%% file: paper.tex
\documentclass[usenatbib]{mn2e}
\usepackage{times}
\usepackage{txfonts}
\usepackage{graphicx}
\usepackage{setspace}
\usepackage{natbib}
\usepackage{sidecap}
\usepackage{color}
\usepackage{subfigure}
\usepackage{floatflt}
\usepackage{rotating}
\usepackage{multirow}
\usepackage{aalongtable, lscape, afterpage}
\newcommand{\comment}[1]{}

\begin{document}

\input{makros}
\input{journals}

\title
{Constraining star formation rates in cool-core brightest cluster galaxies} 
\author[Rupal Mittal, John T.~Whelan and Fran\c{c}oise Combes]{
  {Rupal~Mittal$^{1,2}$, John~T.~Whelan$^{1,3}$, Fran\c{c}oise~Combes$^{4}$}\\
   $^{1}$ Max-Planck-Institut f\"ur Gravitationsphysik
   (Albert-Einstein-Institut), D-30167 Hannover, Germany \\
  $^{2}$ Chester F. Carlson Center for Imaging Science, Rochester
  Institute of Technology, Rochester, NY 14623, USA \\
  $^{3}$ School of Mathematical Sciences, Rochester Institute of Technology, Rochester, NY 14623, USA  \\
  $^{4}$ Observatoire de Paris, LERMA, CNRS, 61 Av. de l'Observatoire, 75014 Paris, France \\
}

\date{Received/Accepted}

\maketitle

\begin{abstract}

  We used broad-band imaging data for 10 cool-core brightest cluster
  galaxies~(BCGs) and conducted a Bayesian analysis using stellar
  population synthesis to determine the likely properties of the
  constituent stellar populations. Determination of ongoing star
  formation rates~(SFRs), in particular, has a direct impact on our
  understanding of the cooling of the intracluster medium~(ICM), star
  formation and AGN-regulated feedback. Our model consists of an old
  stellar population and a series of young stellar components. We
  calculated marginalized posterior probability distributions for
  various model parameters and obtained 68~\% plausible intervals from
  them. The 68~\% plausible interval on the SFRs is broad, owing to a
  wide range of models that are capable of fitting the data, which
  also explains the wide dispersion in the star formation rates
  available in the literature. The ranges of possible SFRs are robust
  and highlight the strength in such a Bayesian analysis.

  The SFRs are correlated with the X-ray mass deposition rates (the
  former are factors of 4 to 50 lower than the latter), implying a
  picture where the cooling of the ICM is a contributing factor to
  star formation in cool-core BCGs. We find that 9 out of 10 BCGs have
  been experiencing starbursts since 6 Gyr ago. While four out of 9
  BCGs seem to require continuous SFRs, 5 out of 9 seem to require
  periodic star formation on intervals ranging from 20 Myr to 200
  Myr. This time scale is similar to the cooling-time of the ICM in
  the central~($<$ 5~kpc) regions.

\end{abstract}

\section{Introduction}
\label{intro}

Brightest cluster galaxies~(BCGs) are luminous early-type galaxies
found at the centers of rich galaxy clusters. These are unique not
only in their special location but also in that they occupy the most
massive end of the galaxy luminosity function. In the hierarchical
model of structure formation, the most massive objects form last
-- referred to as the bottom-up growth of structures.

An important consideration in the hierarchical models is the inclusion
of AGN feedback, which ensures dry mergers, and in the absence of
which, simulations irrevocably produce more luminous elliptical
galaxies than observed and higher star formation rates~(SFRs) than
measured
\citep[e.g.][]{Benson2003,Croton2005,Croton2006,Scannapieco2005,Kormendy2009}. AGN
feedback is also a viable solution to the cooling-flow problem in the
centers of galaxy clusters. Numerous results have indeed confirmed
that (a) radio-loud AGN dwell preferentially in BCGs, compared to
other galaxies of the same stellar mass
\citep{Anja2007,Best2007,Bagchi1994,Valentijn1983} and (b) cool-core
clusters are particularly conducive for cD galaxies which have an AGN
visible at radio wavelengths
\citep[e.g.][]{Burns1990,Birzan2004,Birzan2008,Mittal2009}.

Due to the early formation of the stellar component in massive
early-type galaxies, ellipticals are usually considered to be ``red
and dead'' and are not expected to show any recent star
formation. However, there is compelling observational evidence that is
suggestive of slow but gradual star formation in ellipticals within
the last 1~Gyr. \cite{Kaviraj2007a}, for example, examined 2100
early-type galaxies using SDSS and GALEX photometry data and found
that at least 30\% of the galaxies have experienced recent star
formation (within 1~Gyr), contributing 1\% to 3\% to the total stellar
mass. Similarly, \cite{Pipino2007} and \cite{Liu2012} studied samples
of 7 and 120 early-type BCGs, respectively, and found a recent
starburst superimposed on an old stellar component for all of them.

BCGs at the centers of cooling flows are a special category of massive
ellipticals. \cite{Rafferty2008} investigated the link between star
formation, cooling of the intracluster gas and AGN feedback. They
studied a sample of 47 BCGs in cool-core clusters and found that only
those BCGs with gas cooling times $<0.8$~Gyr show an increase in star
formation towards galaxy centers. A number of studies have indeed
found significant star formation in cool-core BCGs
\citep[e.g.][]{Hansen1995,McNamara1989,Crawford1999,Mittaz2001,ODea2004,McNamara2004,Hicks2005,Rawle2012,Hoffer2012,Liu2012,Fraser-McKelvie2014}
with the implication that the cool gas at the centers of cooling flows
may be feeding the star formation in them. In some systems, it has
been shown that AGN activity may well also trigger star formation in
the central regions \citep[e.g.][]{ODea2004,Tremblay2012b}. While the
general consensus is that the AGN feedback is regulating the cooling
flows, it remains largely unclear as to how the cooling time-scale
relates to the AGN duty cycle and the time-scale of star formation.

In this work, we study the stellar populations of a sample of
cool-core brightest cluster galaxies using broad-band imaging fluxes
to fit the observed spectral energy distributions (SED) with the
best-fit models obtained from stellar population synthesis~(SPS). The
study implements a Bayesian approach to compare the observed SEDs to
models consisting of both old and young stellar populations to
constrain a family of parameters. The cool-core BCGs are taken from a
sample studied in an open time Herschel key programme in which,
Herschel photometry and spectroscopy of 11 strong cool-core BCGs was
conducted. The aim of the Herschel project included determining the
dust temperatures and masses, and understanding the heating mechanisms
of the ionized, molecular and atomic filaments that surround several
of the cool-core BCGs. Note that the Herschel sample is not
complete. It was chosen as to cover a wide range in X-ray, optical and
radio luminosities so that correlations based on the physical
properties of the BCGs could be examined. Results of this ongoing
study can be found in \cite{Edge2010a}, \cite{Edge2010b},
\cite{Mittal2011b} and \cite{Mittal2012}.

One of the goals of the current study is to determine the ongoing star
formation rates~(SFRs) in the BCGs at the centers of strong cool-core
clusters~(SCC) -- defined as those clusters for which the central gas
cooling time is shorter than a Gyr \citep{Mittal2009}. This is of key
importance to understand the factors leading to an order of magnitude
discrepancy between the expected and observed rates of gas mass
condensation in galaxy cluster cores. It is paramount that we
understand the degeneracies among the physical properties
parameterizing the stellar populations, in particular, the young
stellar population, so that we may better understand its connection to
the gas cooling out of the intracluster medium.

\input{table1_log}

Within the framework of AGN-regulated feedback in cool-core clusters,
it is commonly believed that the AGN outbursts are periodically
heating up the in-falling cooling intracluster-gas
\citep{Fabian1994,Peterson2003}.  Since it has been shown that the
cool-core BCGs show enhanced level of star formation compared to other
ellipticals of the same mass \citep[e.g.][]{Edwards2007,Wang2010}, it
is also believed that the star-formation activity may somehow be tied
to the cooling of the ICM gas and the AGN-regulated heating
\citep{Rafferty2008,ODea2008,Donahue2010,Tremblay2012a,Tremblay2012b}. A
rather ambitious goal of this study is to determine if there is any
periodicity in the young stellar component in cool-core BCGs on a
similar time-scale as the cooling time.

A crucial source of concern that provides a further motivation to
conduct this study is a wide range of star formation rates available
in the literature for a given BCG with rather small errorbars. An
important question that we would like to be able to address is the
accuracy with which we may estimate the star formation rates of BCGs
from the data amidst a large subset of unknown model parameters.

The study utilizes HST (far-ultraviolet and optical), GALEX
(near-ultraviolet and far-ultraviolet), SDSS (optical) and 2MASS
(near-infrared) photometric data points to investigate the
best-fitting superposition of old and young synthetic stellar spectra.
In Section~\ref{data} we describe the observations and the various
data used in this work and in Section~\ref{analysis} we discuss the
photometry analysis. In Section~\ref{existingmethods} we describe the
common methods of determining star formation rates and in
Section~\ref{bayesianmethod} we introduce the Bayesian technique of
inferring stellar population parameters, including the star formation
rates. In Section~\ref{params}, we discuss the model parameters and in
Section~\ref{issues} we focus on some of the concerning issues. In
Section~\ref{results} we present the results of the Bayesian inference
technique using galaxy SEDs and SPS. In Section~\ref{discussion} we
correlate the inferred SFRs with the cooling of the ICM and AGN
heating and in Section~\ref{conclusions} we give our conclusions. We
assume throughout this paper the $\Lambda$CDM concordance Universe,
with $H_0 = 71~h_{71}$~km~s$^{-1}$~Mpc$^{-1}$, $\Omega_{\st m} = 0.27$
and $\Omega_{\Lambda} = 0.73$ \citep{Larson2011,Jarosik2011}.

\section{Data Acquisition and Description}
\label{data}

The Herschel sample contains 11 cool-core BCGs, all of which are
located in the centers of clusters with the central gas cooling times
shorter than 1~Gyr.  The BCG of the Centaurus cluster of galaxies,
which we studied in detail in \cite{Mittal2011b}, is the closest BCG
in the Herschel sample with a redshift of 0.01016
\citep{Postman1995}. Owing to its proximity, we were unable to obtain
an accurate estimate of the background in the Hubble
data. Furthermore, it lacks SDSS data as well and so due to the lack
of any reliable photometric constraints we were unable to study
it. Some of the observational characteristics of the remaining 10 BCGs
in the Herschel sample are given in Table~\ref{log} and displayed in
Figures~\ref{optical-images} and \ref{FUV-images} are the optical and
FUV image of the BCGs.

\subsection{Hubble Data}
\label{hubble}

The optical data for Hydra-A and RXC~J1504 and the FUV data for
PKS~0745-191, Hydra-A, A~1068, RXC~J1504 and A~2199 were acquired as
part of the HST Proposal 12220, ``Linking Star Formation with
Intracluster Medium Cooling and AGN Heating in a Sample of Herschel
Galaxy Clusters'' (PI: R.~Mittal, 7 cycles). The rest of the optical
and FUV data were downloaded from the HST archive. All the data
(proprietary and public) were retrieved from the MAST Data Archiving
Distribution System (DADS). The data were reduced through the standard
On-The-Fly-Reprocessing pipeline, kindly provided by the Space
Telescope Science Institute (STScI), which includes tasks such as,
flat-fielding, dark-subtraction, bias-subtraction and non-linearity
correction.

The native units of the final calibrated HST images are
e$^{-}$~s$^{-1}$ (or counts~s$^{-1}$), which were converted into
erg~s$^{-1}$~cm$^{-2}$~$\AA^{-1}$ using the header keyword, {\sc
  photflam}. {\sc photflam} is defined as the mean flux density of a
source in erg~s$^{-1}$~cm$^{-2}$~$\AA^{-1}$ that produces 1 count per
second in the HST observing mode used for the observation.

\subsubsection{HST Optical Data}
\label{opticaldata}

The HST instruments used for the optical data were the Wide-Field
Planetary Camera 2 (WFPC2) with a field of view of
$\p{2.5}\times\p{2.5}$ at a resolution of $\pp{0.1}$~pix$^{-1}$, the
Wide Field Channel (WFC) included in the Advanced Camera for Surveys
(ACS) with a field of view of $\p{3.4}\times\p{3.4}$ at a resolution
of $\pp{0.05}$~pix$^{-1}$ and the Wide Field Camera 3 (WFC3) with a
field of view of $\p{2.7}\times\p{2.7}$ at a resolution of
$\pp{0.04}~$pix$^{-1}$.

All observations involved either dithering (small shifts between
successive exposures) or split exposures (CR-SPLIT) so as to
facilitate the removal of cosmic ray events. Dithering has an added
advantage over CR-SPLIT in that it also makes possible the removal of
hot pixels and gaps between CCDs, hence, yielding better images. The
WFC and WFC3 calibrated data had already been subjected to the
cosmic-ray removal algorithm in the pipeline. This was not the case
for the WFPC2 data. For the WFPC2 data, we extracted the single
flat-fielded, dark-subtracted CR-SPLIT exposures and combined them
using the {\sc iraf} task, {\sc multidrizzle}, with the cosmic-ray
rejection tool switched on. {\sc multidrizzle} is a powerful tool with
the ability to work on dither, mosaic and CR-SPLIT associations. Apart
from masking cosmic-ray events, it calculates dithered offsets and
field distortions (the latter being significant for ACS), which are
then used to correctly register the individual exposures with respect
to one another. The single exposures can be then drizzled onto
separate output frames and later combined.

\input{table2_hst}

\input{images}

The WFC/F500M observations of Perseus were made with the aim of
creating a mosaic of the BCG, NGC~1275 (PI: A.~Fabian). The data
comprised three pointings/visits (north-west, south-east and
south-west), with two orbits per pointing and three-point dithering
per observation. One of the two pointings for the north-west region
had the incorrect World Coordinate System~(WCS) associated with it
(the reasons are not clear), which was fixed via cross-correlation of
the images using the other pointing. The calibrated flat-fielded
exposures were then combined using {\sc multidrizzle}. Even though
{\sc multidrizzle} accurately calculates the dither offsets within a
single visit, some residual offsets on the order of a few pixels may
remain for multiple visits (which often entail guide star
re-acquisitions). Small offsets, visible in the combined image of
NGC~1275, were removed by cross-correlating overlapping stars and
matching them using the {\sc iraf} tasks, {\sc daofind} and {\sc
  xyxymatch}. The results from these tasks were fed into {\sc geomap},
which computes the geometric transformation between the individual
exposures. The residual offsets determined this way were then used as
an additional input to {\sc multidrizzle}.

\subsubsection{HST Far-Ultraviolet Data}
\label{fuvdata}

The far-ultraviolet~(FUV) observations were made with the Solar Blind
Channel~(SBC) included in the ACS, which uses a Multi-Anode
Microchannel Array detector, with a field of view of
$\pp{34.6}\times\pp{30.8}$ at a resolution of $\sim
\pp{0.032}$~pix$^{-1}$. Even though the ACS MAMA detector is not
affected by cosmic rays, all observations included the dithering
technique. This was primarily to eliminate hot and permanently damaged
pixels, and to improve the PSF sampling. 

The MAMA detector has no readout noise and very low detector noise,
with the dark current rate of about
$1.2\times10^{-5}$~e$^{-1}$~s$^{-1}$~pix$^{-1}$, making the dark
correction unnecessary\footnote{The dark correction has since recently
  been switched off in the standard ACS SBC pipeline.} The MAMA dark
rate, though, fairly uniform at operating temperatures below
25~$^{\circ}$C, increases with temperature which increases as the time
elapses during observations. As has been noted in previous studies
\citep[e.g.][]{ODea2004}, this may also give rise to a secondary dark
component in the form of a temperature-dependent glow near the
upper-left quadrant of detector. We checked the detector temperatures
for all observations listed in Table~\ref{hst} and none of the
recorded final temperatures go beyond 25~$^{\circ}$C. Hence there was
no need to apply any primary or secondary dark correction.

The SBC/F140LP observations of A~1795 had two pointings (north and
south) separated by about $\sim \pp{30}$. These were combined in a
similar way as the WFC/F555M data for NGC~1275 using {\sc
  multidrizzle}.

\subsection{Two Micron All Sky Survey~(2MASS), Sloan Digital Sky
  Survey~(SDSS) and Galaxy Evolutionary Explorer~(GALEX) Data}
\label{otherdata}

In addition to the HST optical and FUV observations, we used
broad-band data from Two Micron All Sky Survey~(2MASS), Sloan Digital
Sky Survey~(SDSS) and Galaxy Evolutionary Explorer~(GALEX)
archives. In the following, we briefly describe the data acquisition
and analysis.

The 2MASS is an all-sky survey conducted using the 1.3-m telescopes at
Mt. Hopkins, USA, and CTIO, Chile, in three near-infrared bands $-$
J-band (1.25~{\mm}), H-band (1.65~{\mm}) and K$_s$-band
(2.17~{\mm})~\citep[2MASS][]{Skrutskie2006}. Since the number of
galaxies in this study is small, we preferred to conduct our own
photometry rather than rely on the catalog derivatives. We downloaded
the images of our galaxies using the 2MASS Atlas image service. The
pixel units in the Atlas images are ``data-number'' units, $DN$, which
can be converted into magnitude using,
\begin{equation}
mag_{Vega} = MAGZP - 2.5\times \log10(S)
\end{equation}
where $S$ is the background-subtracted flux in $DN$ determined from
integrating over the desired region and $MAGZP$ is the zero-point
magnitude available in the header of each retrieved image. Note the
default 2MASS magnitudes refer to the Vega magnitude system. We used
the conversions given in \cite{Blanton2005} to obtain the AB
magnitudes ($mag_{AB} = mag_{Vega} + C$, where $C=[0.91, 1.39, 1.85]$
for the J-, H- and K$_s$-bands, respectively), followed by the
conversion, 
\begin{equation}
mag_{AB} = -2.5 \times \log(F_{\nu}) - 48.60, 
\end{equation}
to obtain the flux, $F_{\nu}$, in
ergs~s$^{-1}$~cm$^{-2}$~Hz$^{-1}$. Given a wavelength, $\lambda$, this
can easily be converted into ergs~s$^{-1}$~cm$^{-2}$~\AA$^{-1}$.

\input{table7_2mass}
\input{table8_sdss}

The SDSS is an optical survey covering more than a quarter of the sky
conducted using a 2.5-m telescope at Apache Point Observatory, USA, in
five bands $-$ ultraviolet or {\it u} (3543~\AA), green or {\it g}
(4470~\AA), red or {\it r} (6231~\AA) and near-infrared or {\it i}
(7625~\AA) and infrared or {\it z} (9135~\AA) \citep{York2000}. We
conducted our own aperture photometry. As the first step, we retrieved
the ``corrected'' images using the tenth SDSS data release
\citep{Ahn2014}. The ``corrected'' images are fully-calibrated,
sky-subtracted images with units of nanomaggies~per~pixel (1~nanomaggy
= $3.631\times10^{-6}$~Jansky). However, as noted by
\cite{Anja2007,Bernardi2007,Lauer2007}, the SDSS photometry for nearby
BCGs is unreliable. These studies impute the reason to the level of
sky background that has been over-estimated for large objects,
especially in crowded fields, resulting in the luminosities of such
objects to be under-estimated. Our strategy to overcome this problem
comprised three steps a) we first de-calibrated the image (with the
effect of converting nanomaggies to counts), b) we then obtained an
interpolated sky image and added it to the de-calibrated image
obtained in the first step and c) finally we re-calibrated the
image. We sincerely thank Benjamin~Alan~Weaver from the SDSS help-desk
for helping us obtain calibrated images of our BCGs with the sky
included.

The GALEX is an ultraviolet imaging and spectroscopic survey of the
sky conducted using a 0.5-m space-based telescope in two ultraviolet
bands $-$ far-ultraviolet (FUV) (1350-1780)~AA and near-ultraviolet
(NUV) (1770-2730)~\AA \citep{Martin2005}. We retrieved the images for
all the galaxies in both the bands using GalexView. With the exception
of PKS~0745-191 and A1068, we found data for all the BCGs. While
PKS~0745-191 has no GALEX data, A1068 has only NUV data available. The
images are in units of electrons per second, which can be converted
into ergs~s$^{-1}$~cm$^{-2}$~\AA$^{-1}$ and AB magnitudes using the
conversion factors given in
http://galexgi.gsfc.nasa.gov/docs/galex/FAQ/counts\_background.html.

\section{Data Analysis}
\label{analysis}

\subsection{Photometry}
\label{photo}

Here we describe the details of aperture photometry for the
calculation of the UV, optical and infrared fluxes of the targets. We
invoked a self-written code written in {\sc C} programming
language. We used the {\sc
  cfitsio}\footnote{http://heasarc.gsfc.nasa.gov/fitsio/} library of C
routines for reading and writing {\sc fits} files and the {\sc
  wsctools} package for reading the headers. The code involved placing
an aperture (circular or elliptical) surrounding the emission from the
BCG and integrating the flux over the aperture. Bright sources clearly
not associated with the BCG but which lie within the source aperture
were masked and the corresponding pixels were assigned the average
flux of the neighbouring pixels.

The sky background was estimated and subtracted using the mean of the
sky flux distribution. This was done by placing a circular aperture in
a sky location devoid of visibly bright sources and making a histogram
of the pixel flux values. At optical and infrared wavelengths, the sky
distribution can be well described by a gaussian and so the average of
the pixel values is the same as the mean of the gaussian. Adopting the
mean as the statistic for sky background is useful in cases where
there is a uniform distribution of bright pixels in areas close to the
BCG since such bright pixels are expected to overlap also with the
galaxy emission and so need to be subtracted from the integrated
source flux in addition to the gaussian background. In some cases,
where the background seemed to be non-uniform, we calculated the sky
mean at more than one location and took the average of the values as
the final flux.

The default photometric uncertainties used in this work correspond to
the absolute photometric calibration of the respective
instruments. These are 10\% for HST \citep{Sirianni2005} and GALEX
data, 7\% for 2MASS data \citep{Jarrett2000}, 5\% for SDSS u-band, and
3\% for SDSS g-, r-, i- and z-band
data\footnote{http://www.sdss.org/dr6/algorithms/fluxcal.html}. The
statistical uncertainties are usually very small. These were
calculated using different techniques depending upon the
observations. For example, for HST optical and ultraviolet
observations, the root-mean-square of the background sky distribution
was used. For GALEX observations, the poisson error was used. However,
the errors so calculated vastly underestimate the uncertainty in the
determined flux-densities. The main source of uncertainty arises from
the choice of the aperture size.  It is rather difficult to ascertain
the amount of flux that is lost as the galaxy emission fades into
noise. In order to obtain reliable estimates of the error-bars on the
fluxes, we used several apertures (differing by up to 20\% in size)
and averaged the fluxes estimated therefrom. For some of the 2MASS and
SDSS observations, the flux-uncertainties determined this way could be
as large as 30\% to 40\%. For NGC~1275, the main source of uncertainty
is the AGN emission that may be contributing to the total flux (see
Section~\ref{lines}). In addition, NGC~1275 also includes a foreground
high-velocity system \citep[HVS,
e.g.][]{Burbidge1965,Rubin1977,Hu1983}. However, since the HVS covers
only a small fraction of the northwest part of the BCG nebula, we have
not subtracted its contribution from any of the flux-densities. Note
there are no stellar or CO components detected \citep{Salome2006} from
the HVS, and so its contribution to the star formation rate is likely
to be small. Although \cite{Salome2006} note that the absence of any
stellar components may be due to the edge-on orientation of the disc,
the HVS is undetected in the 2MASS maps as well.

Determining the total fluxes of brightest cluster galaxies is a very
challenging task, especially since BCGs in cool-core galaxy clusters
are often cD galaxies that are embedded in a pool of intracluster
light~(ICL, the origin of which is an issue of open debate), which in
some cases can extend as far out as to a few hundreds of kiloparsecs
\citep[e.g.][]{Lin2004}. Albeit conservative, our approach aims to
obtain reliable uncertainties on the estimated flux-densities of the
galaxies in our sample. The details of aperture photometry are listed
in Tables~\ref{hst}$-$\ref{galex}.

\input{table9_galex}

\subsection{Galactic Extinction}
\label{galext}

The measured fluxes suffer from extinction, arising from dust in both
the interstellar-medium of the Milky Way Galaxy~(Galactic) and the
BCG~(internal). The internal extinction is included in the parameter
set of the models described in Section~\ref{params}. The Galactic
extinction was corrected with the help of the mean extinction law of
the form \citep{Cardelli1989}
\begin{equation}
\left < A(\lambda)/A(V) \right > = a(x) + b(x)/R_V \, .
\label{eqncardelli}
\end{equation}
Here, $A(\lambda)$ is the total extinction (in magnitude) at a given
wavelength, $\lambda$, and $a(x)$ and $b(x)$ are wavelength-dependent
coefficients which assume a polynomial form in $x=(1/\lambda$)\mminv.
$R_V=A(V)/E(B-V)$, where $E(B-V) = A(B) - A(V)$ is the selective
extinction (in magnitude) in set bands, blue ($B$) and visual
($V$). The extinction law is a function of a single parameter, $R_V$,
for which we adopt a value of 3.1, the standard value for the diffuse
ISM \citep[e.g.][]{Schultz1975,Sneden1978}. The polynomial
coefficients, $a(x)$ and $b(x)$, were taken from \cite{Cardelli1989},
those appropriate for the optical (1.1 \mminv $\leq x \leq$ 3.3
\mminv) and the ultraviolet ranges (3.3 \mminv $\leq x \leq$ 8
\mminv). The Galactic $E(B-V)$ values were taken from the NASA/IPAC
Extragalactic Database (NED\footnote{http://nedwww.ipac.caltech.edu}).
and determined $A(\lambda)$ using Eqn.~\ref{eqncardelli}.

\section{Determining Star Formation Rates: Existing Methods}
\label{existingmethods}

There are currently four main diagnostics of determining star
formation rates in galaxies. The first method relies on measuring
emission lines, predominantly, the optical $\ha$ line but often other
recombination lines as well. The underlying assumption is that the
emission lines originate from the ionized gas surrounding hot and
young stars ($30~\ms$ to $40~\ms$) and, hence, serve as an
instantaneous measure of the star formation rate
\citep[e.g.][]{Kennicutt1994}. This method probes only very recent
star formation. The second method relies on the empirically suggested
relation between the FUV$-$NUV colour and the logarithmic ratio of
infrared to UV luminosity (also known as ``IRX-$\beta$'' relation)
\citep[e.g.][]{Meurer1999}. However, this relation exhibits a good
deal of scatter. The third method is to use the FUV luminosity (or
sometimes the UV excess) directly together with synthetic stellar
population models
\citep[e.g.][]{Kennicutt1998,ODea2004,Hicks2005,ODea2010,Mittal2011b,Mittal2012}. This
method, as we will describe below, does not work unless either the
mass or the age of the young stellar population is known. For the
latter two methods, one also assumes a certain initial mass function
(IMF), describing the number density of stars as a function of mass,
for the stellar populations. The fourth method is based on the
far-infrared luminosity (8{\mm} to 1000{\mm}) originating from dust,
assumed to be heated by young and hot stars.

\begin{figure}
    \centering
    \includegraphics[width=0.45\textwidth]{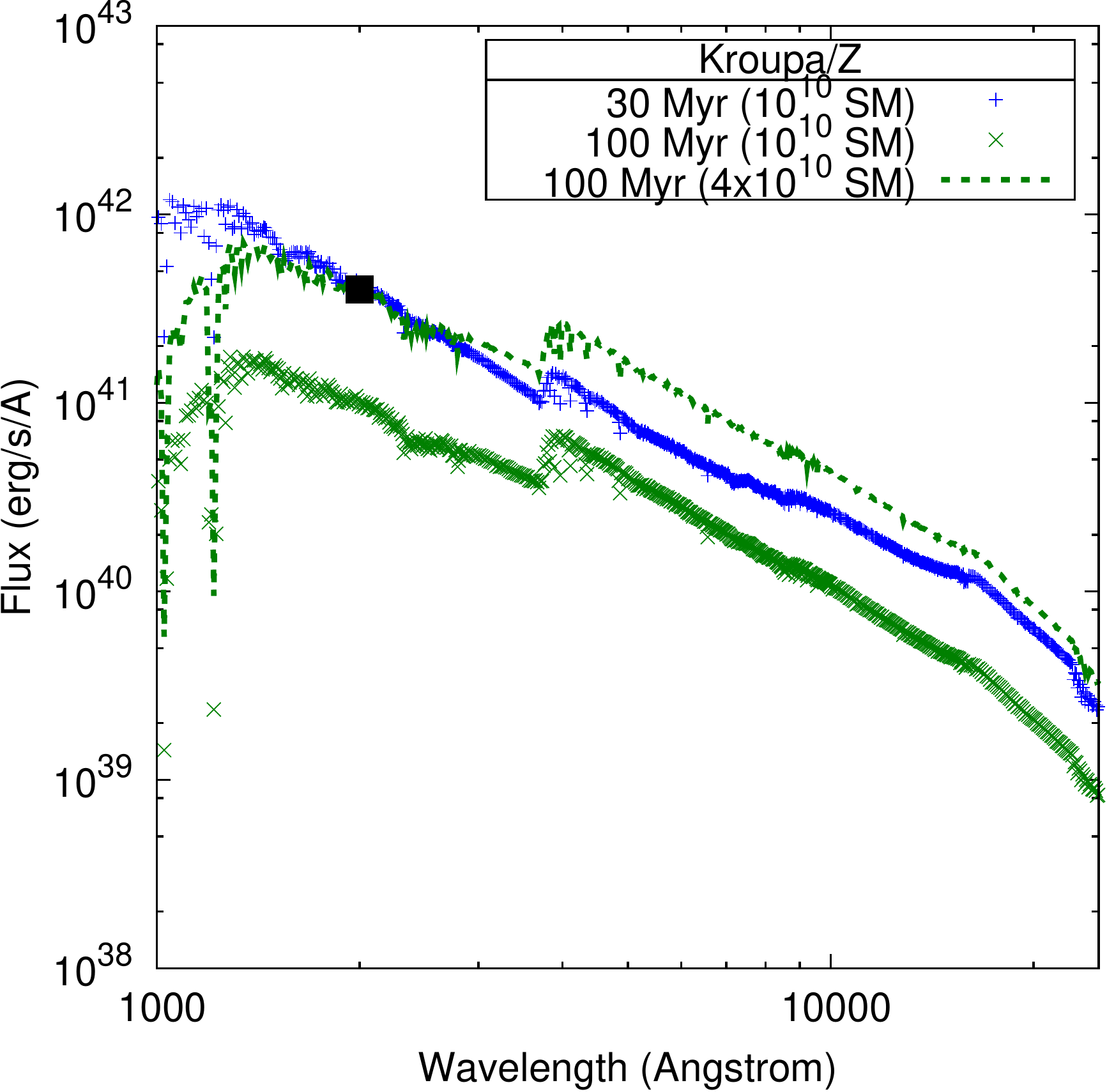}
    \caption{An aging stellar population. A 30~Myr stellar population
      can reproduce the observed FUV luminosity as easily as a 100~Myr
      stellar population four times as massive.}
  \label{ageSP}  
\end{figure}

A major drawback of the first three methods is that these are all
subject to dust extinction. While Galactic extinction is accurately
known for almost all lines of sight, the internal extinction arising
due to dust in the host galaxies remains largely unknown. The fourth
method, based on the FIR luminosity of the BCG, does not suffer from
this drawback, but it is most suitable for dusty circumnuclear
starbursts. For early-type galaxies, the FIR emission is not a very
reliable indicator of young stars since the older stellar component
may also contribute to the heating of dust. In addition, FIR emission
is good calorimeter for measuring ionizing photons from young stars
only if the dust opacity is high everywhere. Lastly, it may be that
some of the IR emission detected is due to heating of dust by an AGN,
and so it is important to use diagnostics such as the Spitzer IRAC
4.5/3.6~{$\mu$m} colour and [O$_{\textsc{iii}}$]~$\lambda5007$/$\hb$
to identify the source of dust heating.

In this work, we use an alternative method to calculate the star
formation rates that precludes many of the above shortcomings. This
method is described in Section~\ref{SEDmethod}.

\subsection*{Mass-Age Degeneracy}
\label{mass-age}

FUV emission, in principle, is indicative of the presence of young and
hot stars and so the FUV luminosity should be directly related to the
active star formation. Synthesis stellar models using a grid of
stellar evolution tracks from one of the well established spectral
libraries can then be used to estimate the parameters of a stellar
population in a galaxy. The simplest model is one where the initial
mass function, metal abundance and internal extinction are known,
which is usually not the case. Even so, that leaves two unknown
parameters -- the age and mass of the stellar population.

The situation may be understood by looking at Figure~\ref{ageSP} which
shows the spectral energy distribution~(SED) created using starburst99
\citep{Leitherer1999} with the IMF set to Kroupa and metallicity set
to solar. We show the SED for a stellar population with a fixed mass
of $10^{10}~\ms$ at two different times during its evolution, 30~Myr
(blue plusses) and 100~Myr (green crosses). It is clear that the
overall normalization of the SED decreases as the stellar population
ages. However, given the FUV luminosity (black filled square), it can
either be fitted by a 30~Myr stellar population with a total mass of
$10^{10}~\ms$ or a 100~Myr stellar population but with a total mass of
$4\times10^{10}~\ms$, i.e.  four times as massive. This is the
mass-age degeneracy. Hence, several studies make use of colour instead
of relying on a single data point. However, since we are interested in
constraining a number of parameters (see Section~\ref{params}), a
well-sampled spectral energy distribution is needed that stretches
across the entire electromagnetic spectrum from the FUV to IR. Under
such circumstances, it is essential to fit both the young and old
stellar population simultaneously.

\section{Determining Star Formation Rates: Bayesian Method}
\label{bayesianmethod}

\subsection{Spectral Energy Distribution and Stellar Population Synthesis }
\label{SEDmethod}

In order to obtain the ongoing star formation rates and explore the
star formation histories of BCGs, we make use of photometry conducted
at several line-free wavebands covering a range from $1500~\AA$~(FUV)
and $25000~\AA$~(IR). The FUV data were obtained from the {\it HST}
telescope and {\it GALEX} \citep{Martin2005}, the NUV data
($2300~\AA$) are take from {\it GALEX} and the IR data~(at three
wavelengths) are all taken from the {\it 2MASS} survey
\citep{Skrutskie2006}. We also used {\it HST} and {\it SDSS} data,
where available. Note that since we are interested in the integrated
galaxy emission, the different resolutions do not complicate handling
of the data. For example, we compared the HST and GALEX FUV fluxes for
BCGs for which both the data were available, and found no systematic
differences despite a smaller point-spread function of the HST
relative to GALEX. We have in total $\geq 8$ data points for all BCGs
except one (PKS~0745-191), which is lacking GALEX and SDSS data
altogether, and for which there are 5 data points available.

Using the SEDs generated with the integrated flux-densities described
above, we aim to fit the data with a model comprising two stellar
populations: an old stellar population~(OSP) and a young stellar
population~(YSP). Each of the two populations has at least two
parameters -- the age and the total mass. Assuming a stellar
population that formed a certain time ago, the normalization of the
SED scales linearly with the total mass in the stars. Hence, the
flux-density at any given frequency, $i$, can be expressed in the
form, $F_i(M, T) = M\times S_i(T)$, where $M$ is the total mass and
$T$ is the age of the population. $S_i(T)$ is the flux-density per
unit mass, which depends on the age. Our model consists of four more
parameters, the IMF
\citep[Kroupa/Salpeter/Chabrier][]{Salpeter1955,Kroupa1993,Chabrier2003},
the metallicity ($0.4~\zs, Z, 2.5~\zs$), the extinction law
(Galactic/extra-Galactic) and the extinction, ($E(B-V) \in [0,
0.6]$). We represent the above four parameters along with the age of
the OSP, $\tosp$, and YSP, $\tysp$, by $\lamvec$. In view of the
findings by \cite{Gao2004} and \cite{deLucia2007}, we assume there
exists a massive old stellar component at least as old as 11~Gyr
(corresponding to a formation redshift of $\sim 3$). Hence, we carried
out two sets of simulations in parallel. For one, we fixed the age of
the OSP to a formation redshift of 3. For the other, the age of the
OSP was left to vary over the range $\sim 10$~Gyr $-13$~Gyr. The
results of these simulations are described in
Section~\ref{results}. The modelled SEDs are convolved with the
instrument bandpasses using the {\sc synphot} library available within
{\sc iraf}.

\subsection{Statistical Method}
\label{stats}

We have a series of flux measurements $\Fis$ with associated weights
$\{w_i\}$, where $1/\sqrt{w_i}$ is the $1\sigma$ uncertainty
associated with $F_i$.  Given a family of models $H$ parametrized by
YSP mass $\mysp$, OSP mass $\mosp$, and some other
(discretely-sampled) parameters $\lamvec$, if the errors on the $F_i$
are assumed to be independent and Gaussian, the likelihood function
may be written as
\begin{equation}
    P(\Fis|\mosp,\mysp,\lamvec,H)
    = \sqrt{\prod_i\frac{w_i}{2\pi}}
    \ \exp\left(-\frac{\chi^2(\mosp,\mysp,\lamvec)}{2}\right)
\end{equation}
where
\begin{equation}
    \chi^2(\mosp,\mysp,\lamvec)
    = \sum_i w_i
    [F_i-\mosp S^{(\st{o})}_{i}(\lamvec)-\mysp S^{(\st{y})}_{i}(\lamvec)]^2 \, ,
    \label{chisq}
\end{equation}
where $S^{(\st{o})}_{i}$ and $S^{(\st{y})}_{i}$ are the flux per unit
mass contributions from the OSP and YSP, respectively. The total
observed flux at a given frequency, $i$, is assumed to be equal to
\begin{equation}
    F^{(\st{o})}_{i}(\lamvec)+ F^{(\st{y})}_{i}(\lamvec) =  \mosp S^{(\st{o})}_{i}(\lamvec)+ \mysp S^{(\st{y})}_{i}(\lamvec) \,
    \label{totalflux}
  \end{equation}
where  $F_i^{(\st{o})}$ and $ F_i^{(\st{y})}$ are the flux contributions from the OSP and YSP, respectively.

If we assume some prior probability distribution\footnote{This is a
  probability density in the continuous parameters $\mosp$ and
  $\mysp$.}  $P(\mosp,\mysp,\lamvec|H)$, Bayes's theorem allows us to
construct a posterior probability distribution
\begin{eqnarray}
  P(\mosp,\mysp,\lamvec|\Fis,H)
  = \frac{P(\Fis|\mosp,\mysp,\lamvec,H)P(\mosp,\mysp,\lamvec|H)}
  {P(\Fis|H)}
  \nonumber
  \\
  \propto P(\mosp,\mysp,\lamvec|H)
  \exp\left(-\frac{\chi^2(\mosp,\mysp,\lamvec)}{2}\right) 
\end{eqnarray}
where we have absorbed everything not dependent on the parameters into
a proportionality constant, which can be easily found from the
requirement that the posterior be normalized, $\int d\mosp\int d\mysp,
\sum_{\lamvec} P(\mosp,\mysp,\lamvec|\Fis,H) = 1$.  To simplify both
the approach and the calculations, we assume uniform priors,
specifically uniform density in $\mysp$ and $\mosp$, with the only
restriction being $0<\mysp<\mosp$, and that each of the discrete
values of extinction, metallicity and $\tosp$, and the different
possibilities for IMF and extinction law, are independently equally
likely.  Since we consider multiple-starburst models with $\nbursts$
bursts evenly spaced in age from $\tysp$ to $\nbursts\tysp$ (see
Section~\ref{SFH}), we consider each of the discrete
$(\nbursts,\tysp)$ combinations sampled to have equal prior
probability.

From the posterior probability, we can calculate useful probability
distributions for various variables, marginalized over the others, such as
the posterior probability density for the YSP mass
\begin{equation}
  \label{pdfmy}
  P(\mysp|\Fis,H) = \sum_{\lamvec} \int_{\mysp}^{\infty} d\mosp
  P(\mosp,\mysp,\lamvec|\Fis,H)
\end{equation}
or the OSP mass
\begin{equation}
  P(\mosp|\Fis,H) = \sum_{\lamvec} \int_{0}^{\mosp} d\mysp
  P(\mosp,\mysp,\lamvec|\Fis,H)
\end{equation}
or the posterior probability distribution $P(x|\Fis,H)$ for a
discretely-sampled or categorical variable $x$ which is among the
parameters $\lamvec$, whose value at some $x=x_0$ is
\begin{equation}
  \label{px}
  P(x_0|\Fis,H) = \sum_{\lamvec:\ x=x_0}
  \int_{0}^{\infty} d\mosp
  \int_{0}^{\mosp} d\mysp
  P(\mosp,\mysp,\lamvec|\Fis,H)
\end{equation}
By similar means, posterior probability densities can be constructed
for derived quantities such as the mass ratio $\mysp/\mosp$ and star
formation rate\footnote{Since $\mysp$ is the total mass in the YSP,
  the mass in each starburst is $\mysp/\nbursts$.} $\mysp/(\nbursts\tysp)$.
Because $\chi^2(\mosp,\mysp,\lamvec)$
is quadratic in $\mosp$ and $\mysp$, the Gaussian integrals over those
parameters in (\ref{pdfmy}--\ref{px}) can be done analytically, with
the sums over the discrete values of the other parameters $\lamvec$
performed numerically. The number of simulations conducted for
different values of the parameters, $\lamvec$, varied from galaxy to
galaxy, depending upon whether the HST FUV data was included or
excluded, and whether the data contained lines or not. The minimum
number of simulations for a given BCG was about $5\times10^5$ and the
maximum was about $2\times10^6$.

The SED-fitting method, like the other methods of estimating star
formation rates, also faces severe challenges due to the existing
uncertainties in stellar evolution, dust properties and attenuation
laws, IMFs and star formation histories \cite[see][for an excellent
review]{Conroy2010}. However, as shown by
\cite{Conroy2010,Pforr2012,Kaviraj2007a,Kaviraj2007b,Pipino2007,Walcher2011},
integrated light from galaxies at well-sampled points of the spectral
energy distribution can be used to constrain the basic parameters,
provided that marginalization techniques are used to incorporate the
uncertainties in the model. The formulation described above does
precisely that and we aspire to yield robust ranges of parameter
values.

In principle, spectroscopic data in addition to photometric data are
highly advantageous and reveal a lot more information about, for
example, the age and metallicity of the various stellar
populations. However, (a) photometric data are readily available for
most of the BCGs whereas spectroscopic data are not and, more
importantly, (b) recent studies claim a different from stellar origin
for the $\ha$ filaments seen in cool-core BCGs \cite[for
e.g.][]{Fabian2011,Ogrean2010}. Many of the cool-core BCGs, including
those being studied in this work, show signs of star formation, and so
spectral lines such as {\niiopt}, $\ha$, {\oiiiopt}, $\hb$ and
{\oiiopt} are expected from the photoionization of the surrounding
gas. However, an unknown fraction of the intensity of the line
emissions may have a non-stellar origin, such as reconnection
diffusion that allows the hot intracluster gas to penetrate the cold
filamentary gas \citep{Fabian2011} or shock heating \cite{Ogrean2010},
making usage of spectral lines problematic for modelling star
formation, (c) spectroscopic data are usually available only for
specific locations of the BCGs and so it is not possible to constrain
their global stellar properties. In this work, therefore, we utilized
only photometric data to constrain the age, mass and metallicity of
both the old and young stellar populations, the initial mass function
and the internal extinction.

There are several other SED-fitting libraries available, such as, {\sc
  starlight} developed by \cite{CidFernandes2005}, {\sc stecmap}
developed by \cite{Ocvirk2006}, {\sc vespa} developed by
\cite{Tojeiro2007}, {\sc ulyss} developed by \cite{Koleva2009}, {\sc
  moped} developed by \cite{Heavens2000} etc. However these are all
spectral synthesis codes that make use of complete spectra of
galaxies. Since these codes make use of a much larger dataset to
obtain the model parameters, they may be able to better constrain
certain stellar parameters by constraining them more tightly. But we
are strongly inclined to using broadband SEDs for this work because of
the reasons given above against using spectral emission lines and
hence we devised our own code. The intention of this work is not to
compete with the existing SED-fitting codes but rather to constrain
the stellar populations of a special category of galaxies $-$
cool-core BCGs. We designed our code to meet two main requirements: a)
The fitting should include only broadband SEDs and b) the star
formation history should include periodicity in star formation to test
if there is a link between periodic cooling, star formation and AGN
heating (Section~\ref{SFH}).

\section{Stellar Populations: Parameters}
\label{params}

Now we will describe in detail the different ingredients of the models
that we used to fit the SED of the BCGs. The various parameters and
their range of variation is given in Table~\ref{paramrange}.

\input{table14_paramrange}

We used two publicly available codes for computing stellar population
evolution synthesis models: {\sc starburst99} \citep{Leitherer1999}
and {\sc galaxev} \citep{Bruzual2003}. While the former code was used
primarily in order to construct simple stellar populations following
the Salpeter and Kroupa initial mass functions, the latter was used to
construct simple stellar populations following Chabrier initial mass
function (see Section~\ref{imfs}). For consistency, we used the Padova
1994 evolutionary stellar tracks for both the codes, containing the
full asymptotic giant branch evolution. 

\subsection{Initial Mass Function~(IMF)}
\label{imfs}

The initial mass function is an important parameter specifying the
mass distribution of the stars in the ISM. There are three generic
IMFs popularly used in literature:
\begin{eqnarray}
{\st {Salpeter}}: \phi(M) \propto M^{-2.35} \quad\st{for}\quad0.1 < M < 100.\nonumber
\end{eqnarray}
\begin{eqnarray}
{\st {Kroupa}:} ~\phi(M) & \propto & M^{-1.3} \quad\st{for}\quad 0.1 < M < 0.5,\nonumber\\
 \phi(M)  & \propto & M^{-2.3} \quad\st{for}\quad 0.5 < M < 100. \nonumber
\end{eqnarray}
\begin{eqnarray}
{\st {Chabrier}:} \phi(M) & \propto & M^{-1} \exp\left[\frac{-(\log~M -
      \log~m_c)^2}{2\sigma^2} \right] ~\st{for}~0.1 < M < 1,\nonumber\\
 \phi(M) & \propto & M^{-2.3} \quad\st{for}\quad 1< M < 100. \nonumber
\end{eqnarray}

Here, $\phi(M)=dN(M)/dM \propto M^{-\alpha}$. The normalization is
determined such that the integral of $\phi(M)$ between the lower
($0.1~\ms$) and upper bounds ($100~\ms$) of mass is unity. While the
\cite{Salpeter1955} and \cite{Kroupa1993} IMFs follow a simple
powerlaw and a broken powerlaw form, respectively, the
\cite{Chabrier2003} IMF follows a simple powerlaw above $1~\ms$ and a
log-normal distribution below $1~\ms$, centered on $m_c = 0.08~\ms$
with the dispersion in logarithmic mass of 0.69.

Whether the IMF is Universal or varies with environment has been a
subject of investigation in many studies
\cite[e.g.][]{Cappellari2013}. The results of these studies indicate
that the IMF may very well indeed be Universal and described by a
power-law with a Salpeter index (2.35) above a few solar masses and
log-normal at low masses \citep[see][]{Bastian2010}. This description
is very close to the Chabrier-type IMF. This is corroborated by a
preliminary analysis that showed that 7 out of 10 BCGs in our sample
show a preference for the Chabrier-type IMF. Thus, in order to reduce
the number of variables and make the results more comprehensible, for
the final simulations we fixed the IMF to Chabrier-type.

\subsection{Metallicity}
\label{metallicity}

For the metallicity, we assigned three different initial chemical
compositions, $Z=0.008$ ($Z=0.4~\zs$), $0.02$ ($Z=\zs$) and $0.05$
($Z=2.5~\zs$), to the old and young stellar populations separately,
yielding 9 different combinations. The three metallicites are denoted
as Z008, Z02 and Z05, respectively. In what follows, we designate the
metallicites of the BCGs with two numbers, ``Z1-Z2'', where $Z1$ is
the metallicity of the OSP and $Z2$ is the metallicity of the
YSP. Even though X-ray observations of the 10 BCGs studied in this
work indicate sub-solar metallicities at the very centers, where the
bulk of the young stars are expected to reside, it is not clear
whether X-ray metallicities are a good measure of the stellar
metallicities. Hence we left this parameter free.

\subsection{Dust: Reddening and Extinction Law}
\label{dust}

The observed BCG spectra do not reflect the intrinsic SED of the
emitting galaxy because of the intervening dust present both in the
inter-stellar medium of the Milky Way and also of the BCG. While the
spectra may be easily de-reddened to correct for the dust absorption
in our Galaxy (Section~\ref{galext}), the internal extinction is very
hard to determine. The Balmer ratios, such as $\ha$/$\hb$, in
principle, can be used to determine the average reddening value
associated with the BCG, $E(B-V)$. However, that line-of-action
entails making assumptions on the exact processes producing the Balmer
lines, such as whether the line emissions are mostly produced by stars
or some other mechanism. A recent study by \cite{Fabian2011} claims a
different origin for the $\ha$ filaments seen in cool-core BCGs
\cite[also see][]{Ogrean2010}. Use of Balmer decrements also depends
on the exact physical properties of the emitting media, such as
whether the intrinsic Balmer decrements are represented by case-A
(optically thin limit) or case-B value (optically thick-limit).

\begin{figure}
    \centering
    \includegraphics[width=0.5\textwidth]{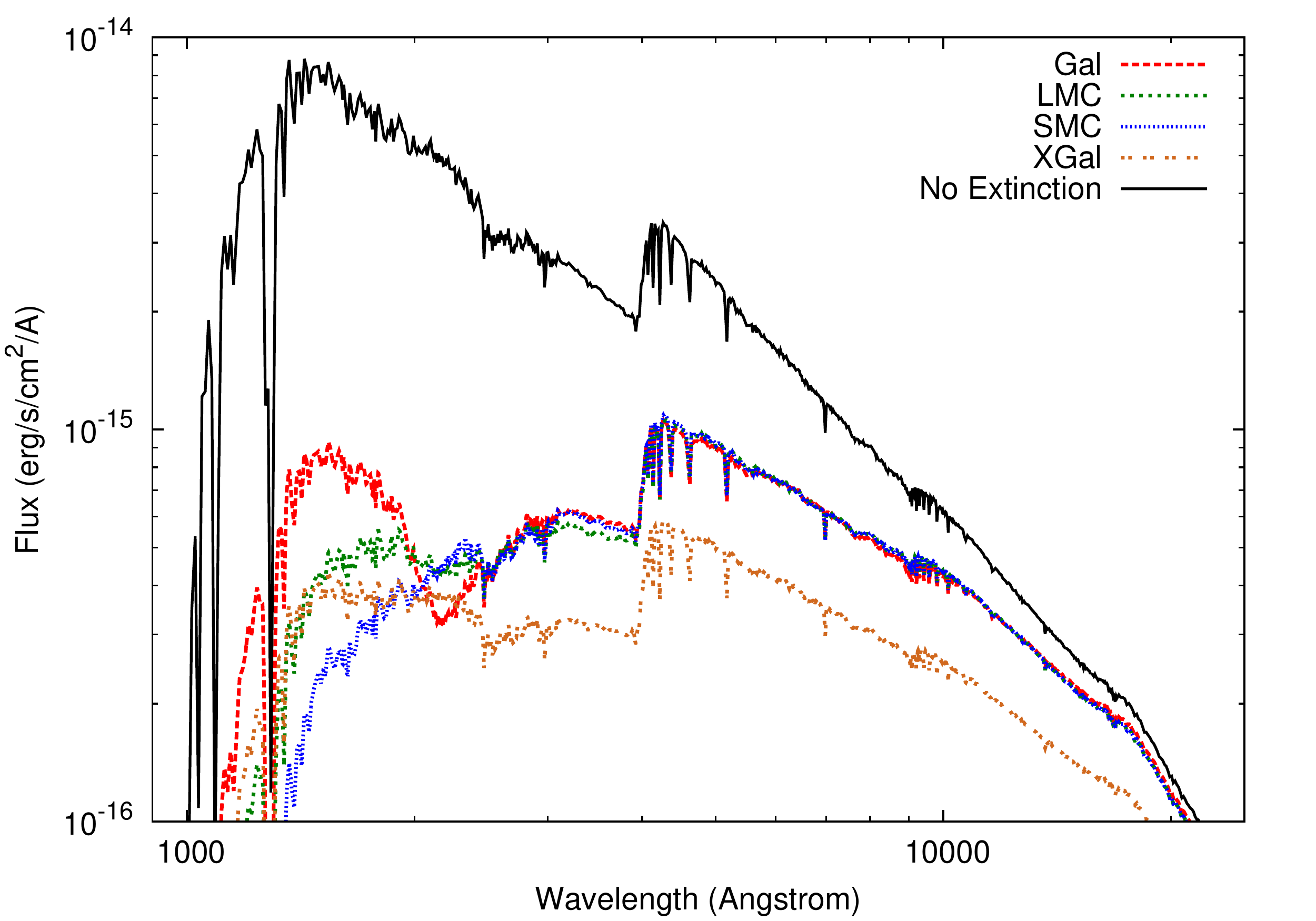}
    \caption{The effect of applying the various extinction
        laws to a SED obtained using the Kroupa IMF, solar metallicity
        and an instantaneous burst of $5\times10^9$~$\ms$ mass stellar
        population $10^8$~yr ago (black curve).}
  \label{extlawsfig}  
\end{figure}

Similarly, there is the issue of which extinction law or curve to
invoke to de-fold the observed spectra to get the true spectra. The
extinction laws depend on chemical composition and physical properties
of the dust grains in the ISM, the metallicity of the ISM, dust
geometry etc. There are four main extinction curves used in the
literature -- the Galactic extinction law (Gal) \citep{Seaton1979},
the Large Magellanic Cloud~(LMC) law
\citep{Koornneef1981,Howarth1983}, the Small Magellanic Cloud~(SMC)
law \citep{Prevot1984,Bouchet1985} and the extragalactic extinction
(XGal) law \citep{Calzetti1994}. These are shown in
Figure~\ref{extlawsfig}. The black solid curve represents the spectrum
from a simple stellar population with no extinction applied to it. The
other curves have the four above-mentioned extinction laws with the
same reddening value, $E(B-V)$, applied to them. The Galactic
extinction law (dashed red) has the well-known $2175~\AA$ dust feature
visible in it. The LMC and SMC laws are very similar to the Galactic
law, although, they differ shortward of $\sim 2500~\AA$ and the
$2175~\AA$ feature becomes weaker from the Galactic to the LMC to the
SMC. The extragalactic extinction law is visibly distinct from the
other three curves in that it is more gray than the other curves but,
additionally, it has no evidence of the $2175~\AA$ dust feature.

In this work, instead of fixing the reddening values to ones available
in literature, we varied it in the range $0.0 \leq E(B-V) \leq 0.6$
and obtained the most-likely value by marginalizing over other
parameters. Similarly, the extinction law was also adopted as a
parameter, and we tried the two extreme extinction laws, the Galactic
and extragalactic. Both the extinction laws are embedded in the {\sc
  synphot} synthetic photometry package distributed as part of the
Space Telescope Science Data Analysis System, {\sc stsdas}. The
extinction laws may be applied to any input spectra using the tool
{\sc ebmvx}. We used the {\sc gal1} option to invoke the Galactic
extinction law, which uses $R_v=3.2$ \citep{Seaton1979}, and the {\sc
  xgal} option to invoke the extragalactic extinction law, which uses
$R_v=2.43$ \citep{Calzetti1994}.

We note that \cite{Calzetti1994} laid out an empirical formulation of
the extragalactic extinction law based on a sample of starburst
galaxies. \cite{Goudfrooij1994}, on the other hand, investigated 10
elliptical galaxies with dust lanes using surface photometry and found
that the most-likely extinction law was close to the Galactic curve,
albeit with a lower $R_v$ (between 2.1 and 3.3). The BCGs studied in
the paper are elliptical galaxies, although with relatively elevated
star formation rates. Letting the extinction law assume one of the two
values (Galactic and extragalactic), we find that 4 out of 10 prefer a
Galactic extinction law, 2 out of 10 prefer the extragalactic
extinction law and 4 out of 10 are compatible with both (see
Table~\ref{extlaw}). Since there is no prior preference for either of
the extinction laws, we let this parameter vary between the two
types. 

\input{table5_extlaw}

\subsection{Star Formation Histories}
\label{SFH}

Star formation history~(SFH) of a galaxy is a crucial ingredient in
stellar population synthesis. Stellar properties, especially, ages and
star formation rages, are sensitive to the stellar population models
adopted in simulations \citep[e.g.][and references
therein]{Pforr2012,Conroy2010}.

N-body and semi-analytic simulations of \cite{Gao2004} and
\cite{deLucia2007} showed an early formation of the constituent stars
of the BCGs and a late formation of the galaxies themselves, such that
most of the stars ($> 80\%$) of the BCGs were already in place in
different smaller number of progenitor galaxies by $z \sim 3$. The
assembly of massive elliptical galaxies and BCGs, in contrast,
occurred relatively late ($z < 1$) through dry (dissipationless)
mergers of massive hosts with other smaller lower-mass halos.

Subsequent studies, however, indicate a model in which the stellar
component in galaxies in cluster centers may have formed as a result
of multiple bursts of vigorous star formation at redshifts as recent
as $z\sim2$, and the galaxies thereafter evolving passively
\citep[e.g.][]{Eisenhardt2008,Mei2009,Mancone2010}. Recently,
\cite{Brodwin2013} studied 16 infrared-selected galaxy clusters in the
redshift range $1 < z < 1.5$ from the IRAC Shallow Cluster
Survey~(ISCS) and found $z\sim1.4$ to be a transition redshift beyond
which the cluster galaxies experienced an unquenched era of star
formation. A similar study of the ISCS clusters in a wider redshift
range $0.3<z<1.5$ was conducted by \cite{Alberts2014} using {\it
  Herschel} 250~{\mm} imaging with the SPIRE instrument. The study
indicates that the cluster galaxies in this redshift range undergo a
monotonic increase in star formation and evolve more rapidly than
field galaxies. However, the results of \cite{Brodwin2013,Alberts2014}
seem to be driven by low-mass galaxies in both cluster cores and
outskirts. High-mass cluster galaxies ($M>6.3\times10^{10}~\ms$) in
the centers, such as the BCGs being studied in this paper, have lower
specific SFRs than field galaxies but show no strong differential
evolution in comparison to the latter \citep{Alberts2014}. This
indicates, in compliance with the findings of \cite{Peng2010}, that
the evolution of the most massive central cluster galaxies is governed
by internal physical mechanisms.

Due to limited number of observational constraints, we assume a model
with an OSP and YSP, both of which are simple stellar populations. In
view of the above studies, it is appropriate that the age of the OSP
be allowed to vary between 10~Gyr~($z\sim1.85$) and 12~Gyr~($z\sim4$)
at 0.5~Gyr intervals. However, our initial results showed that each of
the trial values was equally likely. In other words, our data are not
sensitive to the OSP age. For the work that follows we fixed the age
of the OSP to the formation redshift of $z\sim3$
($\tosp~11.5$~Gyr). Since our BCGs have redshifts in the range
[0.02-0.3], light emitted by the stars take a non-negligible amount of
time to reach us.  So that the OSP has the same formation redshift for
all the BCGs, we need to subtract the light-travel time from the age
of the Universe at the assumed formation redshift. Except for
ZwCl~3146 and A~1835, all the BCGs have a resulting $\tosp$ age fixed
between 9.5~Gyr and 11~Gyr.

For the YSP, we assumed a simple model based on a series of starbursts
of the same mass separated by a constant time interval ranging from
10~Myr to 6~Gyr with a maximum duration of 6~Gyr. Under these
assumptions, namely, that (a) the time separation between bursts is
the same as the age of the most recent burst and (b) the bursts have
the same mass, the number of additional parameters increases only by
one and that parameter is the number of bursts, $\nbursts$. Revisiting
equation~(\ref{totalflux}),
\begin{eqnarray}
    F_i &=&   \mosp S^{(\st{o})}_{i}(\lamvec)+ \mysp \sum_{n=1}^{n=Nbursts} S^{(\st{y})}_{i,n}(\lamvec) \,
    \label{sfhflux}
\end{eqnarray}

\begin{figure*}
   \begin{minipage}{0.33\textwidth}
     \centering
     \includegraphics[width=\textwidth]{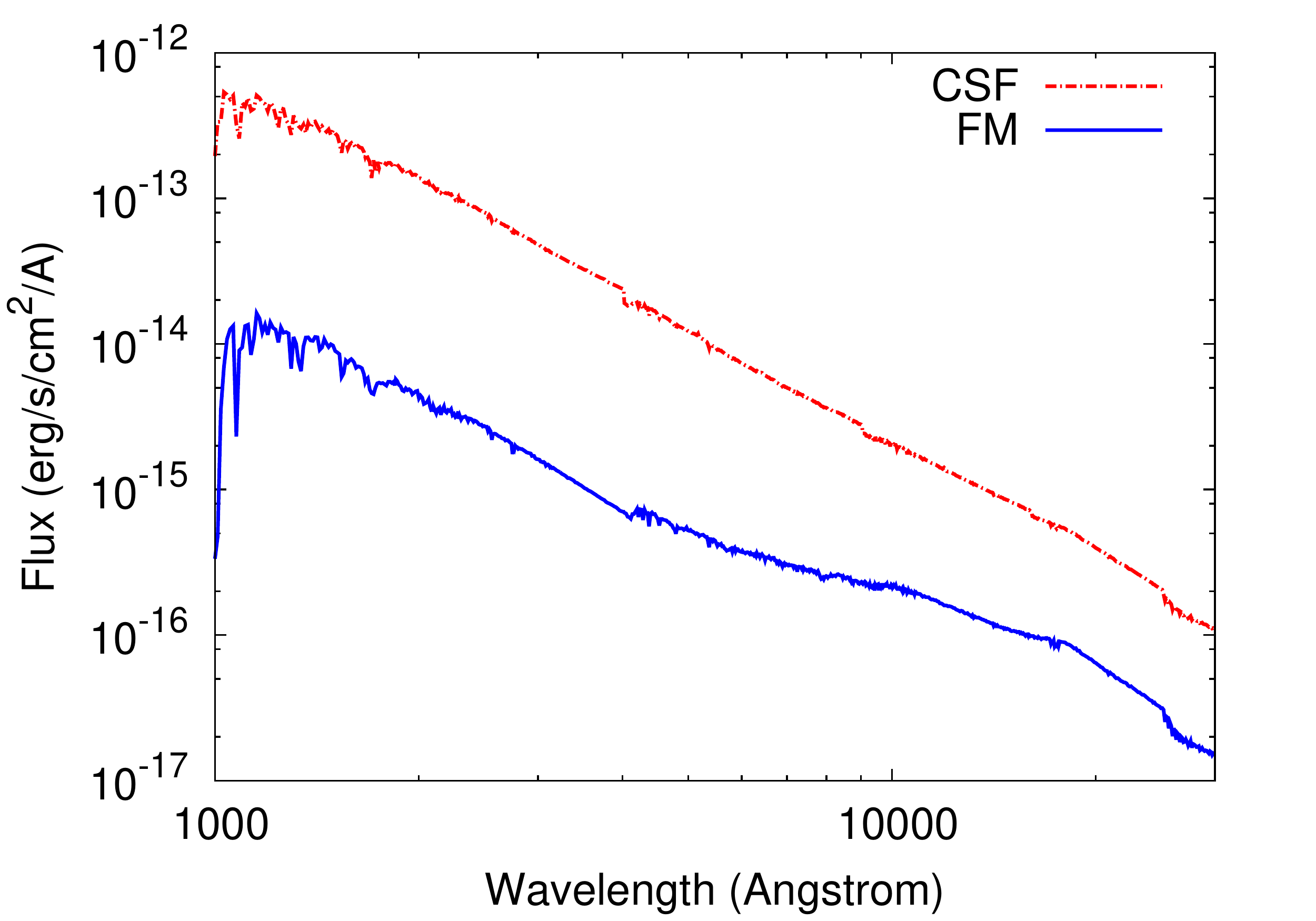}
   \end{minipage}   \begin{minipage}{0.33\textwidth}
     \centering
     \includegraphics[width=\textwidth]{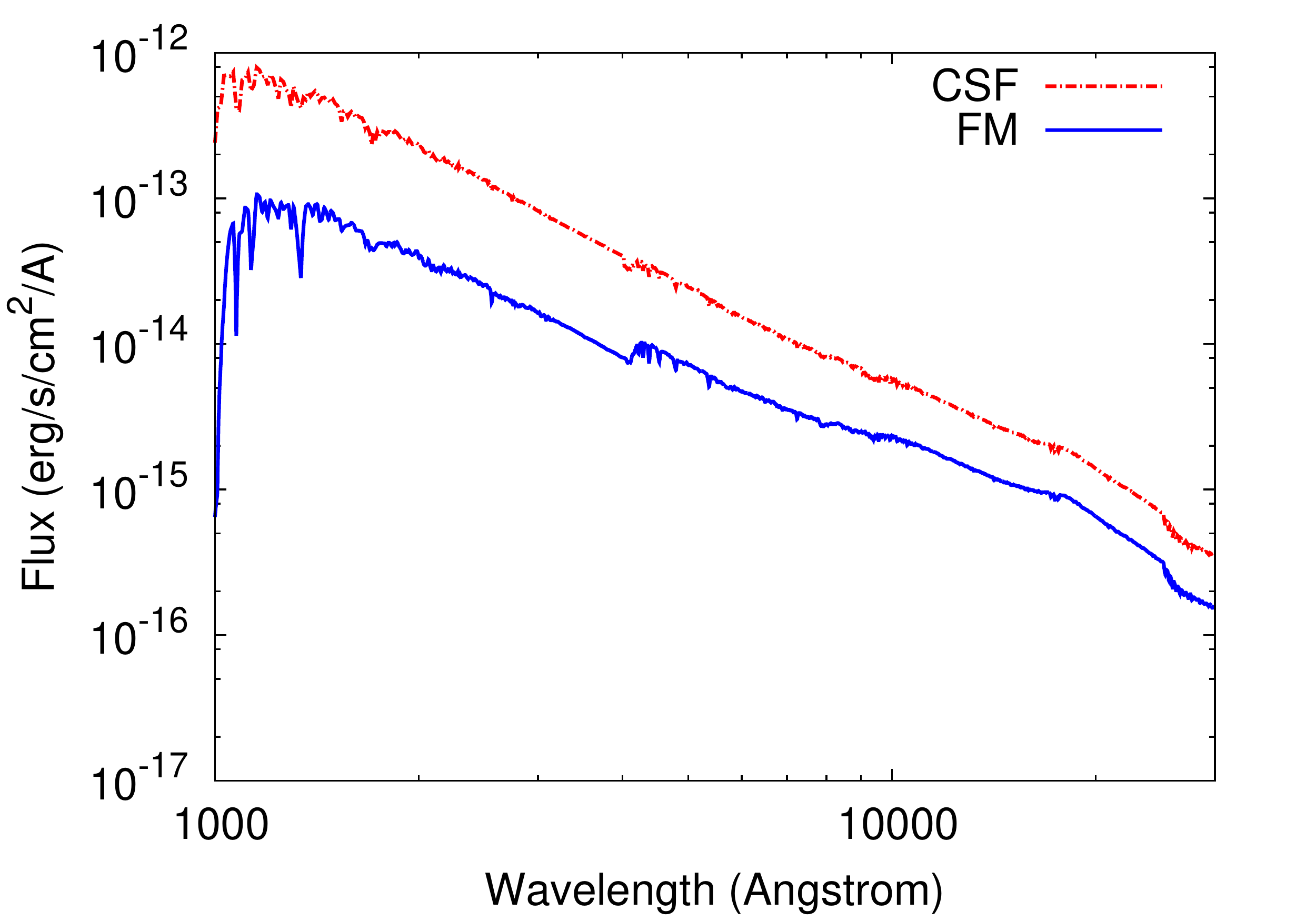}
   \end{minipage}   \begin{minipage}{0.33\textwidth}
     \centering
     \includegraphics[width=\textwidth]{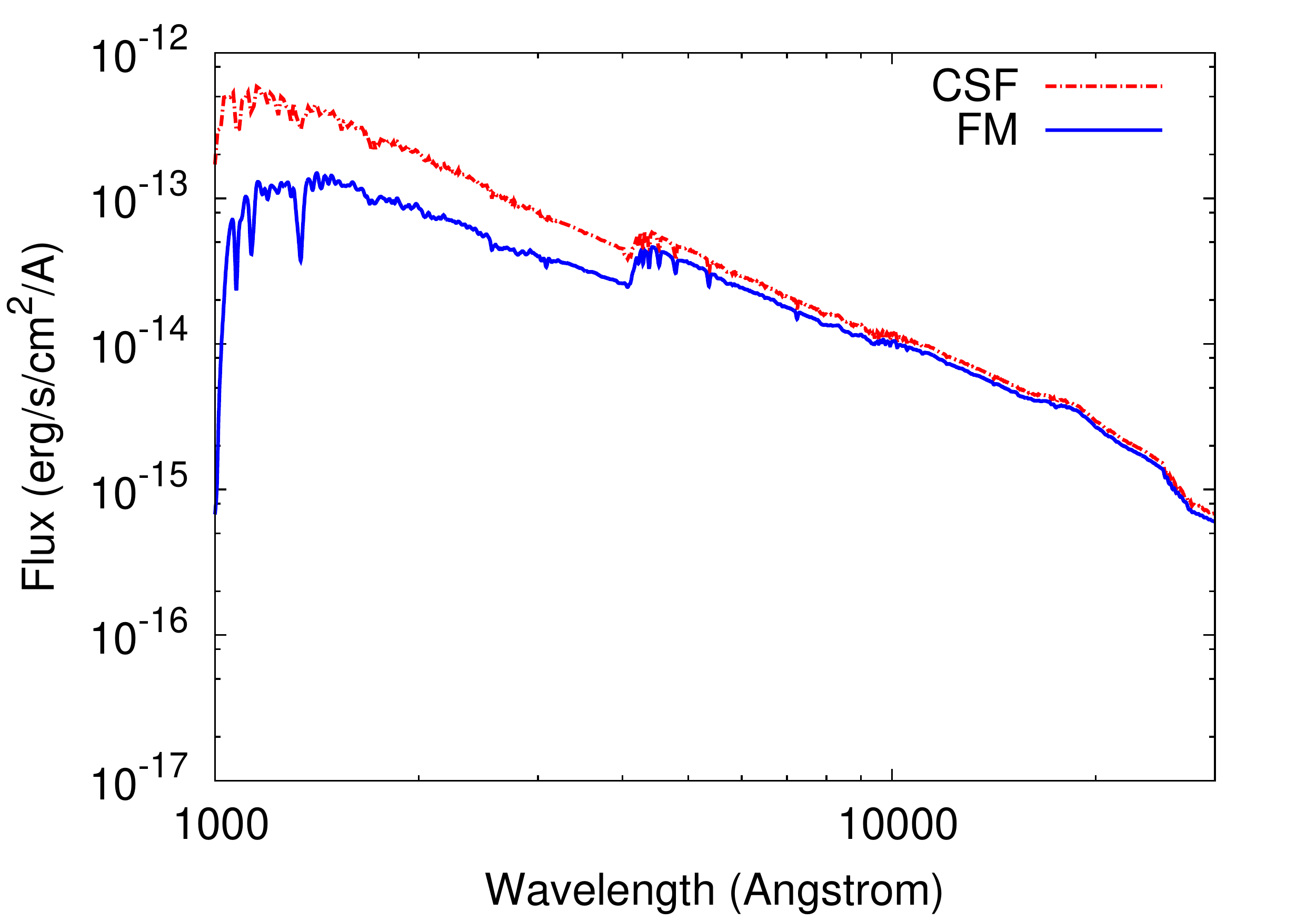}
   \end{minipage}\\
   \begin{minipage}{0.33\textwidth}
     \centering
     \includegraphics[width=\textwidth]{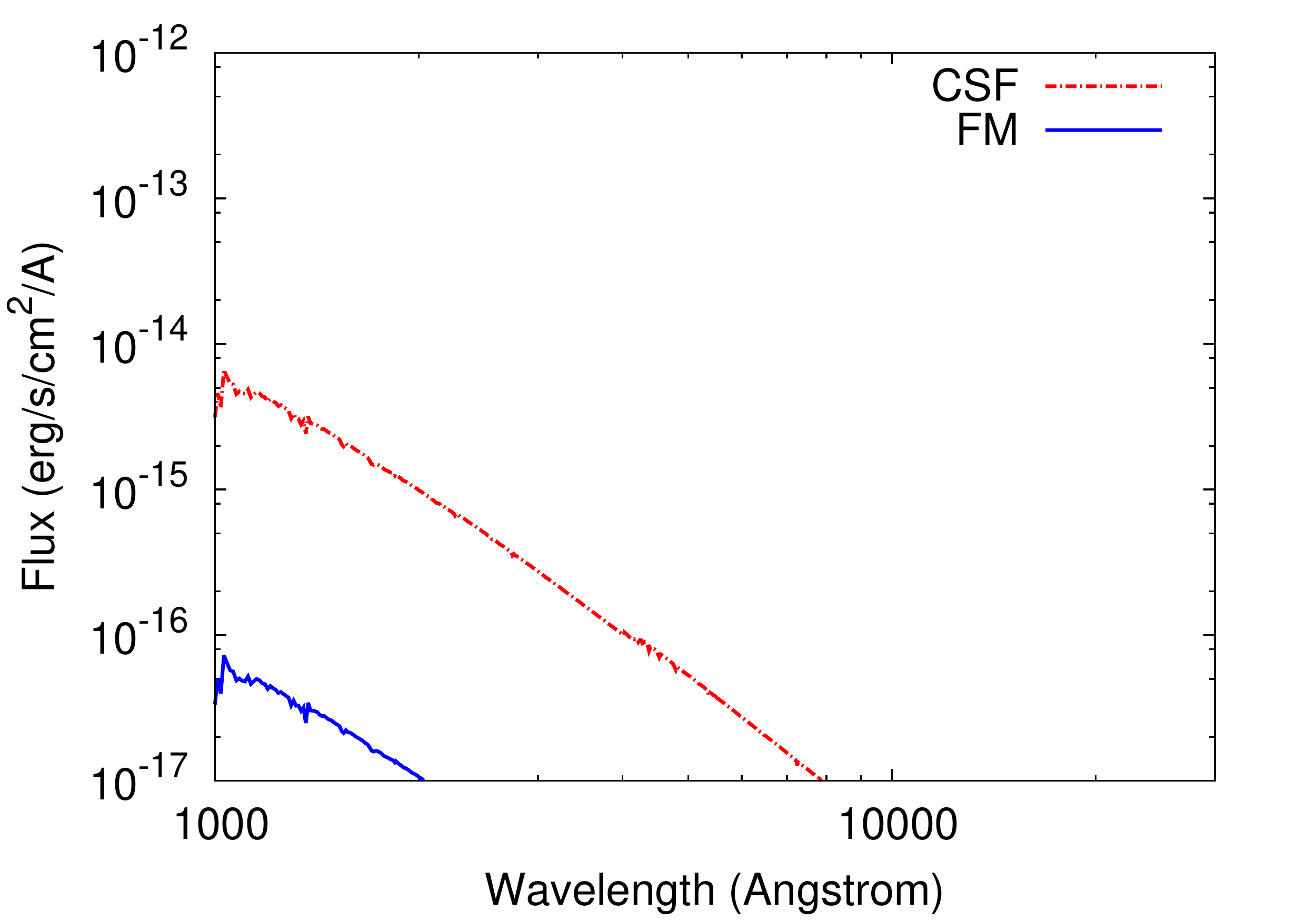}
   \end{minipage}   \begin{minipage}{0.33\textwidth}
     \centering
     \includegraphics[width=\textwidth]{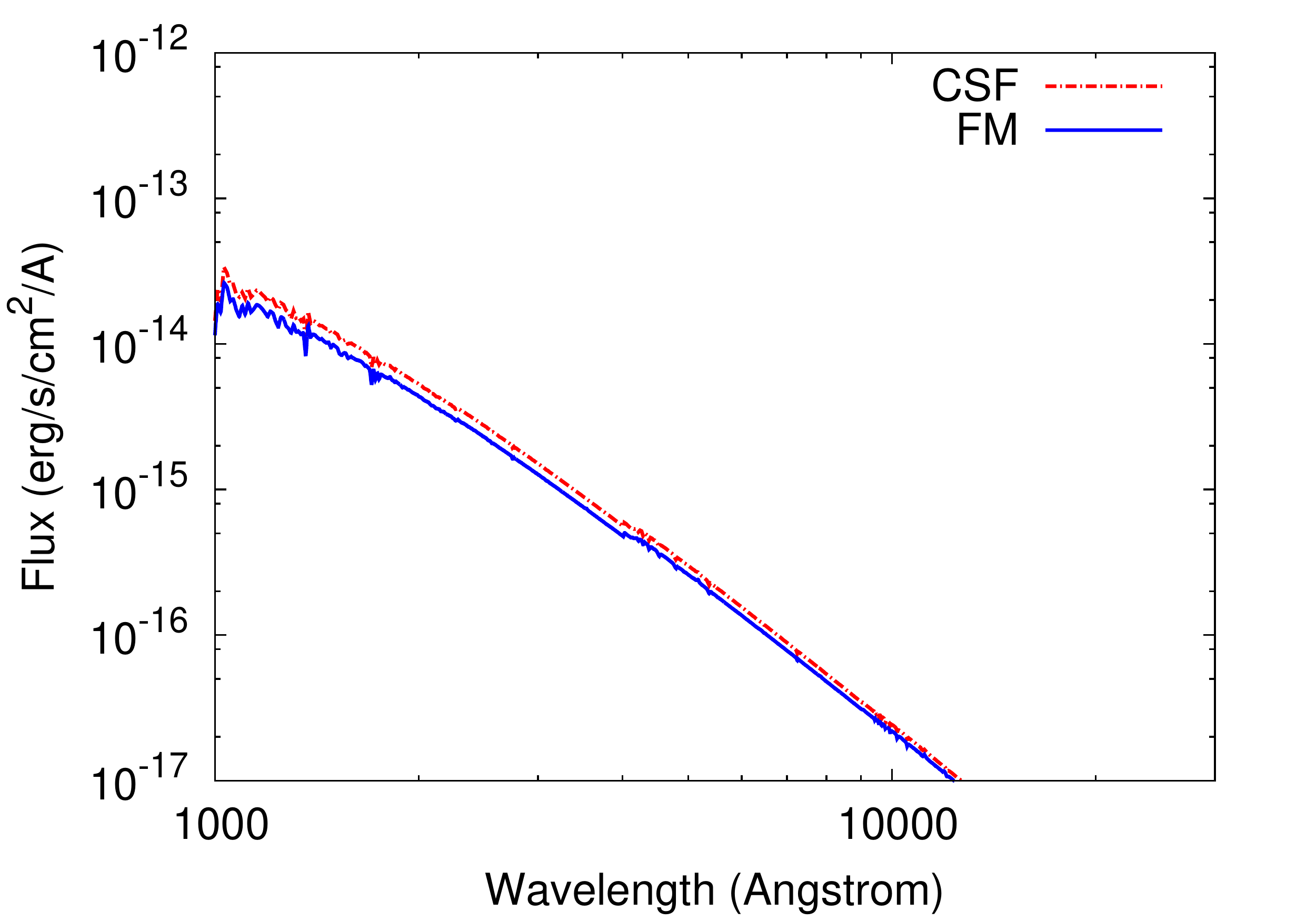}
   \end{minipage}   \begin{minipage}{0.33\textwidth}
     \centering
     \includegraphics[width=\textwidth]{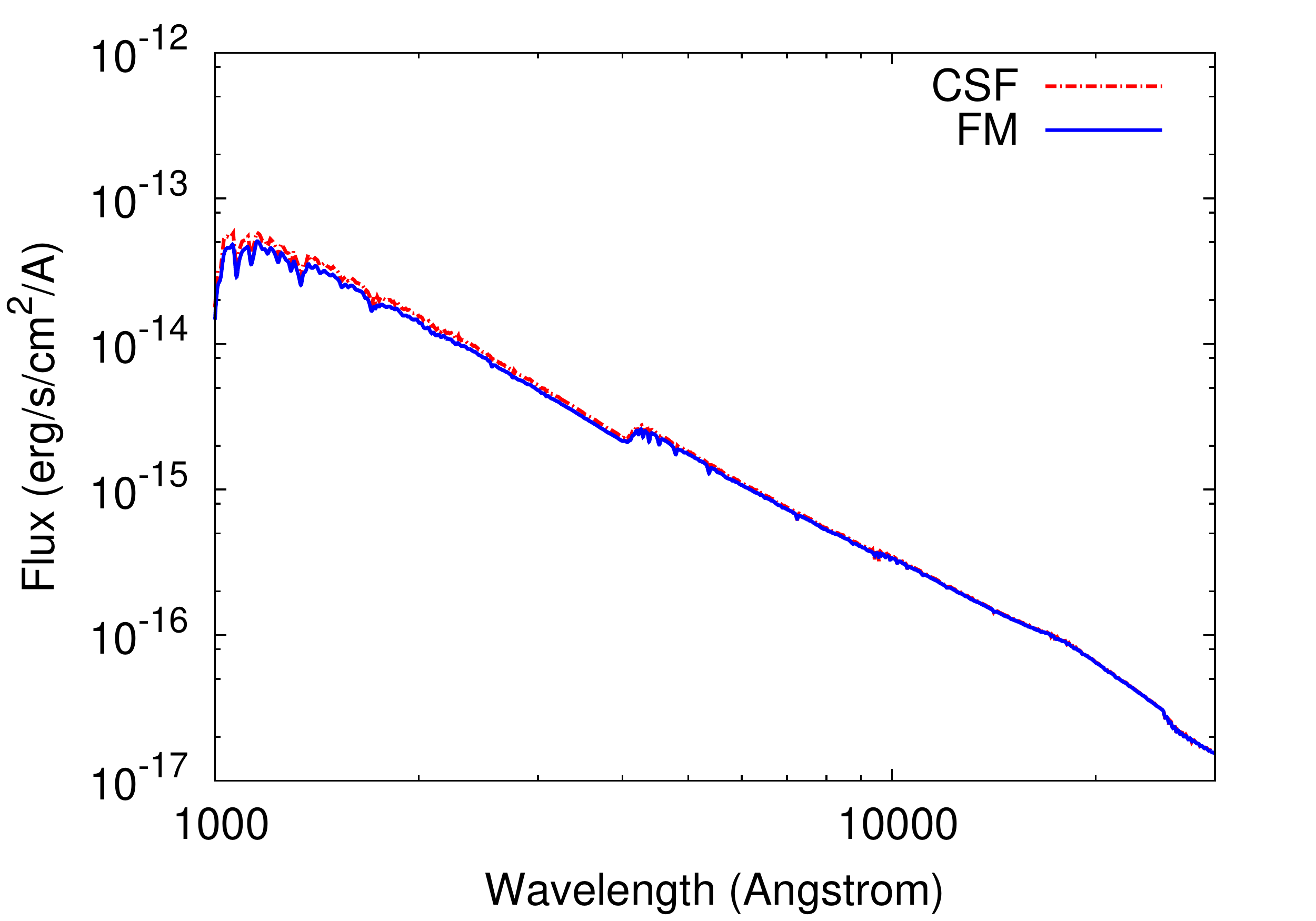}
   \end{minipage}\\
   \caption{Continuous (red dashed curves, ``CSF'') versus
     instantaneous (blue solid curves, ``FM'') star formation models.
     In the top row panels, the separation between bursts is 10~Myr
     and in the bottom row panels, the separation between bursts is
     1~Myr apart. {\it Left:} 1 burst, {\it Middle:} 4 bursts and {\it
       Right:} 50 bursts. For 1~Myr separation, the CSF and FM models
     start coinciding for 4 or greater number of bursts. But for
     10~Myr separation, the two curves remain qualitatively different
     at wavelengths ranging from $1000~\AA$ to $4000~\AA$ by more than
     the typical measurement uncertainties even when the number of
     bursts is 50 or higher. }
   \label{fm_csf}  
 \end{figure*}

An interesting question that arises is whether, within such a
formulation, a model comprising a large number of starbursts occurring
periodically is equivalent to one that has a constant star formation
rate and thus indistinguishable from it. In Figure~\ref{fm_csf}, we
show the two models, continuous (red dashed curves, ``CSF'') versus
instantaneous or fixed mass (blue solid curves, ``FM'') for 10~Myr
separation between bursts (top panel) and 1~Myr separation (bottom
panel). Both the models have the same total stellar mass. Shown is a
comparison between the two models for $\nbursts=1$, $\nbursts=4$ and
$\nbursts=50$. While for 1~Myr separation, the two models start
coinciding for $\nbursts\ge$, for 10~Myr separation the two curves
remain qualitatively different at wavelengths ranging from $1000~\AA$
to $4000~\AA$, by more than the typical measurement
uncertainties. This is true also for a large number of bursts. Hence
our model with the adopted 10~Myr separation between bursts is such
that a multiple burst scenario can {\it not} be equated to a constant
star formation scenario.

 The idea behind including a series of stellar bursts is that we would
 like to determine whether a) there is evidence of any periodicity in
 star formation and b) if so, whether or not it's correlated with
 either the cooling-time of the ICM or AGN activity. Such a
 correlation would imply that the star formation occurs on a similar
 time scale as cooling of the ICM or AGN heating. The cooling times
 within the central few kiloparsec of cool-core clusters can be as
 short as a few to a few tens of Myrs \citep{Voit2014} and also the
 time between AGN outbursts is typically $10^7$~yr to $10^8$~yr
 \citep{Fabian2000,Shabala2008,Birzan2012}. Hence, we designed our
 simulations to include a minimum separation of $10$~Myr between
 starbursts. We believe that nobody has explored such a correlation
 using star formation rates before. While adding a series of periodic
 bursts in our simulations increases the number of parameters by one,
 it is an easy feature to include and allows a better understanding of
 the cooling of the ICM, star formation and AGN-regulated feedback.

\section{Concerns and Issues}
\label{issues}

There are some additional concerns, especially that pertaining to
elliptical galaxies and cooling-flows, that need to be addressed in
the context of modelling stellar populations. These are outlined
below. We also present work-arounds to these problems and, although,
only approximations, we are confident that they help us assess the
parameters more accurately.

\subsection{UV-Upturn}
\label{uvupturn}

The spectrum of some of the elliptical galaxies show an increasing
flux with decreasing wavelength at around 2500~$\AA$
\citep[e.g.][]{Code1979,O'Connell1999,Atlee2009}. This so-called
UV-upturn is believed to be due to hot, extreme horizontal branch
stars. The reason as to the existence of such a population of old hot
stars is not clear. However, a careful modeling of such stars is
required in order to interpret the UV emission from ellipticals and cD
galaxies, in particular, that associated with young stars.

A study conducted by \cite{Yi2011} showed that only 5~\% of cluster
elliptical galaxies (both satellite and BCGs) show the UV-upturn
phenomenon. \cite{Loubser2011} studied, on the other hand, only the
cluster dominant galaxies and found that there are systematic
differences in the UV-colours of BCGs and ordinary ellipticals. To
ascertain whether or not the BCGs in the Herschel sample exhibit a
UV-upturn, we applied the three criteria listed in \cite{Yi2011}
[using the FUV-NUV, FUV-r and NUV-r colours]. We find that of those
that have data in all three bands, none fulfill all three criteria, a
necessary condition for a galaxy to be classified as a UV-upturn
galaxy. Note, the criteria from \cite{Yi2011} are based on a model
that comprises only an old stellar population. It is clear from works
of
\cite{Rafferty2006,Rafferty2008,Hicks2005,ODea2004,ODea2008,McNamara2004,Hansen1995}
that several of the cool-core BCGs are undergoing significant star
formation. So, the contribution from the stars causing a UV-upturn to
the UV emission is probably even lower than the sample studied by
\cite{Yi2011}.  The work of \cite{Donahue2010}, in fact, shows that
once the SFR increases beyond $1~\mpy$, the contribution from UV
upturn stars from the old stellar population of the BCG, despite being
much more massive than the young component, is swamped by that of new
stars.

While we can not completely rule out the existence of a UV-upturn in
our galaxies, strong cool-core BCGs can not be classified as strong
UV-upturn galaxies (since the blue centers are very likely due to
young stars). For this work, we do not apply any correction by
incorporating stellar population models that contain evolved
horizontal branch stars, although such SP models are under
construction (H. Jeong, private communication) and in the future it
will be interesting to directly explore the effects of evolved branch
stars on the SFRs.

\subsection{HST ACS/SBC Redleak}
\label{redleak}

\input{table10_hstfuv}

For a G-type or later star, the ACS/SBC MAMA detector is known to
detect a significant fraction (half or more) of optical and near-UV
photons, i.e photons outside the nominal bandpass. Although this
effect, termed {\it redleak}, is incorporated in {\sc synphot}, recent
investigations carried out by the ACS team have revealed a time and
temperature dependence (not taken into account in the sensitivity
curves), such that the count rates may increase by 10~\% to 20~\%. The
ACS team is currently working to derive the updated throughput curves,
which will return more accurate estimates (Matt McMaster, private
communication). However, since at the time of conducting this work, no
correction factors were available, we decided to carry out simulations
with and without the HST ACS/SBC FUV data. This way we were able to
assess the effect of redleak on the inferred parameters assuming a
worst-case scenario. This exercise was possible only for those
galaxies that have a GALEX FUV measurement so that there is at least
one data point available at FUV wavelengths, needed to constrain the
stellar parameters, especially those corresponding to the young
stellar component.

In Table~\ref{hstfuv} we list the star formation rates obtained from
including the HST FUV data point and excluding it. If the redleak were
a problem, then we would expect the star formation rates inferred from
a dataset including the HST ACS/SBC data to be systematically higher
than those inferred from a dataset excluding it. There is no such
systematic dependance. Aside from the above test, as mentioned in
Section~\ref{SEDmethod}, we also compared the HST FUV to the GALEX FUV
fluxes for the same subset of galaxies for which both the data are
available and did not find any systematic differences.  From this we
conclude, that the ACS/SBC redleak is not a major problem and does not
affect our results in any way.

\subsection{Line Contamination}
\label{lines}

While codes like {\sc starburst99} and {\sc galexev} are able to model
stellar and nebular continua, nebular emission lines are not
calculated. Hence, ideally it would be nice to model the galaxy SEDs
using imaging data devoid of any nebular lines, such as {\oiiopt},
$\ha$, $\hb$, {\niiopt}, {\oiiiopt} etc. However, many of the
available HST and SDSS data contain line emissions and such a
restriction would drastically reduce the number of data points that
may be used to obtain a most-likely model. We attempted to remove the
contribution from various lines by compiling total line fluxes from
the literature, dividing them by the rectangular bandwidth of the
instrument bandpass, and subtracting the resulting line flux-density
from the observed flux-density of the images using aperture
photometry. In some cases the contributions were too high, like for
WFPC2/F439W and WFPC2/F555W observations of PKS~0745-191, the line
contributions calculated using the fluxes given in \cite{Fabian1985}
are 80~\% and 20~\% respectively. Such observations were not used.
Five of the 10 BCGs needed to be corrected for line emission. These
are PKS~0745-191, ZwCl~3146, A~1795, A~1835 and A~2597.

\input{table15_lines}

Note that the line fluxes compiled from the literature usually do not
correspond to the line emission originating from the entire galaxy but
rather to the one-dimensional cross-section of the spectroscopic
slits. We made approximate corrections for this shortcoming in the
following way. \cite{Crawford1999} is a noteworthy repository of data
in this context since they tabulate emission line fluxes of several
lines ({\oiiopt}, $\hb$, {\oiiiopt}, {\niia}, $\ha$, {\niiopt},
{\siia}, {\siib}, {\oia}, {\oib}) originating from 256 dominant
galaxies in 215 clusters. The line fluxes, however, are measured using
a $\pp{6}$ long slit. The $\ha$ line fluxes given in
\cite{Heckman1989}, in contrary, are measured using a narrow-band
image and so the integrated line flux density is expected to be higher
and more accurate. As an example, the Galactic-extinction corrected
$\ha$ flux of A~1795 given in \cite{Heckman1989} is about a factor of
ten more than that given in \cite{Crawford1999}. We used this factor
to correct the other line fluxes obtained by \cite{Crawford1999} that
are present in the HST (WFPC2/F555W and WFPC2/F702W) and SDSS ({\it i}
and {\it r}) observations of A~1795. Similarly, by comparing the $\ha$
flux estimated by \cite{Voit1997} and \cite{Heckman1989}, we found
that the line fluxes determined by \cite{Voit1997} need to be
corrected by a factor of 2.5. The various emission lines, the
bandpasses in which they lie, along with the references used to make
the correction are given in Table~\ref{lineref}.

This correction strategy is crude since it may very well be that the
lines are not co-spatial, and so we may be over-estimating the
correction factors. The difference between the flux-densities with and
without the lines are, however, within the uncertainties. Similarly,
the best-fit parameters, such as the star formation rates, calculated
with and without lines (for NGC~1275: with and without the AGN), are
comparable within the uncertainties. The results quoted from hereon
refer to measurements after the line correction has been made.

NGC~1275 is a special case. It is host to a bright AGN, 3C~84, with a
Seyfert-like spectrum and a radio core-luminosity of order
$10^{44}$~erg~s$^{-1}$ \cite[e.g][]{Pedlar1990}. It is possible that
the AGN contributes to the UV and optical bands and since we are
interested in the flux corresponding only to the stellar component, it
is important to assess the AGN fraction and subtract it from the
measured flux-densities. In \cite{Mittal2012}, we carefully modelled
the SED of NGC~1275 from radio to mid-infrared frequencies and
obtained simultaneous fits for two dust components and an AGN. Taking
the best-fit model at face value, we can use the best-fit parameters
to calculate the flux-density due to the AGN. We tabulate in
Tables~\ref{hst}$-$\ref{galex} both the flux-densities with and
without the AGN contribution. As can be seen, the AGN contribution at
UV wavelengths is high. The uncertainties assumed for NGC~1275 are
conservative and include those due to the AGN contribution. For
comparison, the star formation rates with and without the AGN
contribution are $78^{+165}_{-60}$ and $71^{+136}_{-53}$.

\section{Results}
\label{results}

The marginalized posterior probability distributions for the various
variables in the parameter space are shown in
Figures~\ref{Mo}-\ref{ext} and listed in Table~\ref{spprop}. In
Figure~\ref{bestfitplots}, we show examples of best-fit plots, which
were made by fixing the discrete parameters (metallicity, extinction,
and YSP age) to their most-likely values given in Table~\ref{spprop}
(which are the modes of the individual marginal posteriors) and
choosing the most likely $\mosp$ and $\mysp$ given those choices,
which are the ones that minimize $\chi^2(\mosp,\mysp,\lamvec)$ for
that choice of $\lamvec$.\footnote{Note that this would not generally
  be guaranteed to give a likely combination of all of the parameters.
  If the peaks in the posterior are oddly shaped in the discrete
  parameter space, we could have found find that e.g., the most likely
  metallicity marginalized over $\tysp$ and the most likely $\tysp$
  marginalized over metallicity was not a likely combination.  But in
  practice this effect is only pronounced when considering
  correlations between the masses and other parameters, so as long as
  we choose the most likely masses for a given likely choice of
  discrete parameters.}  The red and blue curves correspond to the
flux contributions from the old and (total) young stellar populations,
respectively, and the black curve corresponds to the total spectrum
energy distribution (sum of the old and young stellar
populations). The green squares correspond to the predicted data and
the orange crosses correspond to the observed data. In the following,
we discuss the most-likely parameters individually and try to draw
generic conclusions.

Since some of the PDFs shown in Figures~\ref{Mo}-\ref{ext} are far
from normal distribution, the mean and standard deviation do not
convey the usual information about the shape of the distribution.  To
provide a more robust quantity which is analogous to a $\pm 1\sigma$
error interval, we construct the narrowest 68~\% plausible interval,
which is the narrowest interval which contains 68\% of the area under
the posterior PDF, along with the median \emph{within that interval}.
Alternative information about the distribution may be captured by the
full-width-half-maximum~(FWHM) interval, where the probability density
at the endpoints is equal to half of the peak value.  For comparison,
we provide in Table~\ref{PI68vsFWHM} both the 68~\% PI and the FWHM
interval for the inferred star formation rates, along with the median
restricted to each interval.  It is encouraging to see that with the
exception of NGC~1275, both the 68~\% PI and FWHM yield similar median
values. For the remainder of the paper, we will use the 68~\% PI.

\input{table16_PI68vsFWHM}

\subsection{Young Stellar Population}
\label{YSPresults}

In Figure~\ref{Ty}, we show the posterior mass density function
corresponding to the YSP age of the BCGs. The X-axis is the age of the
youngest starburst and also the separation between multiple starbursts
and the Y-axis is the number of starbursts for a given separation. The
colourbar represents the probability for a given combination of age
and number of outbursts.  As an example, in case of Hydra-A, the best
fitting combination is given by $\tysp=180$~Myr and
$\nbursts=33$. Hence our results show that the ``young'' stellar
component in Hydra-A comprises about 33 starbursts separated roughly
by about 0.2~Gyr, the youngest being 0.2~Gyr old and the oldest being
6~Gyr old.

Among all the BCGs, A~2199 seems to be an oddball in that the
most-likely model does not call for a ``young'' stellar component at
all. The youngest stellar component is 6~Gyr old. This is in line with
the star formation rate estimates derived by
\cite{Bertola1986,McNamara1989,Crawford1999,Hoffer2011}, the average
being $0.2\pm0.1$ (see Table~\ref{mdrs}), indicating very little
ongoing star formation. Our data, however, suggests a complete absence
of a young stellar population. One reason this may be is that A~2199
is one of the two BCGs (the other being PKS~0745-191) that has only
one data point in the wavelength range [1500~$\AA$ to 4000~$\AA$]. So
it is possible that with our data we are not able to constrain the YSP
as accurately. If we force the age of the YSP to be $<100~$Myr, then
we obtain $\tysp=80~$Myr and a SFR of $\sim 0.3~\mpy$, in consistency
with the previous results. The evidence for such a young SP, however,
is very small given our data.

Even though A~2199 is the weakest cool-core cluster in the sample with
a mass deposition rate of $72^{+2}_{-2}~\mpy$ \citep{paperIII}, the
BCG of A~2199, NGC~6166, harbours a moderately powerful radio source,
3C~338, with symmetric parsec-scale jets \citep[e.g.][ and references
therein]{Gentile2007}. Hence it is possible that the BCG experienced
an enormous AGN outburst in the past such that the cooling has not yet
overcome AGN heating to a degree that is conducive to active star
formation.

Four of the 10 BCGs, ZwCl~3146, A~1795, A~1835 seem to favour the
youngest stars provided by our simulations. While RXC~J1504 is best
fitted by a model with a single burst at 10~Myr, the other three seem
to prefer multiple bursts with the shortest possible spacing (10~Myr),
with ZwCl~3146 and A~1795 going up to the oldest age of 6~Gyr and
A~1835 going up to the oldest age of about 3.5~Gyr. In order to
investigate whether even younger starbursts might fit the SEDs better,
we designed bursts 2~Myr apart for ZwCl~3146 and found that best-fit
results were still favouring the youngest stellar population $-$ 2 Myr
old $-$ with multiple outbursts going up to 6~Gyr. Hence it seems
likely that these four BCGs would be well-fitted by a continuous star
formation model. The median star formation rates inferred from the two
models for ZwCl~3146 based on multiple starbursts 10-Myr~(14.4~$\mpy$)
and 2~Myr~(16~$\mpy$) apart are nearly the same. Note that according
to the bottom middle panel of Figure~\ref{fm_csf}, a model comprising
multiple bursts 1~Myr apart is equivalent to that with a continuous
star formation rate for $nbursts\ge4$.  Similar investigations for
RXC~J1504 (YSPs with 1~Myr separation although with a limited number
of bursts) showed that it might be better-fitted by a continuous star
formation model with $\nbursts\gg1$. Once again, the median star
formation rates inferred for the two models are nearly the same. For
the present work we give results conducted with 10~Myr spacing and
bear in mind that BCGs with a $\chi-2$ minimum at the upper left
corner of the Figures~\ref{Ty} probably favour continuous star
formation model.  Consideration of a broader range of $\tysp$ and/or
simulations with continuous star formation is a promising area for
future investigation.

Note that the stellar property solutions inferred from this work are
not unique. In principle, the SFH could be such that the star
formation rate declines linearly or exponentially or starbursts
instead of being instantaneous may be parameterized by a time-decay
factor. The SFH chosen in this work is motivated by the notion that
cooling of the ICM, star formation and AGN heating may be periodic in
nature. However, choosing a different star formation history, or
initial mass function (e.g bottom-heavy), may lead to different values
of stellar properties. In the future, we will be exploring different
forms of star formation histories and their effects on the inferred
stellar properties.

\subsection{Star Formation Rates}
\label{SFRresults}

\input{table12_ngcpks}

We define the star formation rates as the ratio of the total mass in
the young stellar component (mass in one burst multiplied by the
number of bursts) to the age of the oldest burst. This quantity is the
same as the ratio of the mass in one burst to the age of the most
recent burst, since we are assuming bursts for a given BCG to be of
the same mass separated by a regular interval of time (see
Section~\ref{SFH} for more details).

In Figure~\ref{SFR}, shown are the posterior probability density
function~(PDF) for the star formation rates. This is the probability
distribution for the mass of the YSP divided by its age, marginalized
over all of the other parameters (OSP mass, extinction, etc).  A
distinct feature visible for some of the BCGs is an asymmetric
distribution with a tail extending to the right (towards higher mass)
of the peak SFR. This feature is attributed to the fact that the
masses of both the OSP and YSP are bounded from below, such that
$\mosp$ and $\mysp$ are both greater than zero. In Table~\ref{spprop},
we tabulate the most-likely (mode of the posterior) star formation
rates, defined as the peak of the posterior PDFs, along with the
narrowest 68~\% plausible interval.

It is clear that the star formation rates so derived have a wide range
of plausible values. However, the ranges calculated taking into
account the various input model parameters are robust. Any additional
information, such as priors on the metallicity and/or extinction will
help in narrowing down the range of plausible values of the model
parameters. As an exercise we fixed the extinction in NGC~1275,
PKS~0745 and A~2597 to their best-fit values (the extinction
probability density distributions for these three BCGs is the
broadest). We list in Table~\ref{ngcpks} the 68\% plausible intervals
before and after fixing the extinction. In all three cases, fixing the
extinction results in tighter constraints on the SFRs. This highlights
the importance of having priors on the internal reddening.

\subsection{Chemical Composition - Metallicity}
\label{Zresults}

Since the data used in this study correspond to broadband-imaging
rather than detailed spectra, our results may not be very sensitive to
the metallicity. Moreover, we include in the model only three
candidate metallicities for each stellar population. Therefore we do
not quote errorbars in Table~\ref{spprop} but this is not meant to
imply precise determination of the metallicities. Rather the reader is
encouraged to examine the full posterior distributions. The posterior
probabilities shown in Figure~\ref{metalZ} (where the first and second
strings refers to the OSP and YSP metallicity, respectively) indicate
that for the OSP, 4 out of 10 BCGs favour Z008=$0.4~\zs$, 3 out of 10
favour Z05=$2.5~\zs$, one out of 10 prefers Z=$\zs$ and the remaining
two do not show a preference for any of the models. For the YSP, 7 out
of 10 BCGs favour a low metallicity of Z008=$0.4~\zs$.

As far as the OSP is concerned, theoretical models and simulations of
\cite{Bower2006,deLucia2007,Guo2011} show that the metallicity
distribution of the progenitors peaks in the range $Z=[0.4-0.7]~\zs$,
and is pretty much independent of their redshift of accretion. Massive
galaxies, such as brightest cluster galaxies, are thought to have
formed from smaller progenitors most of whose stellar component was in
existence since $z\sim3$.  As these progenitor galaxies evolve and
form a higher fraction of stars, so should their metallicity.

The low metallicities, as favoured by 4 out of 10 BCGs in our sample,
seem to be in contrast with the results of
\cite{Loubser2009,Loubser2011}, where the author(s) studied 49 and 24
BCGs, respectively, in nearby Universe using long-slit optical data
and found supersolar metallicities \citep[also
see][]{Anja2007}. However, the results based on general samples of
BCGs, which are usually ``red and dead'' elliptical galaxies, need not
apply to BCGs at the centers of cooling flows that have blue
centers. Furthermore, the analysis of \cite{Loubser2009,Loubser2011}
includes only an old stellar population, whereas cool-core BCG spectra
essentially need a YSP component to explain the data. \cite{Liu2012},
on the other hand, analysed a sample of 120 BCGs using SDSS spectral
data of the inner $\pp{3}$ region of each BCG and used the SED fitting
code {\sc starlight} to fit their spectra with a model containing SPs
of three different ages (young, old and intermediate). Although the
average mass-weighted metallicity of their sample is 1.5~$\zs$, the
three cool-core BCGs that overlap between our and their sample --
RXC~J1504, ZwCl~3146 and A~1835 -- have sub-solar metallicities of
around $0.5~\zs$ (Fengshan Liu, private communication). Our results
show that RXC~J1504, on the contrary, seems to clearly prefer a
supersolar metallicity for the OSP.

The work of \cite{Sanderson2009,Leccardi2008} based on {\it
  XMM-Newton} and {\it Chandra} data of 50 and 20 galaxy clusters,
respectively, indicates subsolar metallicities at the centers of
galaxy clusters. Dedicated X-ray {\it XMM-Newton} and {\it Chandra}
observations of cool-core BCGs studied in this work are also
consistent with the overall low metallicity
result. Table~\ref{metalZXray} tabulates the metallicities inferred
from X-ray observations along with the corresponding references. The
central most bins in these observations are a few tens of kiloparsec
at most in size and so reflect the metallicities in the ISM of the
BCG.

\input{table11_metalZXray}

The metallicity inferred from X-ray observations is usually based on
the detected iron lines. Other metals need not necessarily have the
same ratio to iron as in solar and, moreover, the metallicity of the
ISM may not reflect that of the stars. Unfortunately, the metallicity
in the BCGs in our sample do not have a measurement based on direct
optical stellar features.  For A~2597 and NGC~1275, independent
measurements lead to further evidence for low metallicities. In the
case of A~2597, \cite{Voit1997} derived the elemental abundances using
the ratio of forbidden-line fluxes to the hydrogen Balmer lines. The
derived abundances of N{\sc ii} and S{\sc ii} red and blue emission
line doublets led them to conclude that the metallicty in A~2597 is
about 0.5 solar, consistent with the result from X-ray observations
\citep{Morris2005}. The observations carried out by \cite{Voit1997}
are possible but hard to achieve even with modern spectrographs since
not only is the entire coverage from blue to red ends of the spectrum
needed, the spectral resolution needs to be high enough to resolve the
N{\sc ii} and S{\sc ii} line doublets. In \cite{Mittal2012}, we used
optical and infrared line fluxes observed in NGC~1275 and showed that
the fluxes predicted from the radiative transfer code {\sc cloudy}
better fit the data if the metallicty is assumed to be $\sim 0.6~\zs$,
once again, consistent with the result from X-ray observations
\citep{Sanders2007}.

One explanation for low metallicities of the YSP may be embedded in
AGN feedback that is essential in cool-core
clusters. \cite{Khalatyan2008} used the smooth particle hydrodynamics
code to analyse the effect of AGN feedback on the properties of
early-type galaxies, including the impact of AGN winds on the chemical
abundances of the intergalactic medium~(IGM). Their results show that
without AGN feedback, metals remain confined to the galactic center
whereas with AGN feedback, the IGM goes through metal enrichment via
AGN winds that lift the enriched gas from the central regions to outer
regions of the halo. The uplifting of the metal enriched gas leads to
a strong increase in the metal abundance of the IGM \citep[also
see][]{Fabjan2010} and the effects can be seen as early on as at
redshift $z\sim3$. It seems that it is possible that due to AGN
feedback, the ISM in the cool-core BCGs has a low metallicity at all
times and therefore the young stars being born from that ISM also have
low metallicities.

\subsection{Internal Extinction}
\label{extresults}

From Figure~\ref{ext}, it is evident that the internal reddening in
the sample of cool-core BCGs is between 0 and 0.25. This is consistent
with the conclusions drawn by \cite{McDonald2011b}, where the authors
note a slight deviation of the ratio of UV to $\ha$ luminosity from
their assumed model. The authors argue that either an intrinsic
reddening of $E(B-V)\sim0.2$ or a top-heavy IMF can account for such a
deviation. In \cite{McDonald2012}, the authors performed optical
spectroscopy of the $\ha$ filaments observed in several of the
cool-core BCGs \citep[see e.g.][]{Crawford1999} in a sample of 9
cool-core BCGs. Assuming case B recombination ratios for Balmer lines
(an optically thick limit where all photons more energetic than
Ly$\alpha$ are re-absorbed and re-emitted through Ly$\alpha$ and
longer wavelengths), they were able to determine the internal
extinction at more than one location in the BCGs. These locations
included the nucleus as well as the filaments. \cite{McDonald2012}
found that while the reddening, $E(B-V)$, in the nuclei varies from
0.0 to 0.7, the reddening in the filaments peaked at 0 but had a broad
tailed-distribution out to 0.6. These results are in contrast with
those of \cite{Crawford1999}, who found relatively high reddening in
cool core clusters.

We find it very difficult to compare our extinction values to previous
works. The reason is two-fold. First, most of the previous studies
have either focussed on the nuclei or the specific locations of the
filaments (dictated by the spectroscopic slit lengths and positions)
associated with the cool-core galaxies. The likely extinction values
derived in this paper are based on the flux measurements from the
total galaxy. Second, and more importantly, determining internal
extinction based on Balmer lines necessarily entails making
assumptions on the emission mechanism and the physical conditions of
the ISM (such as whether case-A or case-B recombination applies, and
whether processes like shock heating and reconnection diffusion are
prevalent). In this work, we avoid making assumptions in order to
obtain unbiased constraints on the physical parameters of the stellar
populations. This is the reason we subtracted the dominant emission
lines from the observed fluxes (see Section~\ref{lines}).

\section{Discussion}
\label{discussion}

 \begin{figure*}
   \begin{minipage}{0.49\textwidth}
     \centering
     \includegraphics[width=\textwidth]{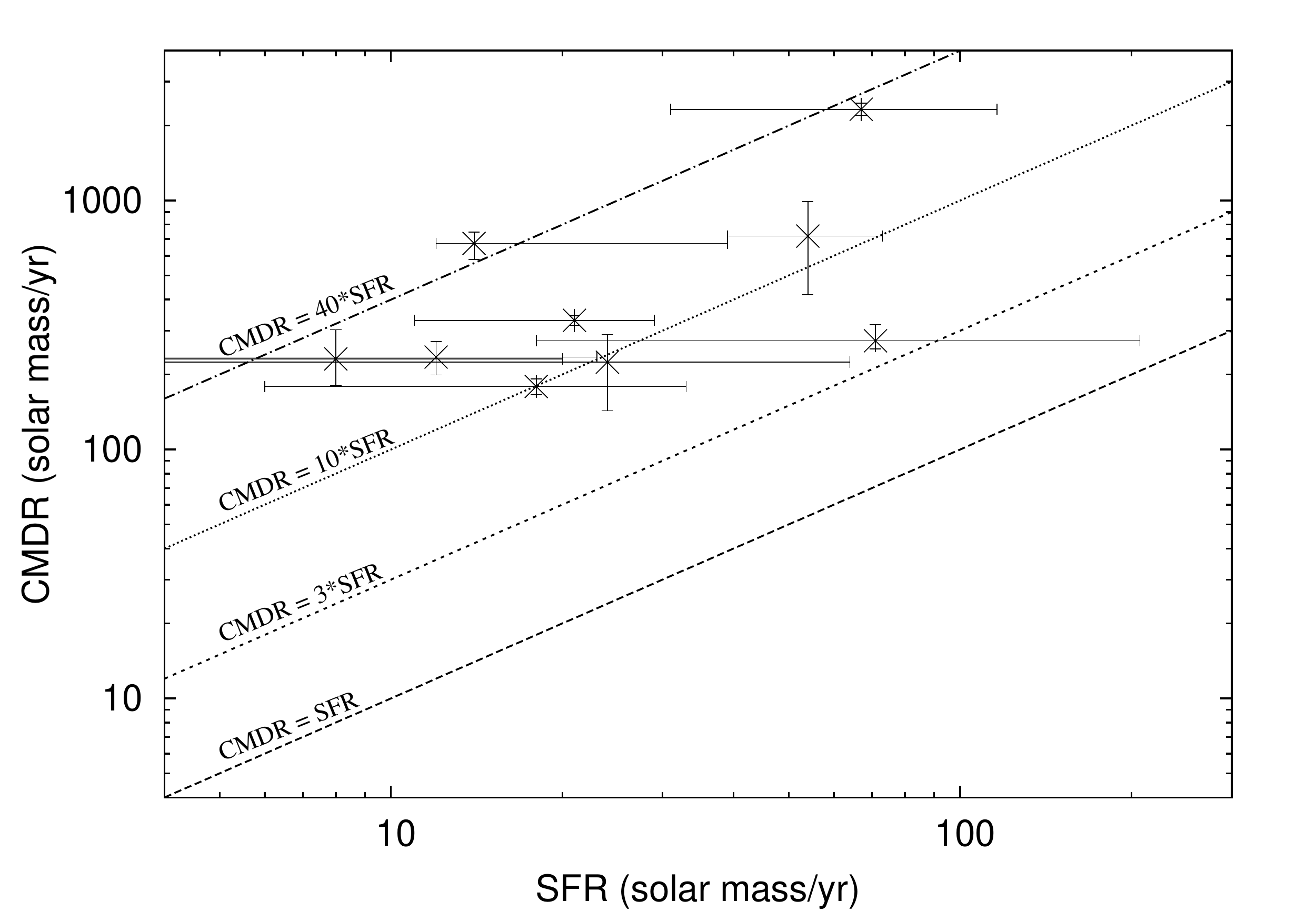}
   \end{minipage}\hfill
   \begin{minipage}{0.49\textwidth}
     \centering
     \includegraphics[width=\textwidth]{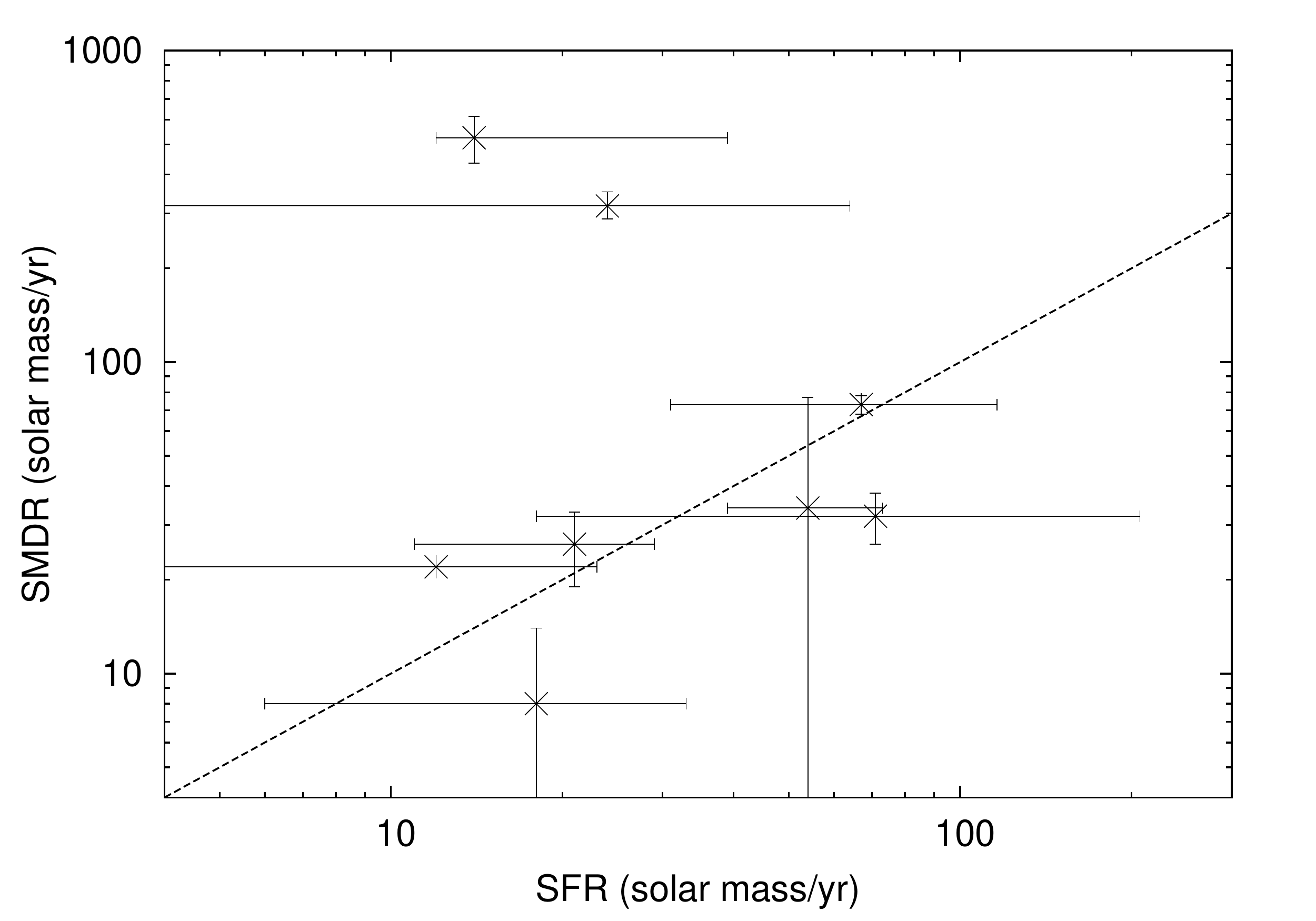}
   \end{minipage}   \caption{The classical~(left) and spectral~(right) mass deposition
     rates~(CMDR) calculated from X-ray observations against star
     formation rates~(SFR) derived from the SPS technique. The
     diagonal line in the right panel represents equality between SMDR
     and SFR.}
   \label{sfr-cmdr}  
 \end{figure*}

In this section we explore the relation between the properties of the
stellar populations in cool-core BCGs, the cooling of the ICM and AGN
heating. The most straight-forward derivative from this work that may
be compared to the cooling and heating properties is the star
formation rate. Note that due to the absence of a young stellar
population in A~2199 (Section~\ref{YSPresults}), we omit this BCG from
all figures that compare the SFRs to cooling and heating properties.

\subsection{Link with Cooling Flows}
\label{CFs}

Cooling flows, mass flowing toward the center of galaxy clusters
coincident with the brightest cluster galaxy, is expected to mainfest
itself in the form of star formation. The SFRs in cooling flow
clusters have been known to be notoriously low in comparison with the
expected mass deposition rate by factors of 10 to 100. We review this
result using the SFRs obtained in this work. However, we caution the
reader that the correlation plots shown in this section present a lot
of scatter, in part due to the fact that the points plotted refer to
the median values, in cases, of rather ill-defined posterior
distributions.

In the left panel of Figure~\ref{sfr-cmdr}, we plot the star formation
rates listed in Table~\ref{spprop} (referred to as ``SPS-SFRs'' from
hereon) against the classical mass deposition rates~(CMDR) derived
from X-ray observations listed in Table~\ref{mdrs}. The CMDRs have
been re-calculated in some of the cases for a $\lambda$CDM cosmology
with $H_0 = 71~h_{71}$~km~s$^{-1}$~Mpc$^{-1}$, $\Omega_{\st m} = 0.27$
(CMDRs scale with luminosity distance inverse square) and correspond
to the mass deposition rate within a radius at which the cooling time
of the gas is 7.7~Gyr to 11~Gyr. Based on the detailed density,
temperature and metallicity radial profiles of the 5 cool-core
clusters belonging to the {\it HIFLUGCS} sample, which were studied in
full extent in \cite{paperIII}, we find that the CMDRs are nearly
constant over radii corresponding to cooling times ranging from
$\sim5~$Gyr to $11~$Gyr (less than 10~\% variation). Hence it is not
necessary to obtain CMDRs at radii corresponding to exactly the same
cooling time. The classical mass deposition rates signify the expected
rate at which the intracluster gas is expected to cool in order to
maintain hydrostatic equilibrium based purely on a cooling-flow model
\citep[e.g.][]{Fabian1994} and in the absence of any kind of heating
\citep[e.g.][]{McNamara2007}. It is typically defined as the ratio of
the gas mass encompassed within a radius to the cooling time at that
radius. A detailed calculation can be found in \cite{paperIII}.

The plot shows a weak positive correlation between the classical mass
deposition rates and the star formation rates, with a Spearman rank
correlation coefficient of 0.43 and a Pearson correlation coefficient
of 0.54. A {\it weak} correlation between the two quantities is not
surprising given the scatter. This correlation has a simple
interpretation that the star formation in cool-core BCGs is in part or
may even largely be due to cooling of the intracluster
medium. \cite{Rafferty2006} and \cite{ODea2008} obtained similar
trends between the two quantities and reached the same
conclusion. Also shown in the figure are diagonal lines representing
various factors needed to equate the two quantities and we find that
the CMDRs are larger than the SPS-SFRs by factors ranging from 4 to
50. This is an expression of the cooling-flow discrepancy
\citep[e.g.][]{paperIII}, wherein the observed cooling rates (as
measured by, for example, star formation rates) are lower than the
predicted rates \citep{Birzan2004,ODea2008}. Our results show that the
SPS-SFRs instead of being lower than the classical CMDRs by factors of
10 to 100 \citep{Rafferty2006,ODea2008} are lower by slightly smaller
factors (4 to 50). One of the reasons our SFRs may be slightly higher
than \cite{Rafferty2006} and \cite{ODea2008} is because we chose
instantaneous bursts as opposed to continuous star formation models
with the former yielding higher masses (left panel of
Figure~\ref{fm_csf}).

Similarly, we compared the SPS-SFRs derived in this work to the
spectral mass deposition rates~(SMDR) derived from X-ray data. SMDRs
reflect the actual rate at which the gas is cooling out of the
intracluster gas and is obtained by fitting a cooling flow model to
the data. The procedure involves technicalities and assumptions that
may vary from study to study resulting in a variance in the
estimates. In Table~\ref{mdrs}, we list the SMDRs from {\it Chandra }
and also other observatories, such as {\it XMM-Newton} and {\it
  FUSE}. Since the sensitivity of {\it Chandra} is lower than that of
{\it XMM-Newton}, the SMDRs obtained from the former are usually
considered as upperlimits. However, {\it XMM-Newton} data for
PKS~0745-191 and RXC~J1504 result in higher SMDRs than {\it Chandra}
data, which merely reflects the spread in the estimates due to
different assumptions.

We plot the SPS-SFRs (this work) against the SMDRs obtained using {\it
  XMM-Newton/FUSE} in the right panel of Figure~\ref{sfr-cmdr}. The
figure shows that with the exception of ZwCl~3146 and PKS~0745-191 the
SPS-SFRs and {\it XMM-Newton/FUSE} SMDRs are similar. It is
interesting that the SPS-SFRs derived in this work are similar to the
spectral MDRs, whereas previously, similar to CMDRs, the SFRs have
been noted to be lower (by factors up to 10) than the SMDRs. We reckon
that the underlying reason for a parity between the SPS-SFRs and SMDRs
is that we have obtained a plausible range of star formation rates
using a grid of models rather than assuming a single model.

 \begin{figure*}
   \begin{minipage}{0.49\textwidth}
     \centering
     \includegraphics[width=1.05\textwidth]{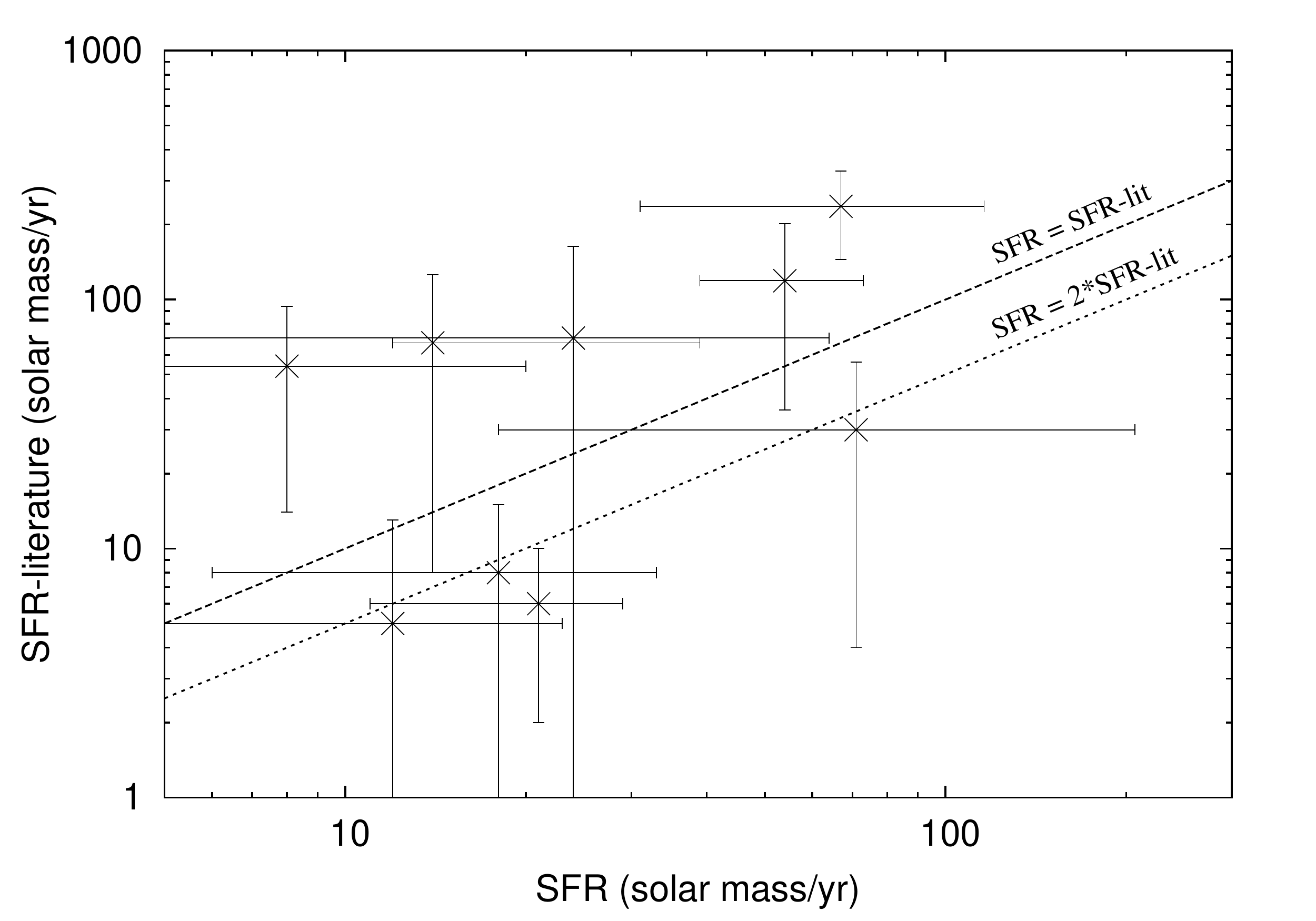}
   \end{minipage}\hfill
   \begin{minipage}{0.49\textwidth}
     \centering
     \includegraphics[width=1.05\textwidth]{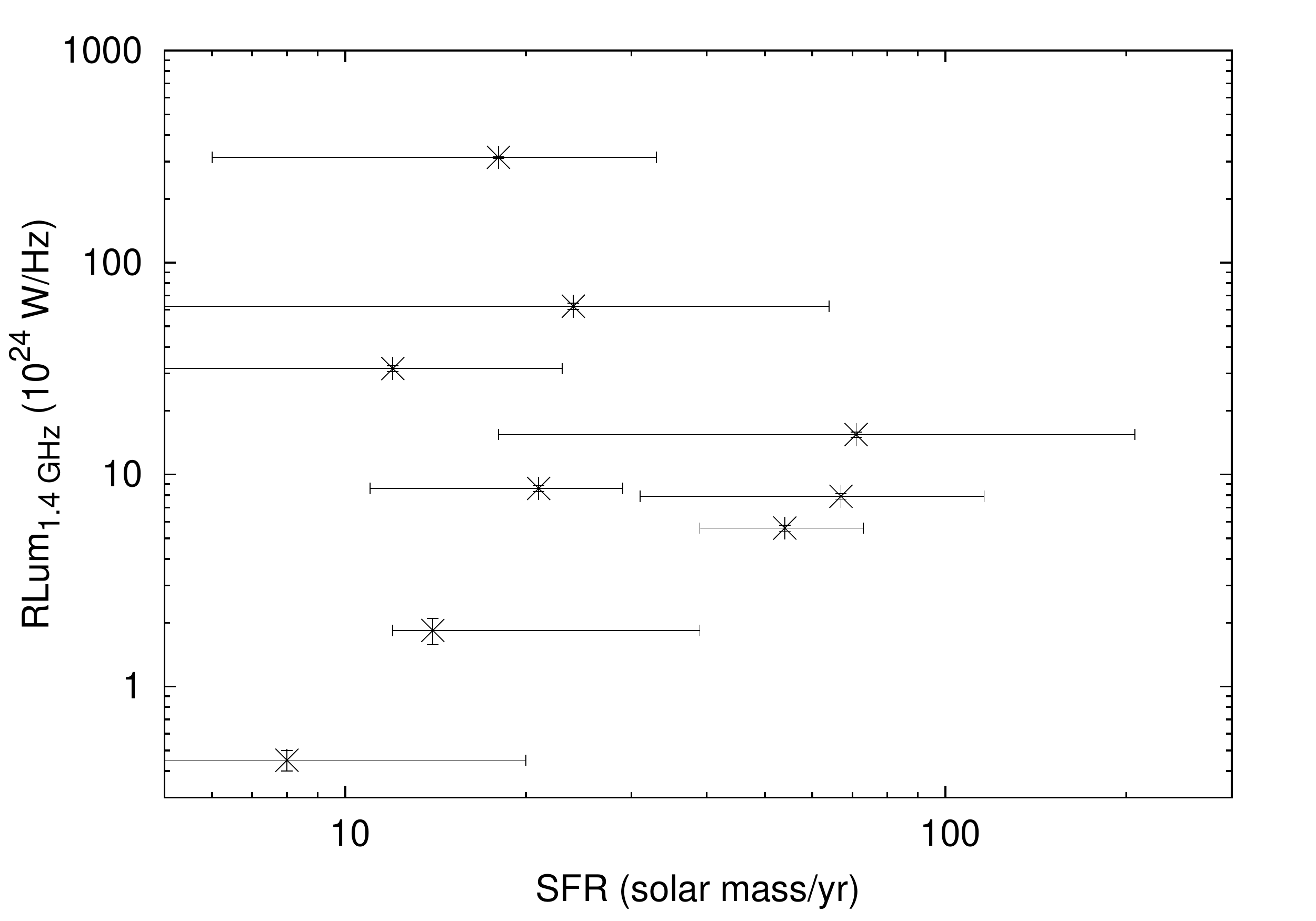}
   \end{minipage}   \caption{{\it Left:} The average star formation rates available in
     the literature versus the star formation rates~(SFR) derived from
     the SPS technique. {\it Right:} 1.4~GHz radio luminosity against
     the star formation rates~(SFR) derived from the SPS
     technique. There seems to be a weak anti-correlation between the
     two quantities. }
   \label{sfr-sfr-rlum}  
 \end{figure*}

 We compiled the star formation rates for the BCGs in our sample from
 the literature to see how they compare to the SPS-SFRs. The SFRs from
 the literature are listed in Table~\ref{mdrs}. The SFRs from
 \cite{Rafferty2006}, compiled in turn from other studies,
 \cite{Hicks2005,ODea2010,Donahue2011} and one of the quoted SFRs in
 \cite{Mittal2012} are based on comparison of data with stellar
 population models. However, these studies usually assume a single
 stellar population with a fixed IMF and/or metallicity. The internal
 extinction has been ignored in some of these studies. The SFRs from
 \cite{Edge2010a,Hoffer2011} and one of the quoted SFRs from
 \cite{Mittal2012} are based on MIR or FIR luminosities. Many of the
 listed SFRs from the literature have small errorbars and are not
 consistent with each other. We also list the average of the SFRs
 available in the literature along with the standard deviation. The
 wide dispersion is clearly visible and comparable to the 68~\%
 plausible intervals for the SPS-SFRs. The reason for such a wide
 dispersion (on both types of SFRs) is clearly due to a large choice
 of models available that are capable of fitting the data. We plot the
 SPS-SFRs against the average SFRs from the literature in left panel
 of Figure~\ref{sfr-sfr-rlum}. We find that the two types of SFRs are
 comparable (A~2199 is omitted because the data are consistent with an
 absence of a YSP).

 From the comparison between the SPS-SFRs and SFRs from the
 literature, we conclude that the star formation rates in BCGs can be
 well-constrained using the SPS technique yielding a robust range of
 possible SFRs. In order to obtain a robust range, a complete range of
 parameters [$\tysp$, $\mosp$, $\mysp$, IMF, $Z$, $E(B-V)$] needs to
 be explored and marginalization techniques need to be employed. To
 this end, we regard our plausible range of SFRs robust in comparison
 with the individual SFRs in the literature. In the future, it will be
 interesting to compare different star formation histories (such as
 instantaneous versus continuous star formation models) and explore a
 wider range of OSP and YSP ages.

\subsection{Link with AGN Heating}
\label{agn}

We briefly inspect whether or not the SPS-SFRs have any correlation
with the properties of the central radio source in the BCGs. In the
generally accepted AGN-regulated feedback framework, the catastrophic
cooling of the ICM is simmered down by AGN heating. As the
intracluster gas cools, a pressure-driven inward flow is established
which serves as fuel for the supermassive black hole residing at the
nucleus of the BCG. AGN outbursts, manifested as giant cavities in the
X-ray images that overlap with radio jets and lobes, transfer some of
the energy back into the ICM via $PdV$ work done by the expanding
cavities, and transmitting weak shocks and sound waves into the
ICM. Observations of multiple cavities at different radii
\citep[e.g. in NGC~1275, Hydra-A,
see][]{Fabian2006,Birzan2008,Birzan2012,Rafferty2006,Rafferty2008}
indicate that the AGN outbursts are likely periodic. This gives rise
to questions as to whether the timescales of AGN activity and cooling
are linked. It has been unambiguously shown that there is a
correlation between the cooling of the ICM and AGN activity
\citep[e.g.][]{Burns1990,Birzan2004,Rafferty2006,Dunn2006,Birzan2008,Mittal2009},
such that every strong cool-core cluster with central cooling time $<
1~$Gyr has a radio AGN at the center, the luminosity of which scales
as the cluster scale (X-ray mass or luminosity).

As discussed in the previous section, star formation is very likely
tied up with cooling, as indicated by a correlation between SPS-SFRs
and the classical mass deposition rate. An interesting aspect we
explore next is a possible correlation between star formation and AGN
activity. Previous studies, such as those by
\cite{Mittal2009,Sun2009}, have found a positive correlation between
the radio luminosity of the AGN and cooling parameters, like, the CMDR
or cooling luminosity (the X-ray luminosity within the cooling
region). Such a correlation may be taken as evidence for AGN feedback,
such that clusters with higher cooling rates harbour stronger AGN
since their ICMs require a greater energy input. As pointed out by
\cite{Sun2009}, although, such a correlation could also be due to the
fact that stronger cool cores are found in more massive clusters which
are also able to better confine radio lobes to stop adiabatic
expansion and reduce radiative losses. The underlying reason for a
positive correlation between the radio luminosity of a centrally
located radio source and cooling activity is not yet clear.

In view of the above results, one would expect a positive correlation
between the SFR and radio luminosity since the gas inflow both fuels
the SFR and the AGN. However, AGN feedback is also believed to quench
cooling activity, and hence also the SFR, so one might also expect to
find an anti-correlation between the SFR and radio luminosity. In the
right panel of Fig.~\ref{sfr-sfr-rlum}, we plot the SPS-SFRs against
the 1.4~GHz radio luminosity \citep[compiled from literature that
included][ and
NED]{Birzan2004,Birzan2008,Mittal2009,Govoni2009}. There is no visible
relation between the two quantities. Considering the uncertainty in
SFRs and a small number of BCGs in the sample, it is difficult to draw
any definite conclusions from this plot. However, given the BCGs in
our sample have very different radio morphologies varying from core
dominated (RXC~J1504), to compact and diffuse (PKS~0745-191 and
A~1835), to small doubles (A~2597 and A~1795), to classic FR-Is
(Hydra-A and A~2199), it is difficult to imagine a scenario where the
power input in the large lobes is matched to the ongoing star
formation. In the future, we will be aiming at expanding the sample so
that we may determine statistical correlations between quantities with
more surety and robustness.

\subsection{Periodicity in Star Formation and its link to Cooling and
  Heating Network}
\label{starbursts}

With the exception of A~2199, all 9 BCGs seem to have had a recent
episode of star formation, requiring a young stellar population with
$\tysp<$ 200~Myr.  Our expectation prior to commencing this work was
for all the BCGs in the cool-core sample to show evidence for multiple
YSPs since the AGN-regulated feedback entails periodicity in both AGN
heating and cooling of the ICM. An important assumption in this
argument is that the cool cores were formed long enough ago
($z\sim1$), that the gas has been cooling for about 7~Gyr $-$
8~Gyr. This assumption was justified by \cite[][]{paperIII}, whose
results ruled out a recent formation of cool-cores ($z\lesssim0.5$)
\cite[also see][]{McDonald2013}.

Our results show that 9 out of 10 BCGs indeed require periodic or
continuous star formation to explain their broad-band SEDs. An
interesting feature for all 9 BCGs is that the star formation seems to
have been ongoing since at least 6~Gyr ago. Hence we find that not
only cooling but also star formation in cool-core BCGs has been a
long-term phenomenon. While ZwCl~3146, A~1795, A~1835 and RXC~J1504
seem to require continuous star formation, NGC~1275, PKS~0745,
Hydra-A, A~1068 and A~2597 seem to require periodic starbursts with
the interval ranging between 20~Myr and 200~Myr. These values
correlate well with both the cooling-time of the hot ICM/ISM gas in
the inner 5~kpc radius of cool-core BCGs \citep{Voit2014} and the
quiescent phase duration of radio sources in massive galaxies
\citep{Shabala2008}. The 5~kpc radius also corresponds to the spatial
extent of the youngest stars seen in the form of FUV emission of most
BCGs shown in Figure~\ref{FUV-images}. The true star formation
history, we note however, is probably much more complex than this
simple hypothesis where the stars form periodically at regular
intervals with a plausible correlation with the cooling and AGN
heating time-scales of the intracluster gas.

In regards to NGC~1275, PKS~0745, Hydra-A, A~1068 and A~2597 seeming
to require periodic starbursts, we note that with the exception of
A~1068, these BCGs are ones with the highest radio luminosities (see
Table~\ref{log}).  From this simple observation, it seems that the
periodic starbursts are somehow connected to the synchrotron output of
the AGN. It may be that cooling of the ICM in these four BCGs is
periodic, such that some of the cooling gas clumps into molecular
clouds leading to star formation and some of it fuels the AGN. There
is, in fact, observational evidence for multiple cavities in NGC~1275
\citep{Fabian2006}, Hydra~A \citep{Wise2007} and A~2597
\citep{Tremblay2012a}. It is not yet clear whether multiple cavities
in cool-core clusters are as a result of continuous jet activity or
due to the episodic nature of the AGN on and off state. While our data
are not capable of distinguishing between the two scenarios, periodic
accretion of cold gas by the supermassive black hole is expected to
have a direct impact on the output of the AGN, and can explain the
correlation between the existence of periodic starbursts and high
radio-luminosity.

As per the results described in Table~\ref{spprop}, A~1068 seems to be
undergoing regular bursts since 6~Gyr ago at intervals of about
80~Myr. The 68~\% plausible range of star formation rate is 2~$\mpy$
to 21~$\mpy$. This BCG has the lowest 1.4~GHz radio luminosity and so
it may be that the AGN is in a quiescent period in the cycle of ICM
cooling-AGN heating, which would imply that the AGN activity in
cool-core clusters is non-continuous and has a less than 100~\%
duty-cycle, consistent with the results of
\cite{Birzan2012,Donahue2010}.

Lastly, we mention another interesting quantity, which is the ratio of
the mass in the young to the mass in the old stellar population,
$\mysp/\mosp$ (column 4 of Table~\ref{spprop}). The ratio ranges from
a few percent to as high as 30~\% suggesting that a significant
fraction of stellar mass accumulated since $z\sim1$.

\section{Conclusions}
\label{conclusions}

We have addressed a crucial issue in the domain of cool-core brightest
cluster galaxies, namely, how accurately can we determine the star
formation rates in these galaxies. This has a direct impact on our
understanding of the cooling of the intracluster-medium, star
formation and AGN-regulated feedback.

We used broad-band imaging of a sample of 10 strong cool-core BCGs
(with central cooling times less than $1~$Gyr) and conducted a
Bayesian analysis using the technique of stellar population
synthesis~(SPS) to determine the most likely stellar populations in
these BCGs and the properties thereof. Our model consists of an old
stellar population and a series of young stellar components.  We
calculated probability distributions for various model parameters,
marginalized over the others. By doing this we were able to obtain
68\% plausible intervals for the model parameters, such as the masses
in the old and young stellar population, and the star formation rates
(SPS-SFRs).

We find that the most-likely SPS-SFRs are factors of 4 to 50 lower
than the classical mass deposition rates, the expected rates of mass
condensation, inferred from the X-ray data. This range of factors is
slightly lower than what was previously thought (10 to 100). The 68\%
plausible interval on the SFRs is broad, owing to a wide range of
models that are capable of fitting the data, and explains the wide
dispersion in the star formation rates available in the
literature. However, the ranges of possible values, despite being
wide, are robust and highlights the strength in Bayesian analyses.

We find that 9 out of 10 BCGs have been experiencing starbursts since
6~Gyr ago. While four out of 9 BCGs seem to require continuous star
formation rates, 5 out of 9 seem to require periodic star formation on
intervals ranging from 20~Myr to 200~Myr. This time scale is similar
to the cooling-time of the intracluster gas in the very central
($<5~$kpc) regions of BCGs.

\nocite{Wright2006}

\section*{Acknowledgments} We thank the anonymous referee for
constructive and detailed feedback. We thank Hyunjin Jeong for the
discussion on the UV-upturn phenomenon, Hans B\"ohringer, Alastair
Edge, Christopher O'Dea, Stefi Baum, Megan Donahue, Gabriella Lucia,
and Brian McNamara for useful comments and Andy Fabian for a critical
feedback. We also thank Fengshan Liu, Shude Mao and Xianmen Meng for
their helpful feedback on their metallicity results. Support for
program number 12220 was provided by NASA through a grant from the
Space Telescope Science Institute, which is operated by the
Association of Universities for Research in Astronomy, Inc., under
NASA contract NAS5-26555. This publication makes use of data products
from the Two Micron All Sky Survey~(2MASS), which is a joint project
of the University of Massachusetts and the Infrared Processing and
Analysis Center/California Institute of Technology, funded by the
National Aeronautics and Space Administration and the National Science
Foundation. This research has made use of the NASA/IPAC Extragalactic
Database (NED) and Galaxy Evolution Explorer (GALEX) data, both of
which are operated by the Jet Propulsion Laboratory, California
Institute of Technology, under contract with the National Aeronautics
and Space Administration, and Ned Wright's cosmological
calculator. STSDAS is a product of the Space Telescope Science
Institute, which is operated by AURA for NASA. We thank the {\sc
  cfitsio} and {\sc wcstools} teams for making available these
software for the astronomical community.

\bibliographystyle{mn2e}
\bibliography{ref}

\input{table6_spprop}
\input{posteriors}
\input{table13_mdrs}

\input{bestfitplots}

\end{document}

%% file: makros.tex

\newcommand{\st}[1]{\mathrm{#1}} 
\newcommand{\pow}[2]{$\st{#1}^{#2}$}
\newcommand{\grad}{\hspace{-0.15em}\r{}}
\newcommand{\lm}{\lambda}
\newcommand{\average}[1]{\left\langle #1 \right\rangle}
\newcommand{\pp}[1]{#1^{\prime\prime}}
\newcommand{\p}[1]{#1^{\prime}}

\newcommand{\fobs}{F$_{\st {obs}}$}
\newcommand{\fintrinsic}{F$_{\st {intr}}$}
\newcommand{\fgal}{F$_{\st {gal}}$}

\newcommand{\oi}{[O{\sc i}]}
\newcommand{\cii}{[C{\sc ii}]}
\newcommand{\nii}{[N{\sc ii}]}
\newcommand{\oiii}{[O{\sc iii}]}
\newcommand{\oii}{[O{\sc ii}]}
\newcommand{\si}{[Si{\sc i}]}
\newcommand{\mm}{~$\mu$m}
\newcommand{\mminv}{\mm$^{-1}$}
\newcommand{\ha}{\st{H}\alpha}
\newcommand{\hb}{\st{H}\beta}
\newcommand{\civ}{[C{\sc iv}]}
\newcommand{\niiopt}{[N{\sc ii}]~$\lambda6583$}
\newcommand{\niia}{[N{\sc ii}]~$\lambda5199$}
\newcommand{\siia}{[S{\sc ii}]~$\lambda6717$}
\newcommand{\siib}{[S{\sc ii}]~$\lambda6731$}
\newcommand{\oiiopt}{[O{\sc ii}]~$\lambda3727$}
\newcommand{\oia}{[O{\sc i}]~$\lambda6300$}
\newcommand{\oib}{[O{\sc i}]~$\lambda6363$}
\newcommand{\oiiiopt}{[O{\sc iii}]~$\lambda5007$}
\newcommand{\ohb}{[\st{O_{~III}}]~\lambda~5007/{\st H}\beta}

\newcommand{\md}{M_{\st{d}}}
\newcommand{\td}{T_{\st{d}}}
\newcommand{\mdc}{M_{\st{d,c}}}
\newcommand{\tdc}{T_{\st{d,c}}}
\newcommand{\mdw}{M_{\st{d,w}}}
\newcommand{\tdw}{T_{\st{d,w}}}
\newcommand{\mg}{M_{\st{g}}}
\newcommand{\tcmb}{T_{\st{cmb}}}
\newcommand{\hextra}{H_{\st{extra}}}
\newcommand{\hcol}{N$_{\st H}$}
\newcommand{\pdv}{$P{\st d}V$}
\newcommand{\nitrogen}{$Z_{\odot}$(N)}

\newcommand{\h}{~h_{71}~}
\newcommand{\hinv}[1]{~h_{71}^{#1}~}
\newcommand{\eq}[1]{(\ref{eq-#1})}
\newcommand{\wrt}{with respect to\ }
\newcommand{\mms}{\frac{M_{\odot}}{M}}
\newcommand{\mpy}{\ms~\st{yr}^{-1}}
\newcommand{\ct}{t_{\st{cool}}}
\newcommand{\rc}{r_{\st{cool}}}
\newcommand{\lc}{L_{\st{cool}}}
\newcommand{\tvir}{T_{\st{vir}}}
\newcommand{\rvir}{R_{\st{500}}}
\newcommand{\mvir}{M_{\st{500}}}
\newcommand{\mbh}{M_{\st{BH}}}
\newcommand{\ms}{M_{\odot}}
\newcommand{\ls}{L_{\odot}}
\newcommand{\zs}{Z_{\odot}}
\newcommand{\lxb}{L_{\st {Xb}}}
\newcommand{\lfir}{L_{\st{FIR}}}
\newcommand{\lfirtot}{L_{\st{FIR,tot}}}
\newcommand{\lx}{L_{\st {X}}}
\newcommand{\lr}{L_{\st {R}}}
\newcommand{\rlum}{L_{\nu~(1.4~\textrm{GHz})}}
\newcommand{\lbcg}{L_{\st{BCG}}}
\newcommand{\lt}{L_{\st{X}}{\st -}\tvir}
\newcommand{\Dl}{D_{\st{L}}}
\newcommand{\Da}{D_{\st{A}}}
\newcommand{\mdr}{\dot{M}_{\st{classical}}}
\newcommand{\smdr}{\dot{M}_{\st{spec}}}
\newcommand{\norm}{$\eta_{\st{OSP}}$}

\newcommand{\hydra}{Hydra-A}
\newcommand{\pks}{PKS~0745-191}
\newcommand{\rxj}{RXC~J1504}
\newcommand{\zw}{ZwCl~3146}
\newcommand{\pers}{NGC~1275}
\newcommand{\cen}{NGC~4696}
\newcommand{\tosp}{\tau_{\st{o}}}
\newcommand{\tysp}{\tau_{\st{y}}}
\newcommand{\mosp}{M_{\st{o}}}
\newcommand{\mysp}{M_{\st{y}}}
\newcommand{\nbursts}{N_{\textrm{bursts}}}
\newcommand{\lamvec}{{\theta}}
\newcommand{\Fis}{\{F_i\}}
\newcommand{\text}[1]{{\rm #1}}

\newcommand{\chandra}{\textit{Chandra}}
\newcommand{\vla}{\textit{VLA}}
\newcommand{\gmrt}{\textit{GMRT}}
\newcommand{\atca}{\textit{ATCA}}
\newcommand{\XMM}{\textit{XMM-Newton}}
\newcommand{\einstein}{\textit{Einstein}}
\newcommand{\asca}{\textit{ASCA}}
\newcommand{\rosat}{\textit{ROSAT}}
\newcommand{\herschel}{\textit{Herschel}}
\newcommand{\iras}{\textit{IRAS}}
\newcommand{\spitzer}{\textit{Spitzer}}
\newcommand{\hiflux}{\textit{HIFLUGCS}}

\newcommand{\combf}[1]{{\bf #1}}

%% file: journals.tex
%
\def\aj{AJ}%
\def\araa{ARA\&A}%
\def\apj{ApJ}%
\def\apjl{ApJ}%
\def\apjs{ApJS}%
\def\ao{Appl.~Opt.}%
\def\apss{Ap\&SS}%
\def\aap{A\&A}%
\def\aapr{A\&A~Rev.}%
\def\aaps{A\&AS}%
\def\azh{AZh}%
\def\baas{BAAS}%
\def\jrasc{JRASC}%
\def\memras{MmRAS}%
\def\mnras{MNRAS}%
\def\pra{Phys.~Rev.~A}%
\def\prb{Phys.~Rev.~B}%
\def\prc{Phys.~Rev.~C}%
\def\prd{Phys.~Rev.~D}%
\def\pre{Phys.~Rev.~E}%
\def\prl{Phys.~Rev.~Lett.}%
\def\pasp{PASP}%
\def\pasj{PASJ}%
\def\qjras{QJRAS}%
\def\skytel{S\&T}%
\def\solphys{Sol.~Phys.}%
\def\sovast{Soviet~Ast.}%
\def\ssr{Space~Sci.~Rev.}%
\def\zap{ZAp}%
\def\nat{Nature}%
\def\iaucirc{IAU~Circ.}%
\def\aplett{Astrophys.~Lett.}%
\def\apspr{Astrophys.~Space~Phys.~Res.}%
\def\bain{Bull.~Astron.~Inst.~Netherlands}%
\def\fcp{Fund.~Cosmic~Phys.}%
\def\gca{Geochim.~Cosmochim.~Acta}%
\def\grl{Geophys.~Res.~Lett.}%
\def\jcp{J.~Chem.~Phys.}%
\def\jgr{J.~Geophys.~Res.}%
\def\jqsrt{J.~Quant.~Spec.~Radiat.~Transf.}%
\def\memsai{Mem.~Soc.~Astron.~Italiana}%
\def\nphysa{Nucl.~Phys.~A}%
\def\physrep{Phys.~Rep.}%
\def\physscr{Phys.~Scr}%
\def\planss{Planet.~Space~Sci.}%
\def\procspie{Proc.~SPIE}%

%% file: table1_log.tex
\begin{table*}
    \centering
    \caption{Basic properties of the sample BCGs taken from NED. The 1.4~GHz radio 
      luminosities have been compiled from literature 
      \citep[][ and NED]{Birzan2004,Birzan2008,Mittal2009,Govoni2009}.}
    \label{log}
    \begin{tabular}{| l | c | c | c | c | c | c|}
    \hline
    
    Cluster		 & RA  (J2000)         &  Dec (J2000)         & Redshift  & $\pp{1}$ (in kpc)& E(B-V) & $L_{\nu~(1.4~\textrm{GHz})}$ (in $10^{24}$~W/Hz)\\
    \hline \hline  									        
    {\pers}              & 03h19m48.16s &  +41d30m42.1s & 0.01756   & 0.352            & 0.144  &$ 15.44\pm0.46$  \\
    {\pks}	         & 07h47m31.35s &  -19d17m39.7s & 0.10280   & 1.868            & 0.463  &$ 62.23\pm2.21$  \\
    {\hydra}		 & 09h18m05.67s &  -12d05m43.9s & 0.05490   & 1.054    	       & 0.036  &$313.73\pm2.78$  \\
    {\zw}		 & 10h23m39.60s &  +04d11m12.0s & 0.28990   & 4.316    	       & 0.026  &$  1.84\pm0.26$  \\
    A~1068		 & 10h40m44.40s &  +39d57m12.0s & 0.13860   & 2.420    	       & 0.020  &$  0.45\pm0.05$  \\
    A~1795		 & 13h48m52.43s &  +26d35m34.0s & 0.06276   & 1.193    	       & 0.012  &$  8.60\pm0.28$  \\
    A~1835		 & 14h01m02.00s &  +02d52m45.0s & 0.25194   & 3.899    	       & 0.026  &$  5.59\pm0.18$  \\
    {\rxj}	 	 & 15h04m07.50s &  -02d48m16.0s & 0.21720   & 3.485    	       & 0.098  &$  7.90\pm0.24$  \\
    A~2199		 & 16h28m38.24s &  +39d33m04.3s & 0.03030   & 0.599 	       & 0.010  &$  7.48\pm0.21$	 \\
    A~2597		 & 23h25m19.82s &  -12d07m26.4s & 0.08300   & 1.542    	       & 0.026  &$ 32.03\pm1.01$  \\
    \hline
    \end{tabular}
\end{table*}

%% file: table2_hst.tex
\begin{table*}
  \centering
  \caption{\small: HST Photometry Details. The columns indicate (1) cluster name, (2) instrument mode, (3) proposal ID, (4) the exposure time, (5) the pivot wavelength, (6) the photflam defined as the flux of the source that produces 1 count per sec, (7) the mean aperture radius (for elliptical apertures, the major and minor axes are given along with the angle of the major axis in brackets), (8) the mean Galactic-extinction corrected flux-density and (9) notes (``..." implies the measurement does not contain any or significant lines, ``w lines" implies the measurement is contaminated by lines, ``wo lines" implies the measurement has been corrected for the lines. For NGC~1275, ``w AGN" implies the measurement contains AGN emission and ``wo AGN" implies the measurement has been corrected for the AGN emission).}
  \label{hst}
  \begin{tabular}{| c | c | c | c | c | c | c | c | c |}
  \hline

  Cluster	 & Aperture/Filter & Prop ID & Exp. Time & Wavelength & PHOTFLAM			     & Aperture   & Mean Flux Density                              & Notes     \\
                 &                 &         &      (s)  & ($\AA$)    & (erg~s$^{-1}$~cm$^{-2}$~$\AA^{-1}$)  & (arcsec)   & (10$^{-16}~$erg~s$^{-1}$~cm$^{-2}$~$\AA^{-1}$) &     \\ 
  \hline \hline
  {\pers}        & WFC/F550M       & 12132   & 12132     & 5581.5     & 3.848E-19 			     & 75	  & 968.7   & w AGN   \\
  		 & ...             & ...     & ...       & ...        & ...                          	     &            & 942.3   & wo AGN  \\
   		 & WFC/F814W       & 12800   & 12900     & 7995.9     & 2.508E-20 			     & 75	  & 827.3   & w AGN   \\
  		 & ...             & ...     & ...       & ...        & ...                           	     &            & 811.1   & wo AGN  \\
  \hline			                         
  {\pks}	 & SBC/F140LP	   & 12220   & 2715      & 1528.0     & 2.713E-17 			     & 3.5	  & 17.9   & ... \\
  		 & WFPC2/F814W	   & 7337    & 2100      & 5439.0     & 3.439E-18			     & 12 (5)	  & 26.5   & w lines \\
  		 & ...             & ...     & ...       & ...        & ...                          	     &            & 25.0   & wo lines \\
      		 
  \hline			                         
  {\hydra}	 & SBC/F140LP	   & 12220   & 2709      & 1528.0     & 2.713E-17 			     & 7.5        & 30.9   & ... \\
  		 & WFC/F814W	   & 12220   & 2367      & 8056.9     & 7.033E-20			     & 23         & 73.5   & ... \\
  		 
  \hline			                         
  {\zw}	 & SBC/F165LP	   & 11230   & 1170      & 1762.5     & 1.359e-16 			     & 8$\times$4.5    (50) & 4.9  & ...      \\
  		 & WFPC2/F606W	   & 8301    & 1000      & 5996.8     & 1.888E-18			     & 8.5$\times$6.5  (50) & 2.6  & w lines  \\
  		 & ...             & ...     & ...       & ...        & ...                          	     &                      & 2.5  & wo lines \\
   		 
  \hline			                         
  A~1068		 & SBC/F150LP	   & 12220   & 2766      & 1612.2     & 4.392E-17 			     & 11$\times$5.5 (61)    & 4.9  & ... \\
  		 & WFPC2/F606W	   & 8301    & 600       & 5996.8     & 1.888E-18			     & 18.5$\times$8.5 (40)  & 16.9 & ...  \\
        	 
  \hline			                         
  A~1795          & $^\dagger$SBC/F140LP      & 11980   & 2394      & 1528.0     & 2.713E-17    		     & 12 + 18$\times$8 (78) & 25.4  & ...      \\
  		 & WFPC2/F555W     & 5212    & 1600      & 5442.9     & 3.483E-18			     & 18$\times$14.5 (95)   & 40.5  & w lines  \\
  		 & ...             & ...     & ...       & ...        & ...                          	     &    	             & 40.2  & wo lines \\
  		 & WFPC2/F702W     & 5212    & 1600      & 6917.1     & 1.872E-18			     & 26$\times$22.5 (95)   & 63.9  & w lines  \\
  		 & ...             & ...     & ...       & ...        & ...                          	     & 	     	      	     & 62.6  & wo lines \\
                 
  \hline			                         
  A~1835		 & SBC/F165LP	   & 11230   & 1170      & 1762.5     & 1.359E-16 			     & 5$\times$4 (65)  & 6.8 & ...      \\
  		 & WFPC2/F702W	   & 8249    & 7500      & 6917.1     & 1.872E-18			     & 7$\times$5 (65)  & 5.6 & w lines  \\
  		 & ...             & ...     & ...       & ...        & ...                          	     &                  & 5.4 & wo lines \\
        	 
  \hline			                         
  {\rxj}	 & SBC/F165LP	   & 12220   & 2700      & 1762.5     & 1.360E-16 			     & 7.0	  & 26.7 & ...   \\
  		 & WFC3/F689M	   & 12220   & 2637      & 6876.3     & 3.714E-19			     & 6.5        & 7.9  & ...   \\
     		 
  \hline			                         
  A~2199		 & SBC/F140LP	   & 12220   & 2767      & 1528.0     & 2.713E-17 			     & 8.3$\times$6.8 (95)   & 5.3    & ... \\
  		 & WFC/F475W	   & 12238   & 5370      & 4745.6     & 1.827E-19			     & 50                    & 283.0  & ... \\
  		 & WFPC/F555W	   & 7265    & 5200      & 5443.0     & 3.483E-18			     & 47.8$\times$35 (300)  & 292.6  & ... \\
		 & WFC/F814W	   & 9293    & 700       & 8059.9     & 6.926E-20			     & 75$\times$47.5 (300)  & 367.3  & ... \\
        	 
  \hline			                         
  A~2597          & SBC/F150LP	   & 11131   & 8141      & 1612.2     & 4.392E-17 			     & 10.5                   & 14.9  & ...      \\
  		 & WFPC2/F450W     & 6228    & 2500      & 4557.3     & 9.022E-18			     & 13$\times$9 (55)       & 16.2  & w lines  \\
  		 & ...             & ...     & ...       & ...        & ...                          	     &                        & 14.8  & wo lines \\
  		 & WFPC2/F702W	   & 6228    & 2100      & 6917.1     & 1.872E-18			     & 20.5$\times$10.5 (55)  & 23.6  & w lines  \\
  		 & ...             & ...     & ...       & ...        & ...                          	     &                        & 21.6  & wo lines \\  		 
  \hline
   \end{tabular}
 \end{table*} 

%% file: images.tex

\begin{figure*}
  \begin{minipage}{0.33\textwidth}
    \centering
    \includegraphics[width=0.95\textwidth]{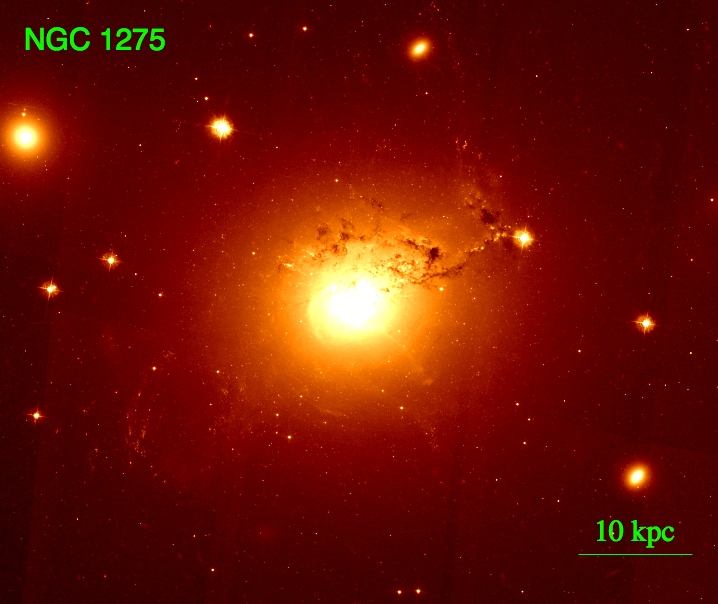}
  \end{minipage}%
  \begin{minipage}{0.33\textwidth}
    \centering
    \includegraphics[width=0.95\textwidth]{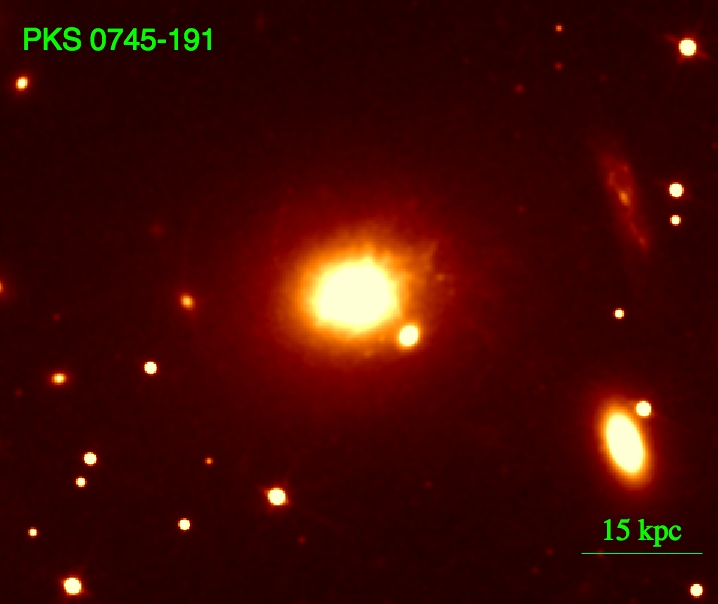}
  \end{minipage}%
  \begin{minipage}{0.33\textwidth}
    \centering
    \includegraphics[width=0.95\textwidth]{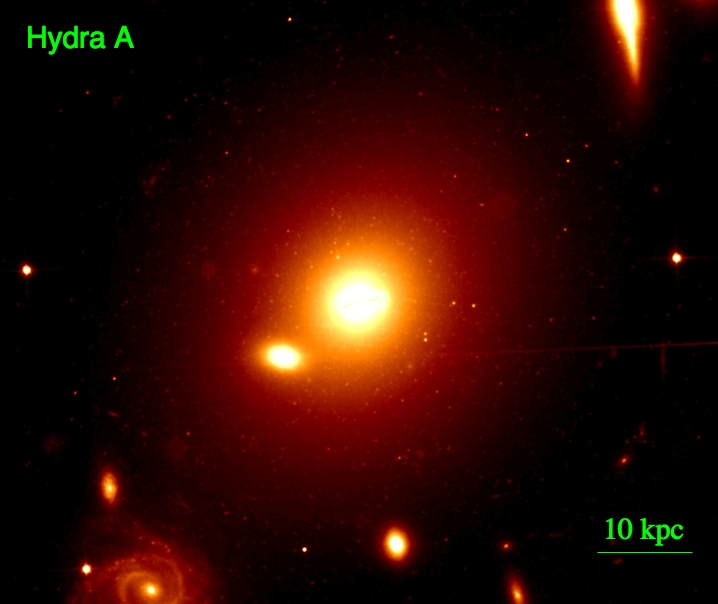}
  \end{minipage}\\[8pt]
  \begin{minipage}{0.33\textwidth}
    \centering
    \includegraphics[width=0.95\textwidth]{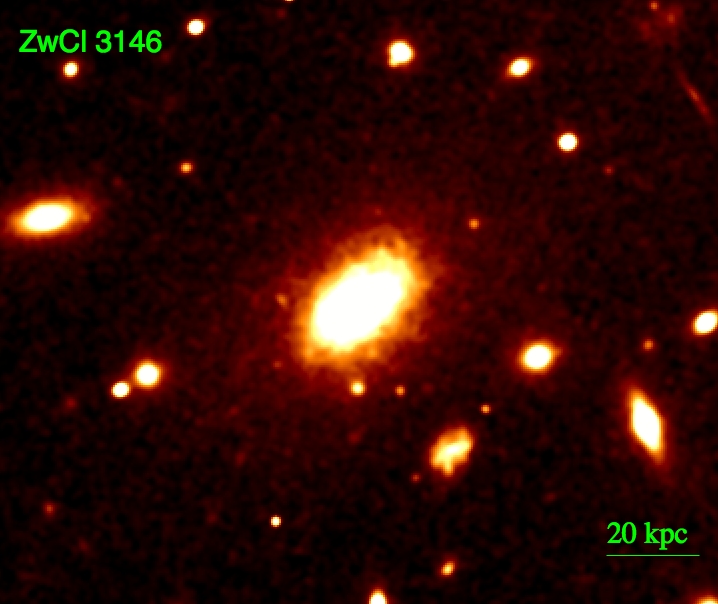}
  \end{minipage}%
  \begin{minipage}{0.33\textwidth}
    \centering
    \includegraphics[width=0.95\textwidth]{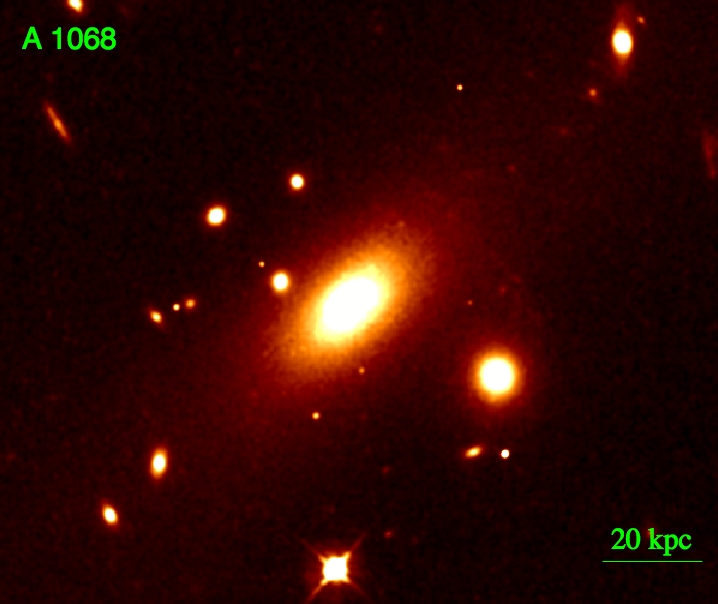}
  \end{minipage}%
  \begin{minipage}{0.33\textwidth}
    \centering
    \includegraphics[width=0.95\textwidth]{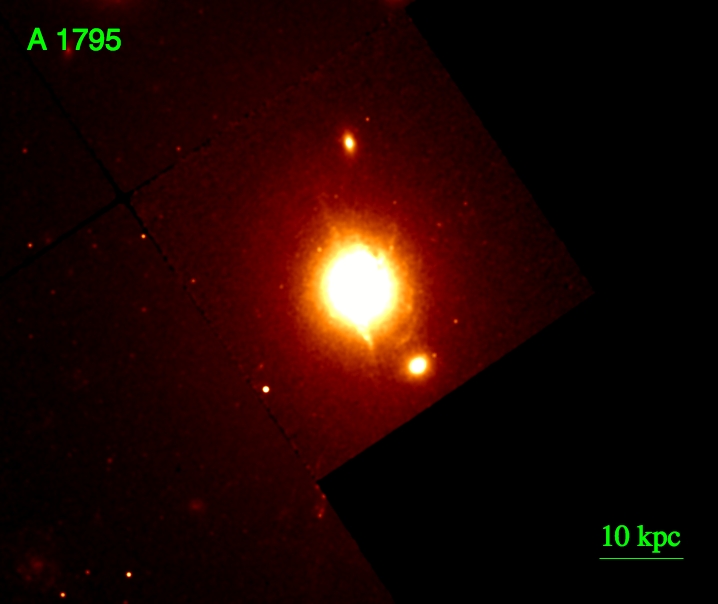}
  \end{minipage}\\[8pt]
  \begin{minipage}{0.33\textwidth}
    \centering
    \includegraphics[width=0.95\textwidth]{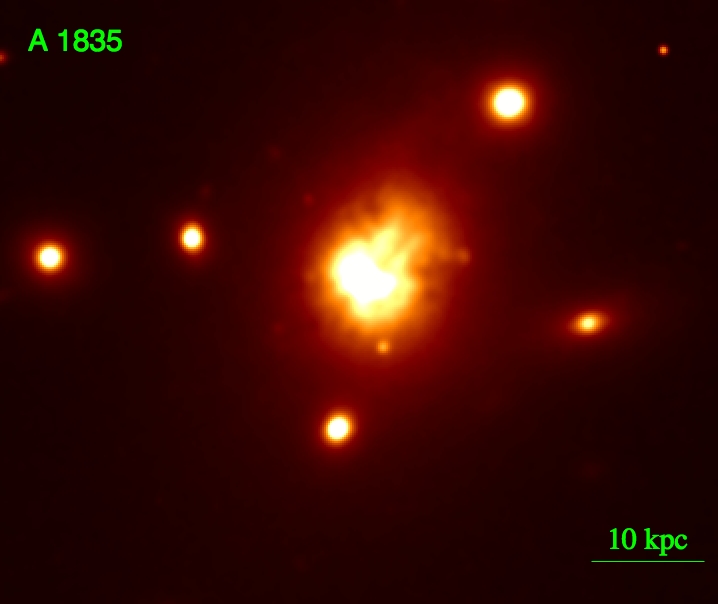}
  \end{minipage}%
  \begin{minipage}{0.33\textwidth}
    \centering
    \includegraphics[width=0.95\textwidth]{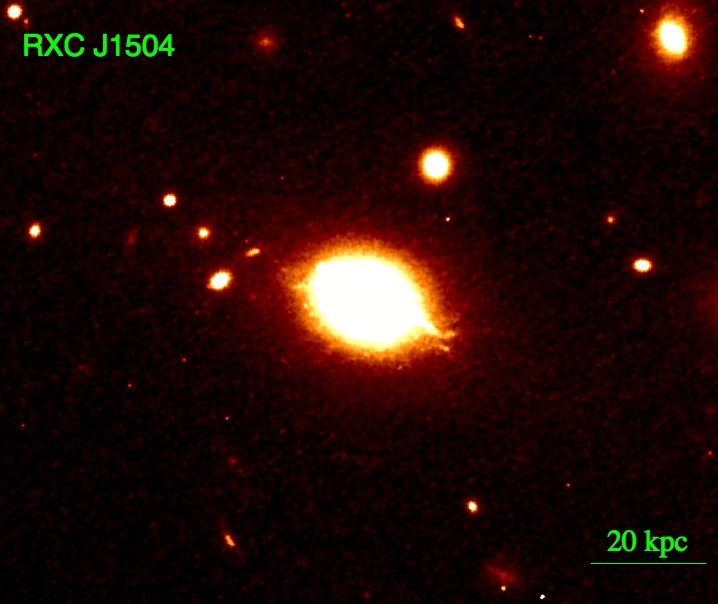}
  \end{minipage}%
  \begin{minipage}{0.33\textwidth}
    \centering
    \includegraphics[width=0.95\textwidth]{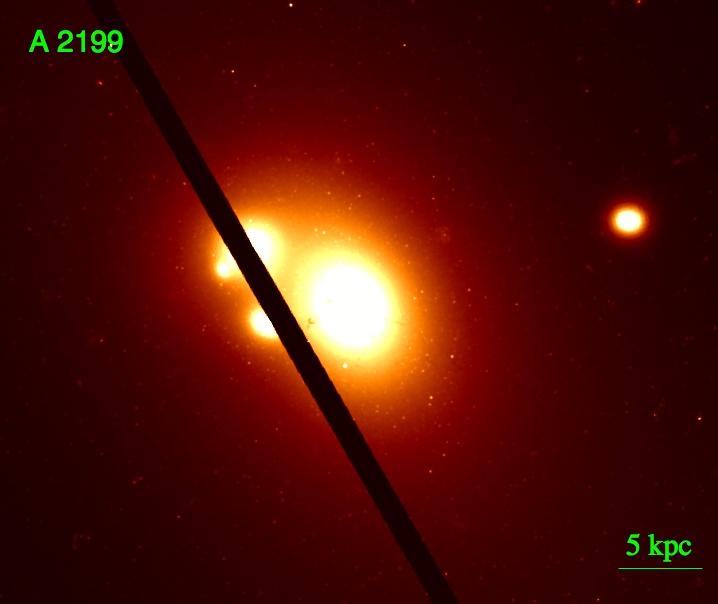}
  \end{minipage}\\[8pt]
  \begin{minipage}{0.33\textwidth}
    \centering
    \includegraphics[width=0.95\textwidth]{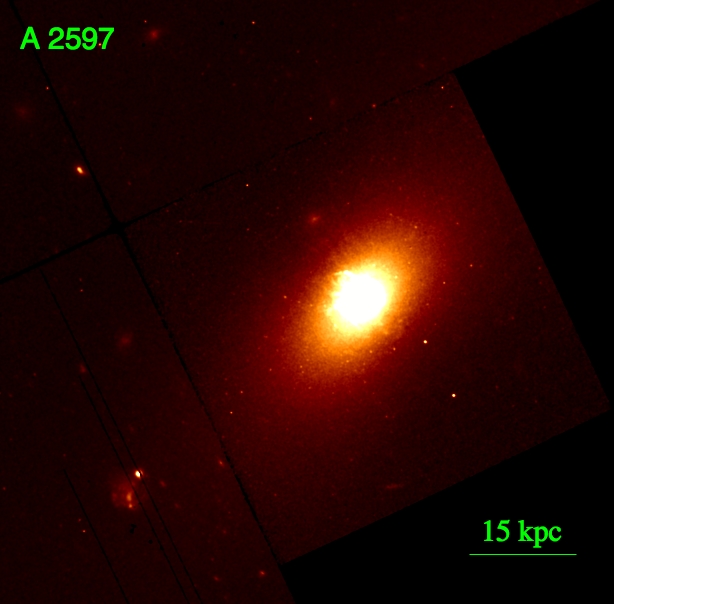}
  \end{minipage}%
  \caption{HST optical images of the BCGs.}
  \label{optical-images}
\end{figure*}


\begin{figure*}
  \begin{minipage}{0.33\textwidth}
    \centering
    \includegraphics[width=0.95\textwidth]{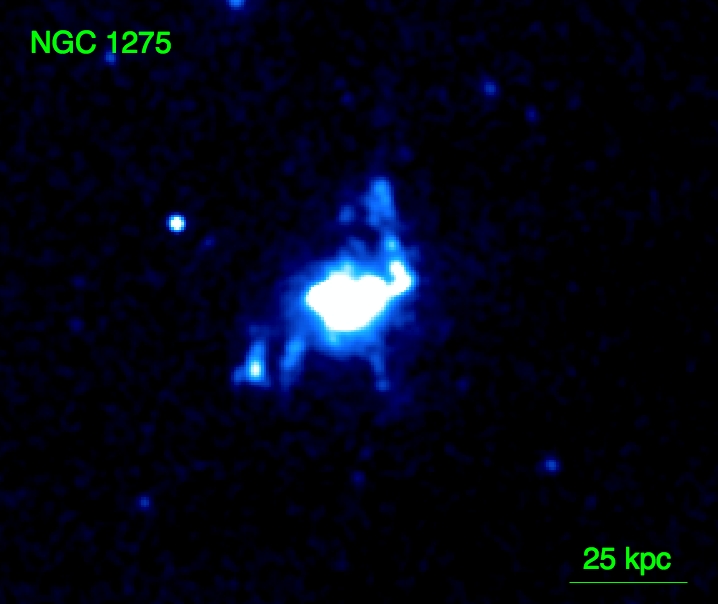}
  \end{minipage}%
  \begin{minipage}{0.33\textwidth}
    \centering
    \includegraphics[width=0.95\textwidth]{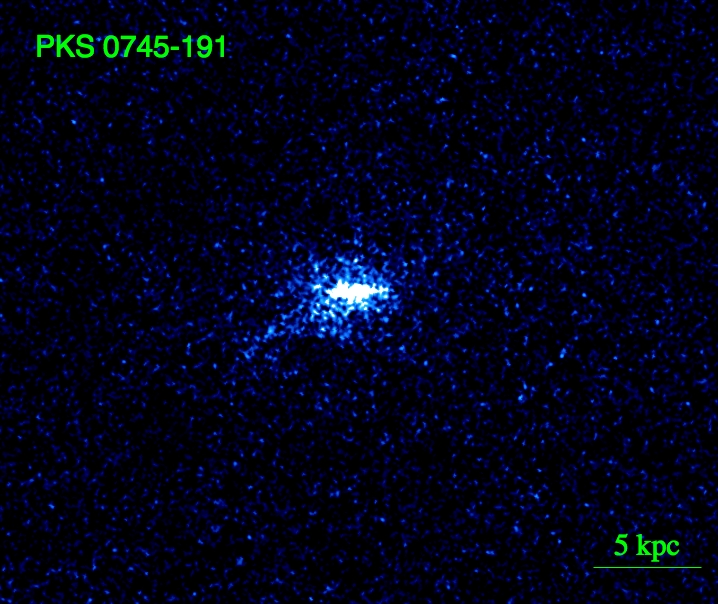}
  \end{minipage}%
  \begin{minipage}{0.33\textwidth}
    \centering
    \includegraphics[width=0.95\textwidth]{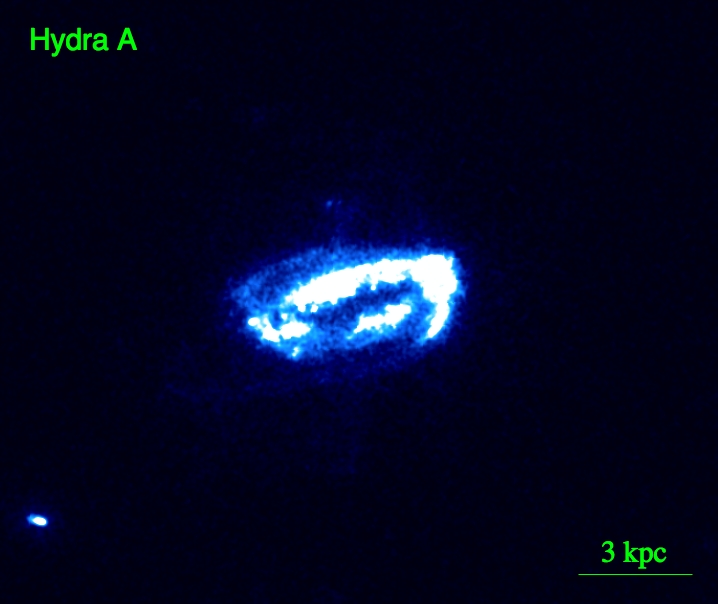}
  \end{minipage}\\[8pt]
  \begin{minipage}{0.33\textwidth}
    \centering
    \includegraphics[width=0.95\textwidth]{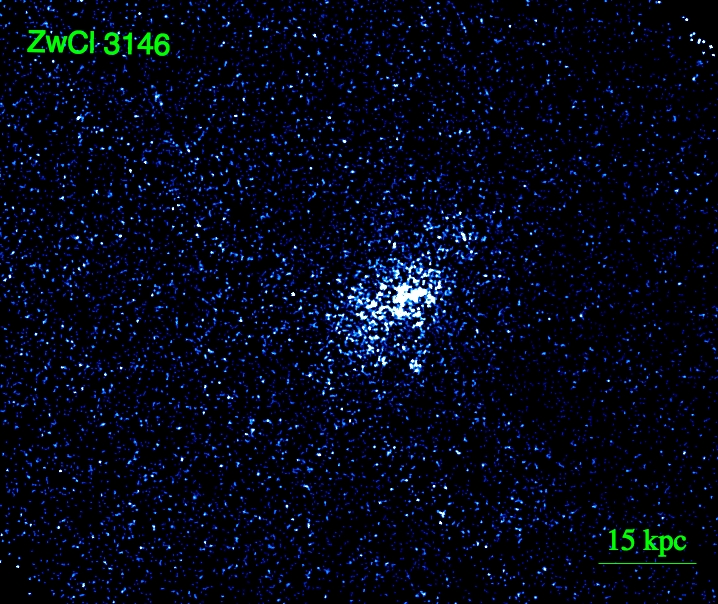}
  \end{minipage}%
  \begin{minipage}{0.33\textwidth}
    \centering
    \includegraphics[width=0.95\textwidth]{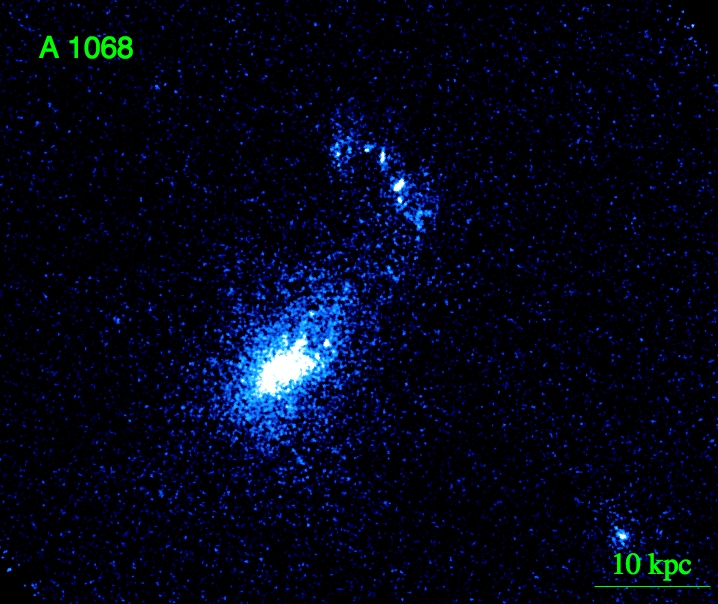}
  \end{minipage}%
  \begin{minipage}{0.33\textwidth}
    \centering
    \includegraphics[width=0.95\textwidth]{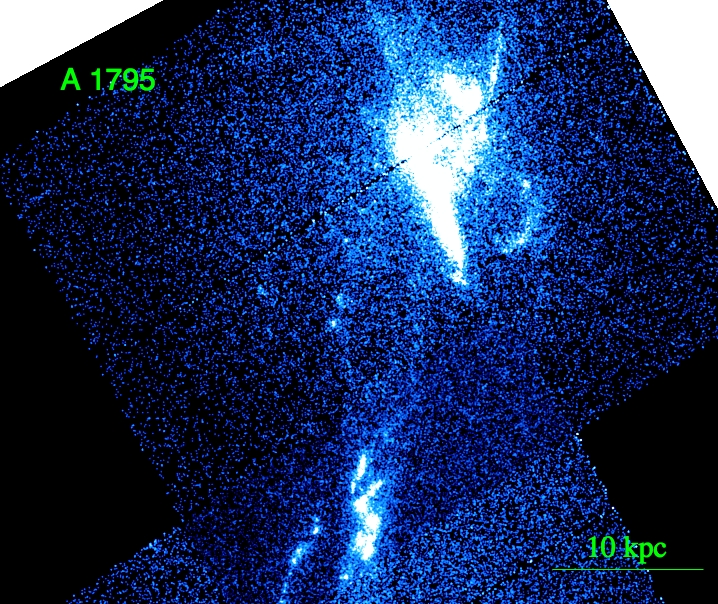}
  \end{minipage}\\[8pt]
  \begin{minipage}{0.33\textwidth}
    \centering
    \includegraphics[width=0.95\textwidth]{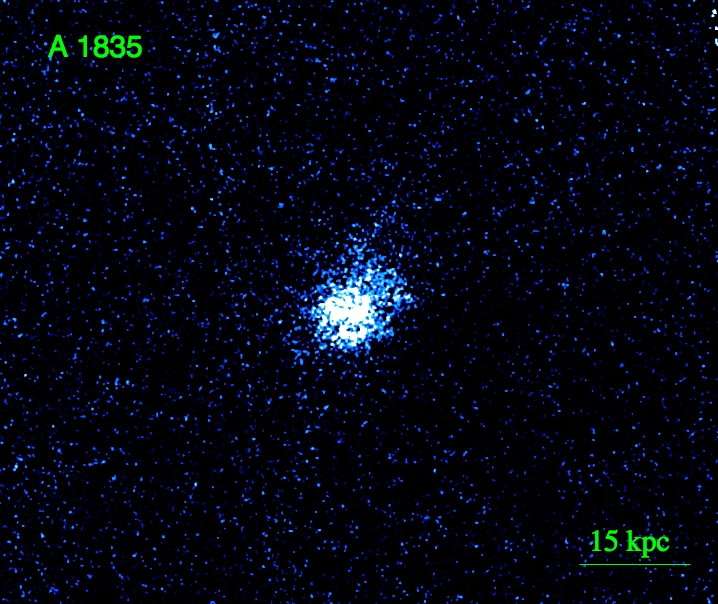}
  \end{minipage}%
  \begin{minipage}{0.33\textwidth}
    \centering
    \includegraphics[width=0.95\textwidth]{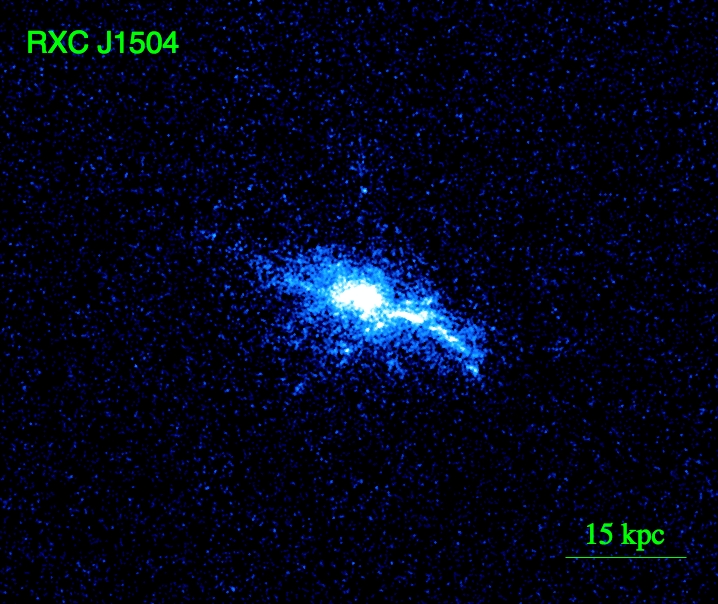}
  \end{minipage}%
  \begin{minipage}{0.33\textwidth}
    \centering
    \includegraphics[width=0.95\textwidth]{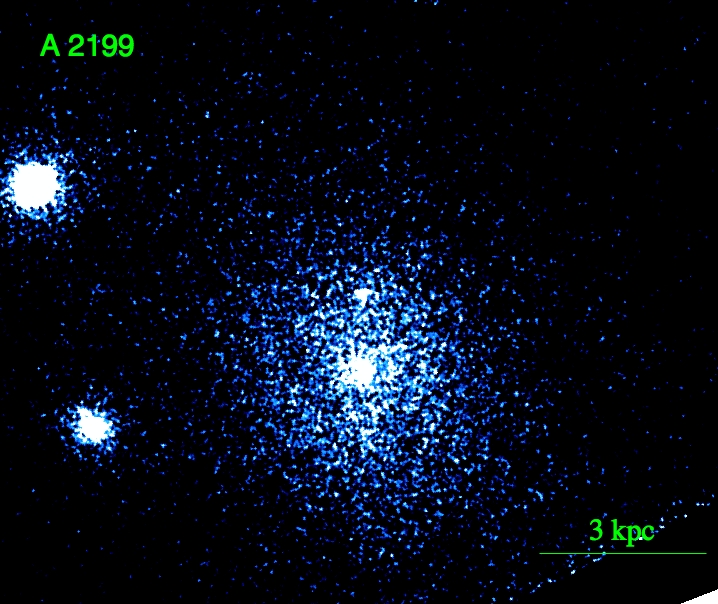}
  \end{minipage}\\[8pt]
  \begin{minipage}{0.33\textwidth}
    \centering
    \includegraphics[width=0.95\textwidth]{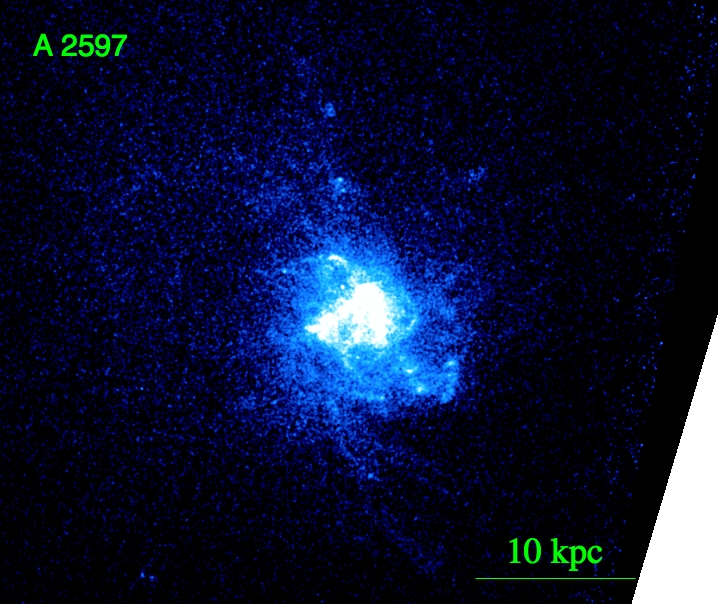}
  \end{minipage}%
  \caption{FUV images of the BCGs. NGC~1275 is from the {\it Galex}
    FUV detector and the others are from {\it HST} ACS/SBC
    FUV detector.}
  \label{FUV-images}
\end{figure*}

%% file: table7_2mass.tex
\begin{table*}
  \centering
  \caption{\small: 2MASS Photometry Details. The columns indicate (1) cluster name, (2) the mean aperture radius in arcsec and (3) the mean Galactic-extinction corrected flux-density in units of 10$^{-16}~$erg~s$^{-1}$~cm$^{-2}$~$\AA^{-1}$. For NGC~1275 the first row indicates the measurements with the AGN emission and the second row without.}
  \label{2mass}
  \begin{tabular}{| c | c | c | c | c | c | c | c }
  \hline

  Cluster	 & Aperture           & \multicolumn{3}{c}{Zero Point}   &  \multicolumn{3}{c}{Mean Flux Density}    \\
                 & (arcsec)    	      & J       & H       & K            &  J       & H        &  K       \\
  \hline \hline							         
  {\pers}        & 69		      & 20.9334 & 20.7121 & 20.0784      &  608.7   &  419.5   &  219.8   \\ 								         
                 & ...		      & ...     & ...     & ...          &  599.2   &  412.5   &  214.5   \\ 								         
  \hline	 						         
  {\pks}	 & 12	              & 20.7320 & 20.3621 & 19.8461      &  22.5    &  13.7    &  7.4     \\ 
  \hline	 	              				         
  {\hydra}	 & 15                 & 20.8561 & 20.4366 & 19.9282      &  42.0    &  30.4    &  15.2    \\   		 	              				         
  \hline	 	              				         
  {\zw	}	 & 4	              & 20.8114 & 20.4221 & 19.9039      &  1.9     &  1.6     &  0.8     \\ 			              				         
  \hline	 	              				         
  A~1068		 & 9$\times$4.5 (40)  & 21.0129 & 20.3648 & 20.0921      &  8.4     &  5.8     &  3.7     \\         	 	              				         
  \hline	 	              				         
  A~1795          & 22                 & 21.0392 & 20.3836 & 20.1139      &  39.0    &  28.8    &  18.0    \\                  	              				         
  \hline	 	              				         
  A~1835		 & 4.5                & 20.8215 & 20.3959 & 19.8729      &  4.0     &  2.7     &  2.0     \\         	 	              				         
  \hline	 	              				         
  {\rxj}	 & 6.5	              & 20.6326 & 20.3051 & 19.7951      &  4.3     &  3.6     &  2.1     \\     		 	              				         
  \hline	 	              				         
  A~2199		 & 40$\times$30 (300) & 20.9123 & 20.6709 & 20.0397      &  211.3   &  147.9   &  74.8    \\        	 	              				         
  \hline	 	              				         
  A~2597          & 12                 & 20.9755 & 20.4543 & 19.9394      &  15.0    &  10.6    &  6.4     \\
  \hline
   \end{tabular}
 \end{table*} 

%% file: table8_sdss.tex
\begin{table*}
  \centering
  \caption{\small: SDSS Photometry Details. The columns
  indicate (1) cluster name, (2) the mean aperture radius in arcsec,
  (3) the mean Galactic-extinction corrected flux-density in units
  of 10$^{-16}~$erg~s$^{-1}$~cm$^{-2}$~$\AA^{-1}$ and (4) notes (same as Table~\ref{hst}).}
  \label{sdss}
  \begin{tabular}{| c | c | c | c | c | c | c | c |}
  \hline

  Cluster	 & Aperture             &  \multicolumn{5}{c}{Mean Flux Density}              &   Notes   \\
                 & (arcsec)    	        &  U       & G        &  R      &  I       &  Z       &           \\
  \hline \hline			        				      		         	    
  {\pers}        & 75		        &  ...     &  ...     &  ...    &  907.6   &  688.7   &  w lines  \\ 								         
         	 &                      &  ...     &  ...     &  ...    &  890.4   &  675.1   &  wo lines \\ 			              				         
  \hline	 		        				      		         
  {\zw	}	 & 8$\times$4.5 (5)     &  1.74    &  ...     &  2.72   &  2.69    &  ...     &  w lines    \\ 			              				         
         	 &                      &  1.74    &  ...     &  2.68   &  2.69    &  ...     &  wo lines \\ 			              				         
  \hline	 	                				      		         
  A~1068		 & 12$\times$6 (42)     &  3.7     &  ...     &  ...    &  ...     &  12.8    &  ...      \\         	 	              				         
  \hline	 	                				      		         
  A~1795          & 26$\times$21 (110)   &  17.0    &  ..      &  67.2   &  66.8    &  60.3    &  w lines    \\                  	              				         
         	 &                      &  17.0    &  ...     &  66.7   &  64.8    &  60.3    &  wo lines \\ 			              				         
  \hline	 	                				      		         
  A~1835		 & 7$\times$5.5 (65)    &  3.61    &  ...     &  6.12   &  5.88    &  5.59    &  w lines    \\         	 	              				         
         	 &                      &  3.61    &  ...     &  6.09   &  5.71    &  5.59    &  wo lines \\ 			              				         
  \hline	 	                				      		         
  {\rxj}	 & 5	                &  6.06    &  ...     &  ...    &  ...     &  6.06    &  ...      \\     		 	              				         
  \hline	 	              				      			         
  A~2199		 & 47.5$\times$30 (300) &  ...     &  267.1   &  ...    &  336.4   &  318.4   &  ...      \\        	 	              				         
  \hline	 	              				      			         
   \end{tabular}
 \end{table*} 

%% file: table9_galex.tex
\begin{table}
  \centering
  \caption{\small: GALEX Photometry Details. The columns indicate (1) cluster name, (2) the exposure time, (3) the mean aperture radius in arcsec and (4) the mean Galactic-extinction corrected flux-density in units of 10$^{-16}~$erg~s$^{-1}$~cm$^{-2}$~$\AA^{-1}$. While PKS~0745-191 has no GALEX data, a nearby galaxy in A2199 is not resolved from the BCG, and A1068 only has NUV data available. For NGC~1275 the first row indicates the measurements with the AGN emission and the second row without.}
  \label{galex}
  \begin{tabular}{| c | c | c | c }
  \hline

  Cluster	 &   Aperture      & \multicolumn{2}{c}{Mean Flux Density}    \\
                 &   (arcsec) 	   & FUV      & NUV          \\
  \hline \hline	               		 	         
  {\pers}        &    97.5    	   &  618.6   &  587.8   \\
                 &    ...    	   &  447.1   &  458.1   \\
  \hline	                  
  {\hydra}	 &    16.5         &  23.4    &  20.6    \\
  \hline	    	           
  {\zw	}	 &    12.5	   &  12.4    &  3.5     \\
  \hline	    	           
  A~1068		 &    10           &  ...     &  3.2     \\
  \hline	    	           
  A~1795          &    30           &  26.5    &  17.5    \\
  \hline	    	           
  A~1835		 &    10.5         &  10.0    &  5.4     \\
  \hline	    	           
  {\rxj}	 &    13.5	   &  44.9    &  13.5    \\
  \hline	    	           
  A~2597          &    13.5         &  10.6    &  6.4     \\
  \hline
   \end{tabular}
 \end{table} 

%% file: table14_paramrange.tex
\begin{table*}
  \centering
   \caption{The range of values for the final set of simulations. The
    star formation history included either single or multiple
    bursts. For the single burst model, the age of the YSP is varied
    between 0.5~Myr and 6~Gyr and for the multiple burst model, the
    age of the most recent burst is varied between 100~Myr to 1~Gyr,
    with the time interval between bursts set to the age of the most
    recent burst.}

  \label{paramrange}
  \begin{tabular}{| l | l |}
  \hline
  Parameter				      & Final Values								 \\
  \hline \hline	      
  Old stellar population age, $\tosp$         & 11~Gyr (10~Gyr for ZwCl~3146 and 10.5~Gyr for A~1835)			 \\ 
  Young stellar population age, $\tysp$       & 10~Myr to 6~Gyr    	       	   	    				 \\
  Initial Mass Function, IMF                  & Chabrier    	       							 \\
  Metallicty      	 		      & 0.4$\zs$, $\zs$, 2.5$\zs$ ($\zs-0.4\zs$ for NGC~1275 and PKS~0745-191)     \\
  Extinction laws 			      & Galactic and extragalactic                  	       	   		 \\
  Reddening, E(B-V)          		      & 0 - 0.6	       							         \\
  Star Formation History, SFH                 & Instantaneous bursts	       						 \\
  \hline
   \end{tabular}
 \end{table*}

%% file: table5_extlaw.tex
\begin{table}
  \centering
  \caption{The posterior mass probablities for the two adopted extinction laws for the final simulations results.}
  \label{extlaw}
  \begin{tabular}{| l | l | l |}
  \hline
  Cluster	 & \multicolumn{2}{c}{Ext-Laws} \\
  		 & gal1 & xgal			\\
  \hline \hline
  {\pers}        & 0.27	     & 0.73     \\
  {\pks}	 & 0.70	     & 0.30	\\
  {\hydra}	 & 0.49	     & 0.51	\\
  {\zw}		 & 0.51	     & 0.49	\\
  A~1068	 & 0.65	     & 0.35	\\
  A~1795         & 0.00      & 1.00	\\
  A~1835	 & 0.34	     & 0.66	\\
  {\rxj}	 & 0.53	     & 0.47	\\
  A~2199	 & 0.51	     & 0.49	\\
  A~2597         & 0.62      & 0.38     \\ 
  \hline
  \end{tabular}
\end{table}

%% file: table10_hstfuv.tex
\begin{table}
  \centering
  \caption{\small: A comparison of star formation rates (defined in Section~\ref{SFRresults}) 
    with the HST ACS/SBC FUV data point included (third column) and excluded (fourth column).}
  \label{hstfuv}
  \begin{tabular}{| c | c | c | }
  \hline

  Cluster    &   \multicolumn{2}{c}{SFR}    \\
             &  FUV incl.            & FUV excl.	\\
  \hline \hline	               		 	         
  {\hydra}   &  $18^{+15}_{-12}$     & $24^{+21}_{-15}$ \\
  \hline	                      
  {\zw}      &  $14^{+25}_{- 2}$     & $37^{+20}_{-22}$ \\
  \hline	                      
  A~1795     &  $21^{+ 8}_{- 10}$     & $20^{+ 9}_{- 10}$ \\
  \hline	                      
  A~1835     &  $54^{+19}_{-15}$     & $40^{+14}_{-15}$  \\
  \hline	                      
  {\rxj}     &  $67^{+49}_{-36}$     & $66^{+49}_{-37}$ \\
  \hline	                      
  A~2597     &   $12^{+11}_{- 9}$    & $15^{+13}_{-10}$ \\
  \hline
   \end{tabular}
 \end{table} 

%% file: table15_lines.tex
\begin{table*}
  \centering
  \caption{\small: The prominent emission lines in the HST and SDSS bands. Given in column 2 are the bandpasses,
   column 3 are the emission lines and column 3 are the references used to correct for the lines.} 
  \label{lineref}
  \begin{tabular}{| p{1cm} | p{4cm} | p {9cm} | p{3cm} | }
  \hline

  Cluster	 &   Bands             &  Lines          & References    \\
  \hline \hline	               	   	      
  PKS~0745	 &   HST/WFPC2 F814W   & $\ha$, {\nii}   & \cite{Fabian1985,Heckman1989}\\ 
  \hline	     	      
  {\zw	}	 &   HST/WFPC2 F606W   & {\oii}, {\niia}, $\hb$, {\oiii}   & \cite{Crawford1999}   \\
  		 &   SDSS R            & {\niia}, $\hb$, {\oiii}   	     	     & \cite{Crawford1999}   \\
  \hline	       
  A~1795         &   HST/WFPC2 F555W, SDSS R  & $\hb$, {\oiii}, {\niia}  & \cite{Crawford1999}   \\
  		 &   HST/WFPC2 F702W, SDSS I  & $\ha$, {\nii}              & \cite{Heckman1989,Crawford1999} \\
  \hline	      
  A~1835	 &   HST/WFPC2 F702W & $\hb$, {\oiii}, {\niia}, $\ha$, {\nii}, {\siia}, {\siib}, {\oia}, {\oib}  & \cite{Crawford1999}   \\
  		 &   SDSS R  & $\hb$, {\oiii}   & \cite{Crawford1999} \\
  		 &   SDSS I  & $\ha$, {\nii}    & \cite{Crawford1999} \\
  \hline	      
  A~2597         &   HST/WFPC2 F450W  & {\oii}, [Ne {\sc iii}]~$\lambda3869$, [H$\eta$], [H$\epsilon$] + [Ne {\sc iii}]~$\lambda3966$, [S{\sc ii}]~$\lambda4069$, [H$\delta$], [H$\gamma$], [He {\sc ii}]~$\lambda4686$  & \cite{Voit1997,Heckman1989}   \\
  		 &   HST/WFPC2 F702W & {\oia}, {\oib}, {\nii}, $\ha$, [He {\sc i}]~$\lambda6678$, {\siia}, {\siib}, [Ca {\sc ii}]~$\lambda7290$, [O{\sc ii}] + [Ca {\sc ii}]~$\lambda7320$   & \cite{Voit1997,Heckman1989} \\  
  \hline
   \end{tabular}
 \end{table*} 


%% file: table16_PI68vsFWHM.tex
\begin{table}
  \centering
  \caption{The posterior mass probablities for the two adopted extinction laws for the final simulations results.}
  \label{PI68vsFWHM}
  \begin{tabular}{| l | l | l | l | l |}
  \hline
   Cluster	 & \multicolumn{2}{l}{SFR~($\mpy$)~68~\%~PI}   & \multicolumn{2}{l}{SFR~($\mpy$)~FWHM}\\
   		 & Median & 68\% PI interval                  & Median & FWHM interval             \\
   \hline \hline
   {\pers}       & 71  &  18 $-$  207    		      & 51  &  24 $-$  90 \\
   {\pks}	 & 24  &   3 $-$  64 	 		      & 20  &   4 $-$  42 \\
   {\hydra}	 & 18  &   6 $-$  34 	 		      & 18  &   7 $-$  30 \\
   {\zw}	 & 14  &  13 $-$  40 	 		      & 14  &  13 $-$  15 \\
   A~1068	 &  8  &   2 $-$  21 	 		      &  6  &   2 $-$  11 \\
   A~1795        & 21  &  11 $-$  28 	 		      & 21  &  11 $-$  27 \\
   A~1835	 & 54  &  39 $-$  73 	 		      & 54  &  38 $-$  74 \\
   {\rxj}	 & 67  &  31 $-$  116	 		      & 68  &  31 $-$  117\\
   A~2199	 & 30  &   3 $-$  57 	 		      & 34  &   0 $-$  72 \\
   A~2597        & 12  &   4 $-$  23     		      & 12  &   5 $-$  21 \\ 
  \hline
  \end{tabular}
\end{table}

%% file: table12_ngcpks.tex
\begin{table*}
  \centering
  \caption{Impact on most likely values and plausible intervals of
  fixing parameters rather than marginalizing. For all three examples,
  the reddening E(B-V) has been fixed to their best-fit values and we
  compare the posterior distributions for other parameters to those
  where we marginalize freely over all extinction values with equal
  prior weights. Quantities listed are defined in the caption of
  table~\ref{spprop}.}
  \label{ngcpks}
  \begin{tabular}{| l | l | l | l | l | l | l |}
  \hline
  Parameter	      &  \multicolumn{2}{c}{NGC~1275}	                       &  \multicolumn{2}{c}{PKS~0745}		                &  \multicolumn{2}{c}{A~2597} 			          \\
		      & E(B-V) free & E(B-V) fixed to 0.12                     &  E(B-V) free & E(B-V) fixed to 0			&  E(B-V) free & E(B-V) fixed to 0.18 			  \\
  \hline \hline	       			      			                                                			                                			  \\		
  Mo ($10^{11}~\ms$)  &	1.8 $-$  4.6 	&   1.6 $-$  4.5                       &  5.0 $-$  8.7     & 6.4 $-$  8.8                       &  2.4 $-$  5.0      & 2.2 $-$  5.2     \\[2pt]
  My ($10^{10}~\ms$)  &	6.3 $-$  16.1 	&   7.3 $-$  15.3                      &  1.3 $-$  11.8    & 0.8 $-$  6.9                       &  1.1 $-$  6.4      & 2.7 $-$  7.6     \\[2pt]
  My/Mo		      &	0.06 $-$ 0.54   &   0.06 $-$  0.52                     &  0.01 $-$  0.21   & 0.01 $-$  0.10                     &  0.01 $-$  0.19    & 0.02 $-$  0.21   \\[2pt]
  Ty (Myr)	      &	110 $-$ 660	&   80 $-$  2160                       &  40 $-$ 6000      & 85  $-$ 5950                       &  20 $-$ 6000       & 20 $-$ 6000      \\[2pt]
  SFR ($\mpy$)        & 18 $-$  207     &   27 $-$  72                         &  3  $-$  64       & 4  $-$  23                         &  4 $-$  23         & 9  $-$  19       \\[2pt]
  E(B-V)	      & 0.03 $-$  0.21  &   ...                                &  0.00 $-$  0.09   & ...      				&  0.03 $-$  0.21    & ... 		\\
  \hline
  \end{tabular}
\end{table*}

%% file: table11_metalZXray.tex
\begin{table}
  \centering
  \caption{\small: The metallicity (within the inner few tens of kpc of 
  the BCGs) inferred from X-ray spectral analysis using either 
  {\it XMM-Newton} or {\it Chandra} data. A range of metallcities, where
  given, corresponds to different models or instruments used, yielding a 
  spread in the average value. (Note PKS~0745 = PKS~0745-191)} 
  \label{metalZXray}
  \begin{tabular}{| l | l | l | }
  \hline

  Cluster	 &   $Z$  & References    \\
  \hline \hline	               	   	      
  {\pers}        &    0.6-0.7 \comment{25 kpc}  & \cite{Sanders2007}   \\
  \hline	                          
  PKS~0745	 &    0.4-0.7 \comment{15 kpc}  & \cite{Chen2003,Hicks2002}\\ 
  \hline	    	                  
  {\hydra}	 &    0.3-0.6  \comment{25 kpc} & \cite{Kirkpatrick2009,Simionescu2009}\\
  \hline	    	                  
  {\zw	}	 &    0.35-0.4 \comment{45 kpc} & \cite{Kausch2007}   \\
  \hline	    	                    
  A~1068		 &    0.7-1   \comment{45 kpc} & \cite{Wise2004}   \\
  \hline	    	                    
  A~1795          &    0.4-0.8 \comment{30 kpc} & \cite{Ettori2002,Gu2012}   \\
  \hline	    	                    
  A~1835		 &    0.3-0.4 \comment{30 kpc} & \cite{Schmidt2001,Majerowicz2002}   \\
  \hline	    	                    
  {\rxj}	 &    0.3 \comment{50 kpc} & \cite{Zhang2012}   \\
  \hline	    	                    
  A~2199	         &    0.3-0.7 \comment{45 kpc}   & \cite{Johnstone2002}   \\
  \hline	    	                    
  A~2597          &    0.5 \comment{30 kpc}  & \cite{Morris2005}   \\
  \hline
   \end{tabular}
 \end{table} 


%% file: table6_spprop.tex
\setlength\LTleft{-2.5in}
\setlength\LTright\fill
\begin{landscape}
  \begin{longtable}{| l | l | l | l | l | l | l | l | l | l | l | l | l |}
  \caption{The most likely physical parameters of the two stellar populations of the BCG sample.}
  \label{spprop}\\
  \hline
  Cluster	 &  \multicolumn{2}{l|}{$\mosp$~($10^{11}~\ms$)}      &   \multicolumn{2}{l|}{$\mysp$~($10^{10}\ms$)}  &   \multicolumn{2}{l|}{$\mysp/\mosp$}  & \multicolumn{2}{l|}{$\tysp$~(Myr)} & \multicolumn{2}{l|}{E(B-V)} & \multicolumn{2}{l|}{SFR~($\mpy$)} \\
  \hline												                                              
  		 &  Median  &  68\% PI                          &   Median & 68\% PI                       &   Median & 68\% PI           & Youngest & Oldest                 & Median & 68\% PI	       & Median & 68\% PI                        \\
  \hline \hline	 											                                              
  \endfirsthead
  \caption{continued.}\\
  \hline	
  Cluster	 &  \multicolumn{2}{l}{$\mosp$~($10^{11}~\ms$)}      &   \multicolumn{2}{l}{$\mysp$~($10^{10}\ms$)}  &   \multicolumn{2}{l}{$\mysp/\mosp$}  & \multicolumn{2}{l}{$\tysp$~(Myr)} & \multicolumn{2}{l}{E(B-V)} & \multicolumn{2}{l}{SFR~($\mpy$)} \\
  \hline												                                  
  		 &  Median  &  68\% PI                          &   Median & 68\% PI                       &   Median & 68\% PI           &	Youngest & Oldest                 & Median & 68\% PI	       & Median & 68\% PI                        \\
  \hline
  \endhead
  \hline
  \endfoot
  NGC~1275       &    3.2  &  1.8 $-$  4.6 			& 11.2 &  6.3  $-$  16.1                   &   0.27  &  0.06 $-$  0.54    &   110    & 660         & 0.14  &  0.03 $-$  0.21     &  71  &  18 $-$  207      \\[2pt]
  PKS~0745-191	 &    7.0  &  5.0 $-$  8.7    			& 6.0  &  1.3  $-$  11.8     		   &   0.08  &  0.01 $-$  0.21    &    40    & 6000        & 0.05  &  0.00 $-$  0.09     &  24  &   3 $-$  64       \\[2pt]
  Hydra-A	 &    3.6  &  2.3 $-$  4.8    			& 6.2  &  2.4  $-$  10.2     		   &   0.14  &  0.02 $-$  0.33    &   180    & 5940        & 0.06  &  0.00 $-$  0.09     &  18  &   6 $-$  34       \\[2pt]
  ZwCl~3146	 &    4.7  &  4.1 $-$  5.3    			& 6.6  &  4.7  $-$  8.2      		   &   0.13  &  0.08 $-$  0.18    &    10    & 6000        & 0.00  &  0.00 $-$  0.00     &  14  &  13 $-$  40       \\[2pt]
  A~1068	 &    6.9  &  5.3 $-$  8.3    			& 2.9  &  0.4  $-$  7.1      		   &   0.04  &  0.00 $-$  0.12    &    80    & 6000        & 0.05  &  0.00 $-$  0.09     &   8  &   2 $-$  21       \\[2pt]
  A~1795         &    8.4  &  7.4 $-$  9.4    			& 6.1  &  1.9  $-$  10.7     		   &   0.06  &  0.01 $-$  0.12    &    10    & 5860        & 0.21  &  0.15 $-$  0.21     &  21  &  11 $-$  28       \\[2pt]
  A~1835	 &    5.7  &  4.0 $-$  7.6    			& 19.7 &  10.8 $-$  28.1     		   &   0.28  &  0.07 $-$  0.51    &    10    & 3560        & 0.15  &  0.09 $-$  0.15     &  54  &  39 $-$  73       \\[2pt]
  RXC~J1504	 &    7.5  &  5.6 $-$  9.3    			& 2.4  &  0.1  $-$  7.9      		   &   0.03  &  0.00 $-$  0.13    &    10    & 10          & 0.00  &  0.00 $-$  0.00     &  67  &  31 $-$  116      \\[2pt]
  A~2199	 &    6.2  &  3.6 $-$  9.7    			& 9.3  &  0.1  $-$  18.8     		   &   0.12  &  0.00 $-$  0.37    &  4130    & 4370        & 0.09  &  0.03 $-$  0.09     &  \ldots &   \ldots	   	    \\[2pt]
  A~2597         &    3.8  &  2.4 $-$  5.0    			& 3.5  &  1.1  $-$  6.4      		   &   0.08  &  0.01 $-$  0.19    &    20    & 6000        & 0.14  &  0.03 $-$  0.21     &  12  &   4 $-$  23       \\		
  \hline
  \multicolumn{13}{p{20cm}}{The most likely physical parameters of the two stellar populations of the BCG sample, 
    defined as the modes of the marginal posterior distributions (shown in figures~\ref{metalZ}--
    \ref{SFR}) for each of the quantities (probability density functions for $\tysp$, $\mosp$, and 
    $\mysp$ and probability mass function for metallicity and extinction).
    The columns are: (1) cluster, (2) mass of the old stellar population (peak and the 68\% plausible interval), 
    (3) the total mass of the young stellar population (peak and the 68\% plausible interval),
    (4) the ratio of the total mass in the YSP to that in the OSP, (5) the youngest and the oldest YSP age (6) the internal 
    extinction and (7) the star formation rate (peak and the 68\% plausible interval); defined as the
    ratio of mass of a single burst to the age of the most recent burst. ``PI 68\%" corresponds to  
    to the endpoints of the narrowest 68\% plausible interval constructed from the marginal posterior 
    (which is shown shaded on the corresponding posterior plot). The SED for A~2199 does not seem
    to require a YSP at all. Therefore there is no ``ongoing'' star formation rate in A~2199. The best-fit model for 
    RXC~J1504 comprises a single YSP burst.}\\
 \hline
 \end{longtable}
\end{landscape}

%% file: posteriors.tex

\begin{figure*}
  \begin{minipage}{0.33\textwidth}
    \centering
    \includegraphics[width=\textwidth]{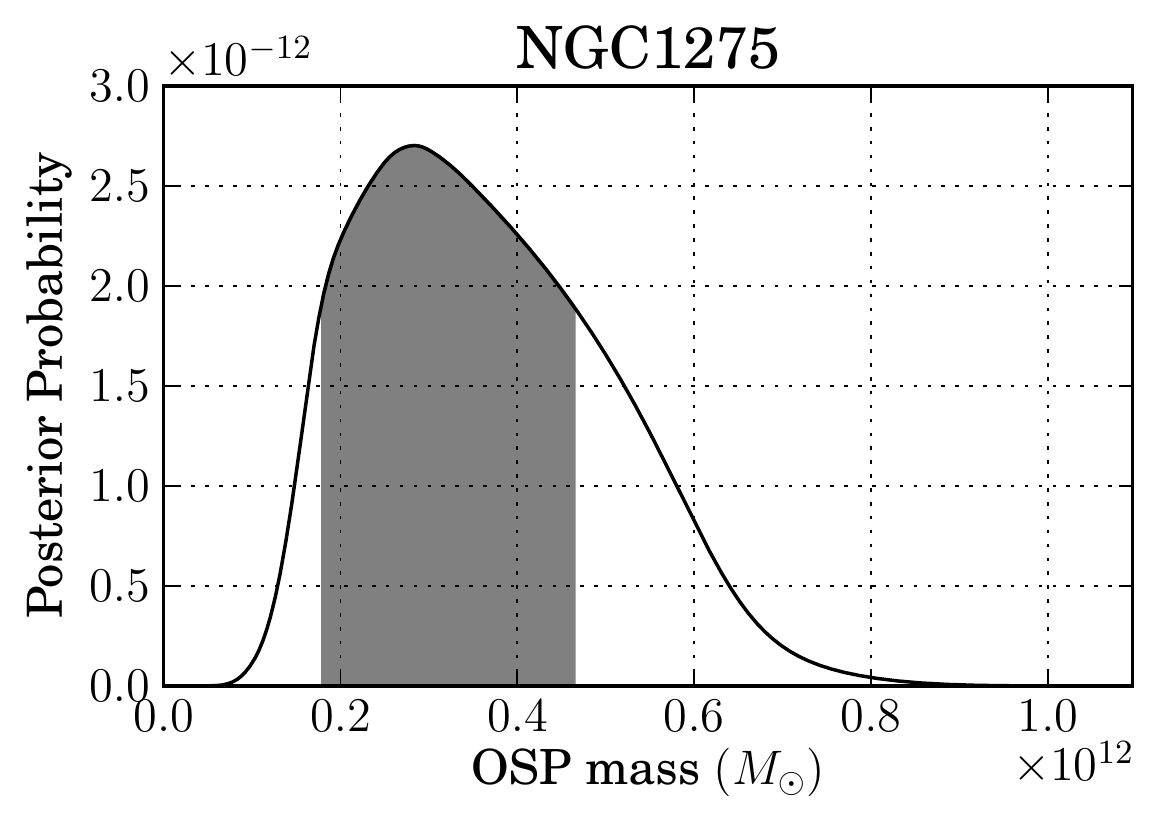}
  \end{minipage}%
  \begin{minipage}{0.33\textwidth}
    \centering
    \includegraphics[width=\textwidth]{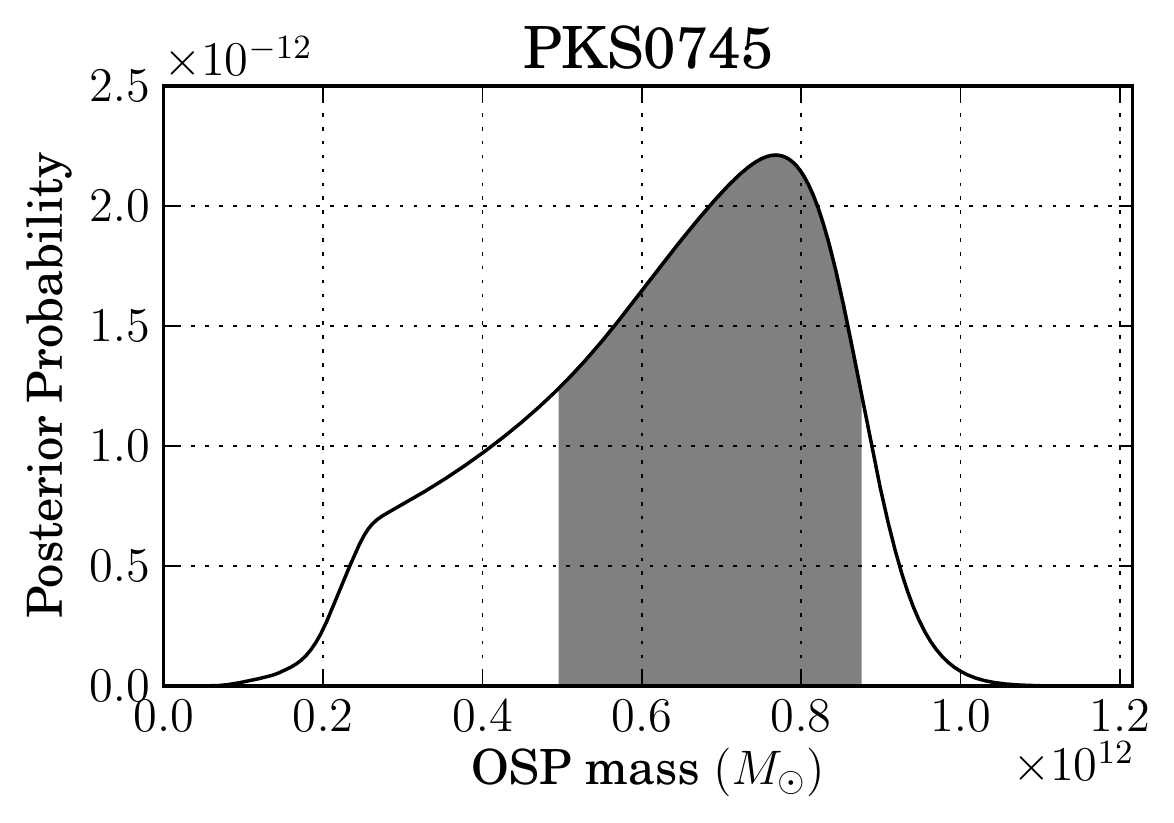}
  \end{minipage}%
  \begin{minipage}{0.33\textwidth}
    \centering
    \includegraphics[width=\textwidth]{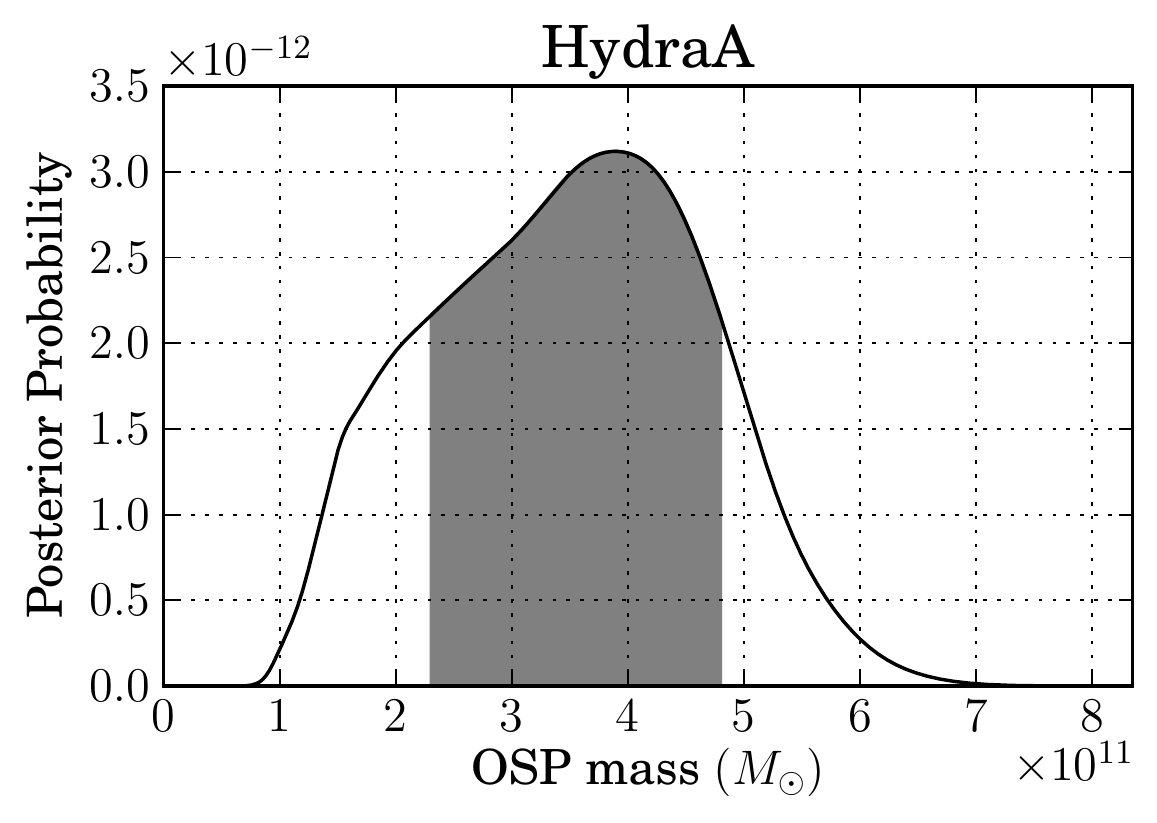}
  \end{minipage}\\
  \begin{minipage}{0.33\textwidth}
    \centering
    \includegraphics[width=\textwidth]{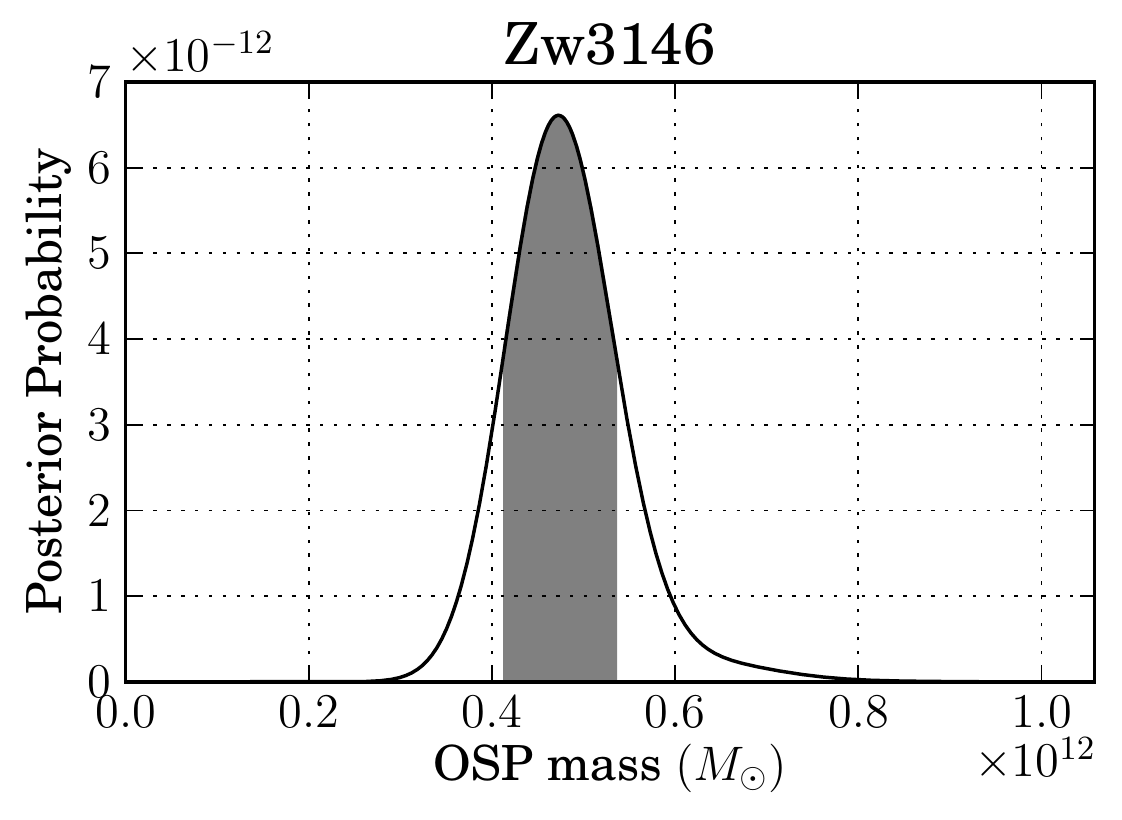}
  \end{minipage}%
  \begin{minipage}{0.33\textwidth}
    \centering
    \includegraphics[width=\textwidth]{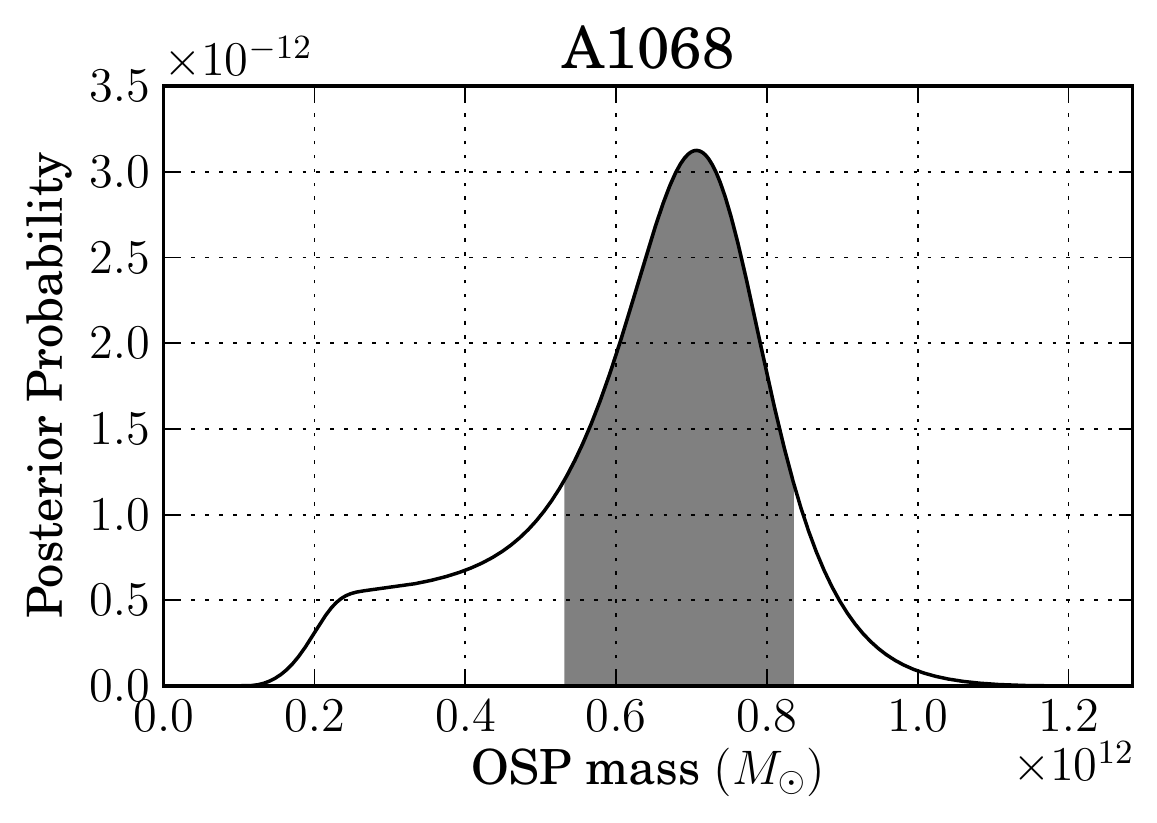}
  \end{minipage}%
  \begin{minipage}{0.33\textwidth}
    \centering
    \includegraphics[width=\textwidth]{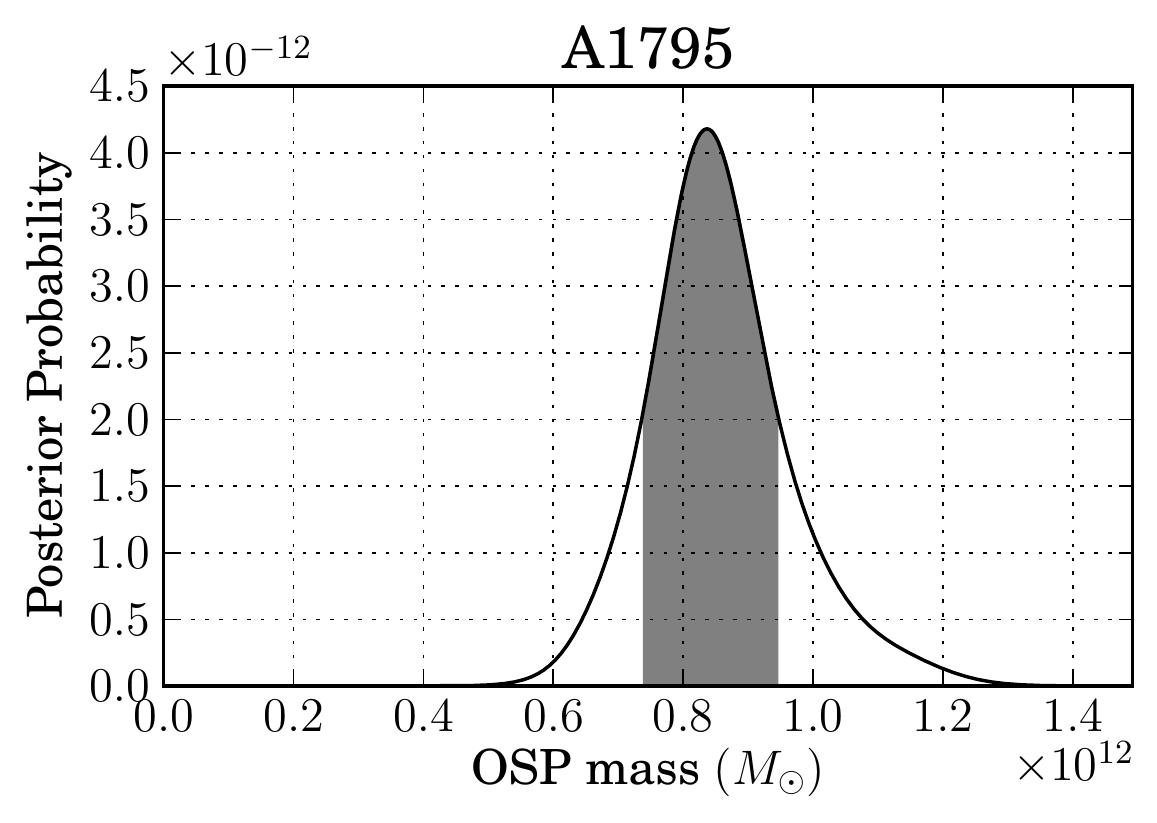}
  \end{minipage}\\
  \begin{minipage}{0.33\textwidth}
    \centering
    \includegraphics[width=\textwidth]{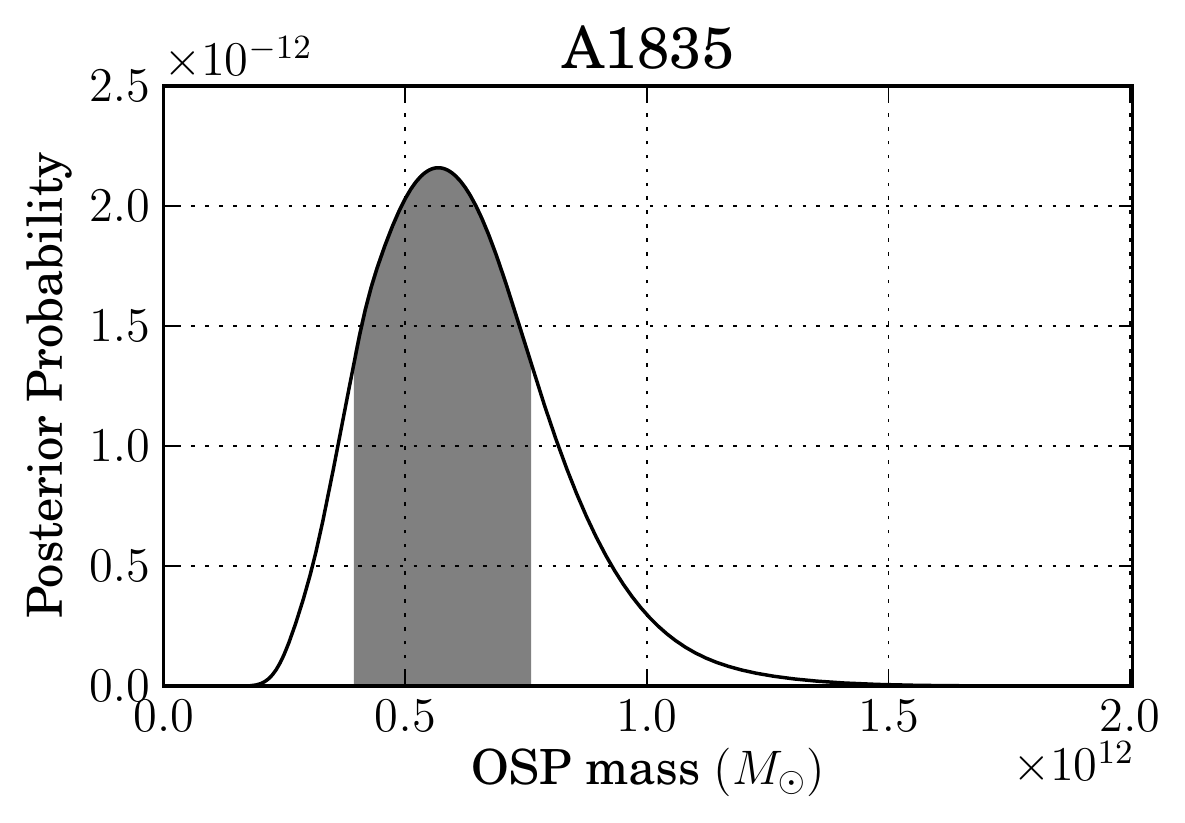}
  \end{minipage}%
  \begin{minipage}{0.33\textwidth}
    \centering
    \includegraphics[width=\textwidth]{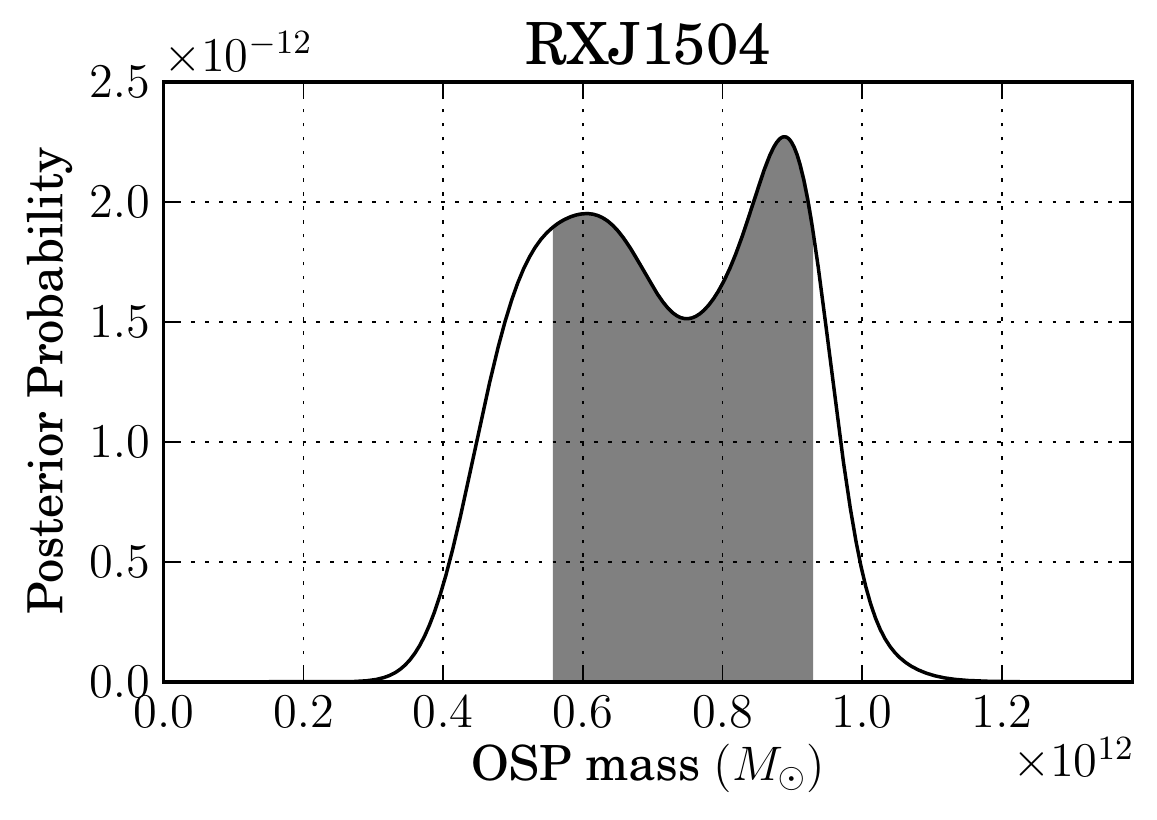}
  \end{minipage}%
  \begin{minipage}{0.33\textwidth}
    \centering
    \includegraphics[width=\textwidth]{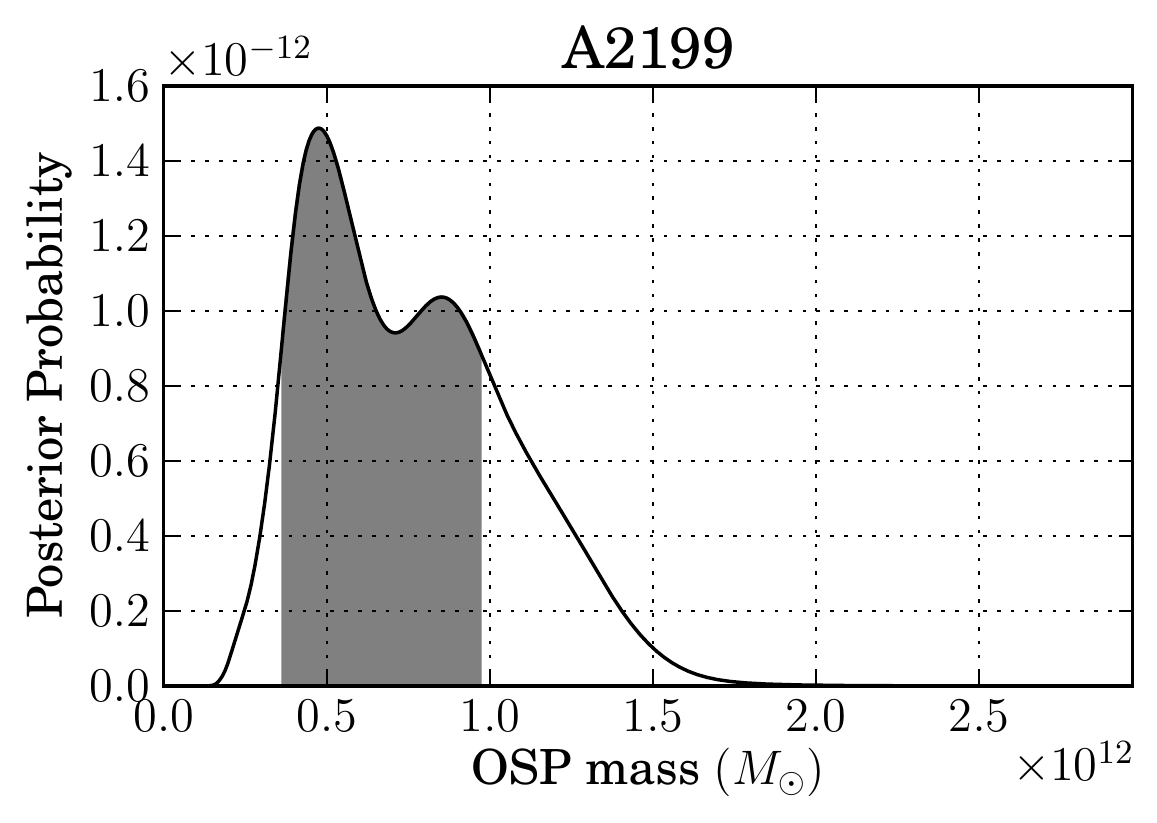}
  \end{minipage}\\
  \begin{minipage}{0.33\textwidth}
    \centering
    \includegraphics[width=\textwidth]{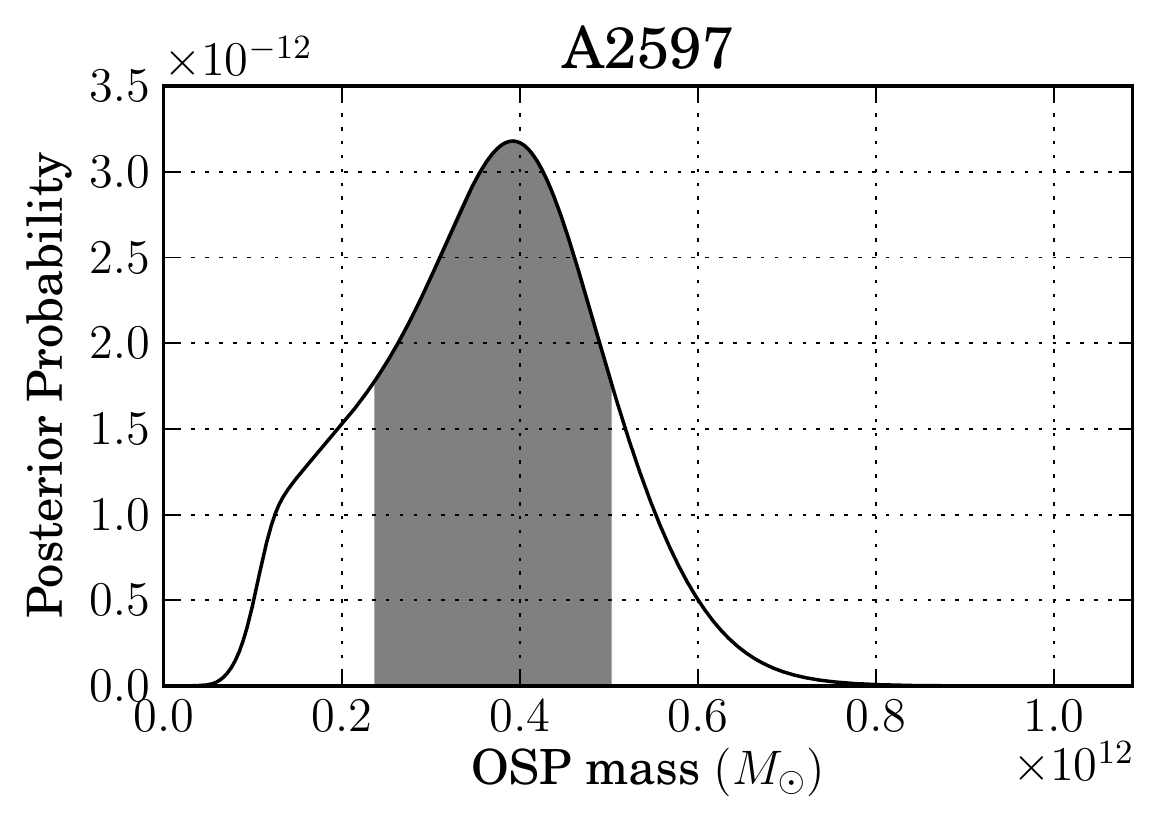}
  \end{minipage}%
  \caption{Posterior probability density functions for the mass of the
    old stellar popluation, $\mosp$, after marginalizing over other
    model parametrs (metallicity, extinction, YSP age and YSP mass).
    The joint prior PDF used for $\mosp$ and $\mysp$ was uniform with
    the constraint that $\mosp>\mysp$.  Since the best fit YSP masses
    for most models were much lower than the corresponding OSP masses,
    this can be thought of effectively as a uniform prior on $\mosp$.
    The shading indicates the narrowest 68\% plausible interval for
    $\mosp$.}
  \label{Mo}
\end{figure*}


\begin{figure*}
  \begin{minipage}{0.33\textwidth}
    \centering
    \includegraphics[width=\textwidth]{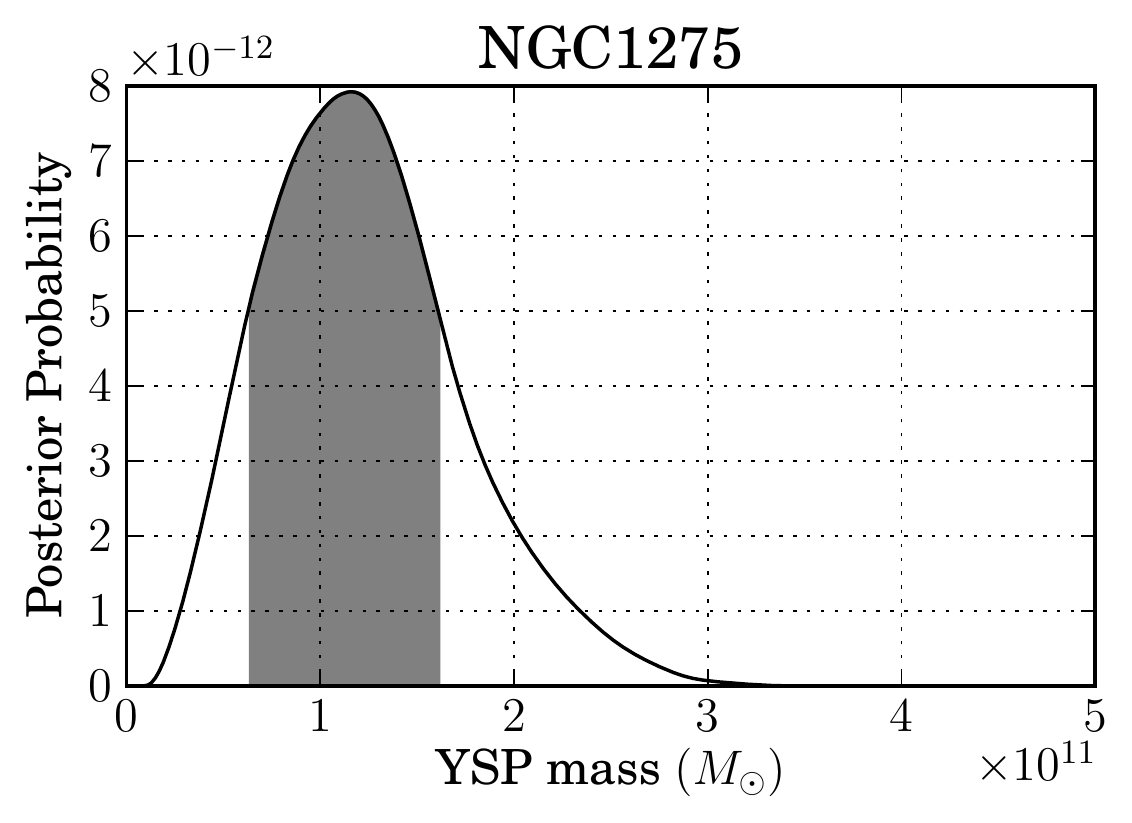}
  \end{minipage}%
  \begin{minipage}{0.33\textwidth}
    \centering
    \includegraphics[width=\textwidth]{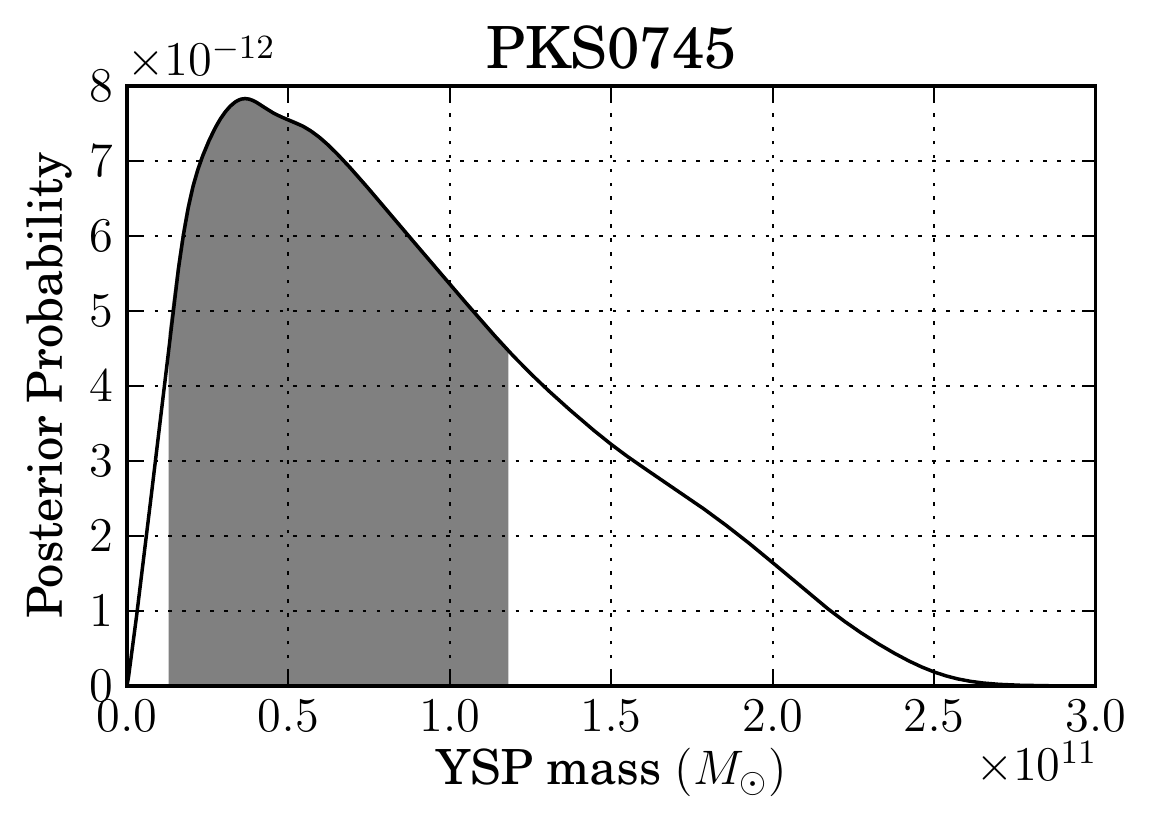}
  \end{minipage}%
  \begin{minipage}{0.33\textwidth}
    \centering
    \includegraphics[width=\textwidth]{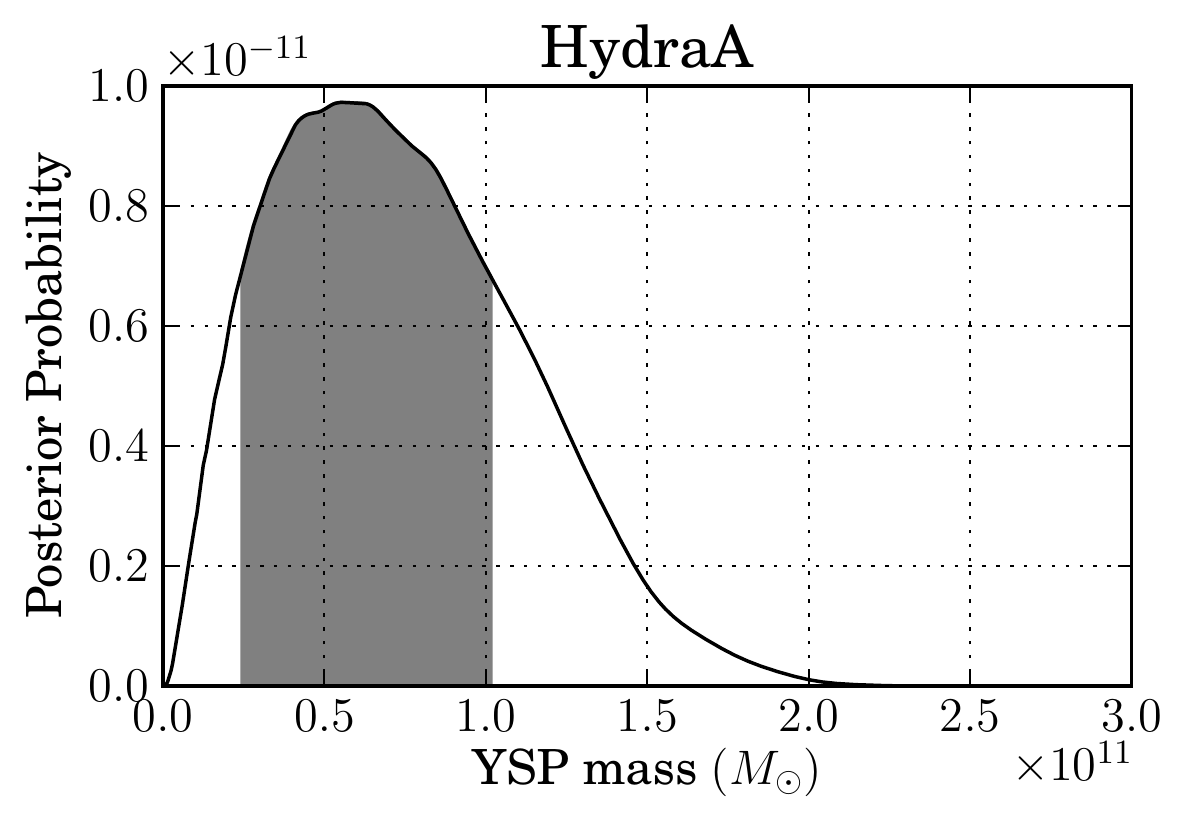}
  \end{minipage}\\
  \begin{minipage}{0.33\textwidth}
    \centering
    \includegraphics[width=\textwidth]{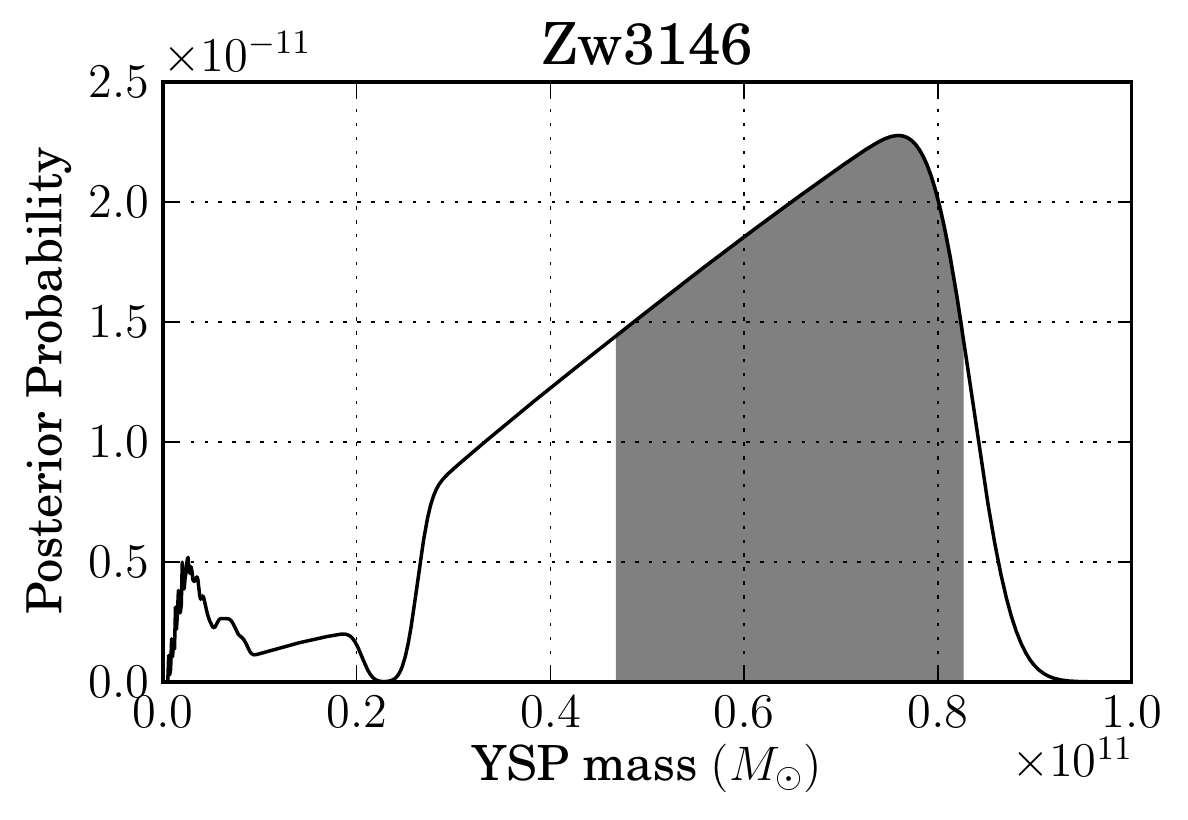}
  \end{minipage}%
  \begin{minipage}{0.33\textwidth}
    \centering
    \includegraphics[width=\textwidth]{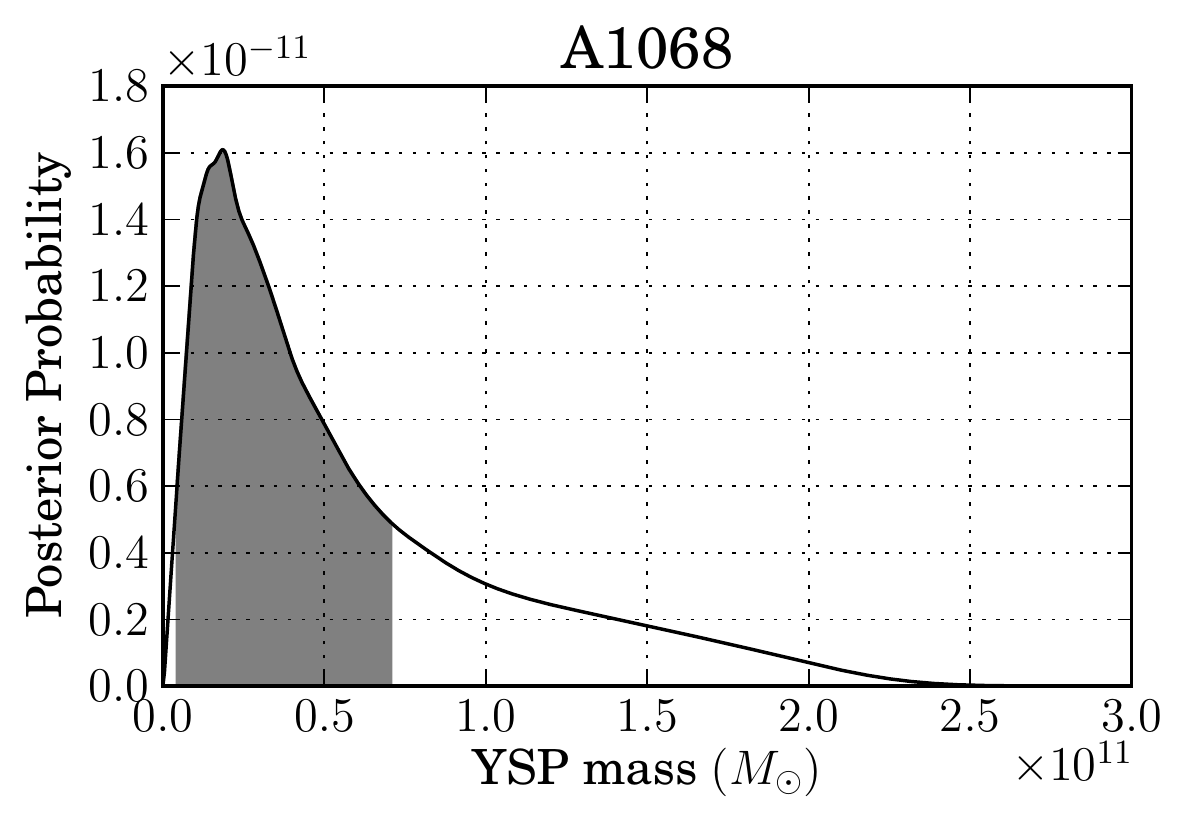}
  \end{minipage}%
  \begin{minipage}{0.33\textwidth}
    \centering
    \includegraphics[width=\textwidth]{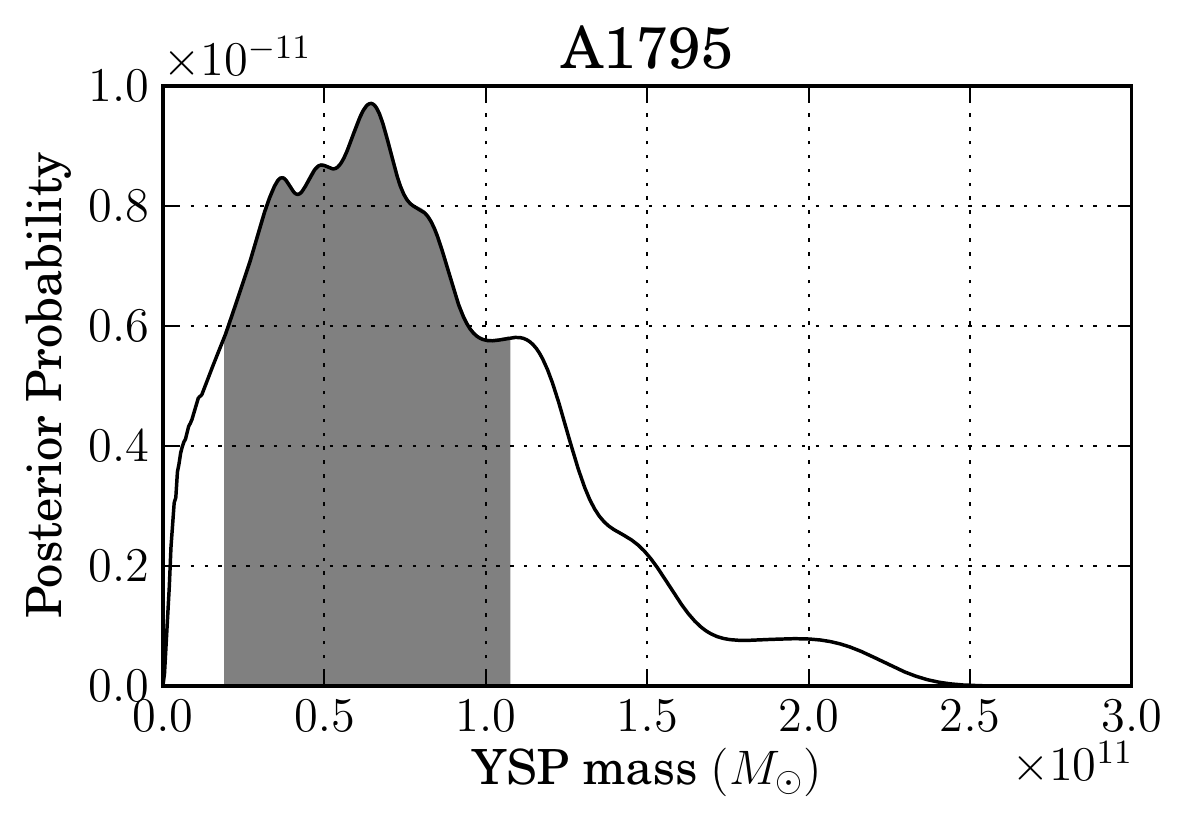}
  \end{minipage}\\
  \begin{minipage}{0.33\textwidth}
    \centering
    \includegraphics[width=\textwidth]{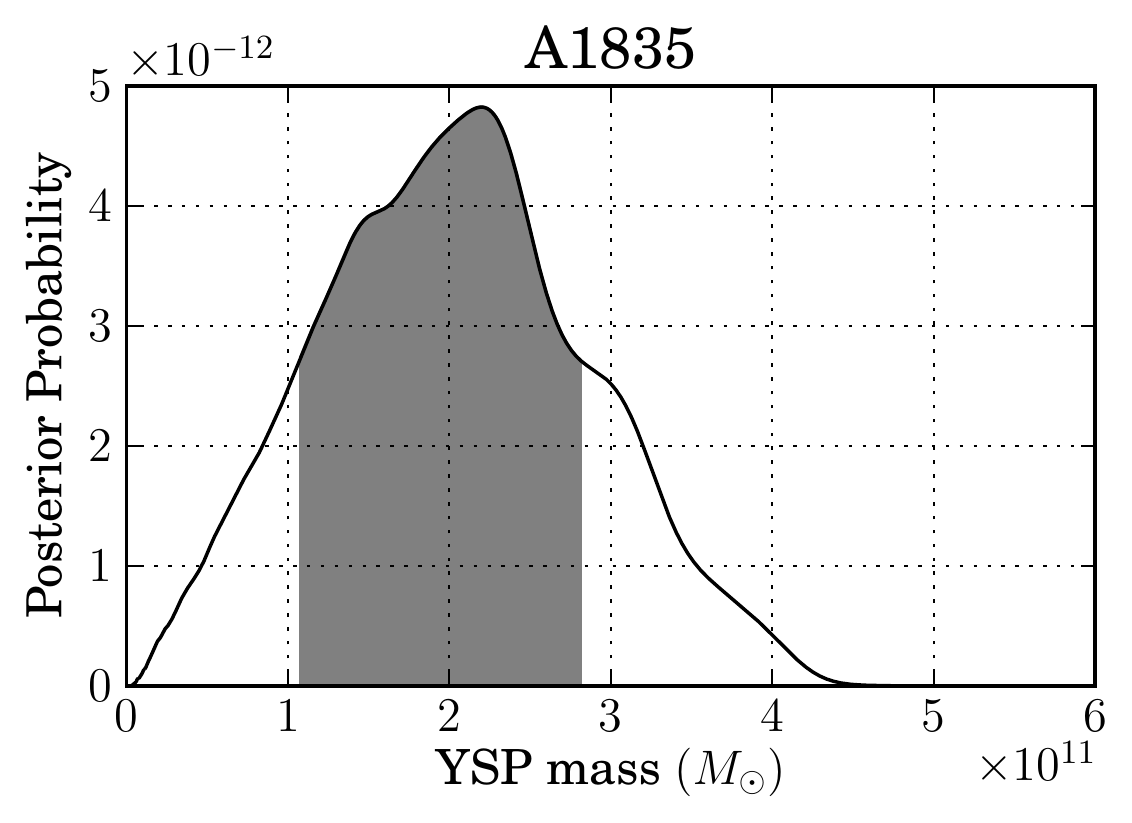}
  \end{minipage}%
  \begin{minipage}{0.33\textwidth}
    \centering
    \includegraphics[width=\textwidth]{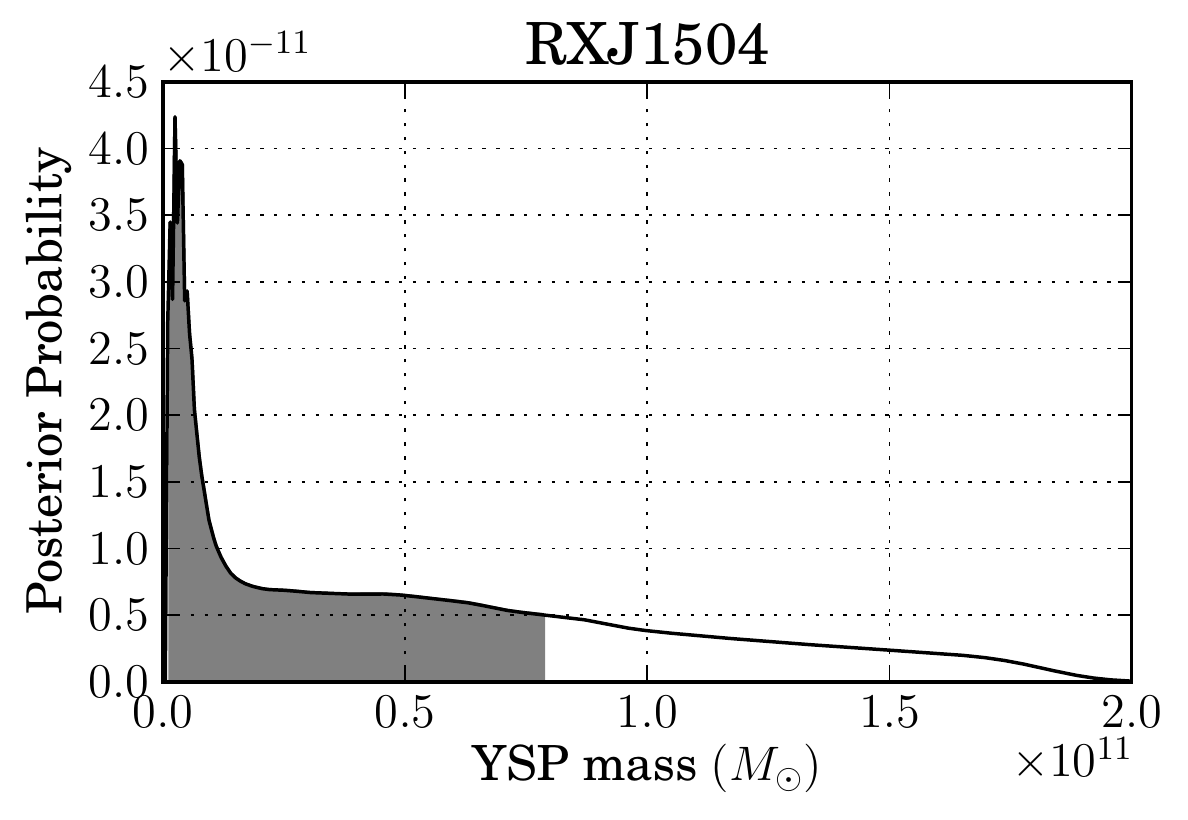}
  \end{minipage}%
  \begin{minipage}{0.33\textwidth}
    \centering
    \includegraphics[width=\textwidth]{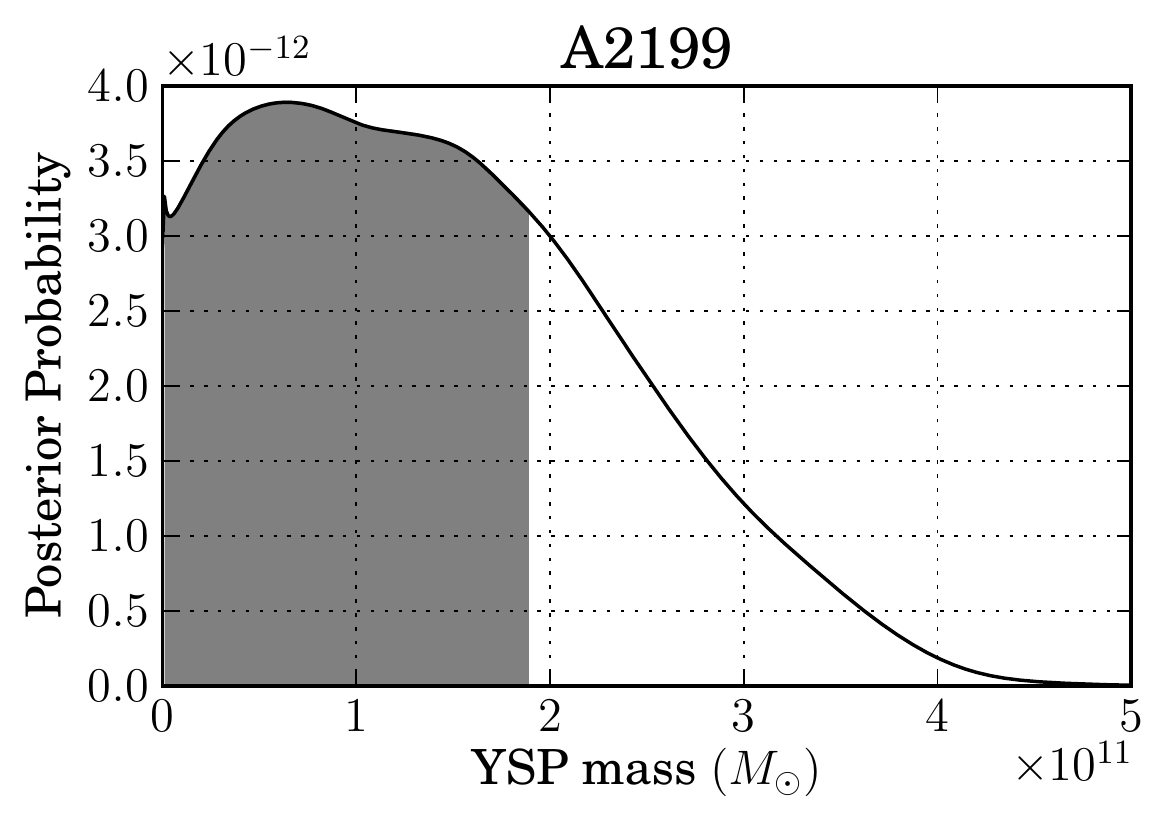}
  \end{minipage}\\
  \begin{minipage}{0.33\textwidth}
    \centering
    \includegraphics[width=\textwidth]{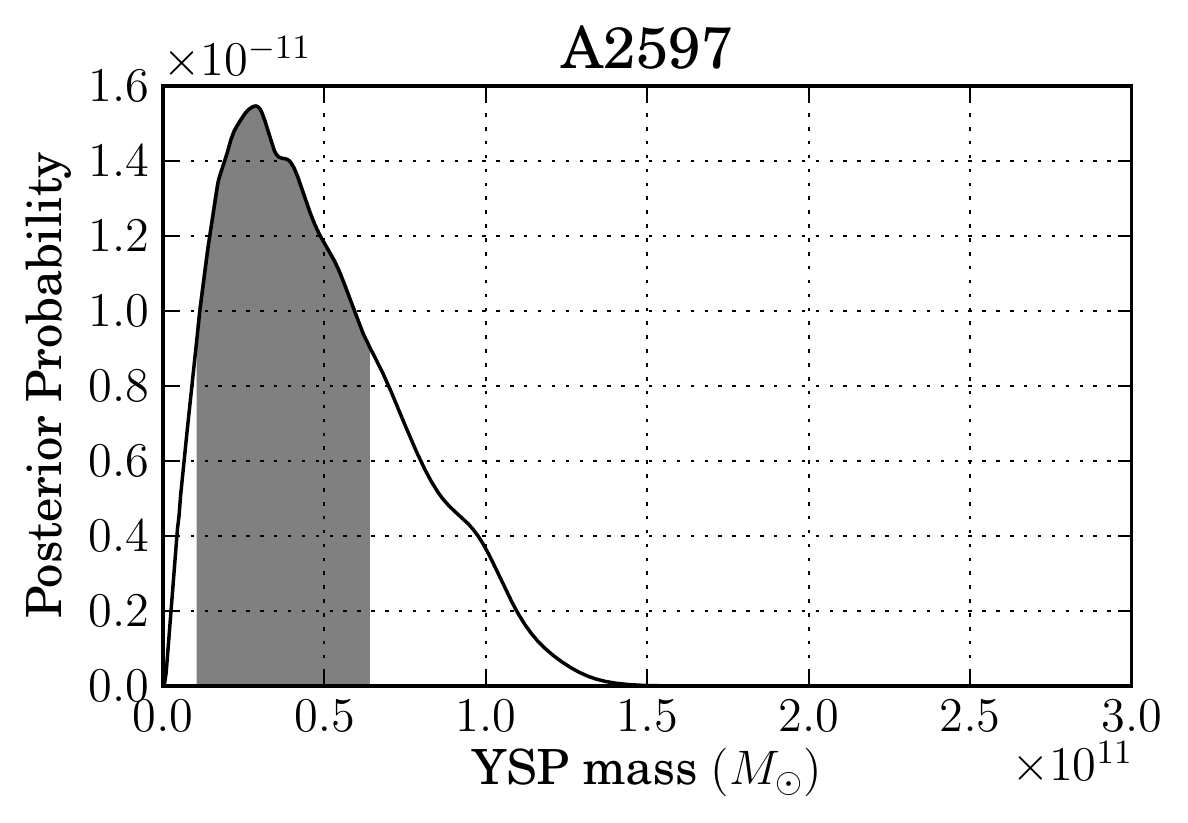}
  \end{minipage}%
  \caption{Posterior probability density distributions for the mass of
    the young stellar popluation, $\mysp$, after marginalizing over
    other model parametrs (metallicity, extinction, YSP age and OSP
    mass).  The joint prior PDF used for $\mosp$ and $\mysp$ was
    uniform with the constraint that $\mosp>\mysp$.  Since the best
    fit YSP masses for most models were much lower than the
    corresponding OSP masses, this can be thought of effectively as a
    uniform prior on $\mysp$.  The shading indicates the narrowest
    68\% plausible interval for $\mosp$.  Note that some of the
    multimodality in the posterior pdf for $\mysp$ is likely not a
    ``real'' feature, but an artifact of the discrete sampling of
    $\tysp$, which is correlated with $\mysp$ in the joint posterior
    distribution.}
  \label{My}
\end{figure*}


\begin{figure*}
  \begin{minipage}{0.33\textwidth}
    \centering
    \includegraphics[width=\textwidth]{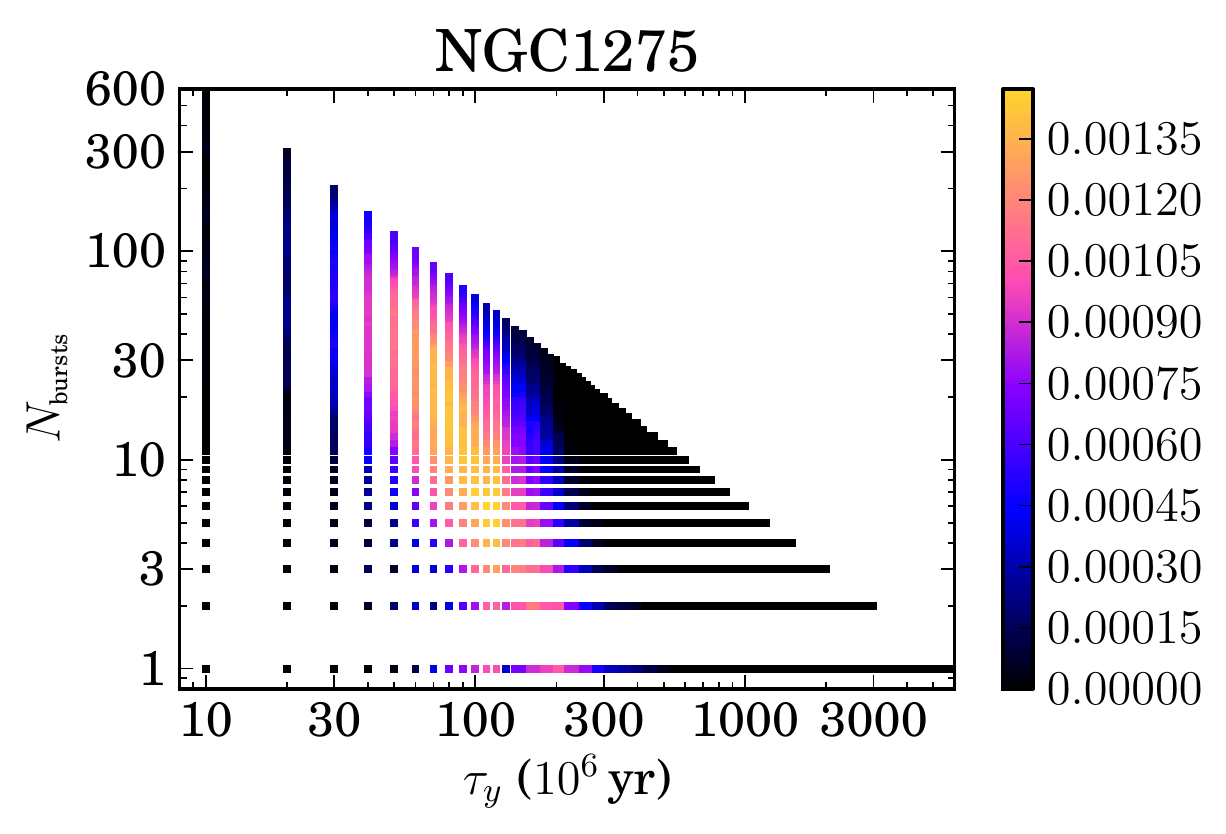}
  \end{minipage}%
  \begin{minipage}{0.33\textwidth}
    \centering
    \includegraphics[width=\textwidth]{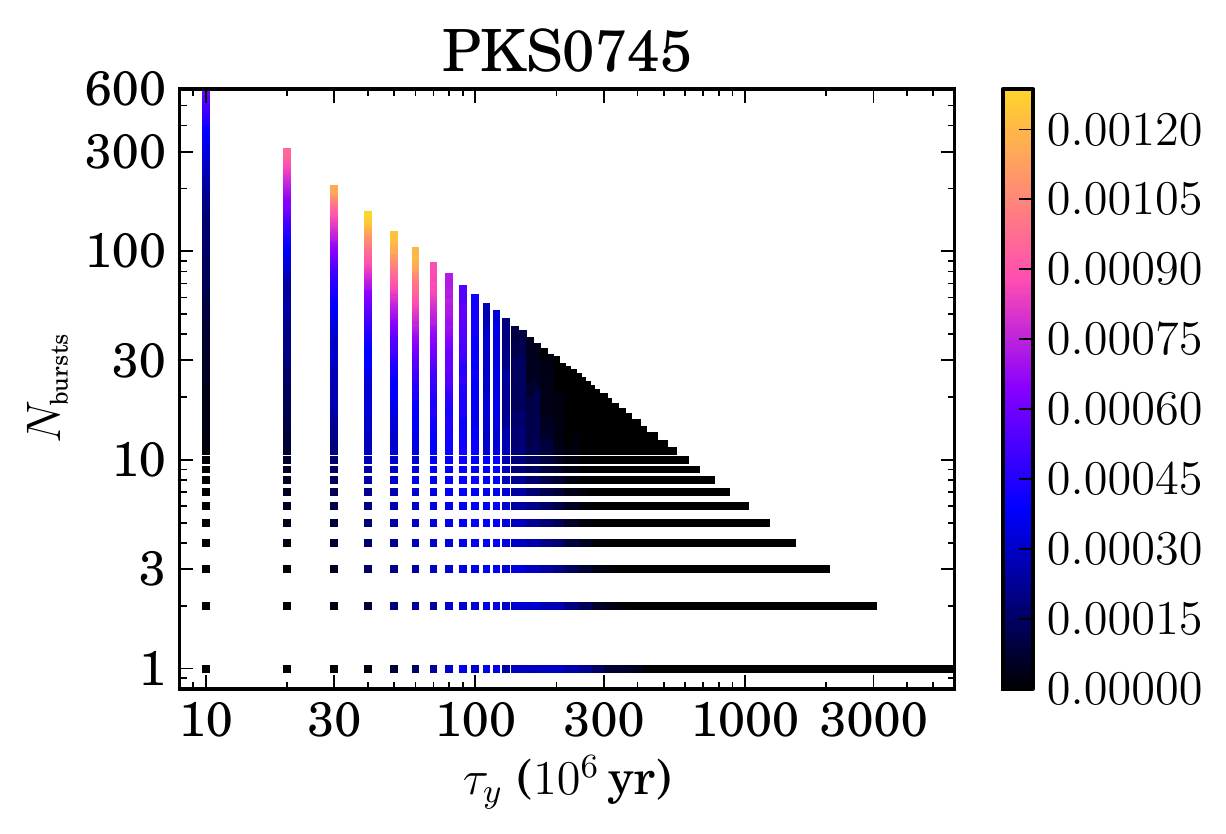}
  \end{minipage}%
  \begin{minipage}{0.33\textwidth}
    \centering
    \includegraphics[width=\textwidth]{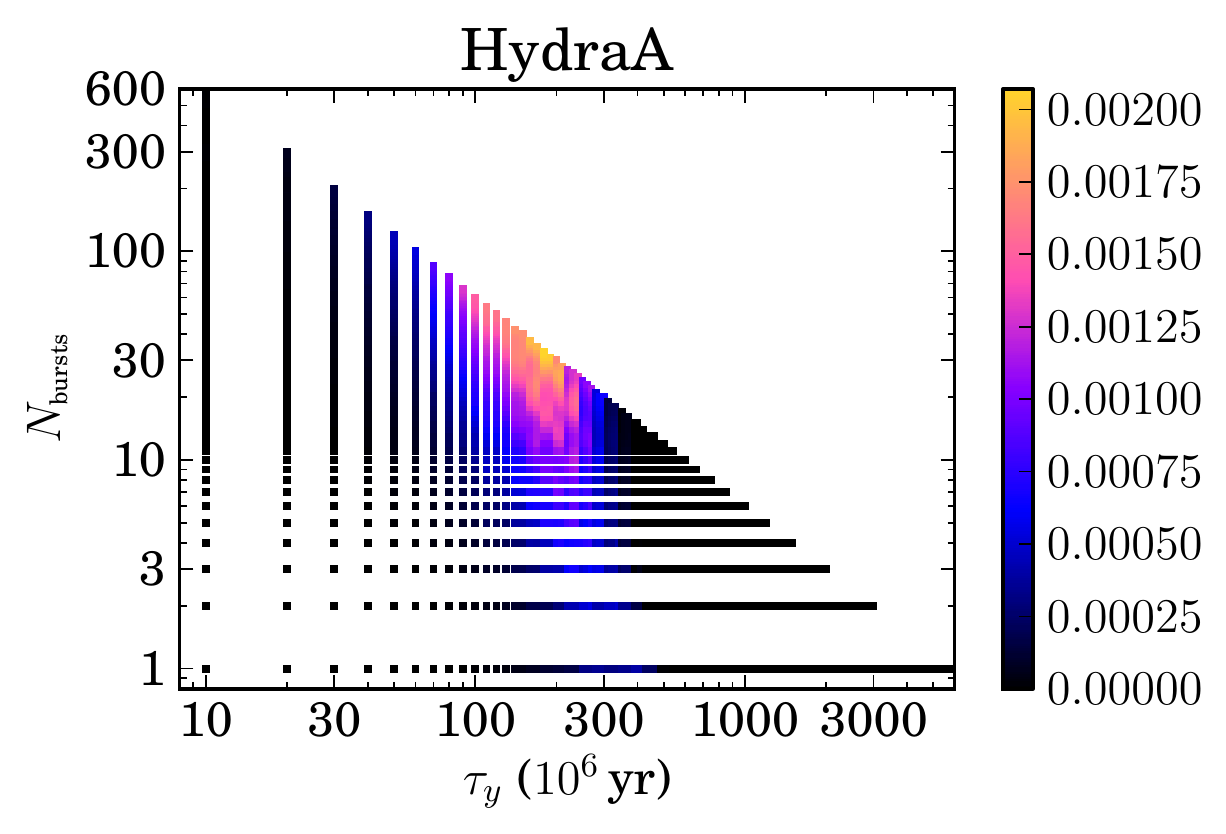}
  \end{minipage}\\
  \begin{minipage}{0.33\textwidth}
    \centering
    \includegraphics[width=\textwidth]{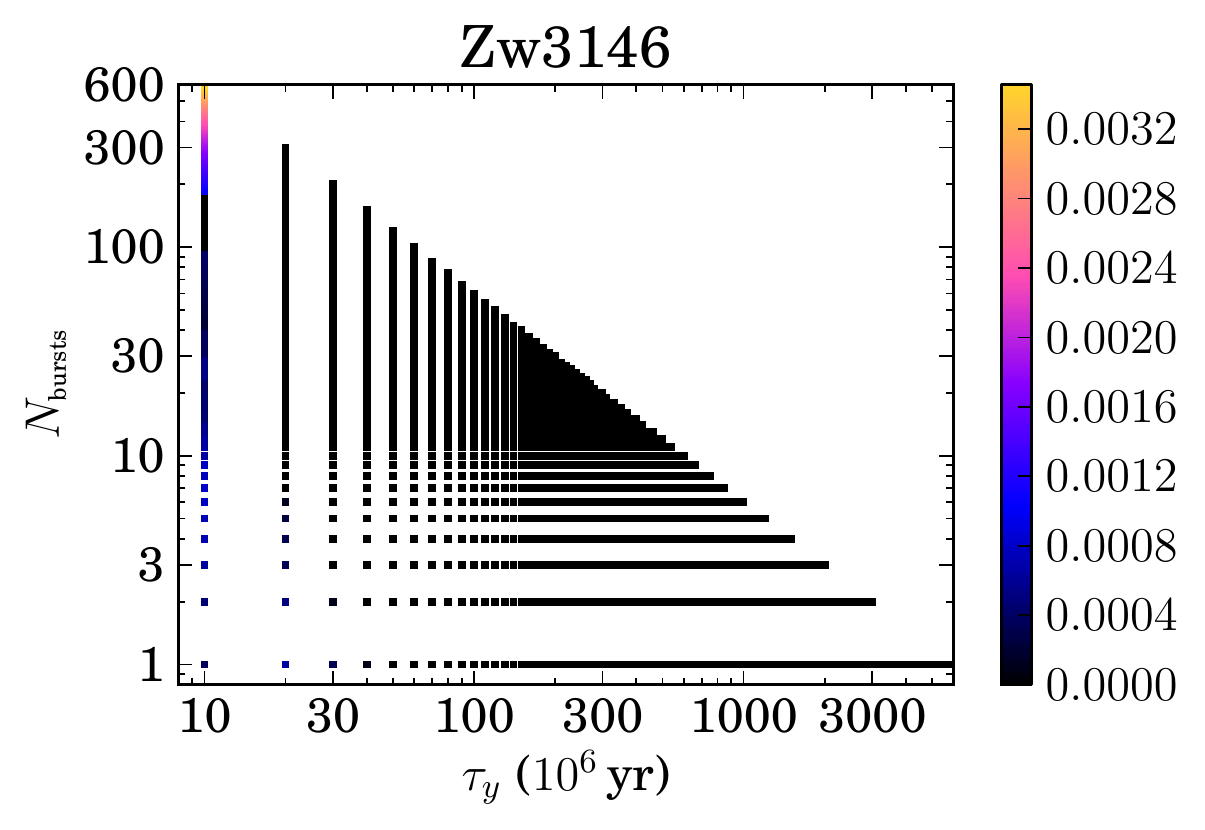}
  \end{minipage}%
  \begin{minipage}{0.33\textwidth}
    \centering
    \includegraphics[width=\textwidth]{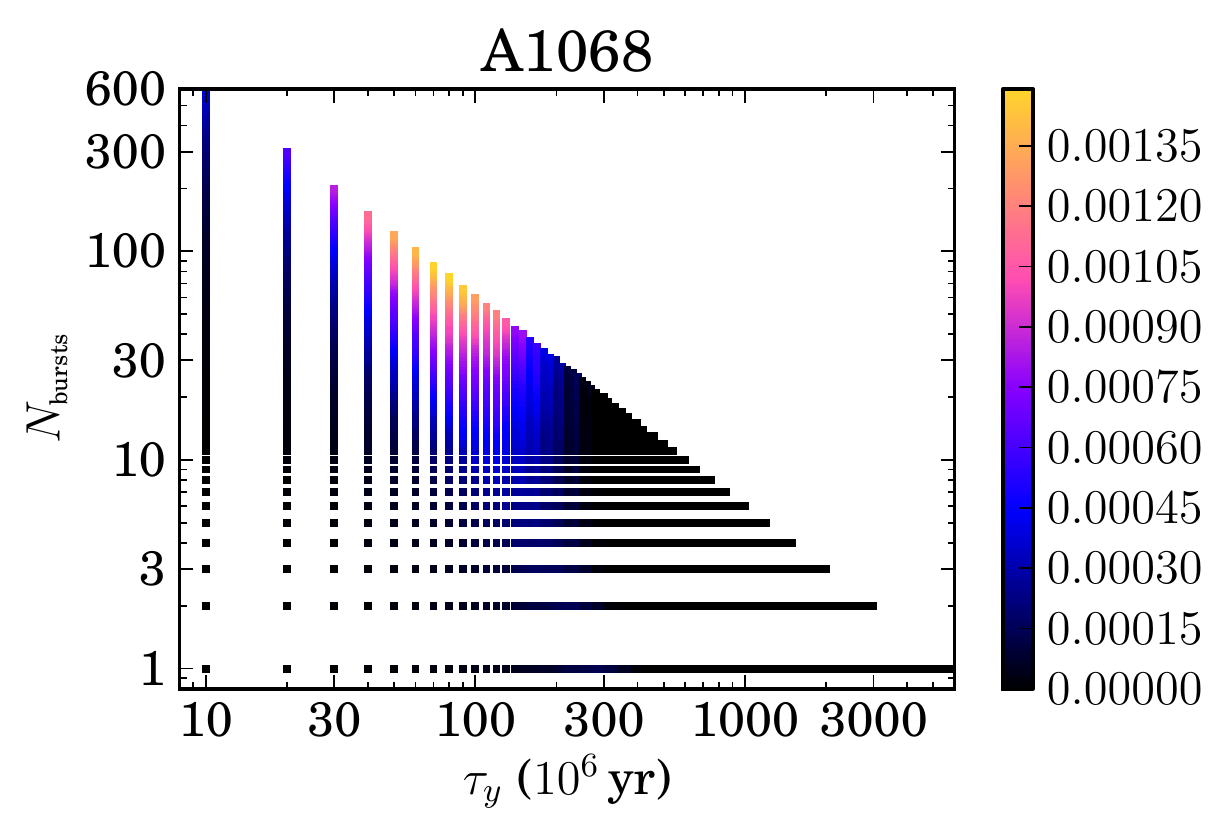}
  \end{minipage}%
  \begin{minipage}{0.33\textwidth}
    \centering
    \includegraphics[width=\textwidth]{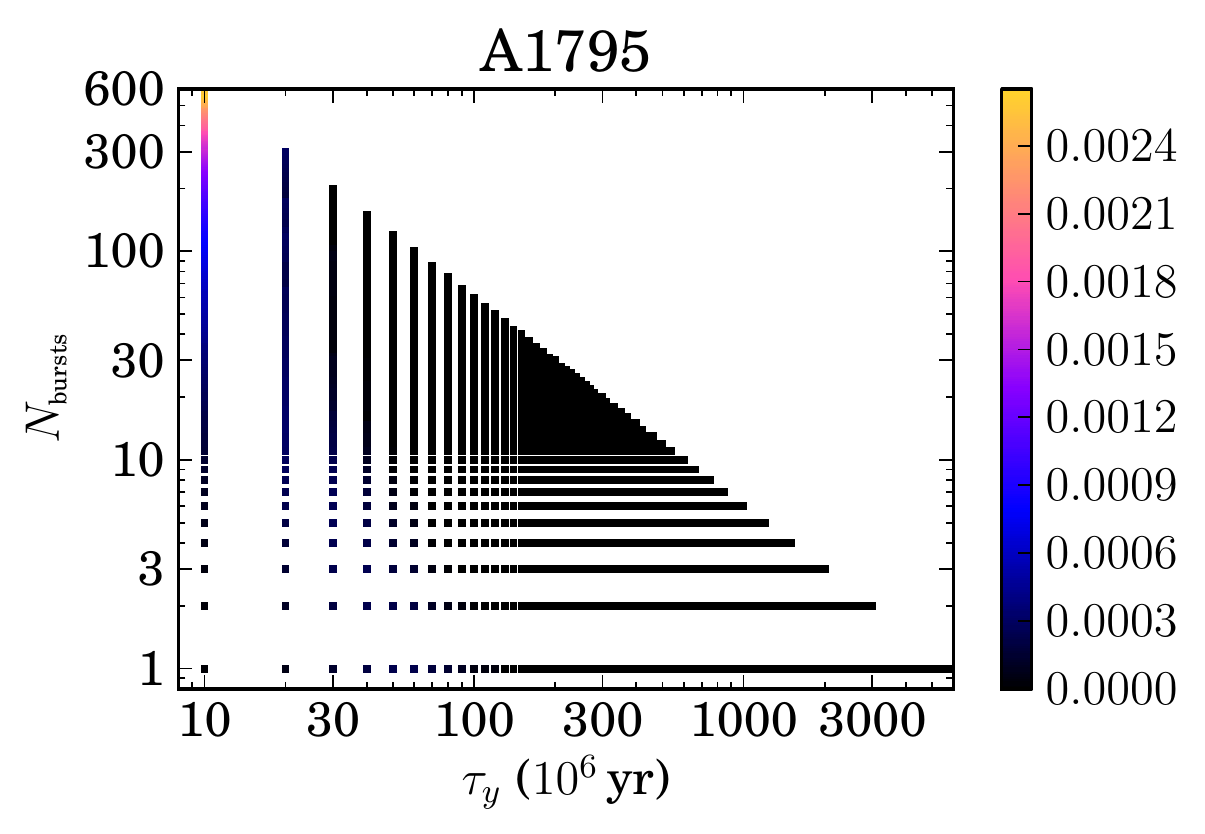}
  \end{minipage}\\
  \begin{minipage}{0.33\textwidth}
    \centering
    \includegraphics[width=\textwidth]{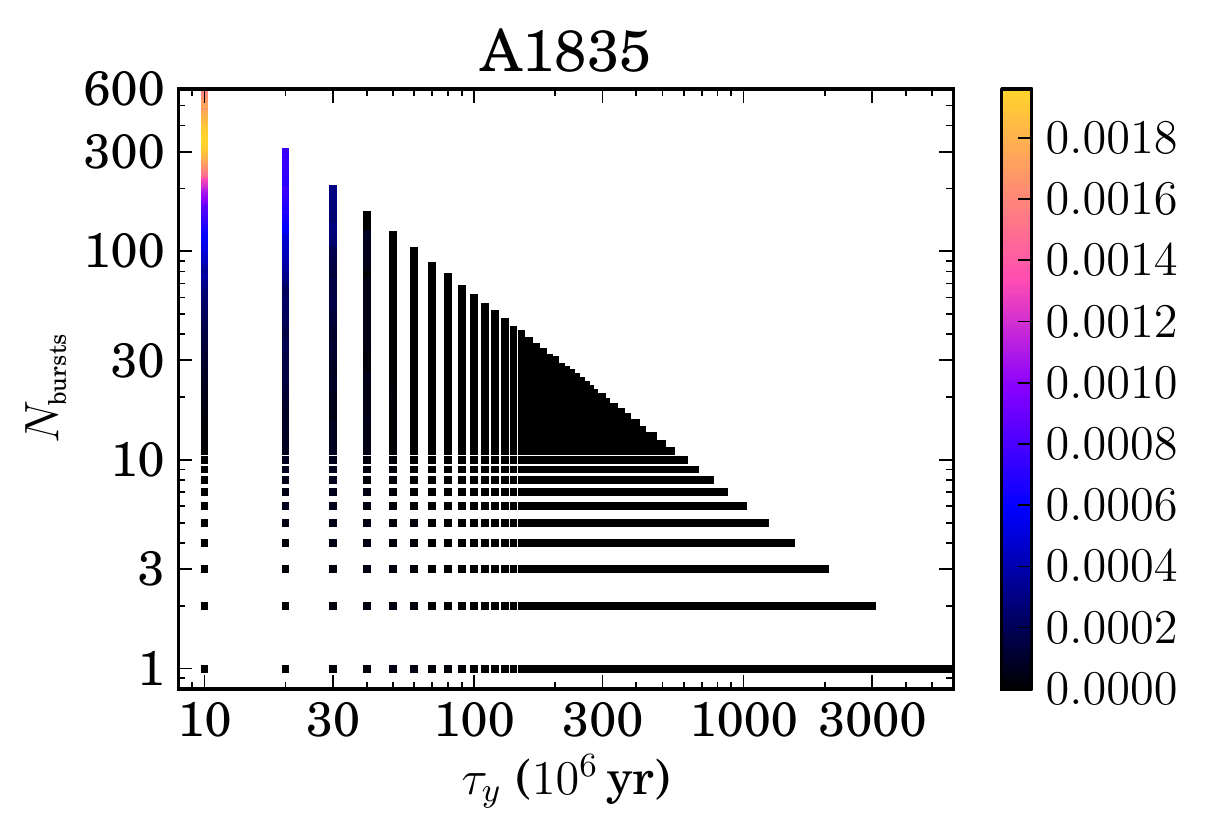}
  \end{minipage}%
  \begin{minipage}{0.33\textwidth}
    \centering
    \includegraphics[width=\textwidth]{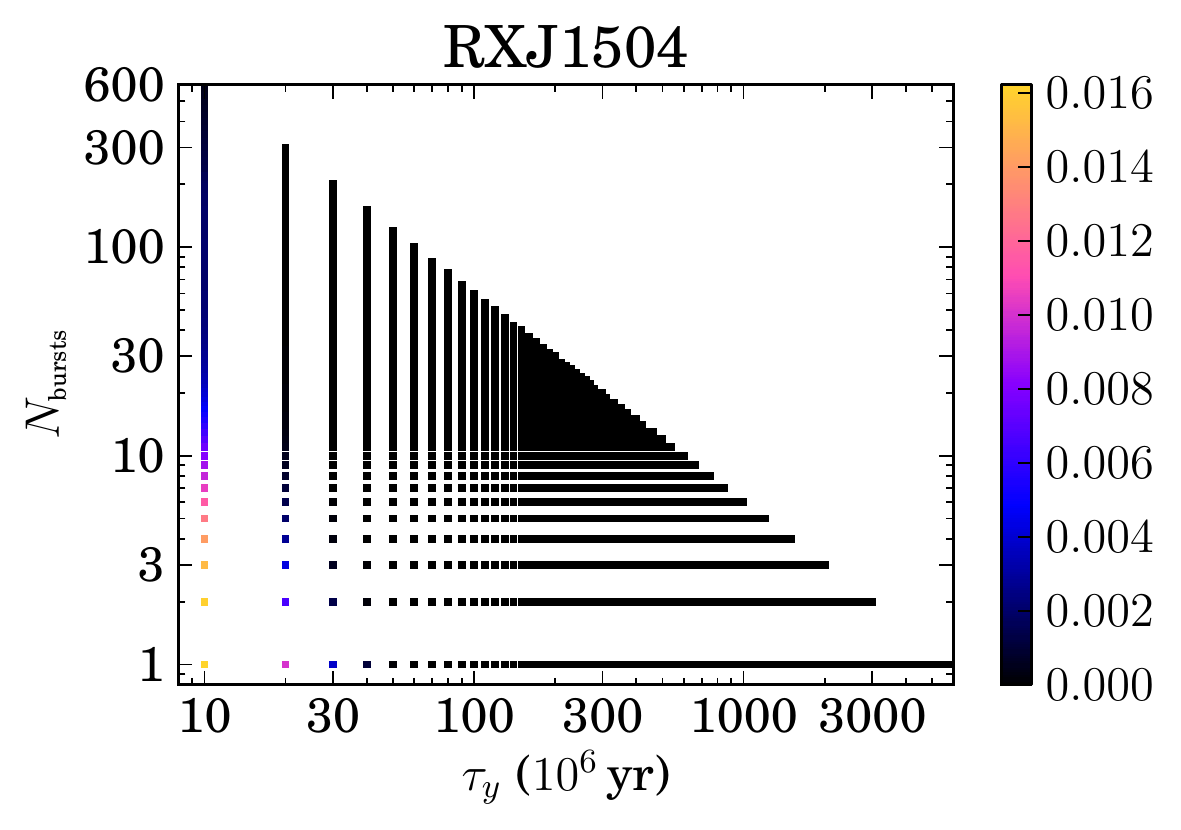}
  \end{minipage}%
  \begin{minipage}{0.33\textwidth}
    \centering
    \includegraphics[width=\textwidth]{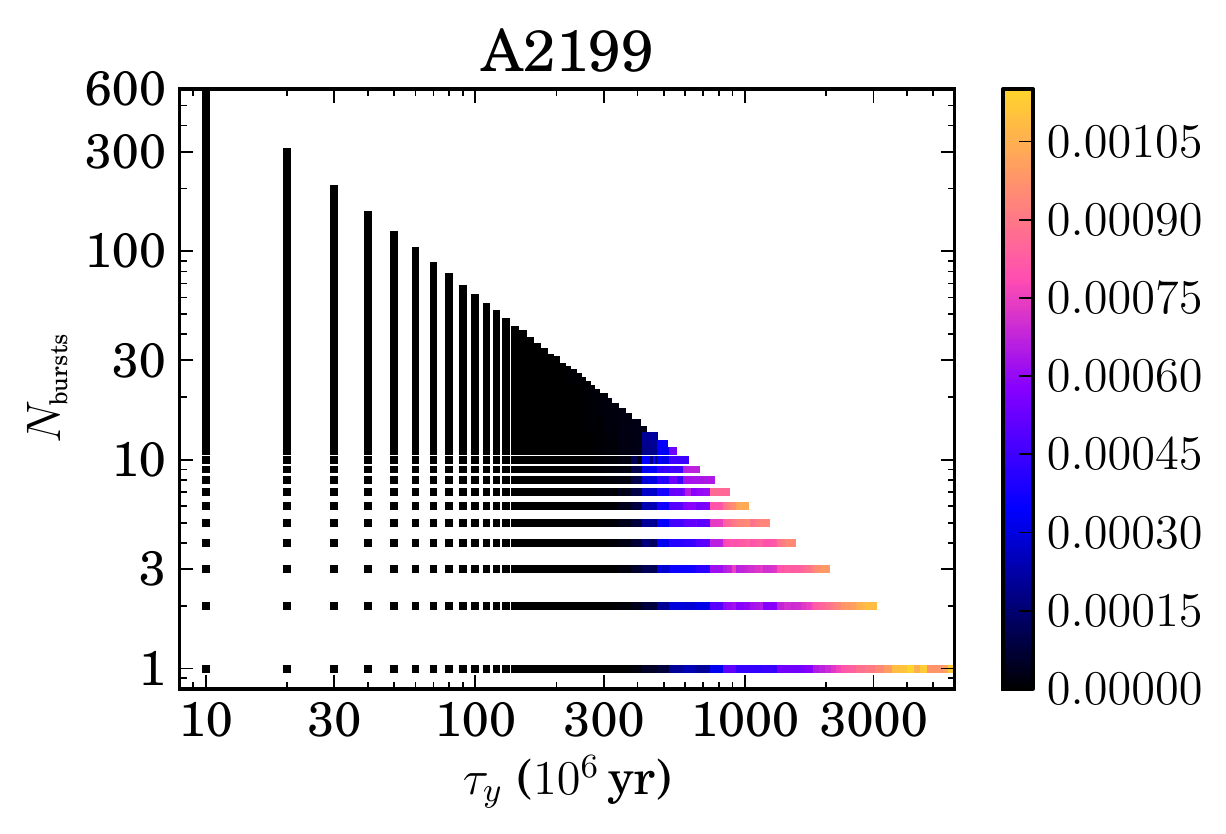}
  \end{minipage}\\
  \begin{minipage}{0.33\textwidth}
    \centering
    \includegraphics[width=\textwidth]{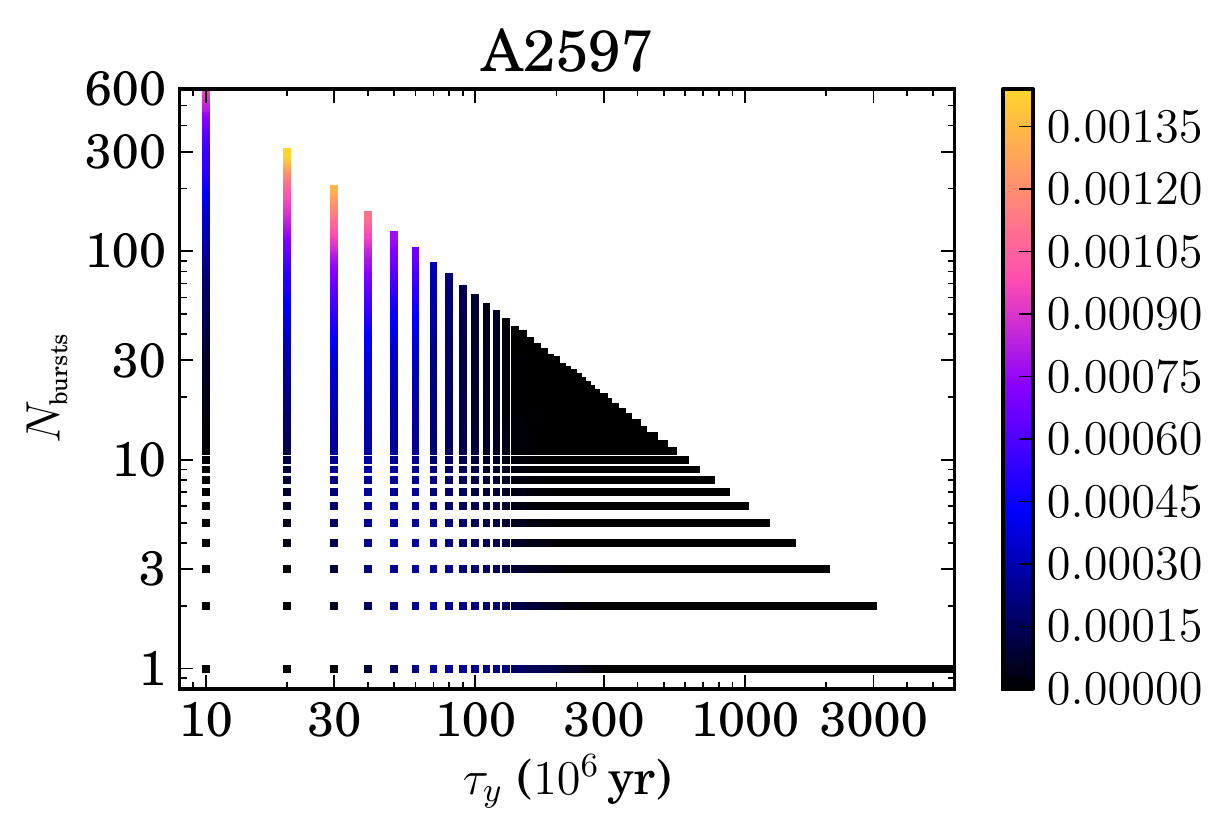}
  \end{minipage}%
  \caption{Posterior probability distributions for the age of the YSP,
    after marginalizing over other model parametrs (metallicity,
    extinction, OSP and YSP).  The X-axis represents the age of the
    most recent starburst and also the separation between multiple
    starbursts and the Y-axis represents the number of outbursts for a
    given separation.  Each of the combinations of these two
    parameters was assigned equal prior probability; The colourbar
    represents the posterior probability for a given combination of
    age and number of outbursts.}
  \label{Ty}
\end{figure*}




\begin{figure*}
  \begin{minipage}{0.33\textwidth}
    \centering
    \includegraphics[width=\textwidth]{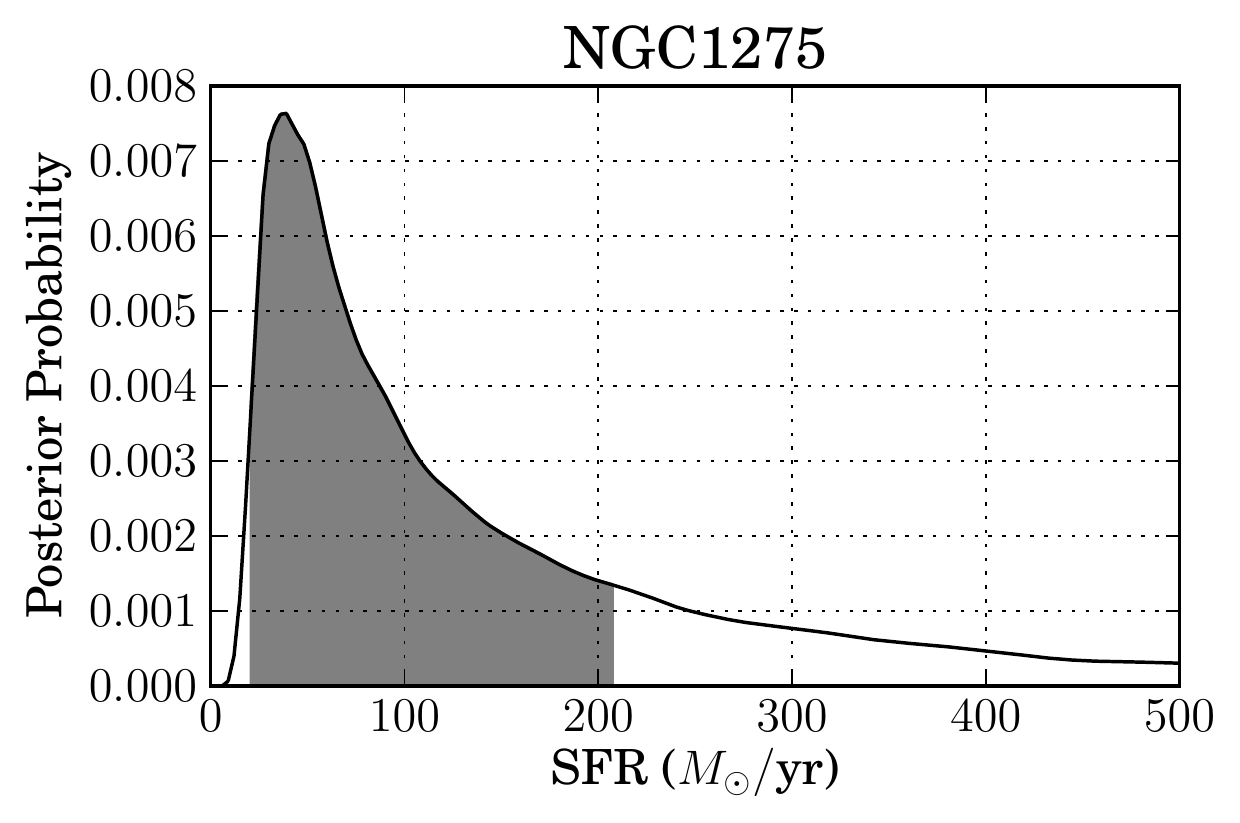}
  \end{minipage}%
  \begin{minipage}{0.33\textwidth}
    \centering
    \includegraphics[width=\textwidth]{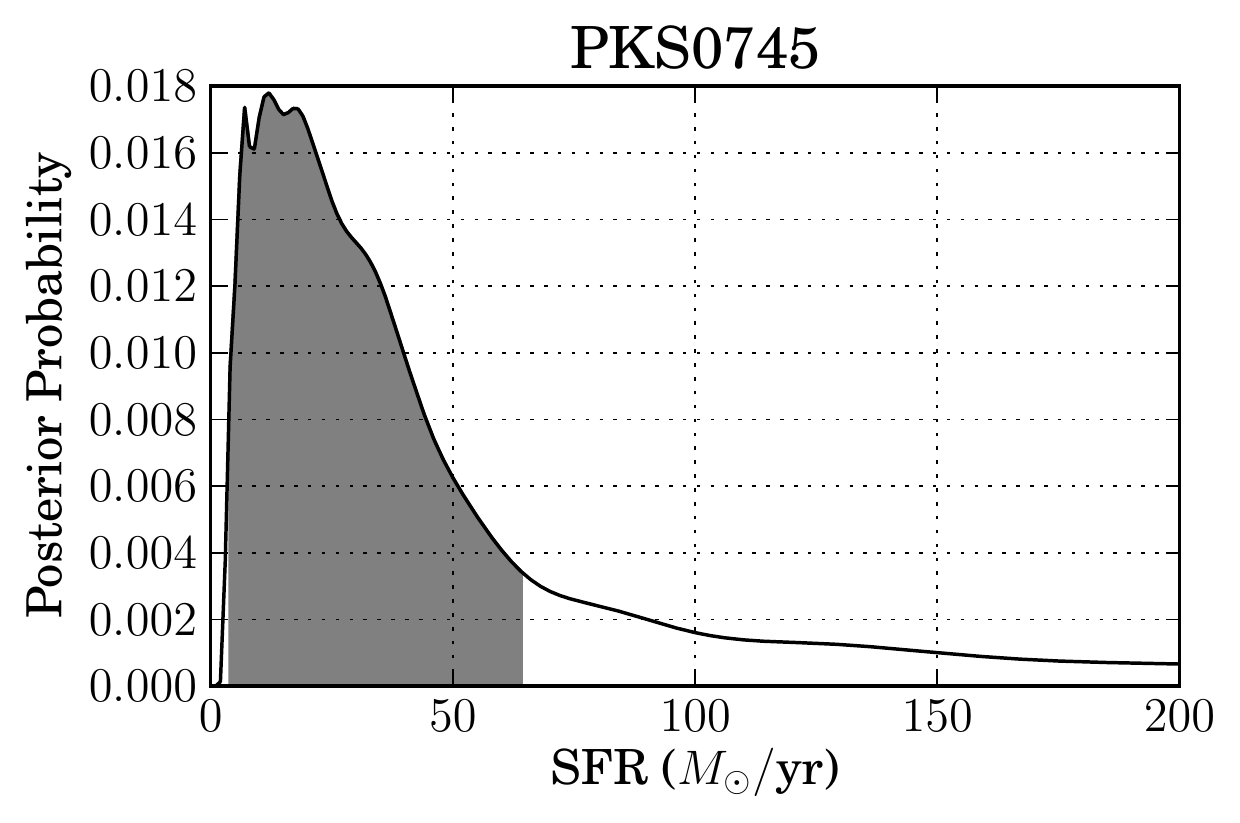}
  \end{minipage}%
  \begin{minipage}{0.33\textwidth}
    \centering
    \includegraphics[width=\textwidth]{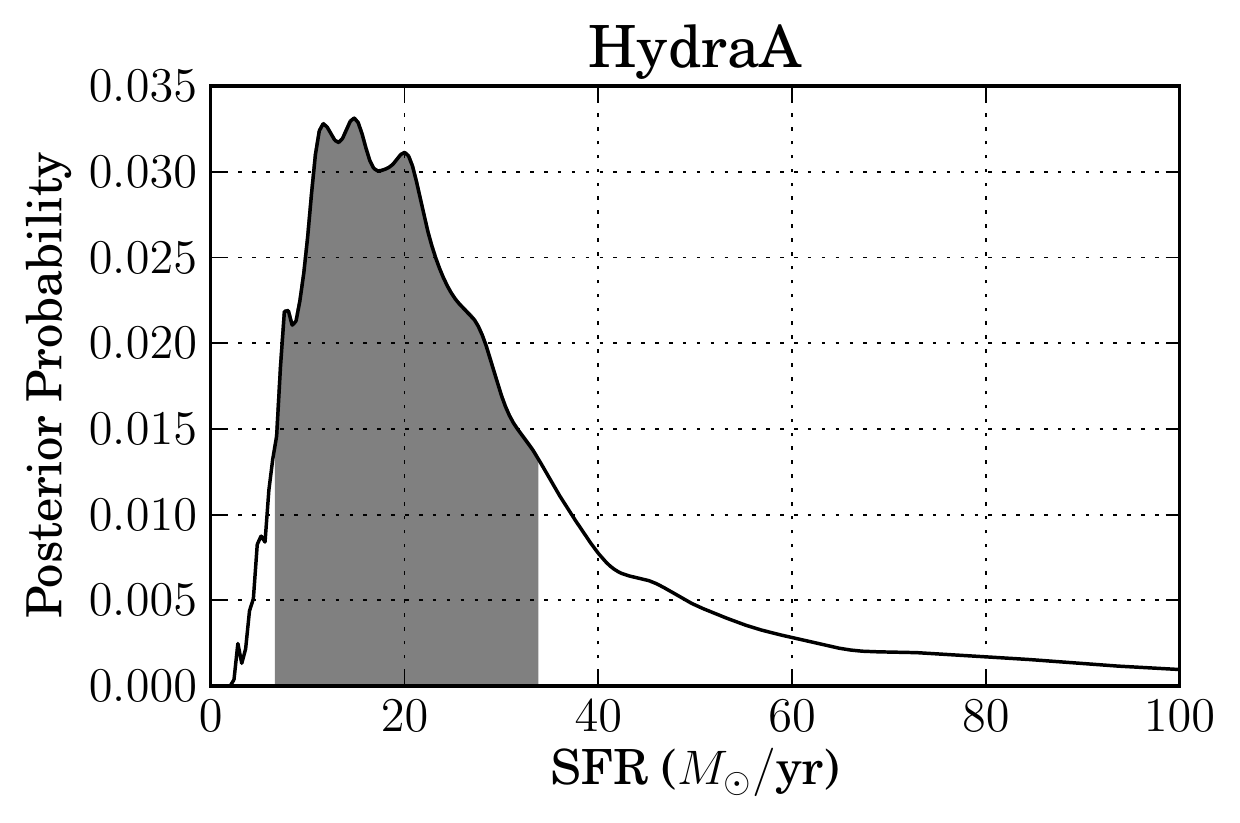}
  \end{minipage}\\
  \begin{minipage}{0.33\textwidth}
    \centering
    \includegraphics[width=\textwidth]{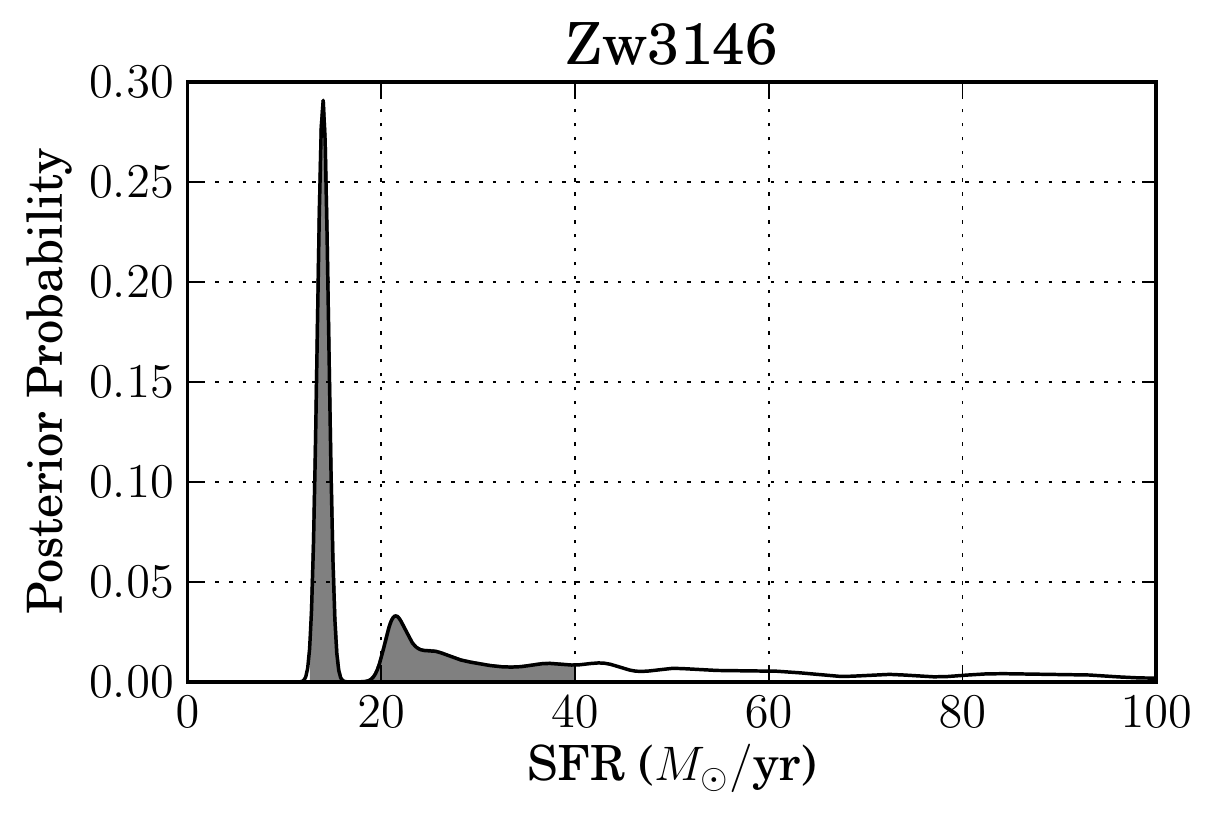}
  \end{minipage}%
  \begin{minipage}{0.33\textwidth}
    \centering
    \includegraphics[width=\textwidth]{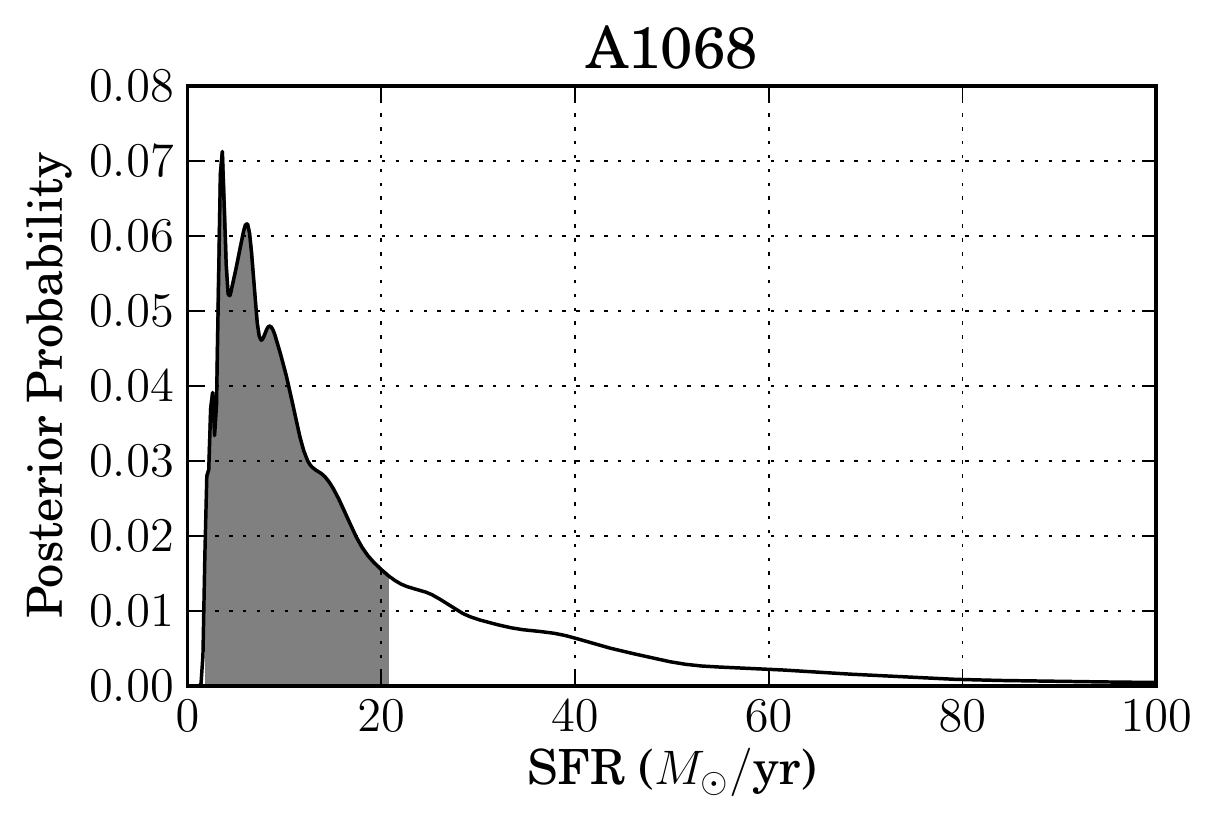}
  \end{minipage}%
  \begin{minipage}{0.33\textwidth}
    \centering
    \includegraphics[width=\textwidth]{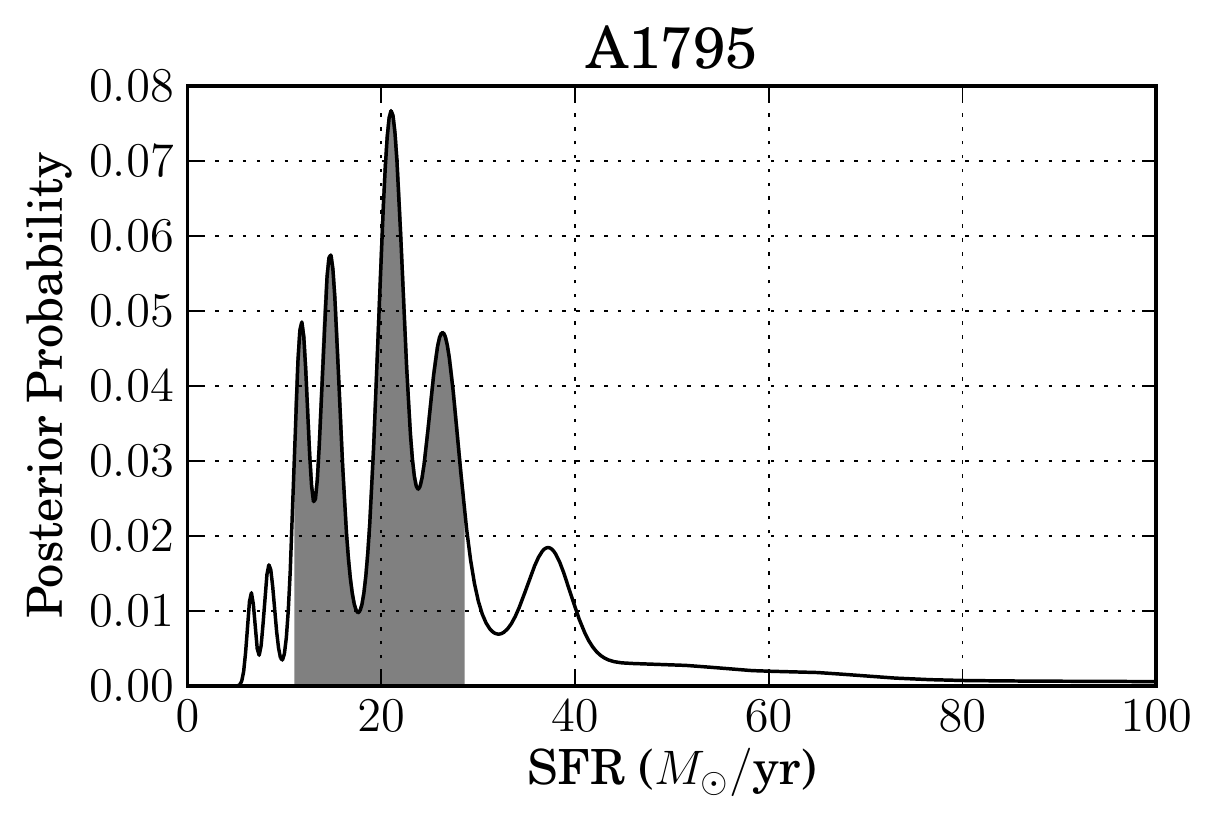}
  \end{minipage}\\
  \begin{minipage}{0.33\textwidth}
    \centering
    \includegraphics[width=\textwidth]{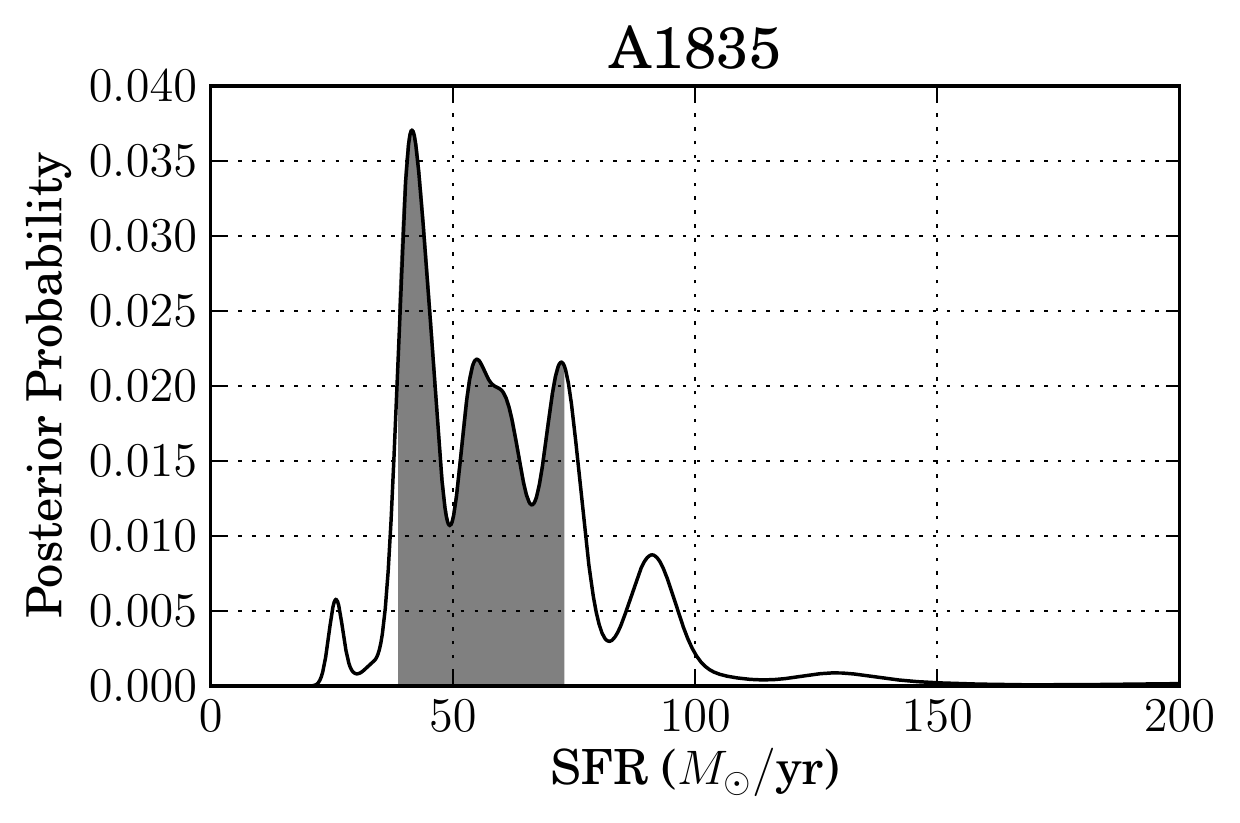}
  \end{minipage}%
  \begin{minipage}{0.33\textwidth}
    \centering
    \includegraphics[width=\textwidth]{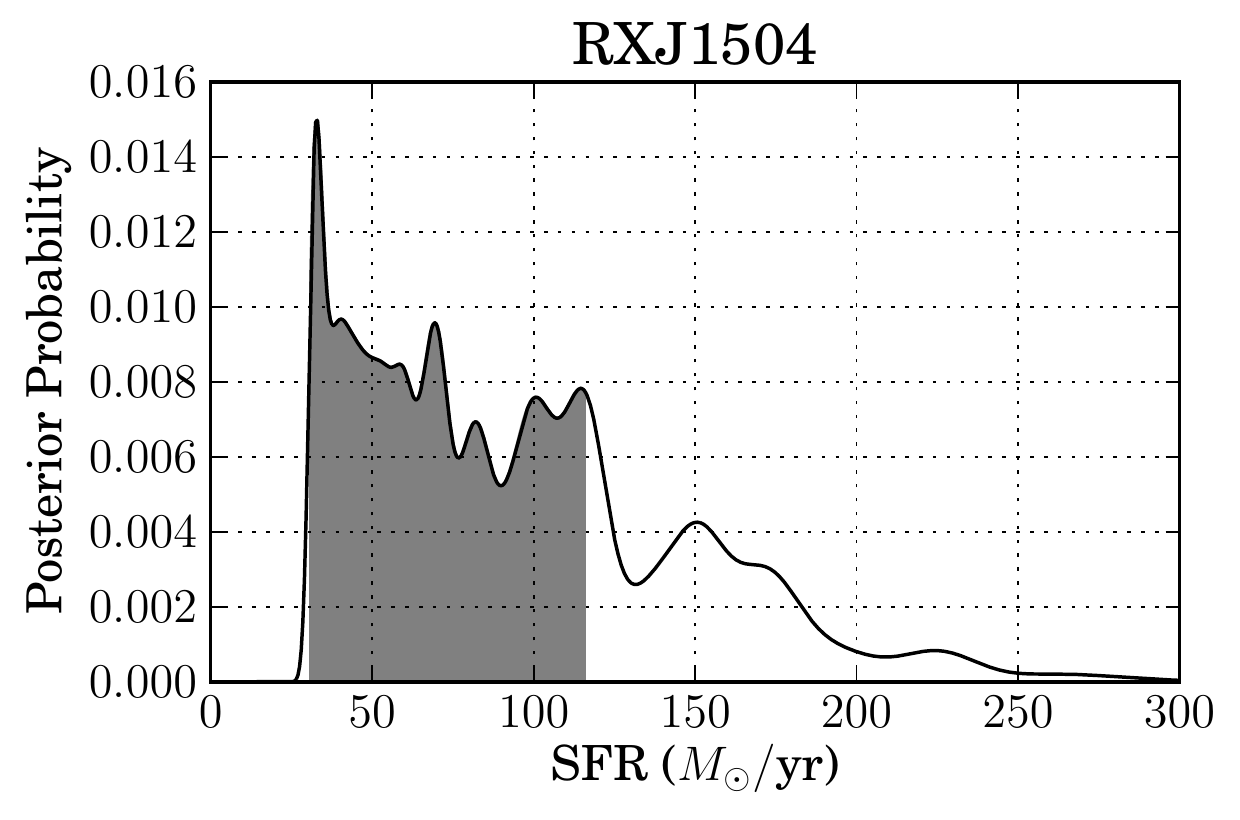}
  \end{minipage}%
  \begin{minipage}{0.33\textwidth}
    \centering
    \includegraphics[width=\textwidth]{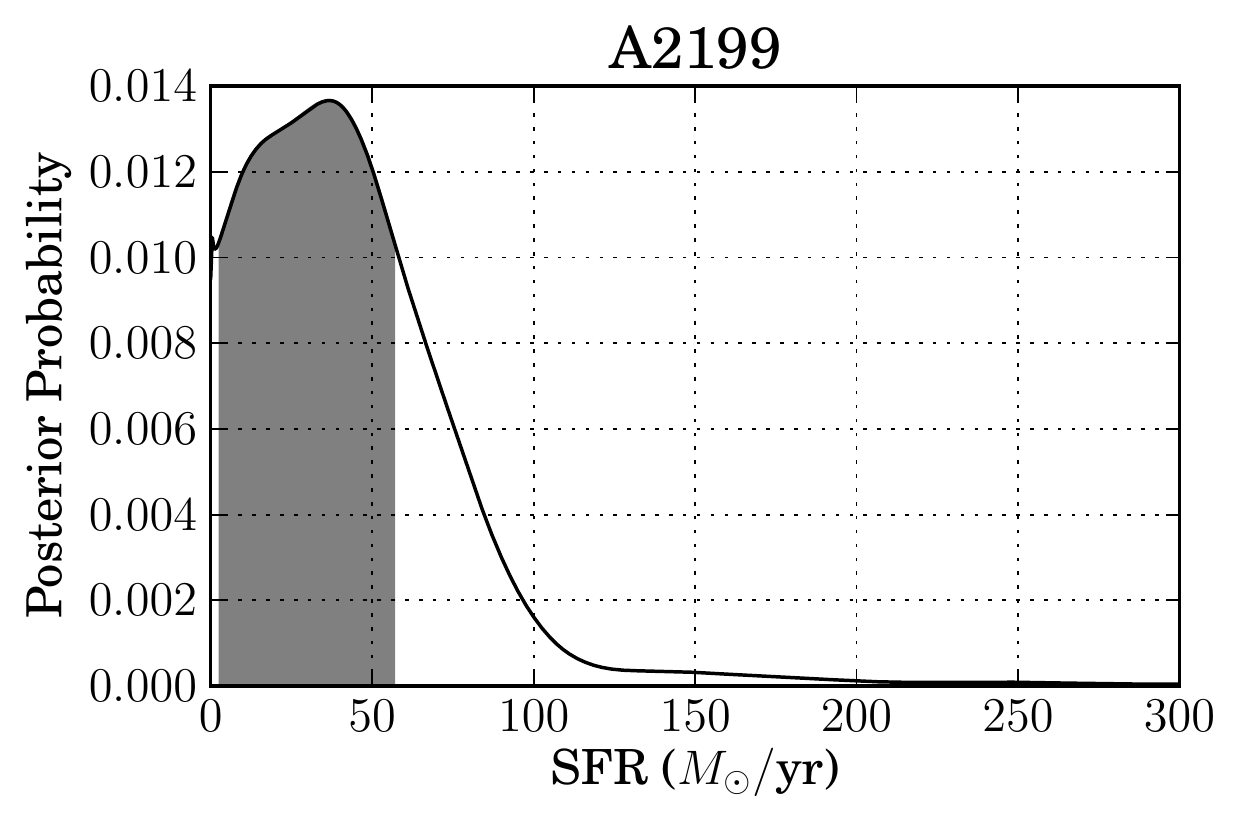}
  \end{minipage}\\
  \begin{minipage}{0.33\textwidth}
    \centering
    \includegraphics[width=\textwidth]{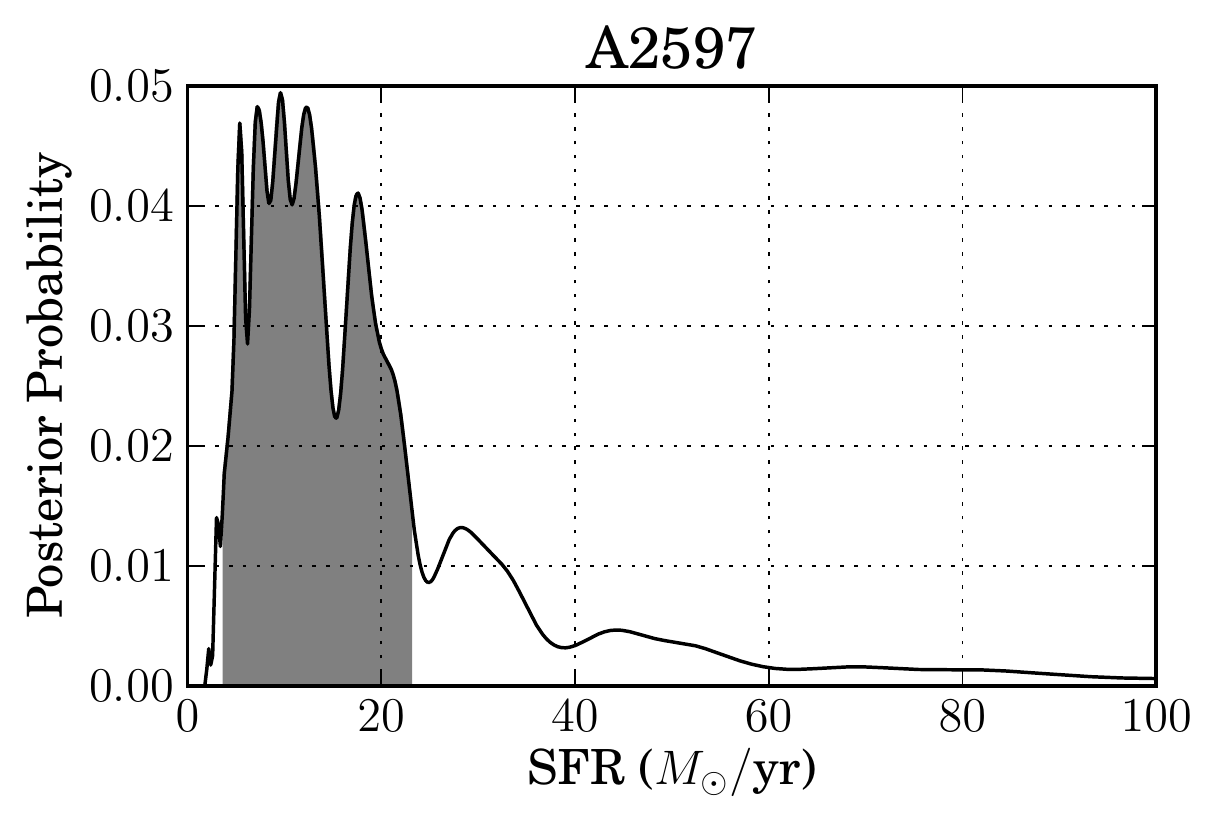}
  \end{minipage}%
  \caption{The posterior probability density distributions for the
    star formation rates, constructed from the joint posterior
    probability distribution for the mass and age of the young stellar
    population, after marginalizing over other model parametrs
    (metallicity, extinction, and OSP mass). The SFR is the total mass
    divided by the age of the oldest burst, or equivalently the mass
    in each burst divided by the time interval between them.  Note
    that some of the multimodality in the posterior pdf for the SFR is
    likely not a ``real'' feature, but an artifact of the discrete
    sampling of $\tysp$.}
  \label{SFR} 
\end{figure*}


\begin{figure*}
  \begin{minipage}{0.33\textwidth}
    \centering
    \includegraphics[width=\textwidth]{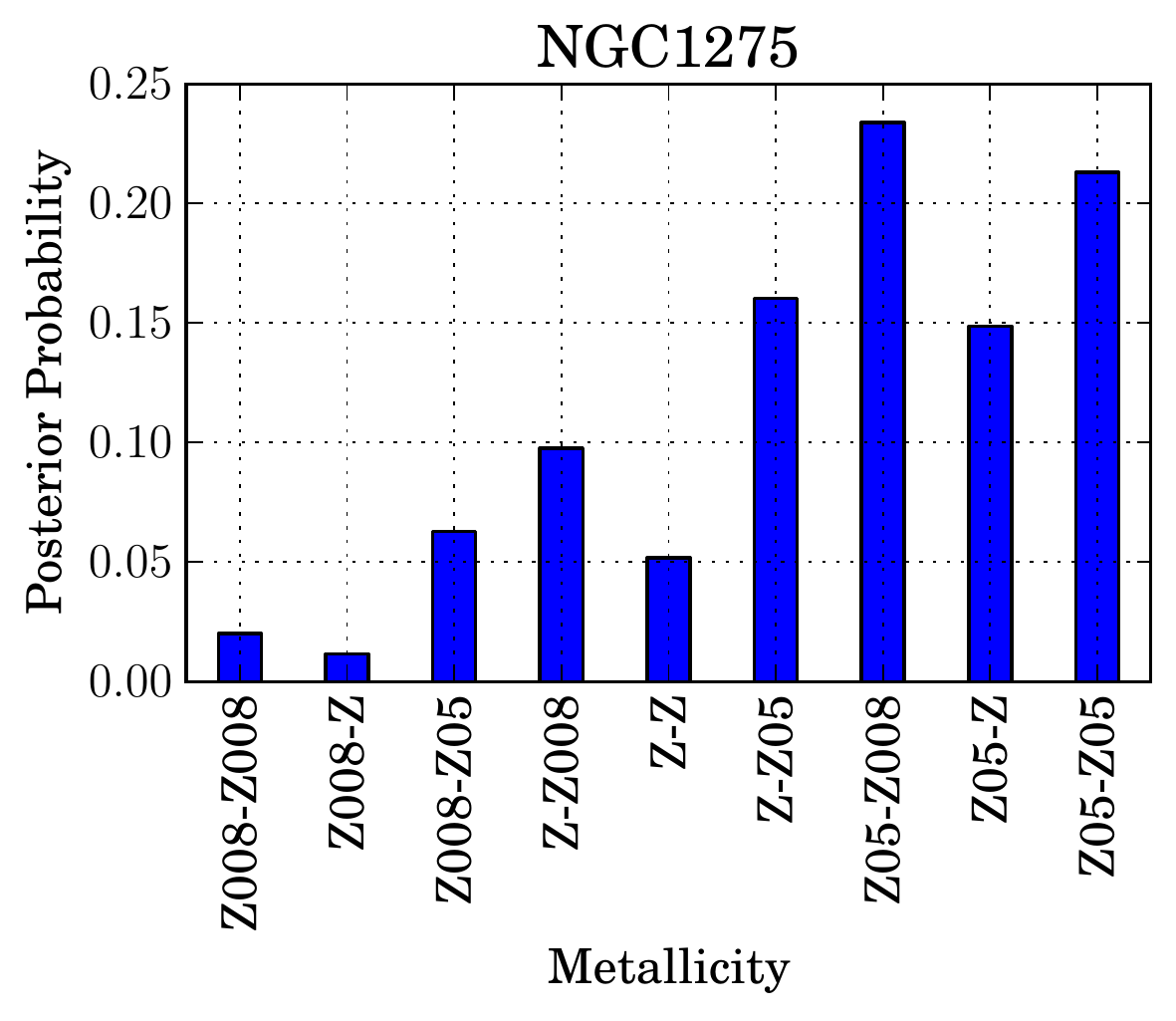}
  \end{minipage}%
  \begin{minipage}{0.33\textwidth}
    \centering
    \includegraphics[width=\textwidth]{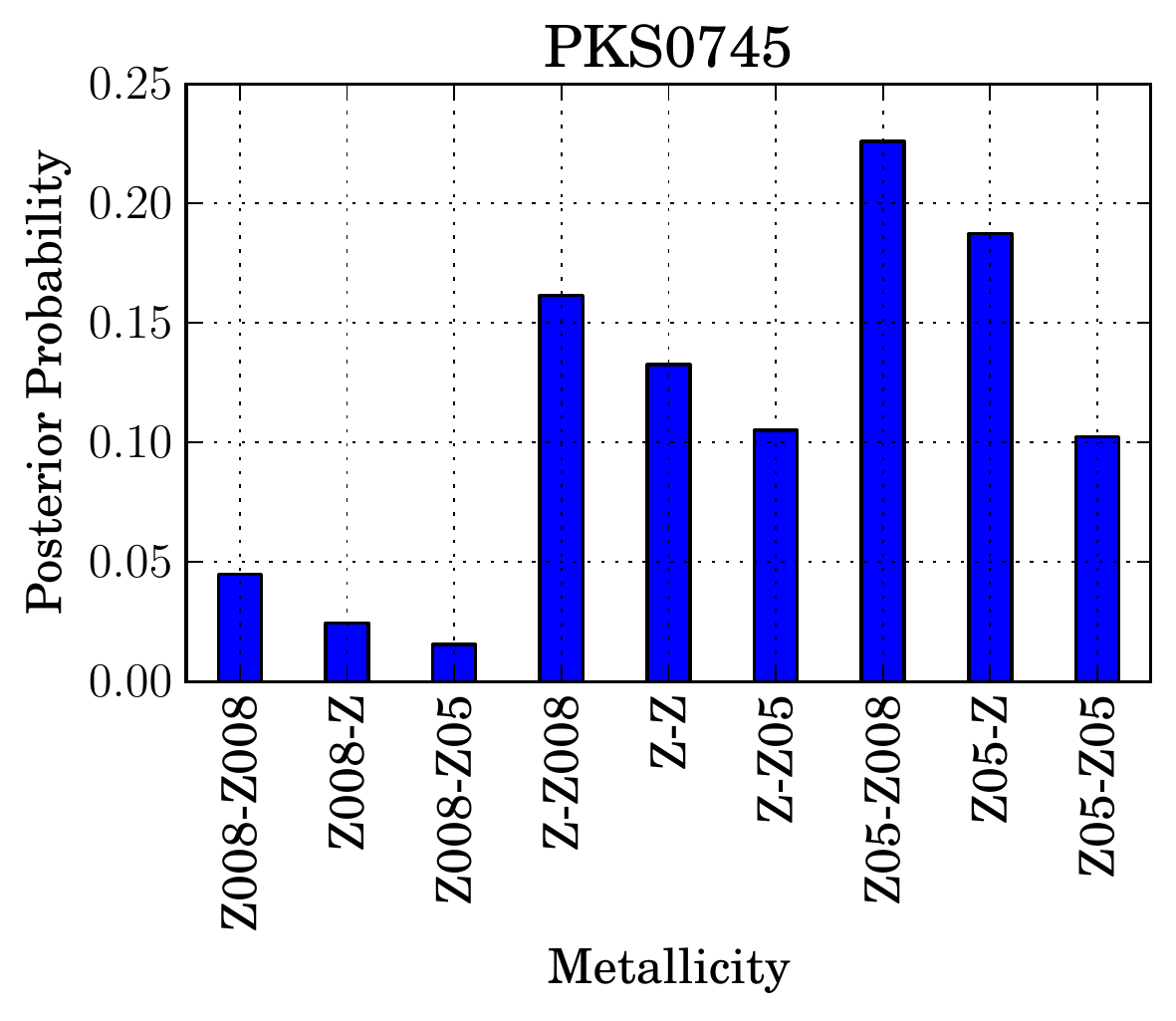}
  \end{minipage}%
  \begin{minipage}{0.33\textwidth}
    \centering
    \includegraphics[width=\textwidth]{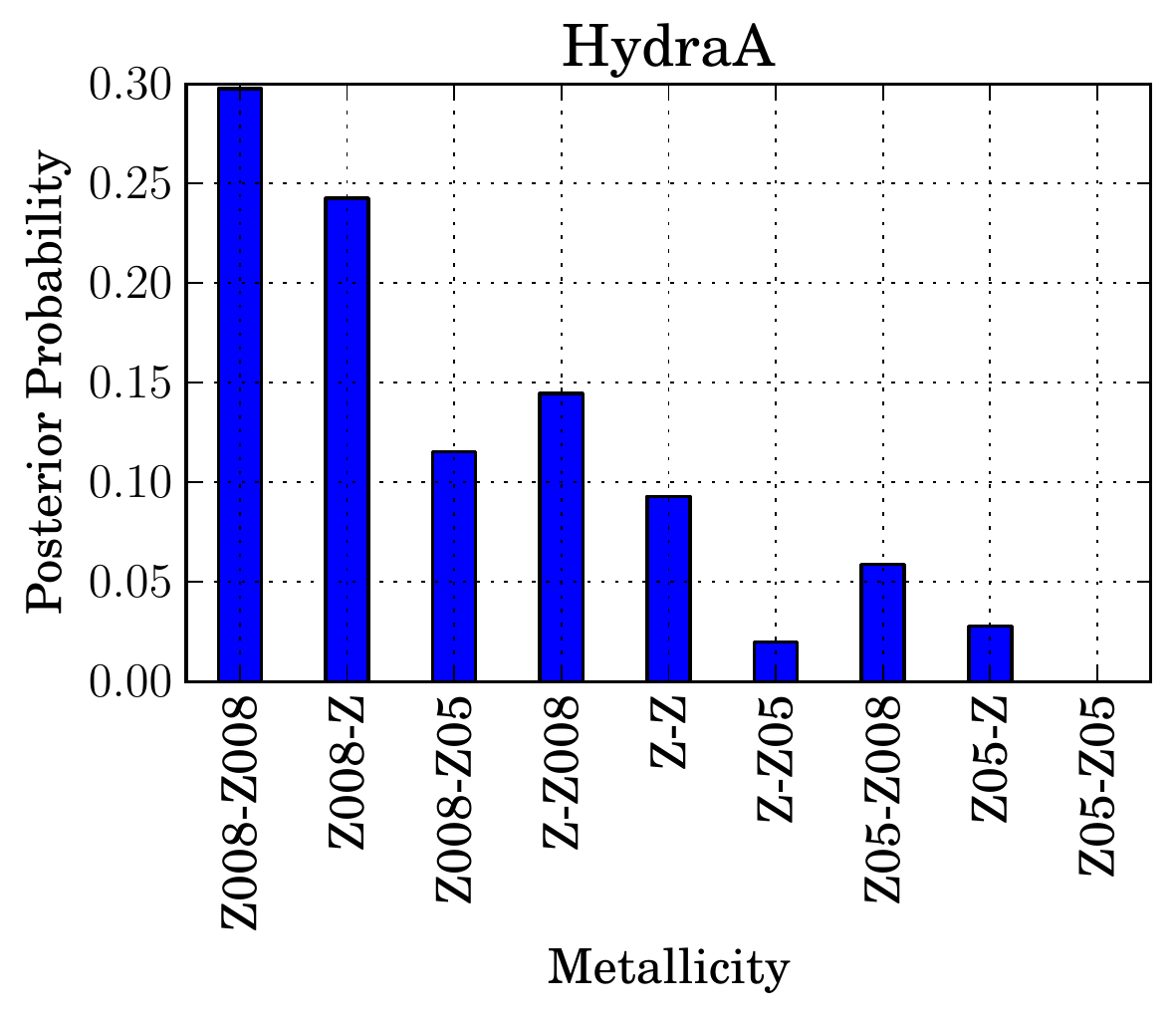}
  \end{minipage}\\
  \begin{minipage}{0.33\textwidth}
    \centering
    \includegraphics[width=\textwidth]{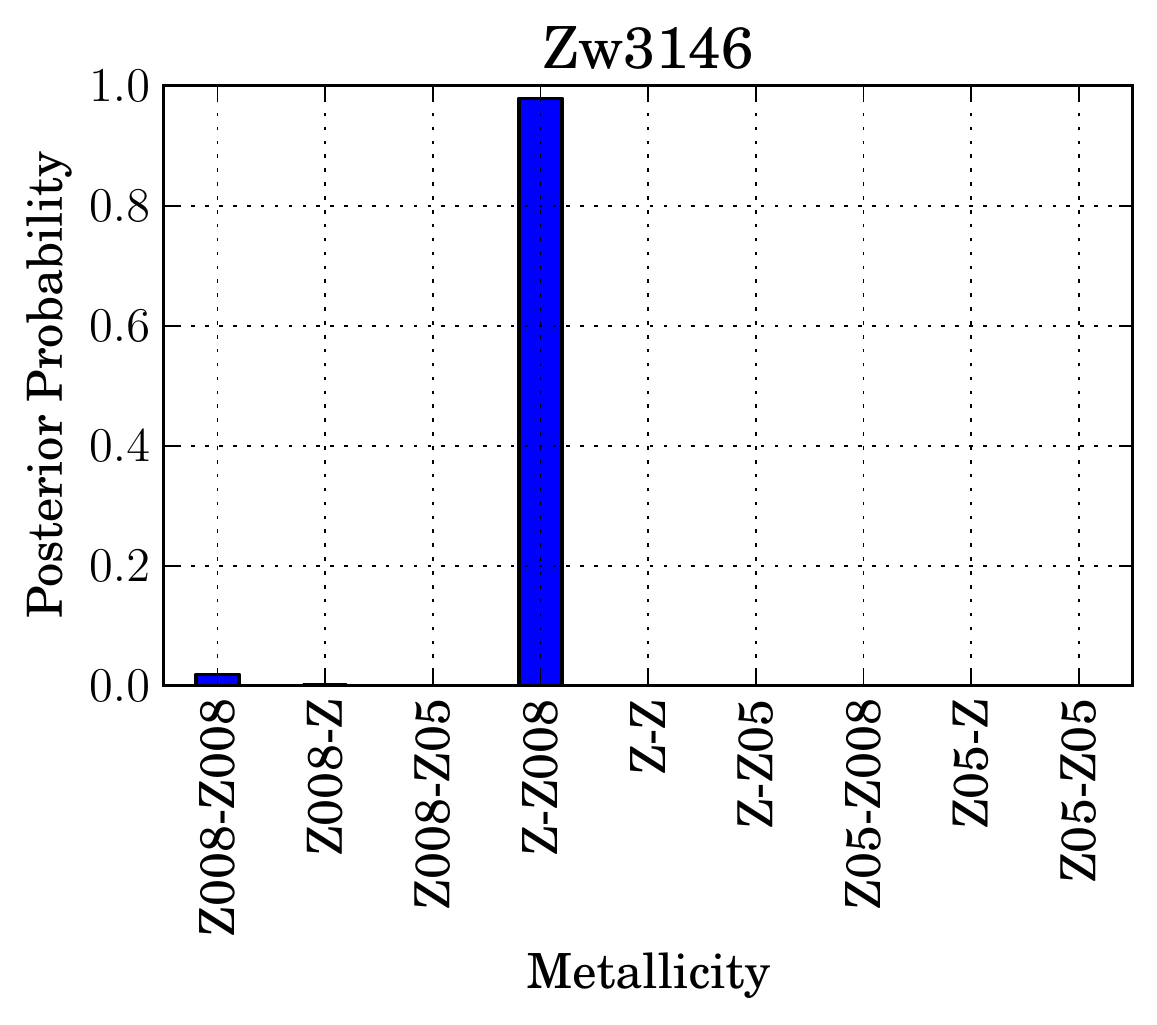}
  \end{minipage}%
  \begin{minipage}{0.33\textwidth}
    \centering
    \includegraphics[width=\textwidth]{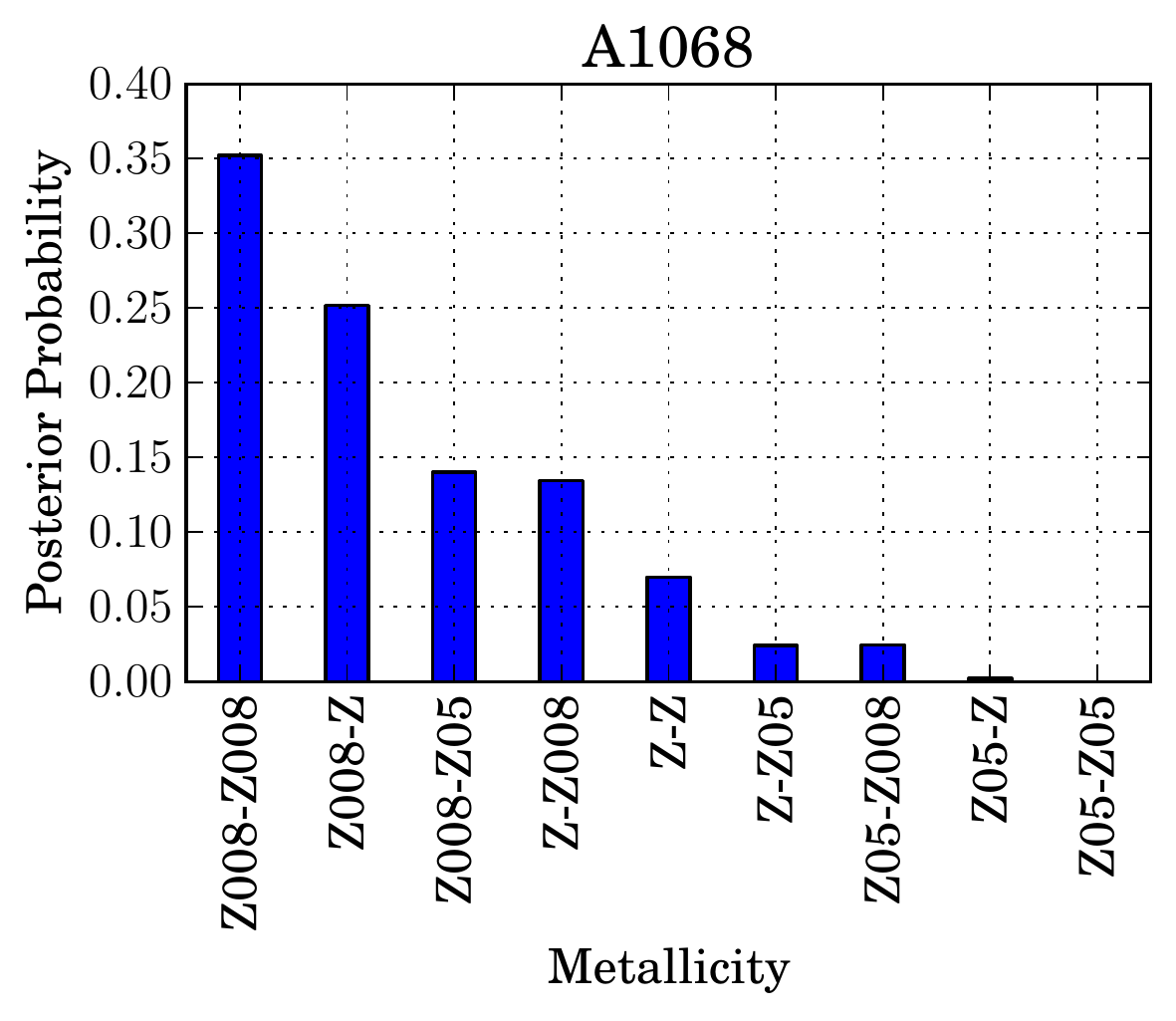}
  \end{minipage}%
  \begin{minipage}{0.33\textwidth}
    \centering
    \includegraphics[width=\textwidth]{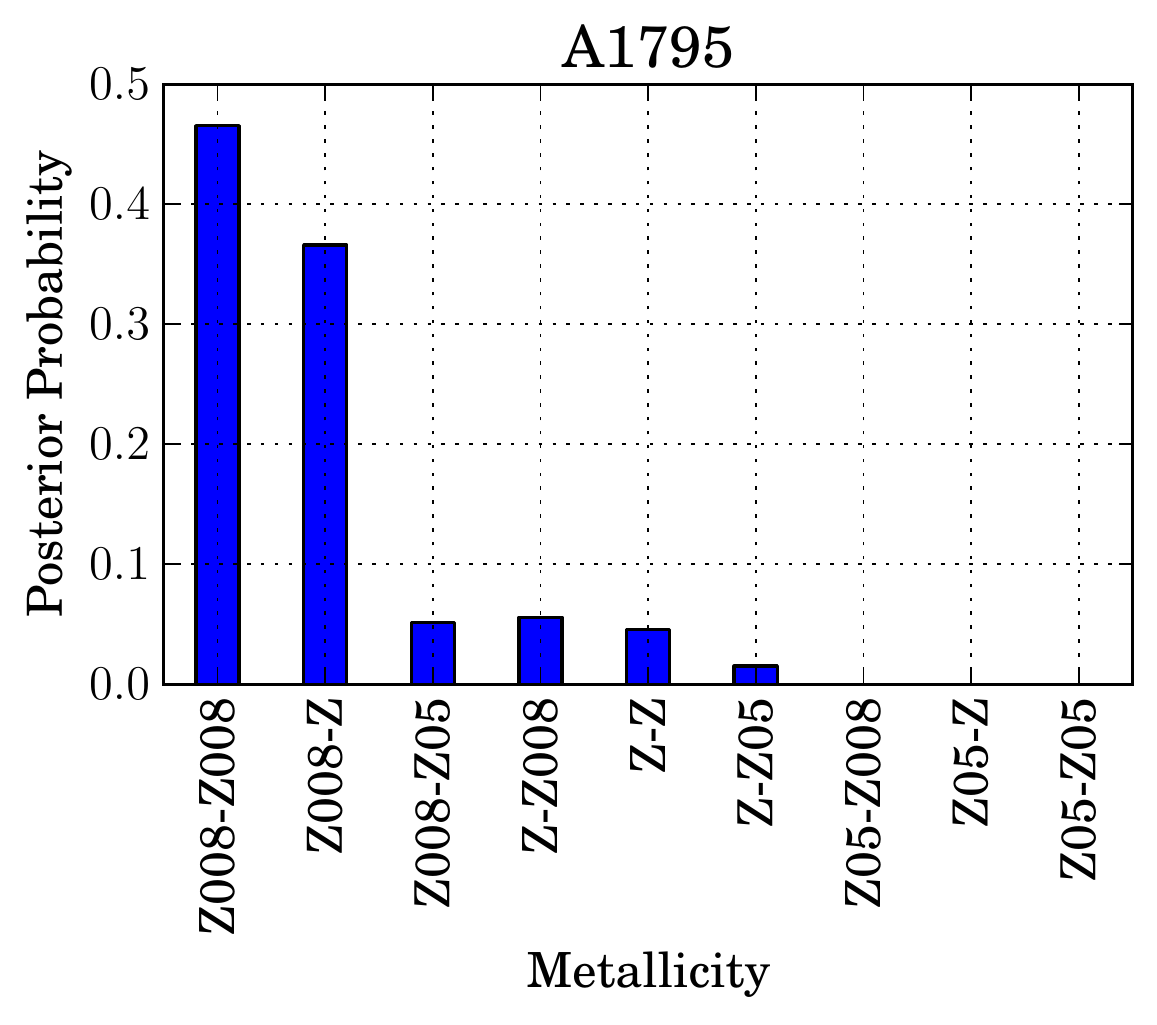}
  \end{minipage}\\
  \begin{minipage}{0.33\textwidth}
    \centering
    \includegraphics[width=\textwidth]{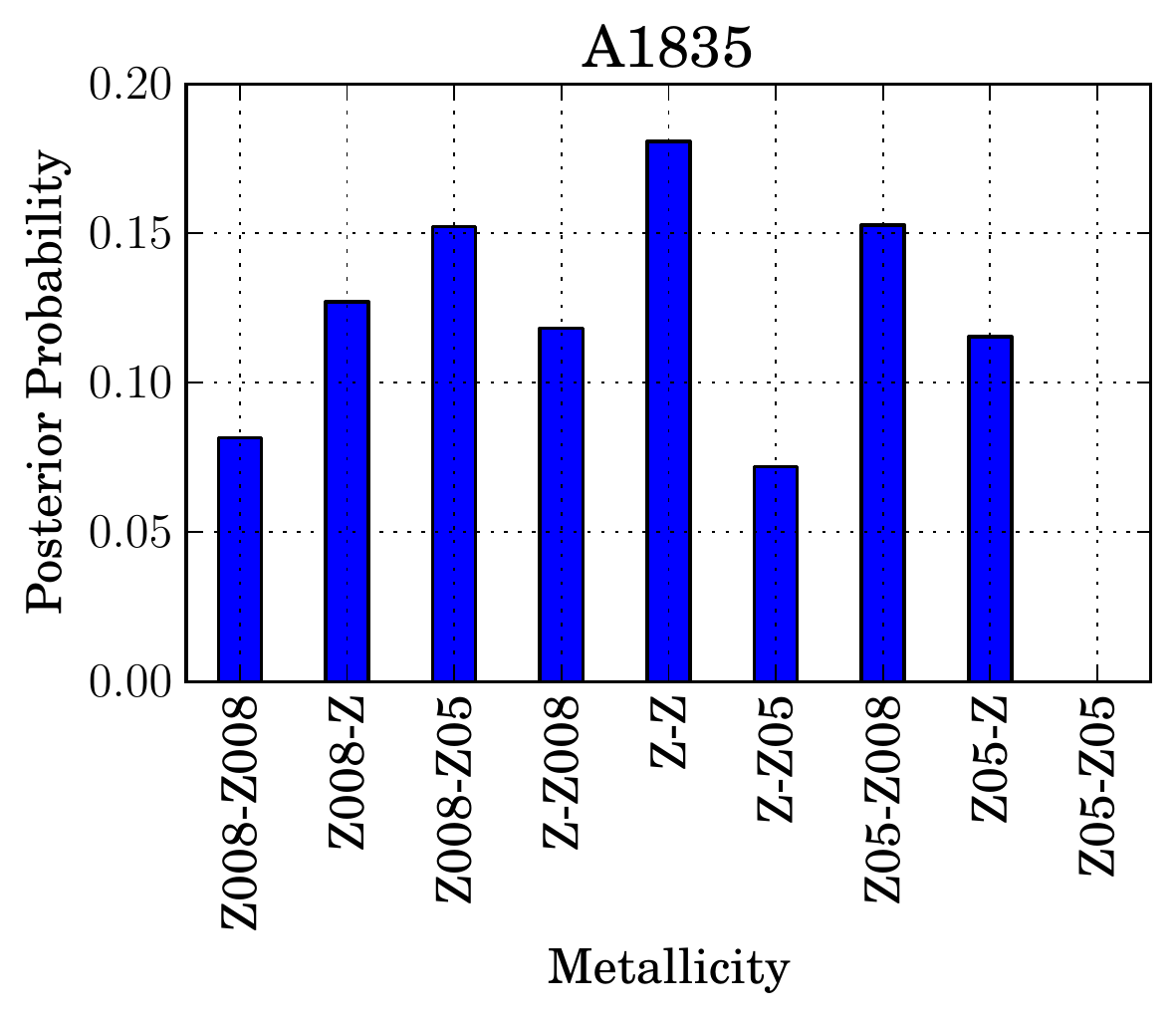}
  \end{minipage}%
  \begin{minipage}{0.33\textwidth}
    \centering
    \includegraphics[width=\textwidth]{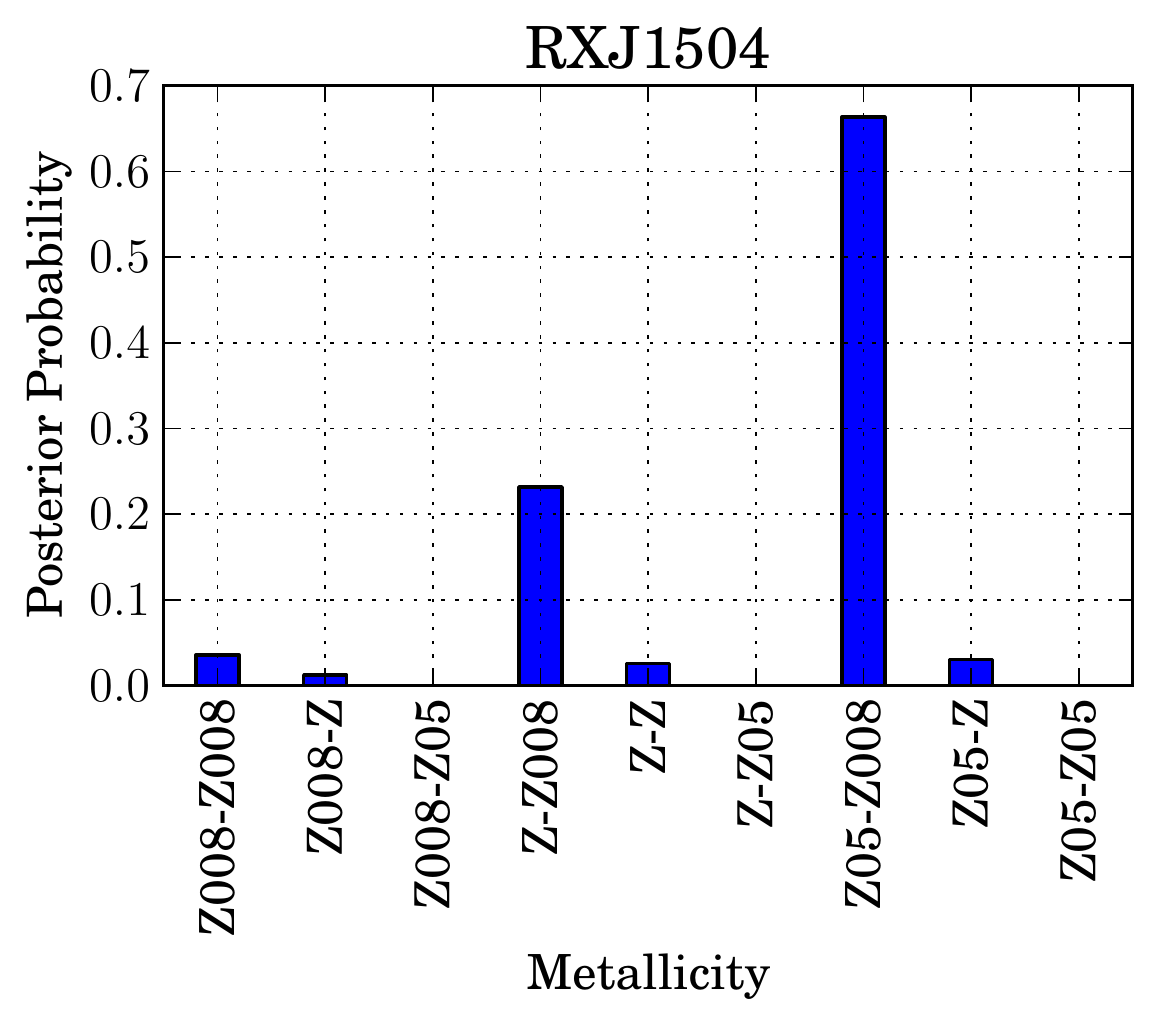}
  \end{minipage}%
  \begin{minipage}{0.33\textwidth}
    \centering
    \includegraphics[width=\textwidth]{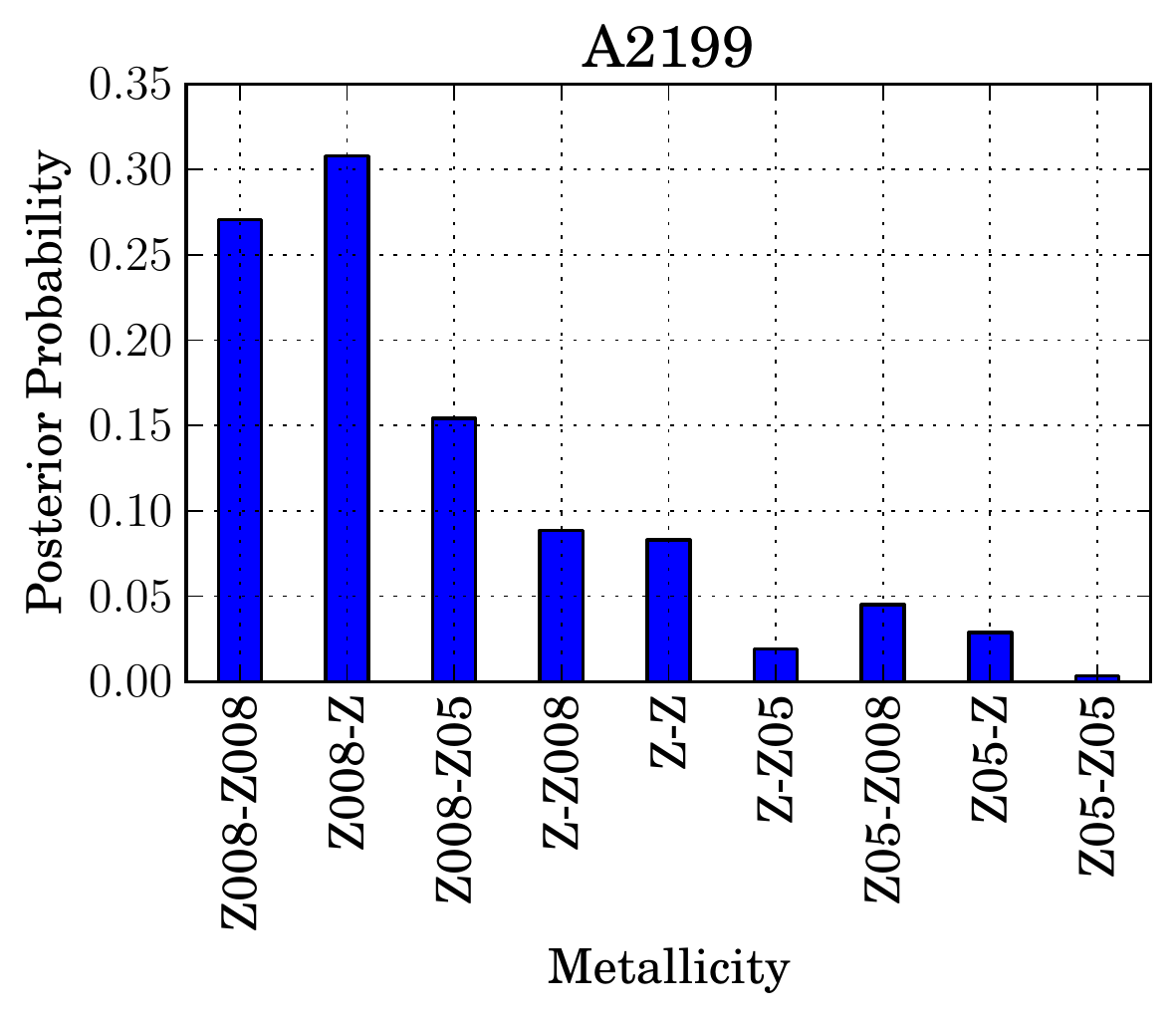}
  \end{minipage}\\
  \begin{minipage}{0.33\textwidth}
    \centering
    \includegraphics[width=\textwidth]{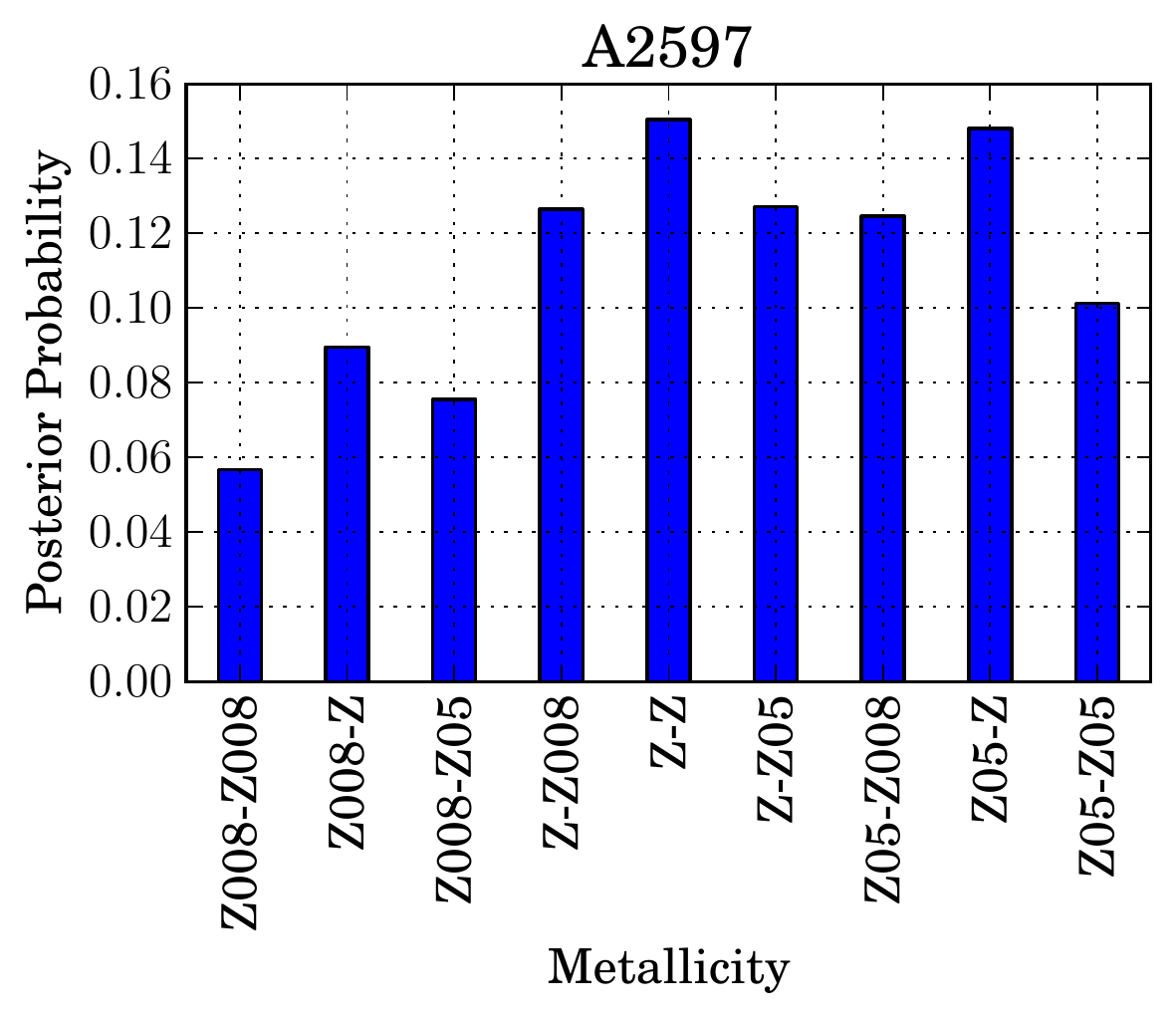}
  \end{minipage}%
  \caption{The posterior probability distributions for the metallicty
    of the young and old stellar populations, marginalized over the
    other model parameters (extinction, YSP age, OSP and YSP masses).
    The labels refer to the metallicity values for the old and young
    stellar populations, respectively.  ``Z008'' means $Z=0.008$,
    ``Z'' means $Z=0.02=Z{\odot}$, and ``Z05'' means $Z=0.05$.  The
    assumed prior probabilities were equal for each of the nine
    combinations.}
  \label{metalZ}
\end{figure*}


 \begin{figure*}
   \begin{minipage}{0.33\textwidth}
     \centering
     \includegraphics[width=\textwidth]{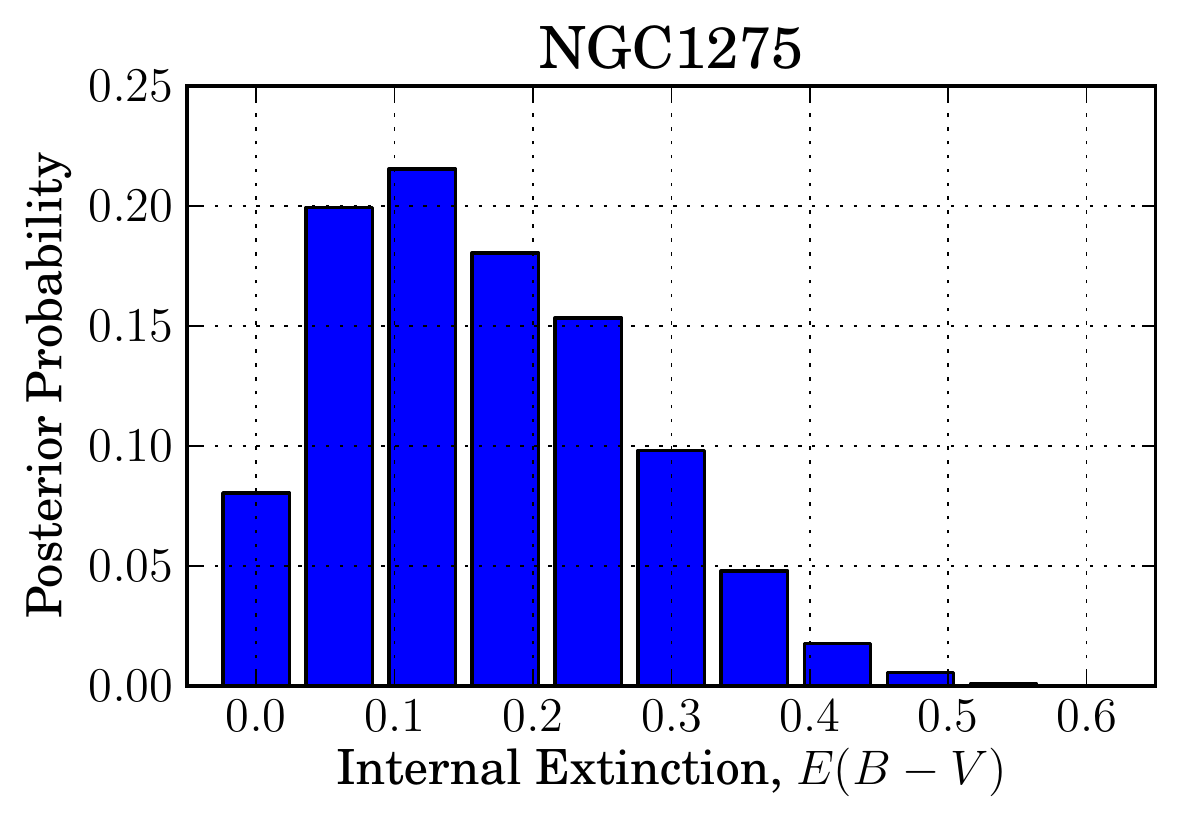}
   \end{minipage}%
   \begin{minipage}{0.33\textwidth}
     \centering
     \includegraphics[width=\textwidth]{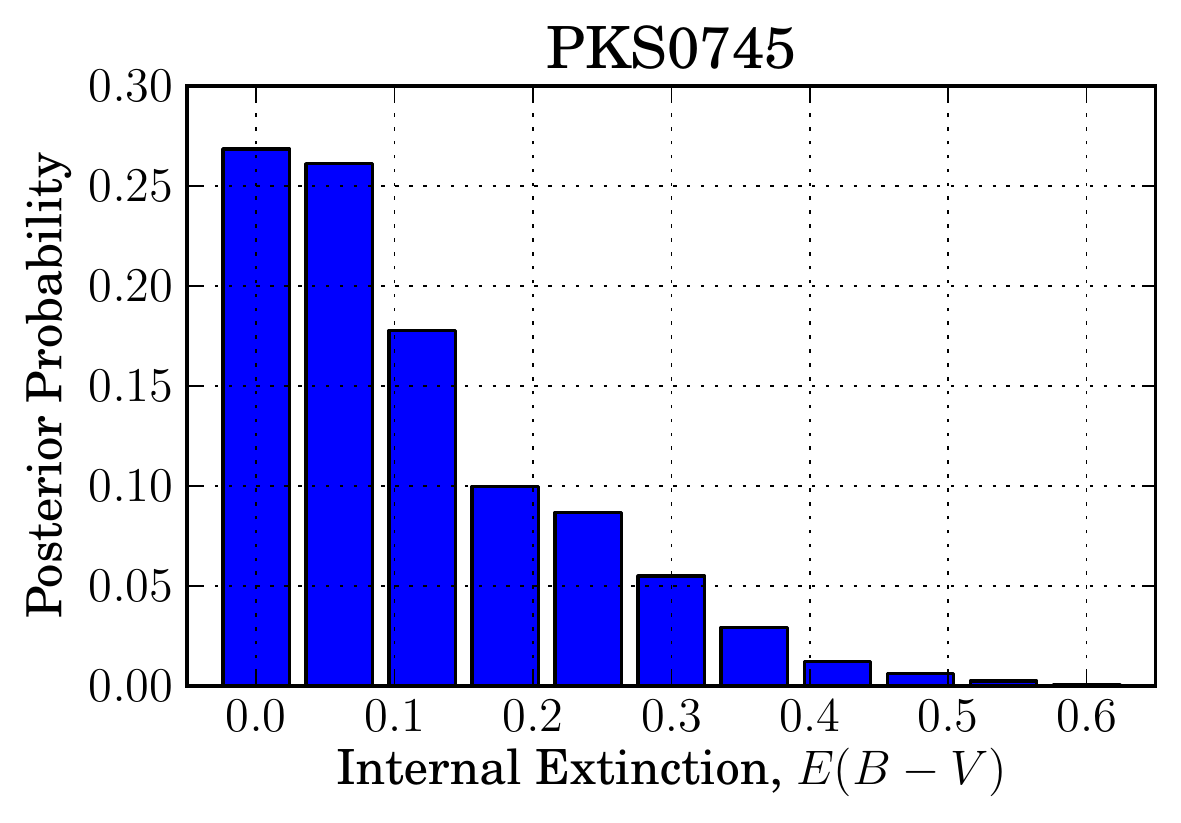}
   \end{minipage}%
   \begin{minipage}{0.33\textwidth}
     \centering
     \includegraphics[width=\textwidth]{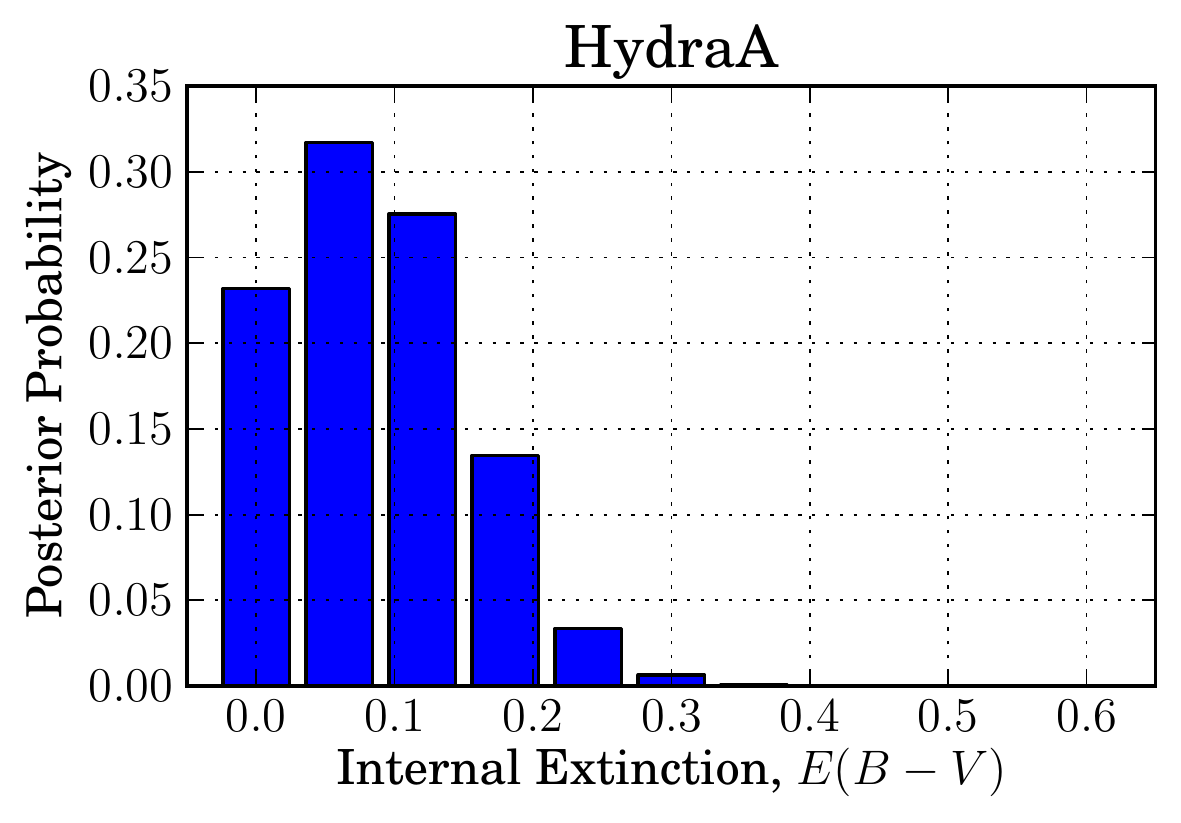}
   \end{minipage}\\
   \begin{minipage}{0.33\textwidth}
     \centering
     \includegraphics[width=\textwidth]{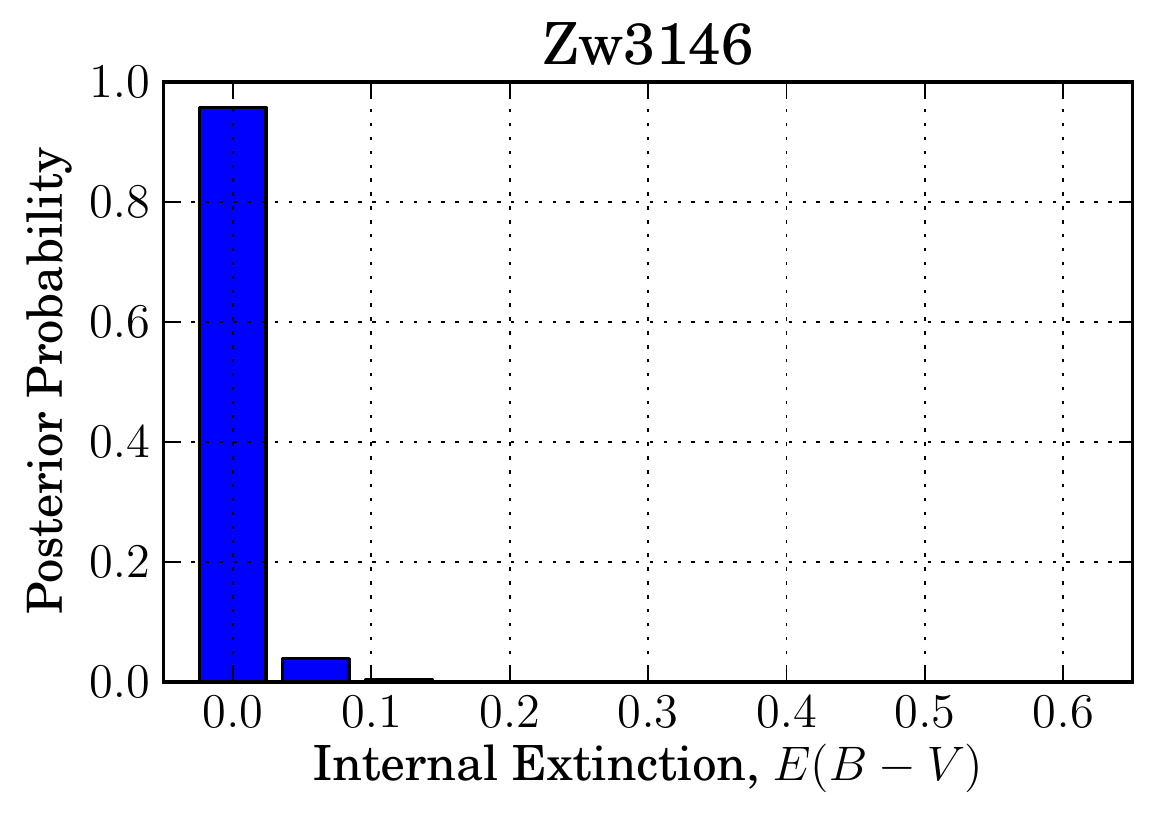}
   \end{minipage}%
   \begin{minipage}{0.33\textwidth}
     \centering
     \includegraphics[width=\textwidth]{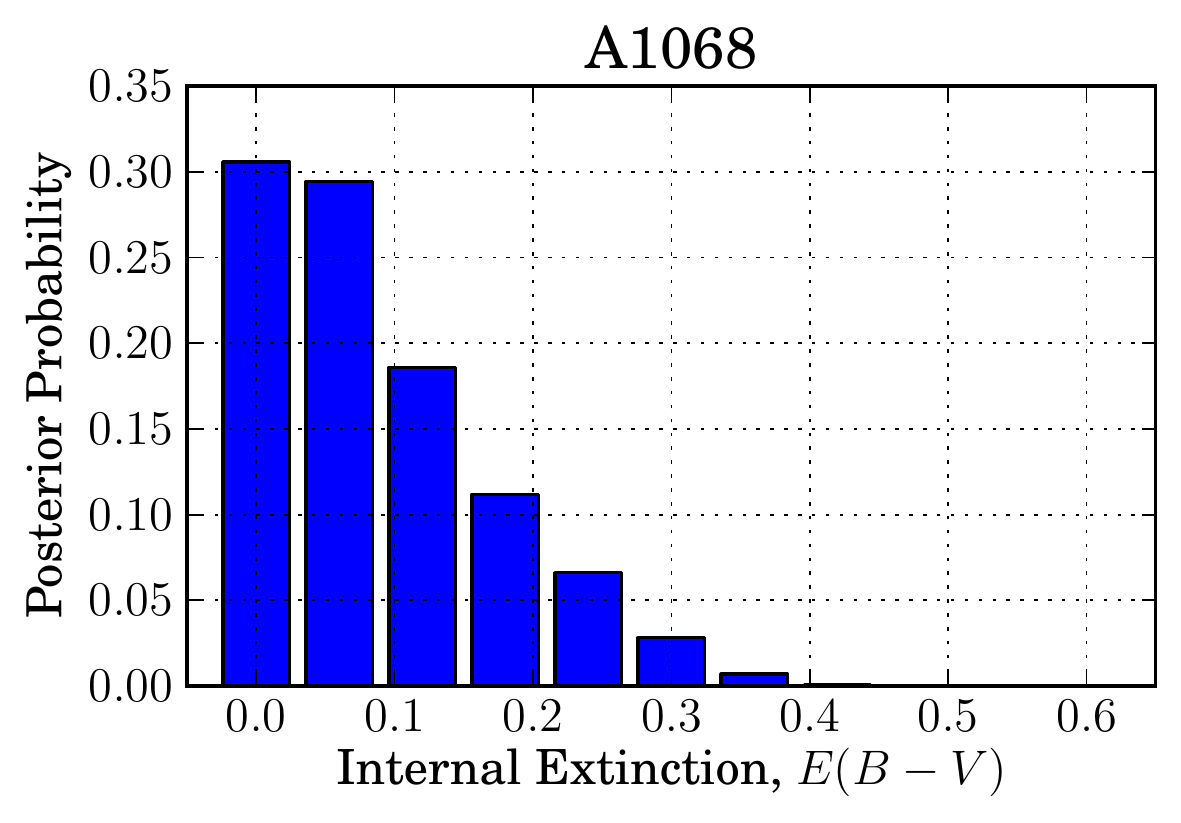}
   \end{minipage}%
   \begin{minipage}{0.33\textwidth}
     \centering
     \includegraphics[width=\textwidth]{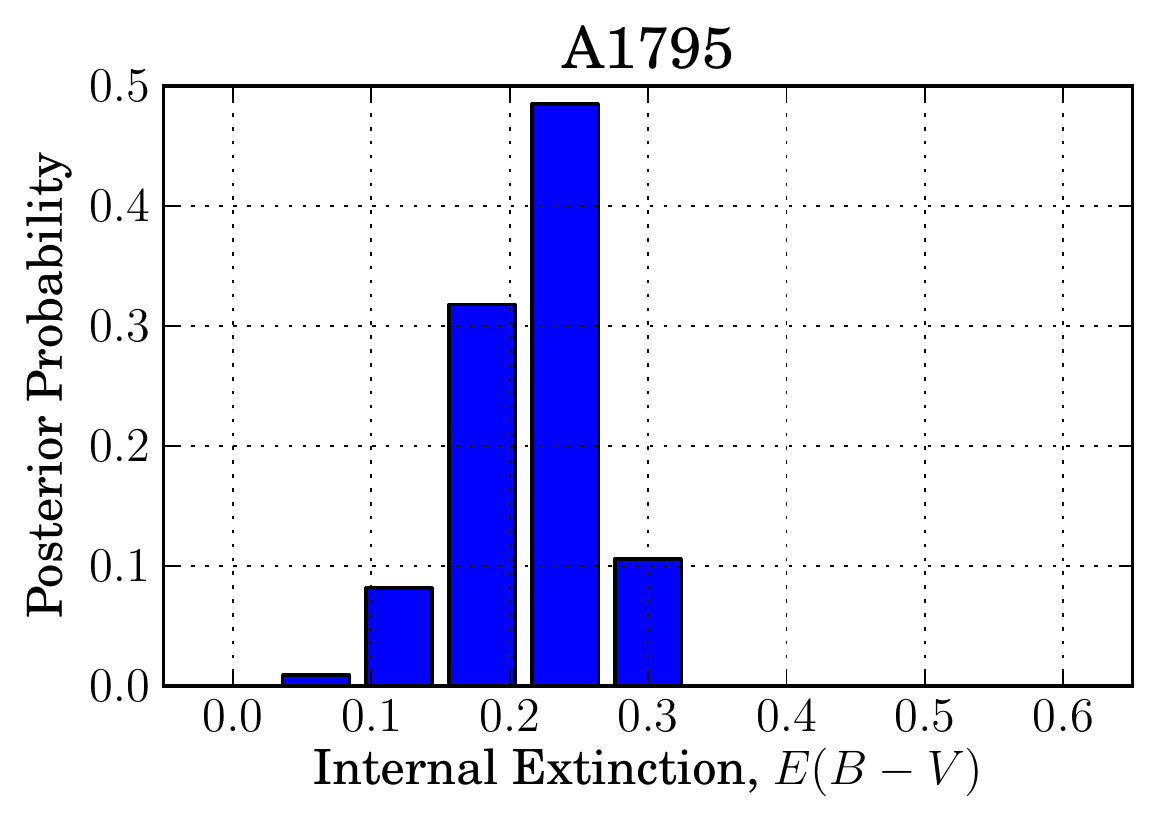}
   \end{minipage}\\
   \begin{minipage}{0.33\textwidth}
     \centering
     \includegraphics[width=\textwidth]{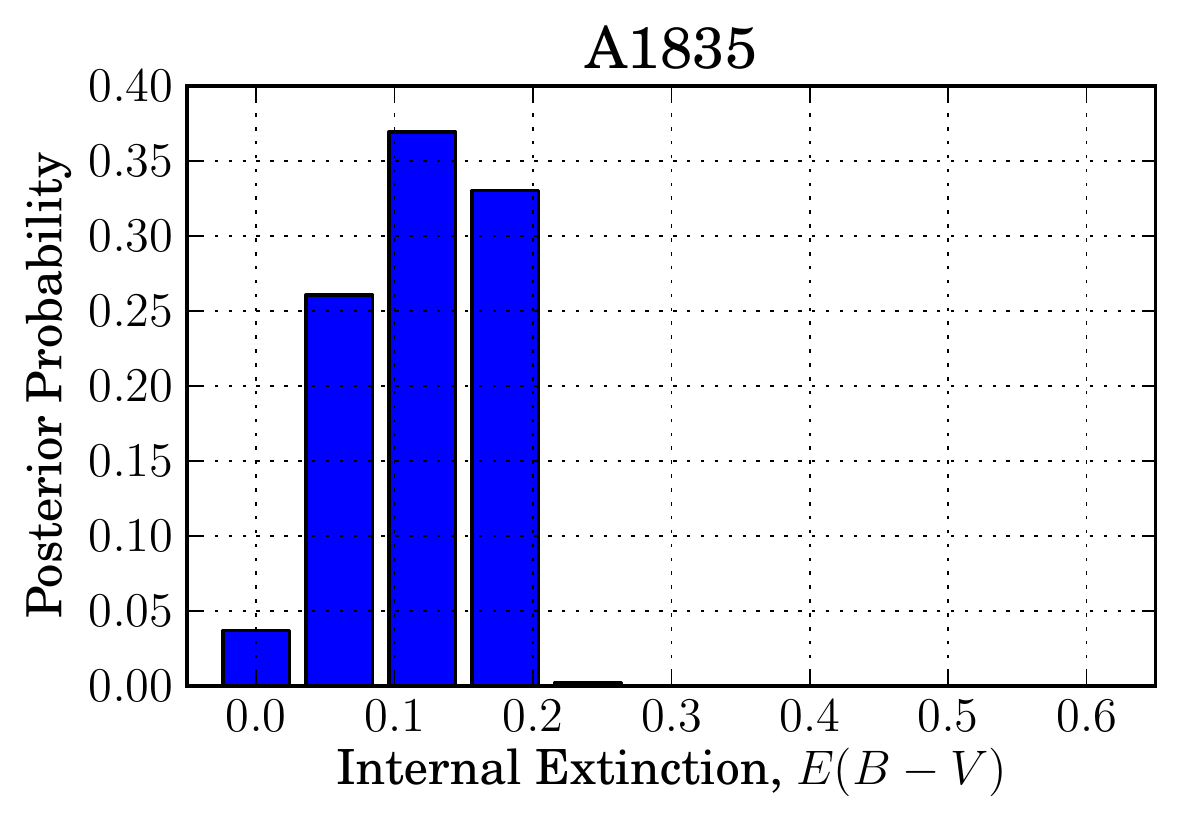}
   \end{minipage}%
   \begin{minipage}{0.33\textwidth}
     \centering
     \includegraphics[width=\textwidth]{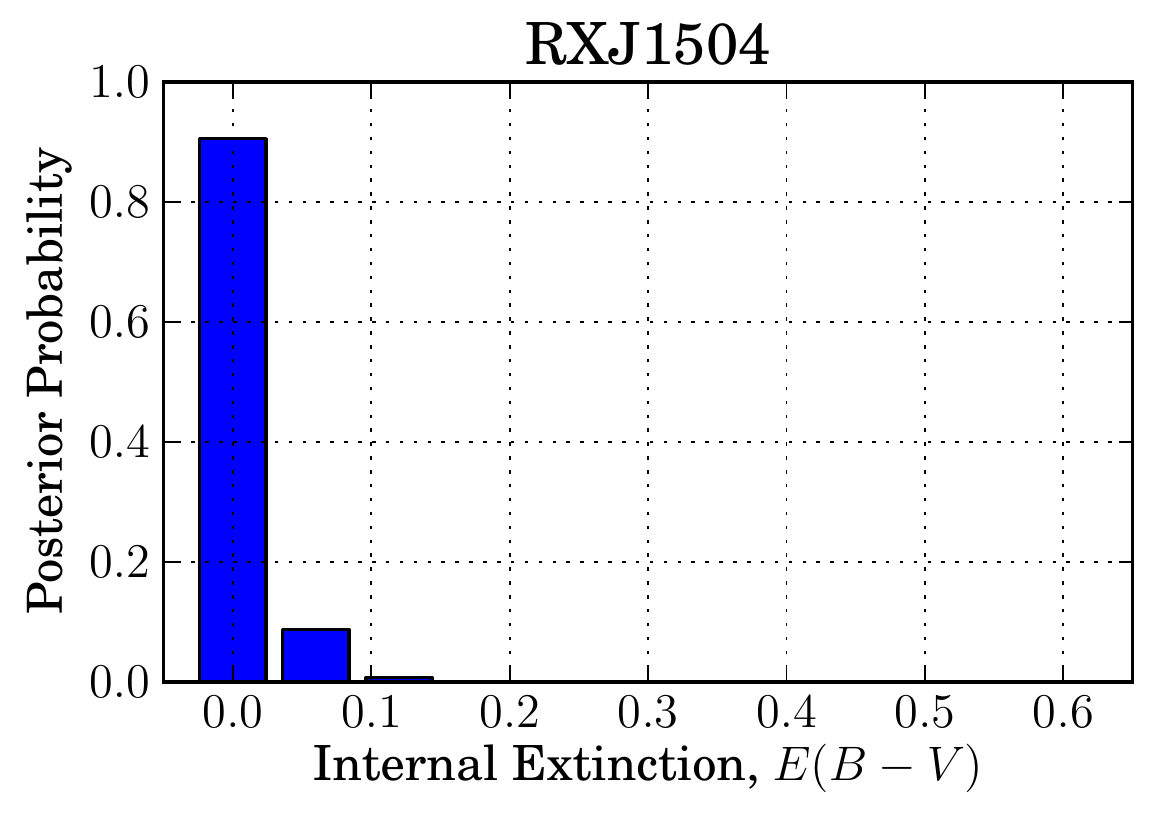}
   \end{minipage}%
   \begin{minipage}{0.33\textwidth}
     \centering
     \includegraphics[width=\textwidth]{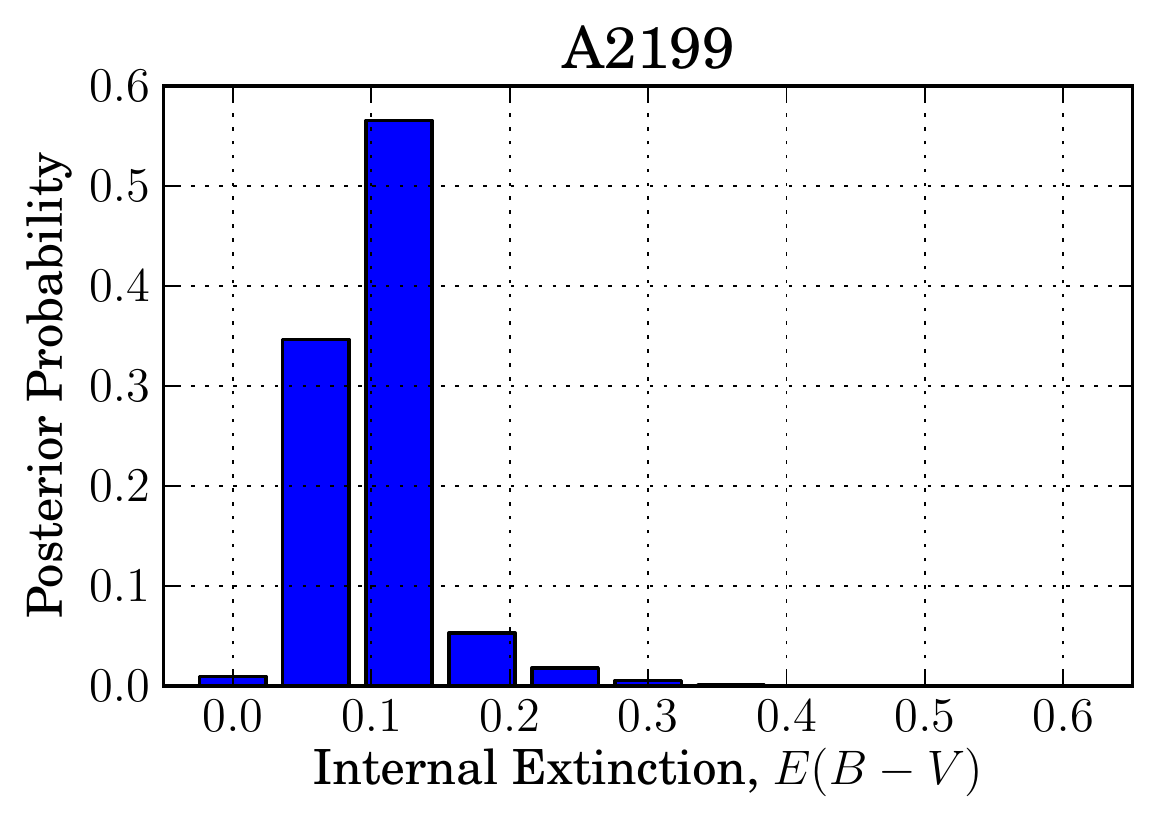}
   \end{minipage}\\
   \begin{minipage}{0.33\textwidth}
     \centering
     \includegraphics[width=\textwidth]{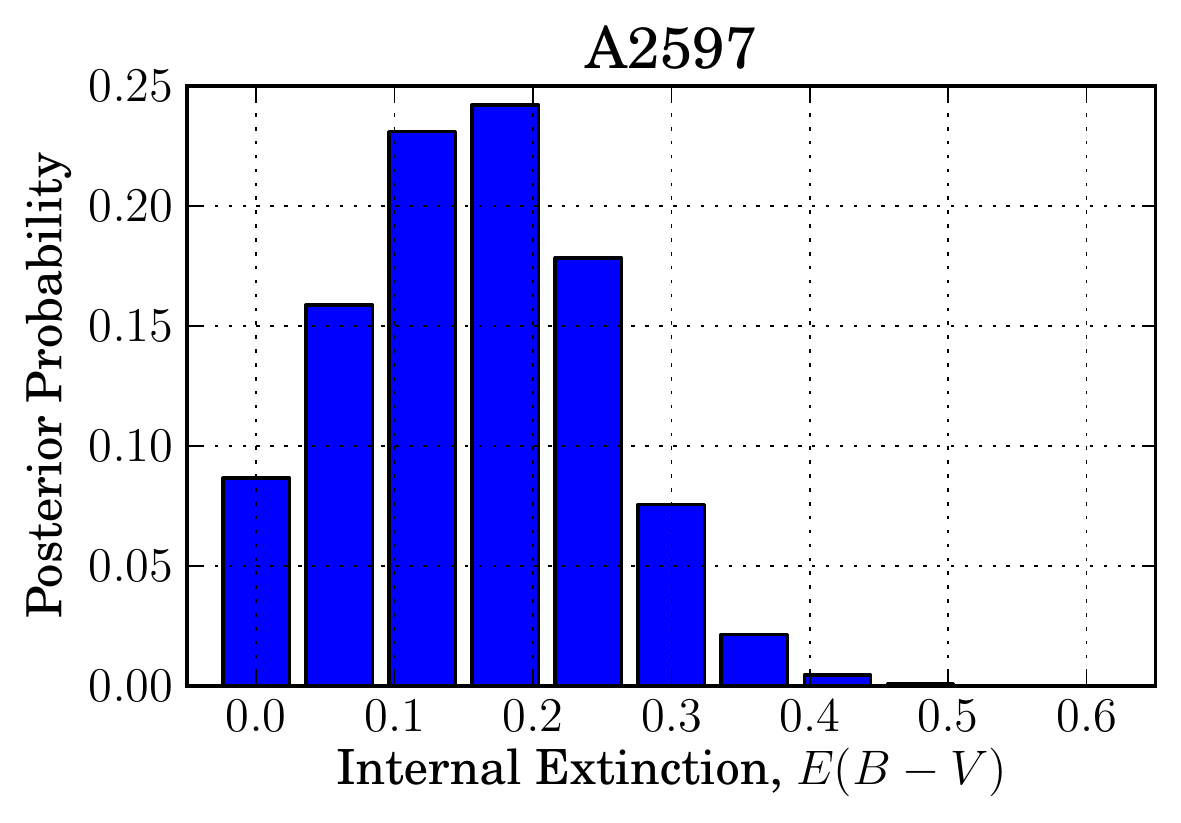}
   \end{minipage}%
   \caption{The posterior probability distributions for the internal
     extinction, $E(B-V)$, after marginalizing over other model
     parametrs (metallicity, YSP age, OSP and YSP masses).  The prior
     probability distribution used for $E(B-V)$ was uniform.}
   \label{ext}
 \end{figure*}

%% file: table13_mdrs.tex
\setlength\LTleft{-2.5in}
\setlength\LTright\fill
\begin{landscape}
  \begin{longtable}{| l | l | l | l | l | l | l |}
    \caption{A comparison of X-ray mass deposition rates and star formation rates.}
    \label{mdrs}\\
    \hline
Cluster      &  $^a$SPS-SFR           &  CMDR                    &  SMDR                    & SMDR                      & Star Formation Rates                                                                                                                                &  Average SFR  \\
             & 		     	      &  {\it Chandra}           &  {\it Chandra}           & {\it XMM/FUSE}            & (literature)	  	   													                      &  (Literature) \\
             & $\mpy~$h$_{71}^{-2}$   & $\mpy~$h$_{71}^{-2}$     & $\mpy$                   & $\mpy$                    & $\mpy~$h$_{71}^{-2}$                                                                                                                                & $\mpy~$h$_{71}^{-2}$\\
 \hline\hline
 \endfirsthead
 \caption{continued.}\\
\hline						         					      	   		     	 									 	       				                        	 
Cluster      &  $^a$SPS-SFR           &  CMDR                    &  SMDR                    & SMDR                      & Star Formation Rates                                                                                                                                &  Average SFR  \\
             & 		     	      &  {\it Chandra}           &  {\it Chandra}           & {\it XMM/FUSE}            & (literature)	  	   													                      &  (Literature) \\
             & $\mpy~$h$_{71}^{-2}$   & $\mpy~$h$_{71}^{-2}$ 	 &  $\mpy$                  & $\mpy$                    & $\mpy~$h$_{71}^{-2}$                                                                                                                                & $\mpy~$h$_{71}^{-2}$\\
\hline
\endhead
\hline
\endfoot
NGC1275      &  $71^{+136}_{-53}$     &  $273^{+44}_{-20}$ (d)   &  $54^{+48}_{-18}$    (f) &   $32^{+6}_{-6}$     (e)  & $2.3^{+0.2}_{-0.2}$ (e), $15.5^{+5.2}_{-5.2}$ (e), $37$ (e), $24^{+1}_{-1}$ (i), $27$ (i), $70^{+7}_{-7}$ (n)  	                              &  $30 \pm 23$    \\[4pt]    
PKS0745      &  $24^{+40}_{-21}$      &  $224^{+66}_{-81}$ (c)   &  $170^{+90}_{-90}$   (e) &   $317^{+35}_{-29}$  (g)  & $16.9^{+5.6}_{-5.6}$ (e), $129^{+7}_{-7}$ (g), $237^{+13}_{-13}$ (g), $17$ (k), $11$ (m), $8.5^{+0.9}_{-0.9}$ (n) 	   	                      &  $70 \pm 94$     \\[4pt]    
HydraA       &  $18^{+15}_{-12}$      &  $179^{+13}_{-13}$ (b)   &  $16^{+5}_{-5}$      (e) &   $8^{+6}_{-4}$      (g)  & $\le 0.5$ (e), $\sim 16$ (e), $9.5^{+0.2}_{-0.2}$ (g), $17.5^{+0.4}_{-0.4}$ (g), $4.3$ (m), $2.1^{+0.2}_{-0.2}$ (n)		                      &  $8  \pm 7$     \\[4pt]    
ZwCl3146     &  $14^{+25}_{-7}$       &  $673^{+74}_{-93}$ (c)   &  $590^{+190}_{-170}$ (e) &   $525^{+90}_{-90}$  (g)  & $10.7$ (e), $\le 110$ (e), $91^{+4}_{-4}$ (g), $168^{+8}_{-8}$ (g), $44^{+14}_{-14}$ (j), $12.4$ (l), $31.3^{+3.1}_{-3.1}$ (n)                      &  $67 \pm 59$    \\[4pt]    
A1068	     &  $8^{+12}_{-10}$       &  $231^{+72}_{-51}$ (c)   &  $\le 48$            (e) &   ...	      	        & $18.1$ (e), $28^{+12}_{-12}$ (e), $46^{+21}_{-21}$ (e), $60^{+20}_{-20}$ (j), $188$ (k), $100$ (m), $119.4^{+11.9}_{-11.9}$ (n)                     &  $80 \pm 60$    \\[4pt]    
A1795	     &  $21^{+8}_{-15}$       &  $330^{+15}_{-15}$ (b)   &  $8^{+13}_{-7}$      (e) &   $26^{+7}_{-7}$     (e)  & $1^{+0.1}_{-0.1}$ (e), $1.1$ (e), $2.1^{+0.9}_{-0.9}$ (e), $6.3$ (e), $23.2$ (e), $9.1^{+0.2}_{-0.2}$ (g), $16.7^{+0.4}_{-0.4}$ (g), $2.3$ (m)      &  $8 \pm 8$      \\[4pt]
A1835	     &  $54^{+19}_{-36}$      &  $720^{+270}_{-301}$ (c) &  $100^{+100}_{-70}$  (h) &   $34^{+43}_{-34}$   (g)  & $48.9$ (e), $79$ (e), $79.5$ (e), $140^{+40}_{-40}$ (e), $123^{+5}_{-5}$ (g), $226^{+9}_{-9}$ (g), $11.7$ (l), $270$ (m), $97.2^{+9.7}_{-9.7}$ (n)  &  $119 \pm 83$   \\[4pt]
RXC~J1504    &  $67^{+49}_{-27}$      &  $2327^{+130}_{-130}$ (b)&  $\le7.4$            (b) &   $73$ (o)	        & $136$ (o), $262$ (o), $314$ (o) 				 				 			     	       	   	      &  $237 \pm 92$   \\[4pt]    
A2199	     &  ...      	      &  $72^{+12}_{-1}$ (b)     &  $2^{+1}_{-1}$       (b) &   ...		        & $0.10^{+0.03}_{-0.03}$ (e), $0.10$ (e), $0.4^{+0.1}_{-0.1}$ (n) 			 					   		      &  $0.2 \pm 0.1$  \\[4pt]    
A2597        &   $12^{+11}_{-9}$      &  $235^{+36}_{-36}$ (b)   &  $30^{+30}_{-20}$    (b) &   $22$		   (e)  & $2.3^{+1.3}_{-1.3}$ (e), $6.4$ (e), $22.3$ (e), $2^{+1}_{-1}$ (j), $5.4$ (m), $1.4^{+0.2}_{-0.2}$ (n)  	     	  	   		      &  $5 \pm 8$       \\
\hline
\multicolumn{7}{|p{22cm}|}{a,~this work; b,~Hudson et al. (2010), unpublished; c,~Allen (2000); d,~Allen et al. (2001); e,~Rafferty et al. (2006), and references therein; f,~Birzan et al. (2004);  
g,~Hicks et al. (2005); h,~Egami et al. (2006); i,~Mittal et al. (2012); j,~Edge et al. (2010); k,~O'Dea et al. (2008); l,~O'Dea et al. (2010); m,~Donahue et al. (2011); n,~Hoffer et al. (2012);
o, Ogrean et al. (2010).}\\
    \hline
 \end{longtable}
\end{landscape}

\nocite{Allen2000} 
\nocite{Allen2001}  
\nocite{Edge2010a} 
\nocite{Birzan2004} 
\nocite{Rafferty2006} 
\nocite{paperIII} 
\nocite{Hicks2005} 
\nocite{Egami2006b} 
\nocite{Mittal2012} 
\nocite{ODea2008}
\nocite{ODea2010}
\nocite{Donahue2011}
\nocite{Hoffer2011}
\nocite{Ogrean2010}

%% file: bestfitplots.tex

\begin{figure*}
  \begin{minipage}{0.33\textwidth}
    \centering
    \includegraphics[width=1.1\textwidth]{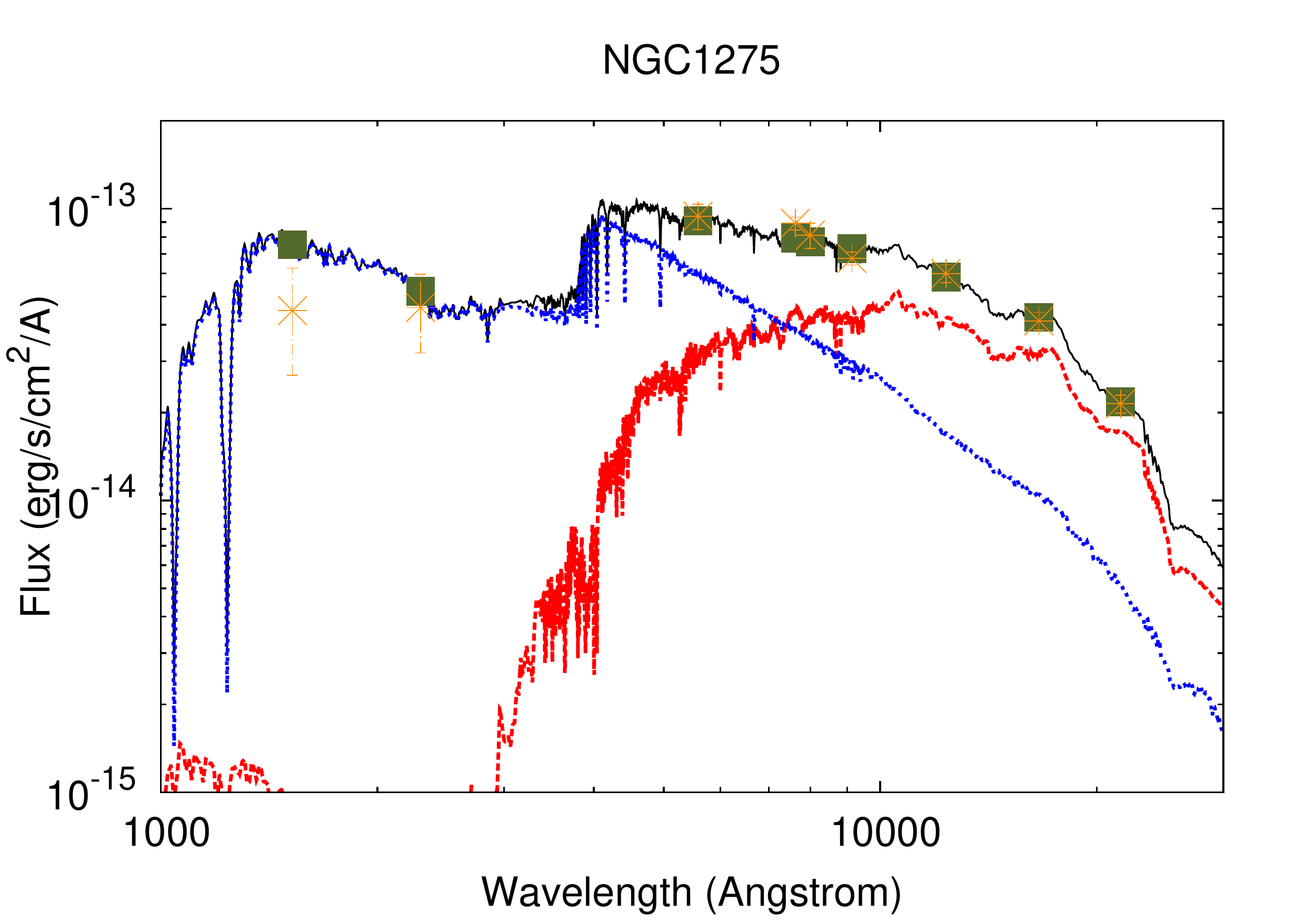}
  \end{minipage}\hfill
  \begin{minipage}{0.33\textwidth}
    \centering
    \hspace*{0.1cm}
    \includegraphics[width=1.1\textwidth]{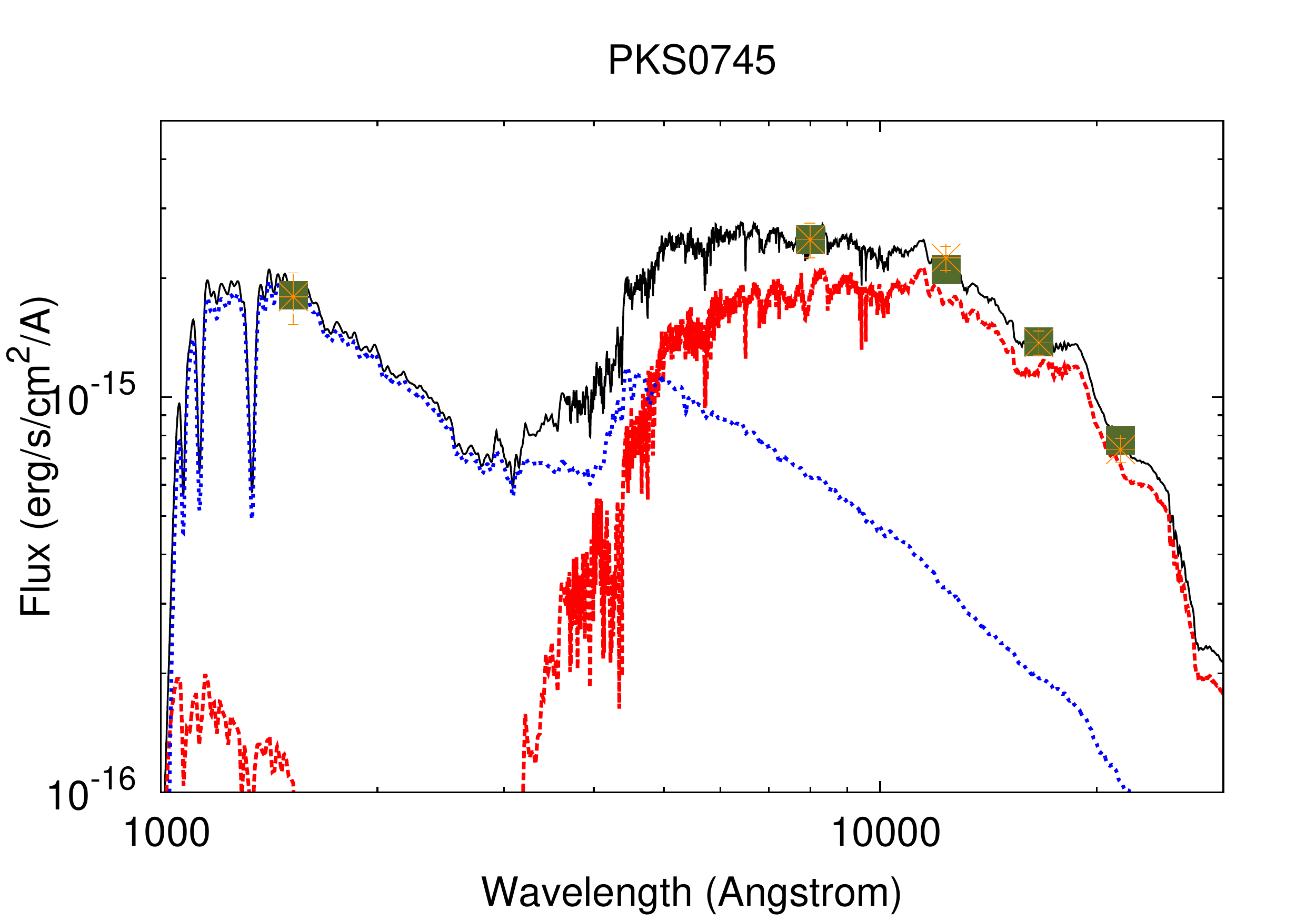}
  \end{minipage}\hfill
  \begin{minipage}{0.33\textwidth}
    \centering
    \hspace*{0.2cm}
    \includegraphics[width=1.1\textwidth]{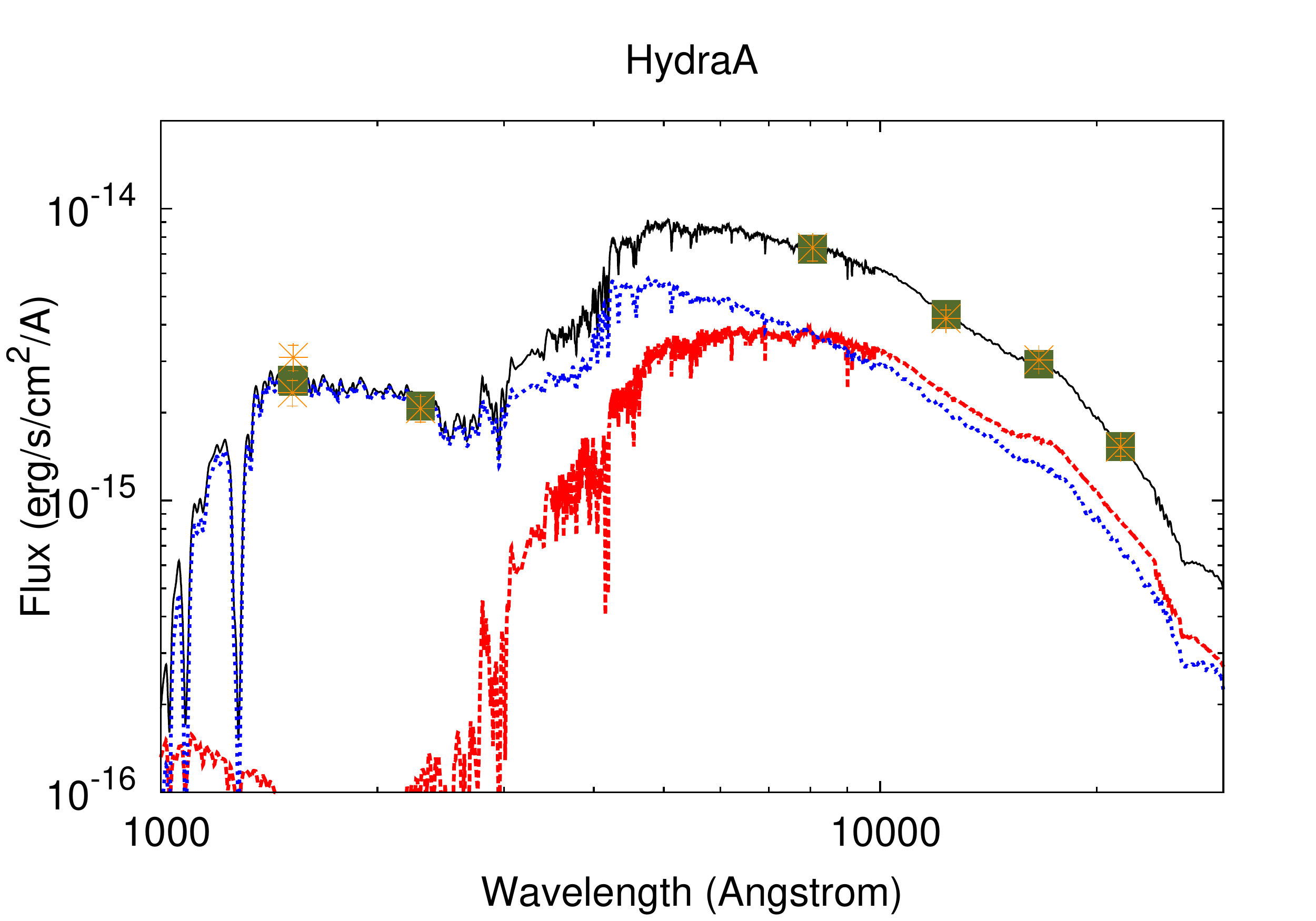}
  \end{minipage}\\[8pt]
  \begin{minipage}{0.33\textwidth}
    \centering
    \includegraphics[width=1.1\textwidth]{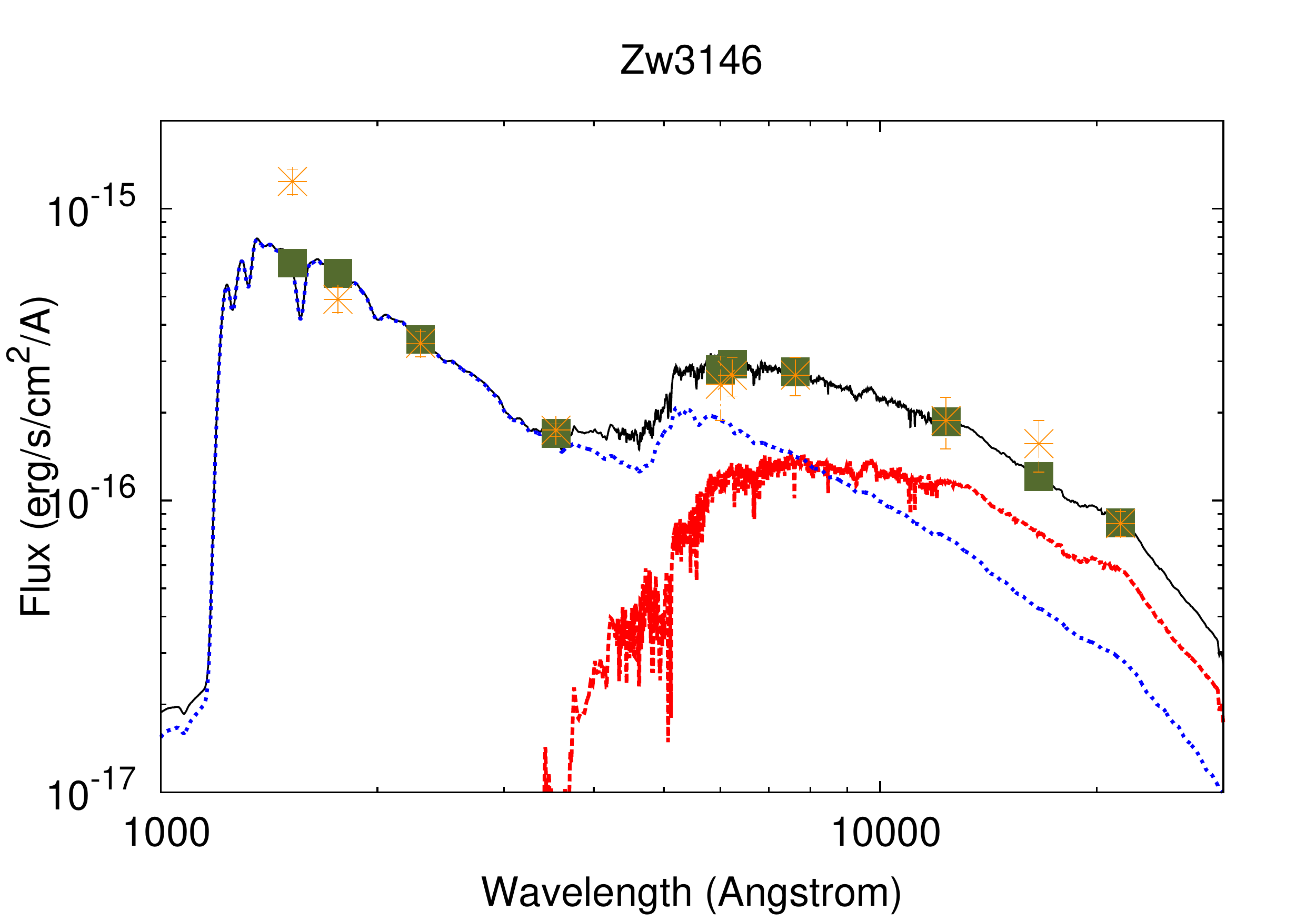}
  \end{minipage}\hfill
  \begin{minipage}{0.33\textwidth}
    \centering
    \hspace*{0.1cm}
    \includegraphics[width=1.1\textwidth]{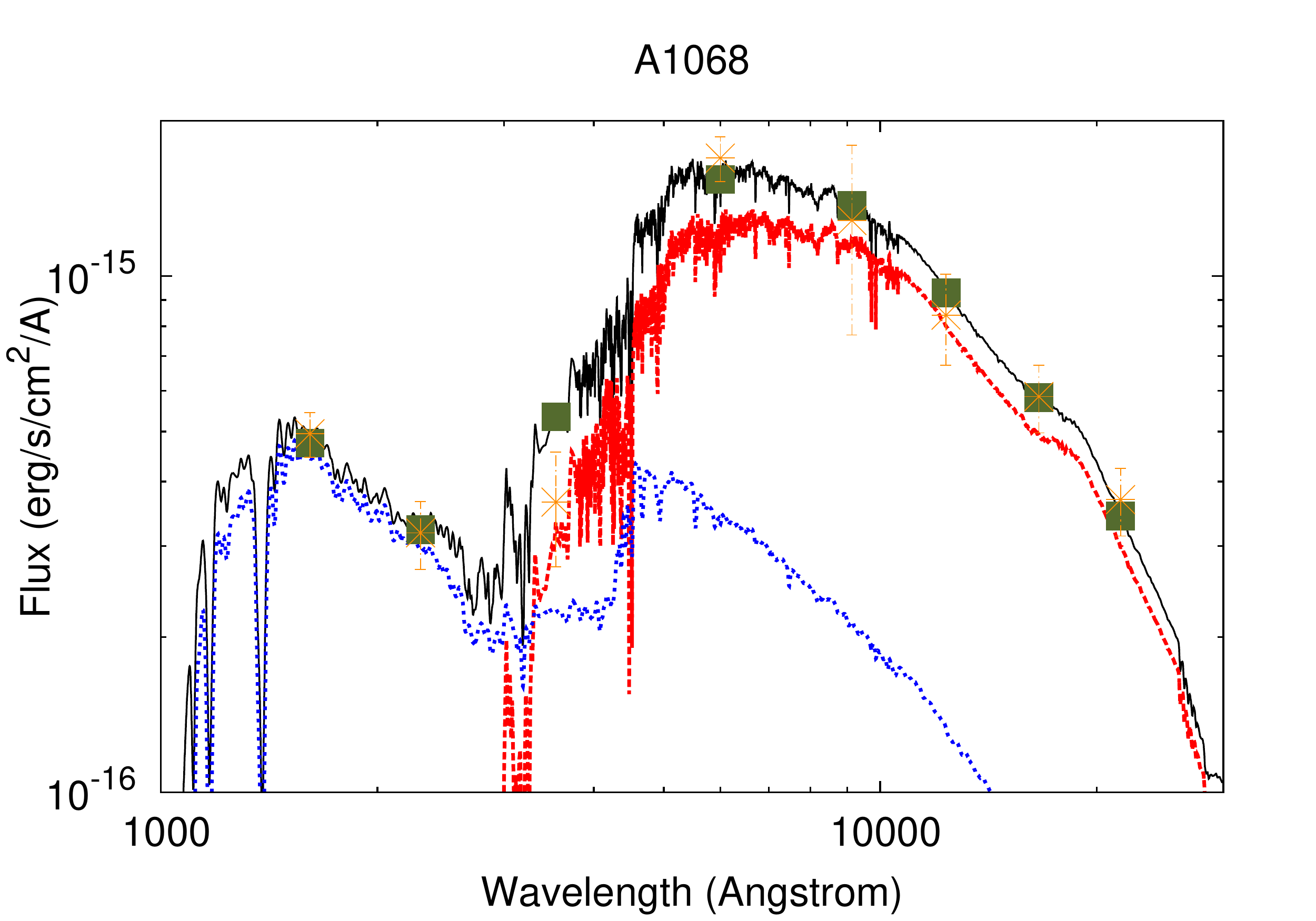}
  \end{minipage}\hfill
  \begin{minipage}{0.33\textwidth}
    \centering
    \hspace*{0.2cm}
    \includegraphics[width=1.1\textwidth]{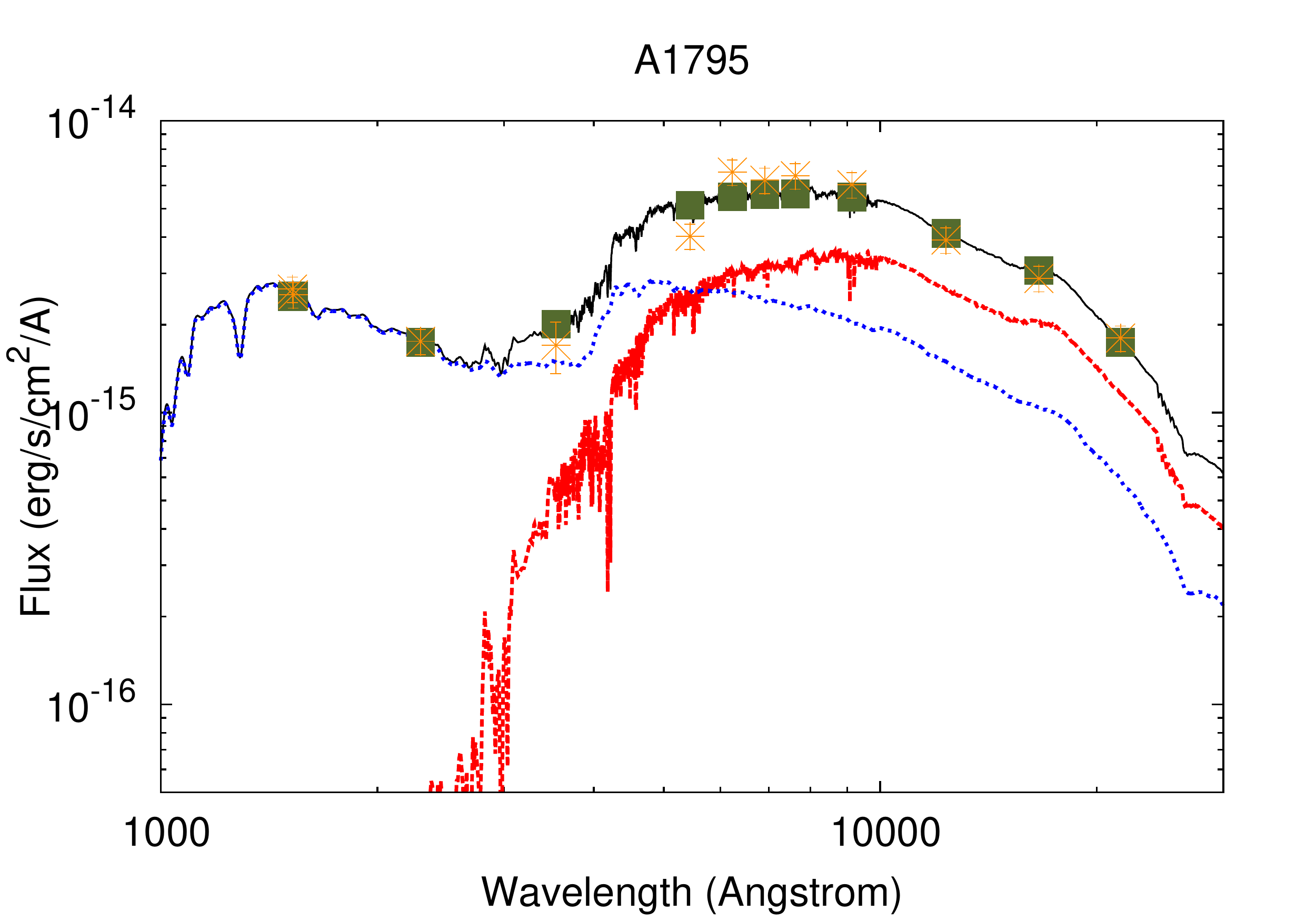}
  \end{minipage}\\[8pt]
  \begin{minipage}{0.33\textwidth}
    \centering
    \includegraphics[width=1.1\textwidth]{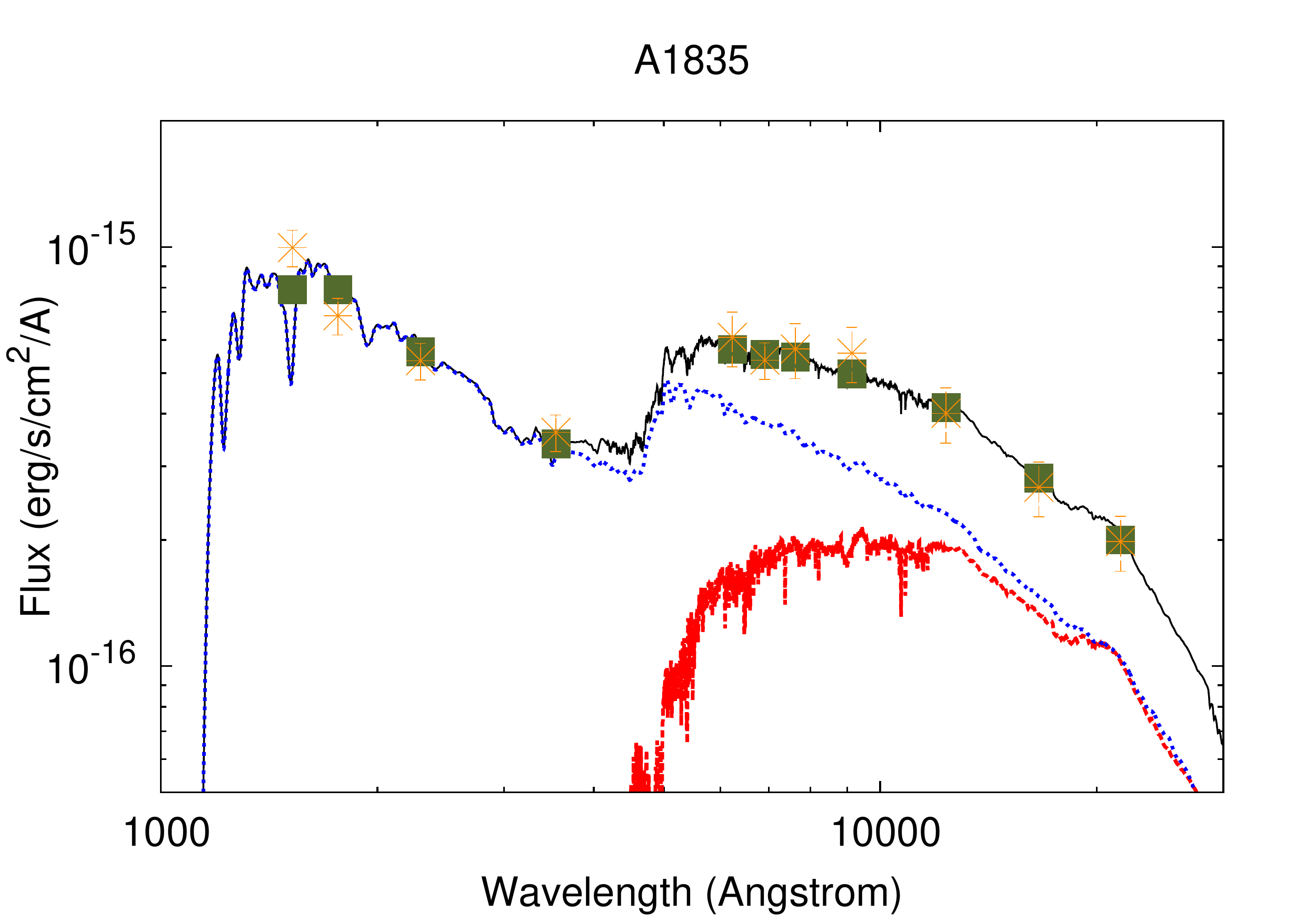}
  \end{minipage}\hfill
  \begin{minipage}{0.33\textwidth}
    \centering
    \hspace*{0.1cm}
    \includegraphics[width=1.1\textwidth]{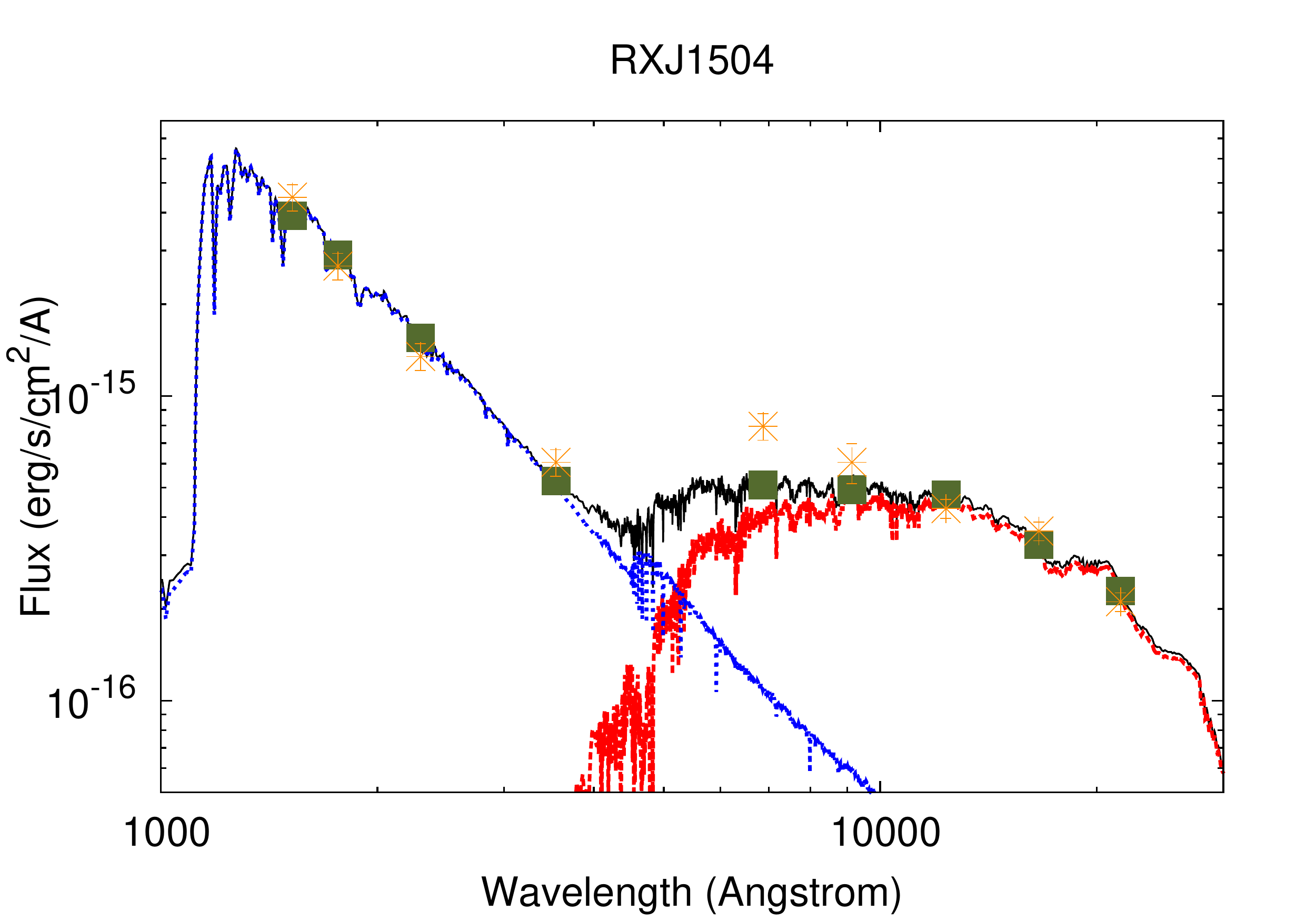}
  \end{minipage}\hfill
  \begin{minipage}{0.33\textwidth}
    \centering
    \hspace*{0.2cm}
    \includegraphics[width=1.1\textwidth]{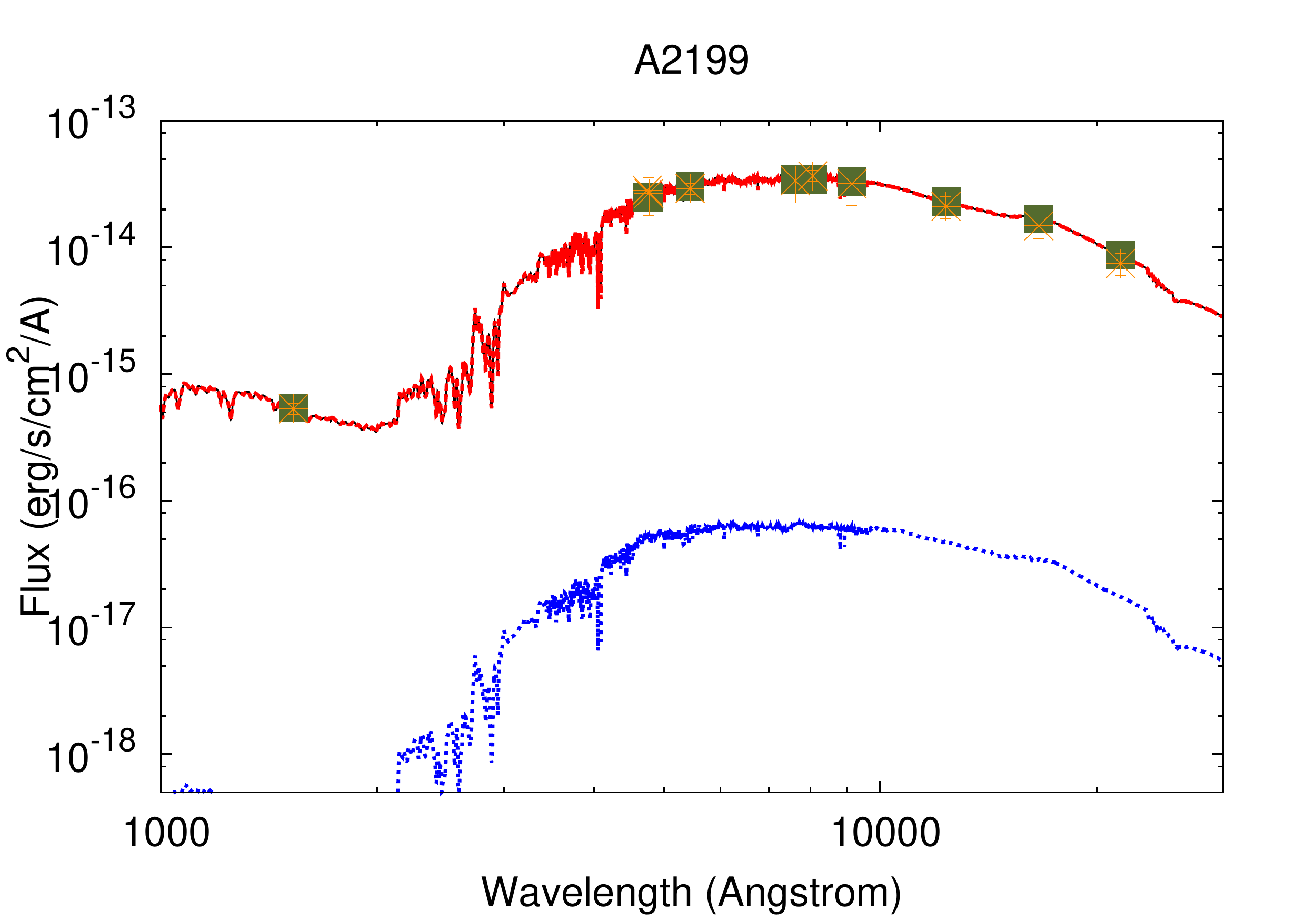}
  \end{minipage}\\[8pt]
  \begin{minipage}{0.33\textwidth}
    \centering
    \includegraphics[width=1.1\textwidth]{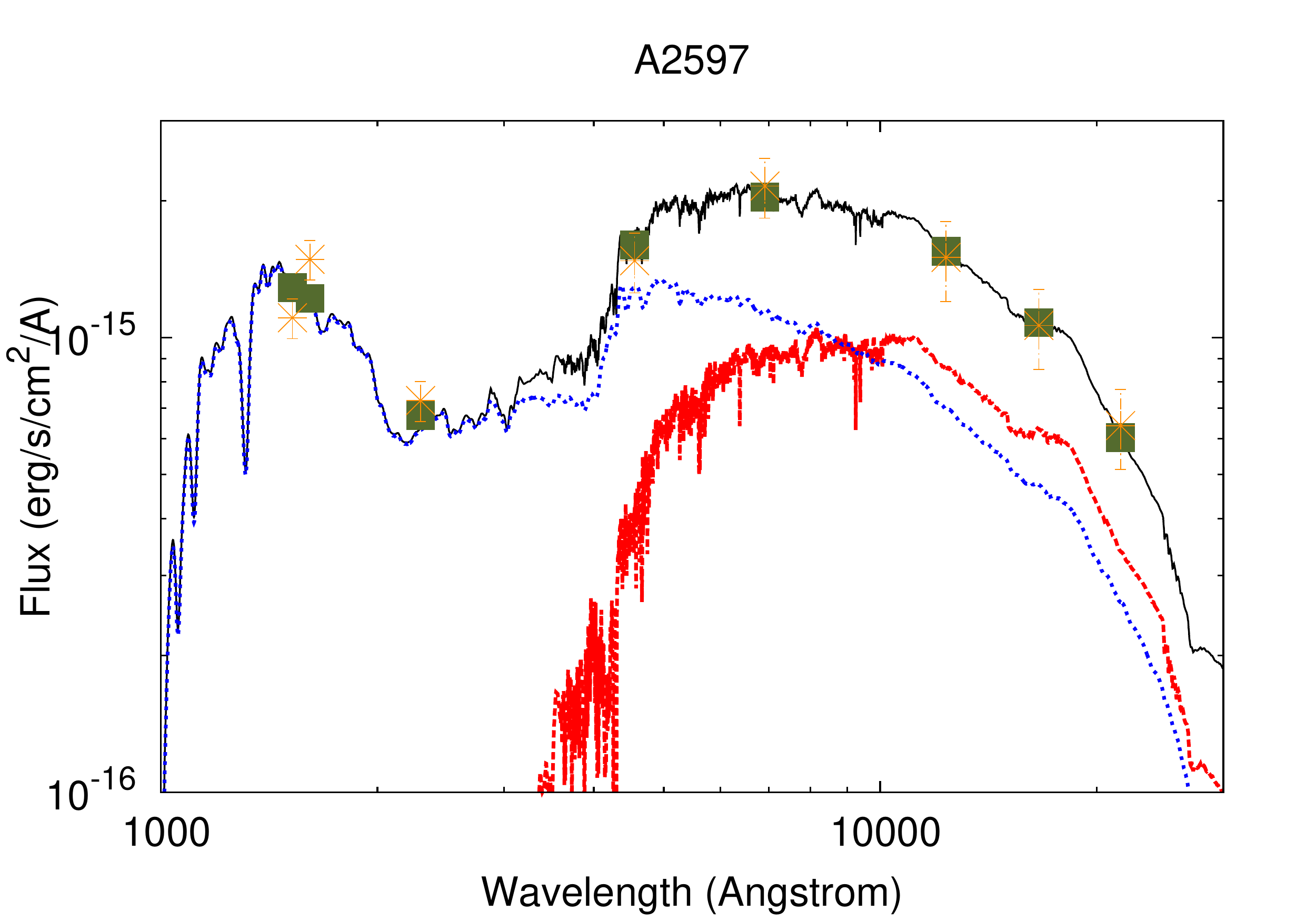}
  \end{minipage}\hfill
  \caption{Example best-fit plots shown by fixing all the discrete
    parameters (metallicity, extinction, and YSP age) to their
    most-likely values (the modes of their individual marginal
    posteriors) and then choosing the most likely YSP and OSP masses
    given that choice (which occurs at the minimum $\chi^2$). The red
    and blue curves correspond to the flux contributions from the old
    and (total) young stellar populations, respectively, and the black
    curve corresponds to the total spectrum energy distribution (sum
    of the old and young stellar populations). The green squares
    correspond to the predicted data and the orange crosses correspond
    to the observed data.}
  \label{bestfitplots}  
\end{figure*}